\pdfoutput=1
\documentclass{article}

\usepackage[final]{neurips_2026}

\makeatletter
\renewcommand{\@noticestring}{%
  Accepted at the AI for Accelerated Materials Design (AI4Mat) and AI for Science (AI4Science) workshops, NeurIPS 2026.%
}
\makeatother

\usepackage{amsmath}
\usepackage{amssymb}
\usepackage{booktabs}      %
\usepackage{graphicx}
\usepackage[table]{xcolor}   %
\usepackage{microtype}
\usepackage{url}
\usepackage{etoolbox}
\usepackage{enumitem}   %
\usepackage{needspace}   %

\usepackage{caption}
\usepackage{graphicx}
\graphicspath{{./}}
\setkeys{Gin}{width=\linewidth,height=0.92\textheight,keepaspectratio}
\usepackage{float}
\usepackage{longtable,array,calc}
\usepackage[colorlinks=true,linkcolor=xrefred,citecolor=citegreen,urlcolor=urlblue]{hyperref}
\definecolor{xrefred}{rgb}{0.55,0,0}
\definecolor{citegreen}{rgb}{0,0.15,0.55}   %
\definecolor{urlblue}{rgb}{0,0.2,0.6}
\newcommand{\citelink}[2]{{\hypersetup{linkcolor=citegreen}\hyperlink{#1}{#2}}}
\providecommand{\real}[1]{#1}          %
\providecommand{\tightlist}{}

\usepackage{newunicodechar}
\newunicodechar{σ}{\ensuremath{\sigma}}\newunicodechar{μ}{\ensuremath{\mu}}
\newunicodechar{λ}{\ensuremath{\lambda}}\newunicodechar{κ}{\ensuremath{\kappa}}
\newunicodechar{γ}{\ensuremath{\gamma}}\newunicodechar{ρ}{\ensuremath{\rho}}\newunicodechar{Σ}{\ensuremath{\Sigma}}
\newunicodechar{Δ}{\ensuremath{\Delta}}\newunicodechar{≈}{\ensuremath{\approx}}\newunicodechar{∼}{\ensuremath{\sim}}
\newunicodechar{≤}{\ensuremath{\leq}}\newunicodechar{≥}{\ensuremath{\geq}}
\newunicodechar{≠}{\ensuremath{\neq}}\newunicodechar{×}{\ensuremath{\times}}
\newunicodechar{·}{\ensuremath{\cdot}}\newunicodechar{−}{\ensuremath{-}}
\newunicodechar{√}{\ensuremath{\surd}}\newunicodechar{‖}{\ensuremath{\|}}
\newunicodechar{∑}{\ensuremath{\sum}}\newunicodechar{∈}{\ensuremath{\in}}
\newunicodechar{ℓ}{\ensuremath{\ell}}\newunicodechar{Å}{\AA}
\newunicodechar{↔}{\ensuremath{\leftrightarrow}}\newunicodechar{→}{\ensuremath{\rightarrow}}
\newunicodechar{²}{\textsuperscript{2}}\newunicodechar{³}{\textsuperscript{3}}
\newunicodechar{⁴}{\textsuperscript{4}}\newunicodechar{⁻}{\textsuperscript{-}}
\newunicodechar{₀}{\textsubscript{0}}\newunicodechar{₁}{\textsubscript{1}}
\newunicodechar{₂}{\textsubscript{2}}\newunicodechar{₃}{\textsubscript{3}}
\newunicodechar{₄}{\textsubscript{4}}\newunicodechar{₅}{\textsubscript{5}}
\newunicodechar{₋}{\textsubscript{-}}\newunicodechar{ₐ}{\textsubscript{a}}

\title{Out-of-training properties in frozen interatomic potential embeddings
are decodable and ordered by physical scale}

\author{%
  Natalija Stepurko\thanks{Primary contact.} \\
  Altrove \\
  25 rue de Ponthieu, 75008 Paris, France \\
  \texttt{natalija.stepurko@gmail.com} \\
  \And
  Joonatan E.M. Laulainen \\
  Altrove \\
  25 rue de Ponthieu, 75008 Paris, France \\
  \texttt{joonatan.laulainen@altrove.ai} \\
}

\begin{document}

\maketitle
\newif\ifmaintable                            %
\AtBeginEnvironment{tabular}{\ifmaintable\small\else\scriptsize\fi}  %
\AtBeginEnvironment{longtable}{\scriptsize}

\begin{abstract}
Machine-learned interatomic potentials (MLIPs) are trained on energies, forces and stresses, yet their frozen representations can predict other materials properties. We probe ORB-v3 and UMA-S across layers, read-outs and equivariant channels for nine properties of 154,875 Materials Project crystals. Both models outperform composition-only descriptors on all targets, including five outside their training objectives; errors in physical units help assess potential screening applications. Optimal depth varies by property, consistent with a physical-scale ordering. Random-weight controls reveal strongly property-dependent benefits of pretraining. UMA-S's rotation-invariant angular channels improve crystal-system classification, while higher scores from its full tensor disappear under rotation; matched-model controls do not isolate equivariance as the explanation. Accuracy varies across different probes, but nearest-neighbour prediction often approaches XGBoost performance, even though materials do not form distinct clusters according to their properties. We also test whether a linear predictor trained on one model's embeddings can predict properties using the other model's embeddings after a learned linear transformation. Predictive accuracy is largely preserved, despite differences in which atoms the two models place close together. These results guide the reuse of frozen MLIP representations for materials prediction.
\end{abstract}

\hypertarget{introduction}{%
\section{Introduction}\label{introduction}}

Universal machine-learned interatomic potentials (MLIPs) are pretrained on large datasets of density-functional-theory (DFT) energies, forces and stresses to model diverse atomic systems. Examples include M3GNet {[}\citelink{cite:Chen2022}{Chen \& Ong 2022}{]}, CHGNet {[}\citelink{cite:Deng2023}{Deng 2023}{]}, MACE-MP-0 {[}\citelink{cite:Batatia2024}{Batatia 2024}{]}, ORB-v3 {[}\citelink{cite:Rhodes2025}{Rhodes 2025}{]} and UMA {[}\citelink{cite:Wood2025}{Wood 2025}{]}.

If a frozen network reproduces energies and forces across all of chemistry, its internal representation must capture composition, local bonding and long-range structure. We ask what each layer encodes and where in the network using a standard interpretability technique {[}\citelink{cite:Alain2016}{Alain \& Bengio 2016}{]}: freeze the network, take the activations at a given layer, and train a probe --- gradient-boosted trees {[}\citelink{cite:Chen2016}{Chen \& Guestrin 2016}{]} and linear models {[}\citelink{cite:Pedregosa2011}{Pedregosa 2011}{]} --- to predict a target property (\textbf{\hyperlink{fig:1}{Fig.~1}}).

\begin{figure}[t]
  \hypertarget{fig:1}{}%
  \centering
  \includegraphics{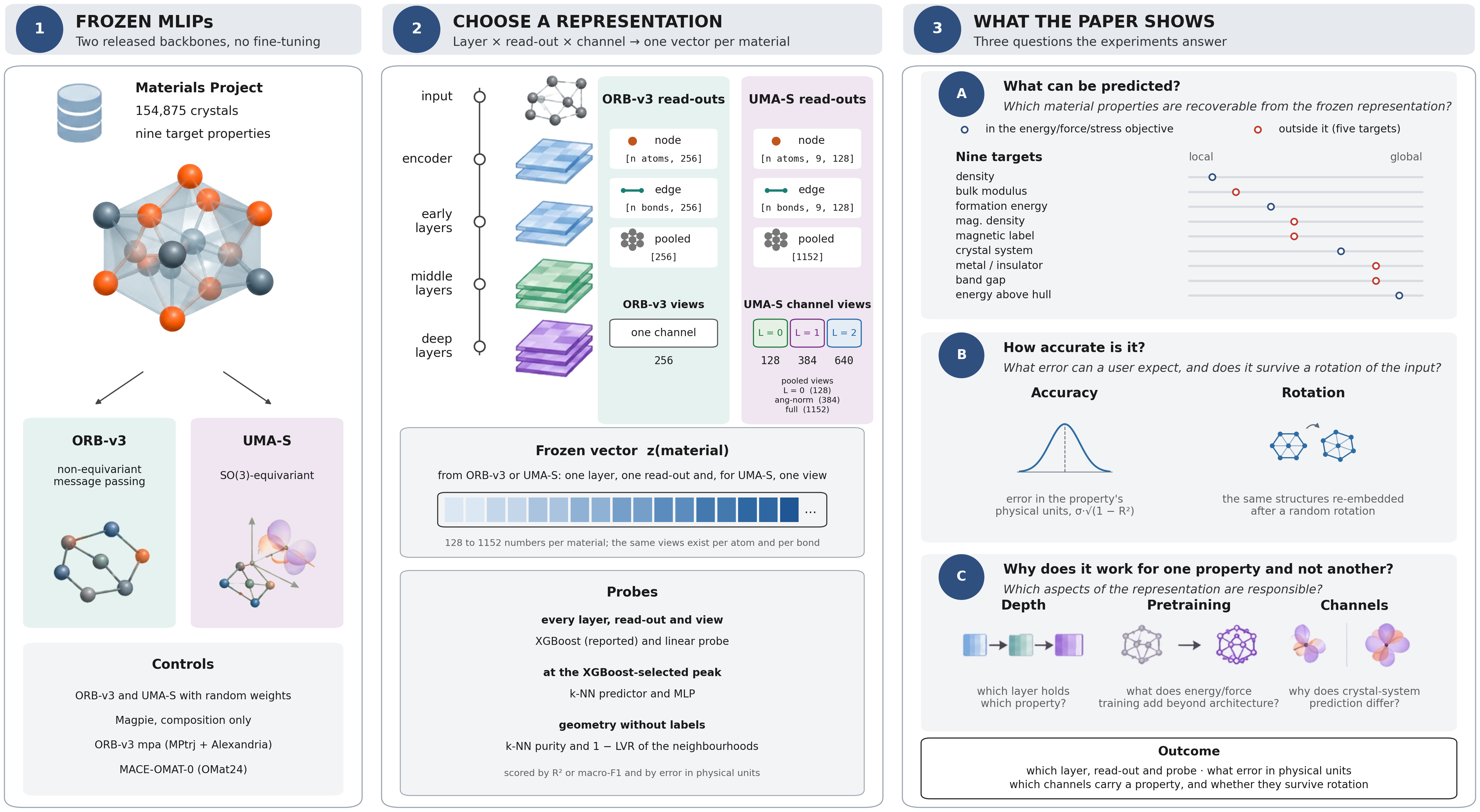}
  \caption*{\textbf{\emph{Figure~1 \textbar{} From frozen MLIP representations to downstream materials use.}} \emph{(\textbf{1}) Frozen ORB-v3 and UMA-S representations are extracted for 154,875 Materials Project crystals, with composition, random-weight and matched-model controls. (\textbf{2}) Layer-, read-out- and channel-resolved vectors are probed for nine material properties. (\textbf{3}) The experiments ask what the representations predict (\textbf{A}), how accurate they are in physical units and under rotation of the input (\textbf{B}), and how their usefulness depends on depth, pretraining and representation channel (\textbf{C}).}}
\end{figure}

This transfer-learning paradigm is established for foundation models in other domains trained on a single masked-token objective: protein language models expose secondary structure and inter-residue contacts {[}\citelink{cite:Rives2021}{Rives 2021}; \citelink{cite:Rao2021}{Rao 2021}; \citelink{cite:Lin2023}{Lin 2023}{]}, the Nucleotide Transformer encodes regulatory elements from raw DNA {[}\citelink{cite:DallaTorre2025}{Dalla-Torre 2025}{]}, and single-cell models transfer to perturbation and disease-state prediction {[}\citelink{cite:Theodoris2023}{Theodoris 2023}; \citelink{cite:Cui2024}{Cui 2024}{]}.

Recent work has begun to characterise MLIP latent representations directly. Li and Walsh {[}\citelink{cite:Li2026}{Li \& Walsh 2026}{]} found substantial agreement in global chemical and structural organisation across seven MLIPs, alongside differences in local neighbourhoods; Chorna et al.~{[}\citelink{cite:Chorna2026}{Chorna 2026}{]} also identified differences in latent representations associated with architecture, training data and evaluation protocol. Layer-dependent predictive performance has been reported by Pinto {[}\citelink{cite:Pinto2025}{Pinto 2025}{]} for molecular ADMET tasks and by Kim et al.~{[}\citelink{cite:Kim2025}{Kim 2025}{]} using mean-pooled embeddings for crystalline materials. Building on these findings, we systematically examine how the decodability of nine crystal properties varies jointly across network depth, node and edge read-outs, and equivariant channels. Random-initialisation, matched-model and rotation controls further distinguish contributions from pretraining, architectural priors and representation choice, while neighbourhood analyses test whether predictive information is organised into property-aligned clusters.

We compare two MLIPs with contrasting architectures: ORB-v3 {[}\citelink{cite:Neumann2024}{Neumann 2024}; \citelink{cite:Rhodes2025}{Rhodes 2025}{]}, a non-equivariant message-passing GNN, and UMA-S (FairChem) {[}\citelink{cite:Wood2025}{Wood 2025}{]}, an SO(3)-equivariant network whose explicit L=0/L=1/L=2 decomposition exposes scalar, directional and rank-2 angular channels at each layer. Because the two also differ in training data and capacity, their performance differences are descriptive comparisons of two released models rather than controlled estimates of equivariance; matched-data and matched-architecture controls (\textbf{\hyperlink{uma-channel-views-and-the-rotation-invariant-crystal-system-gap}{§3.4}}, \textbf{\hyperlink{random-initialisation-and-matched-training-data-controls}{§A.9}}) constrain this interpretation. We pass the same 154,875 Materials Project {[}\citelink{cite:Horton2025}{Horton 2025}; \citelink{cite:Jain2013}{Jain 2013}{]} structures through each model and, at every layer, ask what can be predicted from the activations and how they are geometrically organised.

Both models' frozen embeddings outperform a Magpie {[}\citelink{cite:Ward2016}{Ward 2016}{]} composition-only baseline across all nine properties, including five absent from their training objectives (\textbf{\hyperlink{frozen-mlip-embeddings-predict-properties-outside-the-training-objective}{§3.1}}). Predictive performance varies with depth, with random-weight controls revealing particularly large pretraining gains for energy above hull, band gap and crystal system (\textbf{\hyperlink{property-decodability-depends-on-network-depth}{§3.3}}). UMA-S substantially outperforms ORB-v3 on crystal-system classification. Random rotations remove the additional gain from its full-tensor features, but the advantage of rotation-invariant angular channels persists; matched-model controls suggest that equivariance alone does not explain it (\textbf{\hyperlink{uma-channel-views-and-the-rotation-invariant-crystal-system-gap}{§3.4}}). Prediction based solely on nearest-neighbour geometry in the frozen embedding space approaches XGBoost performance for most properties, despite the absence of distinct property-aligned clusters (\textbf{\hyperlink{property-information-is-locally-predictable-without-property-aligned-clusters}{§3.5}}).

\hypertarget{methods}{%
\section{Methods}\label{methods}}

Full methodological details, including formulas, probe hyperparameters, controls, compute resources and asset licences, are provided in Supplementary Methods (\textbf{\hyperlink{read-outs}{§A.1}--A.13}).

\hypertarget{models-and-layer-wise-embedding-extraction}{%
\subsection{Models and layer-wise embedding extraction}\label{models-and-layer-wise-embedding-extraction}}

We extract and cache activations from every layer of two frozen foundation MLIPs. ORB-v3 {[}\citelink{cite:Neumann2024}{Neumann 2024}; \citelink{cite:Rhodes2025}{Rhodes 2025}{]} is a non-equivariant GNN with an encoder and five message-passing layers (gnn\_0--4), each producing 256-dimensional node and edge features. UMA-S (FairChem) {[}\citelink{cite:Wood2025}{Wood 2025}{]} is SO(3)-equivariant, with an eSCN-MD backbone comprising an embedding block and four interaction blocks (block\_0--3). Its final backbone read-out aggregates node features but provides no edge features. For each structure, we retain node and edge embeddings wherever available (\textbf{\hyperlink{fig:1}{Fig.~1}}).

\hypertarget{supervised-probes}{%
\subsection{Supervised probes}\label{supervised-probes}}

We evaluate XGBoost {[}\citelink{cite:Chen2016}{Chen \& Guestrin 2016}{]} and linear probes {[}\citelink{cite:Pedregosa2011}{Pedregosa 2011}{]} across all layers and read-outs, including each equivariant channel view of UMA-S. XGBoost determines the reported scores and optimal configurations, while linear probes assess the influence of decoder flexibility on layer and read-out trends. We predict six continuous properties (formation energy, energy above hull, band gap, density, bulk modulus and magnetic-moment density) using R², and three categorical properties (crystal system, metal/insulator and magnetic label) using macro-F1. All configurations use the same fixed 80/20 split of 154,875 Materials Project structures, except for bulk modulus, which has fewer available labels (12,853 structures); the targets are defined in \textbf{\hyperlink{targets-and-labels}{§A.3}}. Magpie {[}\citelink{cite:Ward2016}{Ward 2016}{]} provides the composition-only baseline.

At each XGBoost-selected peak, we additionally evaluate a parameter-free k-NN predictor {[}\citelink{cite:Cover1967}{Cover \& Hart 1967}{]} and an MLP {[}\citelink{cite:Rumelhart1986}{Rumelhart 1986}{]}, both as implemented in scikit-learn {[}\citelink{cite:Pedregosa2011}{Pedregosa 2011}{]}. The four probes are compared at the same configuration and on the same training and test materials in \textbf{\hyperlink{probe-configuration}{§A.2}} and \textbf{\hyperlink{tab:S4}{Table~S4}}; the validation-split selection, grouped-split leakage and permuted-label controls are described in \textbf{\hyperlink{robustness-controls}{§A.4}}. We also evaluate retrospective screening: how well the probes' predictions recover the best materials, those within stability or band-gap criteria, and members of each class (\textbf{\hyperlink{screening-enrichment-results}{§B.6}}), and label efficiency, by retraining the probes on 50 to 20,000 labelled materials (\textbf{\hyperlink{label-efficiency-results}{§B.7}}). The label-efficiency curves include a small message-passing network trained from scratch on the same crystal structures at matched training sizes, to test whether a frozen foundation model plus a probe outperforms a small task-specific model trained directly on the labels; both methods are described in \textbf{\hyperlink{screening-enrichment-and-label-efficiency}{§A.11}}.

\hypertarget{unsupervised-label-organisation-metrics-main}{%
\subsection{Unsupervised label-organisation metrics}\label{unsupervised-label-organisation-metrics-main}}

A property is decodable if a trained probe predicts it from the embedding, and geometrically organised if the raw embedding neighbourhoods already align with it. We measure organisation with chance-corrected k-NN purity (categorical) and 1−LVR (continuous), the inverted Local Variance Ratio; these quantify how strongly a point's nearest neighbours match it in label or value. We also cluster with HDBSCAN {[}\citelink{cite:McInnes2017}{McInnes 2017}{]}, scored by ARI and NMI {[}\citelink{cite:Pedregosa2011}{Pedregosa 2011}{]}, which tests whether the embedding fragments into label-consistent clusters without supervision. Prediction and organisation scores are different quantities on different scales and are compared only qualitatively; the quantitative comparison between a trained probe and the raw neighbourhood structure is the k-NN predictor of \textbf{\hyperlink{supervised-probes}{§2.2}}, scored in the same units as the probe (\textbf{\hyperlink{property-information-is-locally-predictable-without-property-aligned-clusters}{§3.5}}). UMAP projections {[}\citelink{cite:McInnes2018}{McInnes 2018}{]} are used for visual inspection only (\textbf{\hyperlink{unsupervised-label-organisation-metrics}{§A.5}}).

\hypertarget{controls-and-external-transfer}{%
\subsection{Controls}\label{controls-and-external-transfer}}

To assess the contribution of pretraining, we randomly reinitialise both models and repeat the probing pipeline. We examine training-data and architecture effects on a 20,000-material subsample stratified by crystal system, comparing ORB-v3 and UMA-S with an ORB-v3 variant trained on MPtrj and Alexandria and the equivariant MACE-OMAT-0 {[}\citelink{cite:Batatia2024}{Batatia 2024}{]}, which shares ORB-v3's OMat24 training data but differs in architecture (\textbf{\hyperlink{random-initialisation-and-matched-training-data-controls}{§A.9}}). Two headline scores are additionally checked at a different scale: bulk modulus is re-scored on a log\textsubscript{10} scale, the convention of the Matbench elastic benchmark, to check that its raw-GPa score is not carried by the few very stiff materials, and band gap is re-scored on the non-metal test materials only, since a pooled R² partly rewards separating the 47~\% of metals with zero gap from the rest (\textbf{\hyperlink{robustness-controls}{§A.4}}). Finally, we test rotational robustness by re-extracting embeddings from randomly rotated structures and measuring changes in both features and probe predictions (\textbf{\hyperlink{robustness-controls}{§A.4}}).

\hypertarget{statistics}{%
\subsection{Statistics}\label{statistics}}

Unless stated otherwise, scores are reported on the fixed seed-42 test split, using the configuration (read-out, channel view and layer) selected on that same split. Headline scores (\textbf{\hyperlink{tab:1}{Table~1}}; intervals shown as error bars in \textbf{\hyperlink{fig:2}{Fig.~2a}}) and paired bootstrap comparisons (\textbf{\hyperlink{fig:2}{Fig.~2b--c}}, \textbf{\hyperlink{fig:3}{Fig.~3e}}) have 95~\% confidence intervals from 2,000 test-set resamples; supplementary controls are reported as point estimates unless indicated otherwise. Statistical tests are chosen according to the source of uncertainty (\textbf{\hyperlink{statistics-1}{§A.8}}). Comparisons that require no retraining (UMA vs ORB, XGBoost vs linear, node vs edge vs concat) use paired bootstraps with Benjamini--Hochberg FDR correction across nine properties {[}\citelink{cite:Benjamini1995}{Benjamini \& Hochberg 1995}{]}. Because the test set contains approximately 31,000 materials, even small differences can be statistically significant; we therefore also report predefined practical-effect bands (\textbf{\hyperlink{statistics-1}{§A.8}}). Layer comparisons instead require retraining and use Nadeau--Bengio corrected repeated-split tests {[}\citelink{cite:Nadeau2003}{Nadeau \& Bengio 2003}{]}. These tests use node read-outs only, so their means may differ from the \textbf{\hyperlink{fig:3}{Fig.~3}} curves, which use each target's selected read-out, including concatenated features.

\hypertarget{results-and-discussion}{%
\section{Results and discussion}\label{results-and-discussion}}

\hypertarget{frozen-mlip-embeddings-predict-properties-outside-the-training-objective}{%
\subsection{Frozen MLIP embeddings predict properties outside the training objective}\label{frozen-mlip-embeddings-predict-properties-outside-the-training-objective}}

Frozen representations from both MLIPs predict all five properties outside their energy/force/stress training objective: band gap, bulk modulus, magnetic-moment density, metal/insulator character and magnetic ordering. At their best layer and read-out, the three continuous targets reach R² = 0.84--0.96 and the two categorical targets macro-F1 = 0.89--0.91 (\textbf{\hyperlink{fig:2}{Fig.~2a}}, \textbf{\hyperlink{tab:1}{Table~1}}; error metrics by read-out in \textbf{\hyperlink{tab:S5}{Table~S5}}). Both models outperform the composition-only Magpie baseline on all nine properties using the same XGBoost probe and data split (\textbf{\hyperlink{probe-configuration}{§A.2}}; \textbf{\hyperlink{fig:2}{Fig.~2b}}, \textbf{\hyperlink{tab:1}{Table~1}}, \textbf{\hyperlink{tab:S6}{Table~S6}}). Because Magpie contains no structural information, polymorphs with the same formula receive identical descriptors. Across six of the nine targets, ORB reduces the baseline error by 35--73~\% and UMA by 41--76~\%, with the largest gains for energy above hull and the smallest for the magnetic properties. These improvements can be more substantial than changes in R² suggest: increasing formation-energy R² from 0.89 to 0.99 reduces RMSE by about 70~\% (\textbf{\hyperlink{from-ruxb2-to-expected-downstream-accuracy}{§3.2}}).

Apart from crystal system, the models perform similarly. UMA has statistically significant but practically small advantages (+0.005 to +0.016) for energy above hull, band gap, metal/insulator, magnetic label and bulk modulus; formation energy and density are ties, while ORB narrowly leads on magnetic-moment density (\textbf{\hyperlink{fig:2}{Fig.~2c}}). Crystal system is the clearest exception: UMA removes 62~\% of Magpie's error, compared with 21~\% for ORB. Node features outperform edge features in 17 of 18 model--target comparisons, and concatenating edges with nodes changes scores by less than 0.015 except for UMA crystal system, where the gain is 0.14 (\textbf{\hyperlink{fig:S1}{Fig.~S1}}, \textbf{\hyperlink{tab:S5}{Table~S5}}). For the remaining targets, the largest gains over composition occur for properties that also improve most across layers (\textbf{\hyperlink{property-decodability-depends-on-network-depth}{§3.3}}); UMA's crystal-system advantage is examined in \textbf{\hyperlink{uma-channel-views-and-the-rotation-invariant-crystal-system-gap}{§3.4}}.

UMA's larger configuration search (96 versus 18 configurations; \textbf{\hyperlink{probe-configuration}{§A.2}}) does not explain these differences: selecting the best configuration on a held-out validation split limits score inflation to ≤~0.002 for every target and changes the UMA−ORB differences by at most 0.002 (\textbf{\hyperlink{tab:S9}{Table~S9}}). Grouping the split so that related compounds cannot span train and test reduces scores by at most 0.04, so near-duplicates do not drive the result (\textbf{\hyperlink{robustness-controls}{§A.4}}, \textbf{\hyperlink{tab:S10}{Table~S10}}). Because ORB and UMA differ in architecture and training data, the matched-model controls in \textbf{\hyperlink{uma-channel-views-and-the-rotation-invariant-crystal-system-gap}{§3.4}} (\textbf{\hyperlink{fig:2}{Fig.~2d}}) further constrain, but do not isolate, the sources of their differences.

At each XGBoost-selected configuration, we also evaluate an MLP and a k-NN predictor on the same training and test materials (\textbf{\hyperlink{probe-configuration}{§A.2}}, \textbf{\hyperlink{tab:S4}{Table~S4}}, \textbf{\hyperlink{fig:S2}{Fig.~S2}}). The MLP scores within 0.035 R² or macro-F1 of XGBoost in 17 of 18 cases while reducing the error in physical units further; the reported XGBoost peaks are therefore decoder-dependent. Finally, we test the transfer of Materials Project-trained probes to JARVIS-DFT 3D {[}\citelink{cite:Choudhary2020}{Choudhary 2020}{]} and OQMD {[}\citelink{cite:Kirklin2015}{Kirklin 2015}{]} (\textbf{\hyperlink{external-transfer-results}{§B.4}}). Because these databases differ in both their materials and DFT calculations, the transfer results reflect changes in reference labels as well as prediction on new crystals.

\begin{figure}[t]
  \hypertarget{fig:2}{}%
  \centering
  \includegraphics{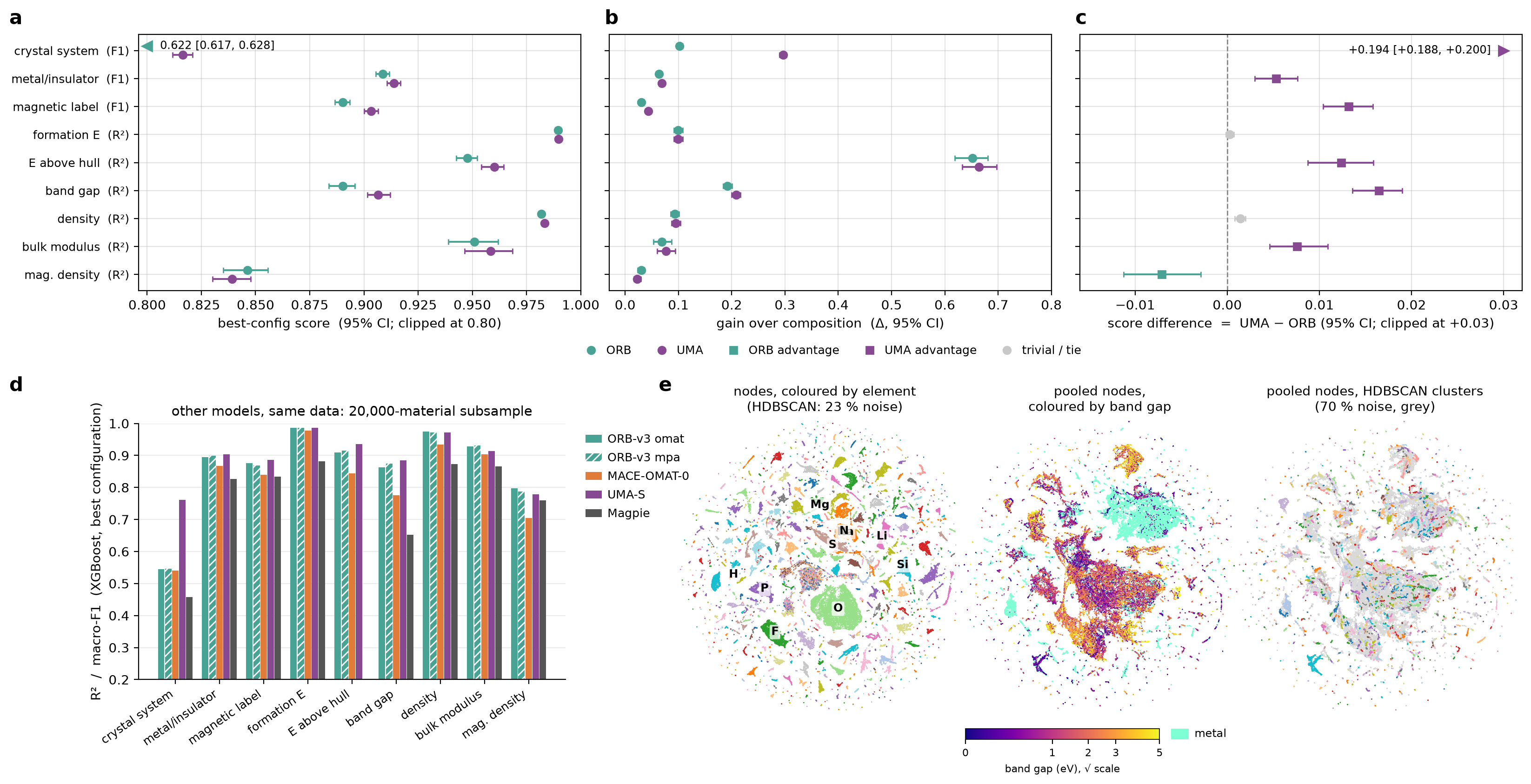}
  \caption*{\textbf{\emph{Figure~2. Property prediction and organisation in frozen MLIP embeddings.}} \emph{\textbf{(a)} Best XGBoost score per property across all layers, read-outs and channel views (axis clipped at 0.80). Confusion matrices and parity plots are shown in \textbf{\hyperlink{fig:S3}{Fig.~S3a--b}} and \textbf{\hyperlink{fig:S3}{Fig.~S3c--d}}, respectively. \textbf{(b)} Improvement over the composition-only Magpie baseline. \textbf{(c)} UMA − ORB score differences, with coloured squares indicating a practical advantage for the corresponding model and grey dots indicating negligible differences. \textbf{(d)} Best scores on a common 20,000-material subset. ORB-v3 mpa is trained on MPtrj and Alexandria rather than OMat24; MACE-OMAT-0 is equivariant and trained on OMat24, like ORB-v3 omat (\textbf{\hyperlink{random-initialisation-and-matched-training-data-controls}{§A.9}}, \textbf{\hyperlink{tab:S18}{Table~S18}}). \textbf{(e)} UMAP projections of UMA-S angular-norm node features at block 0, showing atoms coloured by element and pooled material embeddings coloured by band gap or HDBSCAN cluster (noise in grey). Error bars in (a--c) show 95~\% bootstrap confidence intervals (2,000 resamples; paired in b and c). UMA-S crystal-system scores use rotation-invariant features (\textbf{\hyperlink{uma-channel-views-and-the-rotation-invariant-crystal-system-gap}{§3.4}}).}}
\end{figure}

\hypertarget{from-ruxb2-to-expected-downstream-accuracy}{%
\subsection{From predictive scores to practical accuracy}\label{from-ruxb2-to-expected-downstream-accuracy}}

The high R² scores in \textbf{\hyperlink{frozen-mlip-embeddings-predict-properties-outside-the-training-objective}{§3.1}} do not necessarily mean that the predictions are accurate enough for practical use. To assess this, we examine errors in physical units: by how many GPa, eV or meV/atom the predictions typically differ from their DFT reference values. For continuous targets, RMSE ≈~σ·√(1 − R²), where σ is the target's standard deviation, so the same R² can correspond to very different errors (RMSE, MAE and σ in \textbf{\hyperlink{tab:S5}{Table~S5}}; parity plots in \textbf{\hyperlink{fig:S3}{Fig.~S3c--d}}).

For bulk modulus (σ = 77~GPa), ORB and UMA reach R² = 0.951 / 0.958 and RMSE = 17 / 16~GPa, sufficient to distinguish stiff from compliant materials, whose moduli differ by far more than 17~GPa {[}\citelink{cite:MansouriTehrani2018}{Mansouri Tehrani 2018}{]}. Although a few very stiff outliers inflate the raw-GPa R² (\textbf{\hyperlink{fig:S3}{Fig.~S3c--d}}), performance remains high on the Matbench log\textsubscript{10}(K\textsubscript{VRH}) scale (R² = 0.935 / 0.939; \textbf{\hyperlink{tab:S11}{Table~S11b}}, \textbf{\hyperlink{robustness-controls}{§A.4}}). For band gap (σ = 1.52~eV), R² = 0.890 / 0.907 corresponds to RMSE = 0.50 / 0.46~eV, comparable to ML band-gap models used for screening {[}\citelink{cite:Zhuo2018}{Zhuo 2018}{]} and enough to tell metals, semiconductors and wide-gap insulators apart. Restricted to the non-metal test materials, R² is 0.859 / 0.881 (\textbf{\hyperlink{tab:S11}{Table~S11a}}), so the pooled score is not carried by the metal/non-metal separation alone. These errors are measured against GGA/PBE reference values, which underestimate experimental gaps by roughly 30--50~\% {[}\citelink{cite:Borlido2019}{Borlido 2019}{]}, so the probe reproduces DFT gaps, bias included.

\begin{table}[t]\maintabletrue
  \hypertarget{tab:1}{}%
  \caption*{\textbf{Table~1 --- Per-property prediction from frozen embeddings: peak score and error removed against a composition-only baseline.} \emph{The five targets absent from the energy/force/stress objective are listed first; the four rows with a shaded label are reference targets. All targets use 154,875 Materials Project structures (80/20 seed-42 split) except bulk modulus (12,853, the Materials Project elastic-tensor subset); target definitions in \hyperlink{targets-and-labels}{§A.3}. For each model we give the peak configuration (read-out · channel view · layer), its score (R² for regression, macro-F1 for classification), and \textbf{cut}, the percentage of the baseline's error removed, calculated from RMSE for regression and 1 − macro-F1 for classification. Scores are those of the XGBoost probe; the linear, nearest-neighbour and multilayer-perceptron probes at the same configurations are in \textbf{\hyperlink{tab:S4}{Table~S4}}. Per-score confidence intervals and the UMA − ORB differences with their intervals are the error bars of \textbf{\hyperlink{fig:2}{Fig.~2a}} and \textbf{\hyperlink{fig:2}{Fig.~2c}}.}}
  \centering

{\setlength{\tabcolsep}{5pt}\begin{tabular}{@{}
  >{\raggedright\arraybackslash}p{(\columnwidth - 16\tabcolsep) * \real{0.2632}}
  >{\raggedright\arraybackslash}p{(\columnwidth - 16\tabcolsep) * \real{0.0947}}
  >{\raggedright\arraybackslash}p{(\columnwidth - 16\tabcolsep) * \real{0.1368}}
  >{\raggedright\arraybackslash}p{(\columnwidth - 16\tabcolsep) * \real{0.0632}}
  >{\raggedright\arraybackslash}p{(\columnwidth - 16\tabcolsep) * \real{0.0526}}
  >{\raggedright\arraybackslash}p{(\columnwidth - 16\tabcolsep) * \real{0.2737}}
  >{\raggedright\arraybackslash}p{(\columnwidth - 16\tabcolsep) * \real{0.0632}}
  >{\raggedright\arraybackslash}p{(\columnwidth - 16\tabcolsep) * \real{0.0526}}@{}}
\toprule\noalign{}
\begin{minipage}[c]{\linewidth}\raggedright
Target
\end{minipage} & \begin{minipage}[c]{\linewidth}\raggedright
Comp.-only
\end{minipage} & \begin{minipage}[c]{\linewidth}\raggedright
ORB config
\end{minipage} & \begin{minipage}[c]{\linewidth}\raggedright
ORB score
\end{minipage} & \begin{minipage}[c]{\linewidth}\raggedright
ORB cut
\end{minipage} & \begin{minipage}[c]{\linewidth}\raggedright
UMA config
\end{minipage} & \begin{minipage}[c]{\linewidth}\raggedright
UMA score
\end{minipage} & \begin{minipage}[c]{\linewidth}\raggedright
UMA cut
\end{minipage} \\
\midrule\noalign{}
band gap (R²) & 0.698 & concat · g4 & 0.890 & 40~\% & concat · ang-norm · b2 & 0.907 & 44~\% \\
bulk modulus (R²) & 0.882 & concat · g0 & 0.951 & 35~\% & concat · ang-norm · emb & 0.958 & 41~\% \\
metal/insulator (F1) & 0.845 & concat · g4 & 0.909 & 41~\% & node · ang-norm · bb & 0.914 & 44~\% \\
magnetic label (F1) & 0.860 & concat · g4 & 0.890 & 22~\% & node · ang-norm · bb & 0.903 & 31~\% \\
mag. density (R²) & 0.816 & concat · g0 & 0.846 & 9~\% & concat · ang-norm · emb & 0.839 & 6~\% \\
\rowcolor{black!8}crystal system (F1) & 0.520 & concat · g2 & 0.622 & 21~\% & concat · ang-norm · b0 & 0.817 & 62~\% \\
\rowcolor{black!8}density (R²) & 0.888 & concat · g0 & 0.982 & 60~\% & concat · ang-norm · emb & 0.983 & 61~\% \\
\rowcolor{black!8}energy above hull (R²) & 0.296 & concat · g4 & 0.948 & 73~\% & concat · L0 · b3 & 0.960 & 76~\% \\
\rowcolor{black!8}formation E/atom (R²) & 0.890 & node · g0 & 0.990 & 69~\% & concat · ang-norm · b3 & 0.990 & 69~\% \\
  \bottomrule\end{tabular}}
\end{table}

Energy above hull illustrates why a high R² need not mean sufficient accuracy for the intended task. Despite R² = 0.948 / 0.960, errors of 103 / 90~meV/atom exceed the 67~meV/atom 90th percentile of observed inorganic metastability {[}\citelink{cite:Sun2016}{Sun 2016}{]}, making near-hull stability decisions unreliable without further validation. Conversely, magnetic-moment density has the lowest R² of any target (0.846 / 0.839) but a small absolute error (RMSE 0.0115 / 0.0118~μ\_B~Å⁻³; σ = 0.029~μ\_B~Å⁻³), potentially useful for identifying magnetic candidates. Prediction errors must therefore be assessed in physical units, from σ·√(1 − R²), against the accuracy the intended application requires, and not from R² alone. Retrospective screening and label-efficiency tests of the same probes are reported in \textbf{\hyperlink{screening-enrichment-results}{§B.6}}--\textbf{\hyperlink{label-efficiency-results}{B.7}}.

\hypertarget{property-decodability-depends-on-network-depth}{%
\subsection{Property decodability varies across network layers}\label{property-decodability-depends-on-network-depth}}

Different properties favour different layers in both models (\textbf{\hyperlink{fig:3}{Fig.~3a--b}}). Formation energy, density, bulk modulus and magnetic-moment density are already well predicted before message passing (R² ≥~0.82). ORB's first message-passing layer improves formation energy by 0.11 R² and density, bulk modulus and magnetic-moment density by 0.01--0.03, while UMA gains little (\textbf{\hyperlink{fig:3}{Fig.~3a--b}}). Formation energy then plateaus; density and bulk modulus decline from their early peaks, while magnetic-moment density peaks early in ORB but remains nearly flat in UMA. Early representations are dominated by element identity: atom-level k-NN element purity is 1.0 in both models (\textbf{\hyperlink{fig:S4}{Fig.~S4}}), and HDBSCAN strongly recovers elemental clusters (\textbf{\hyperlink{fig:S5}{Fig.~S5a--c}}), the element structure that the atom-level UMAP of \textbf{\hyperlink{fig:2}{Fig.~2e}} shows directly. With depth, element purity declines (\textbf{\hyperlink{fig:S4}{Fig.~S4}}) and these clusters dissolve (\textbf{\hyperlink{fig:S5}{Fig.~S5a--c}}). Density and bulk modulus depend mainly on composition and packing, consistent with their declining scores as the representations shift towards structural context while electronic targets gain (\textbf{\hyperlink{tab:S7}{Table~S7}}).

By contrast, band gap, metal/insulator character and magnetic label benefit from deeper layers. Band gap gains 0.20 R² in ORB and 0.14 in UMA from the initial embedding to its peak. Energy above hull also peaks deep, but 0.30 of ORB's total 0.31 gain (0.10 of UMA's 0.13) occurs at the first message-passing layer (\textbf{\hyperlink{fig:3}{Fig.~3a--b}}). Moreover, hull energy is defined relative to competing phases rather than by the spatial extent of a single crystal. These depth profiles are consistent with a physical-scale interpretation, in which local and additive properties require less integration than electronic ones. However, depth also changes non-linear processing, so the experiments do not isolate spatial range. The finding that final layers are not uniformly optimal also echoes results from genomic {[}\citelink{cite:DallaTorre2025}{Dalla-Torre 2025}{]} and molecular {[}\citelink{cite:Pinto2025}{Pinto 2025}{]} encoders.

The depth contrasts survive five retrainings under the Nadeau--Bengio corrected test on the node read-out (\textbf{\hyperlink{fig:3}{Fig.~3c--d}}, \textbf{\hyperlink{tab:S7}{Table~S7}}), while permuted labels give chance-level scores (\textbf{\hyperlink{tab:S12}{Table~S12}}). Linear probes select the same best layer as XGBoost in 10 of 18 cases and a layer within one step in 16, preserving the broad early-versus-late pattern (per-layer curves for all nine properties and both probes in \textbf{\hyperlink{fig:S6}{Fig.~S6}}). The XGBoost--linear gap is negligible for formation energy but substantial for band gap and crystal system (\textbf{\hyperlink{fig:3}{Fig.~3e}}), indicating greater dependence on non-linear decoding. Embedding-health checks also show that the depth trends are not explained by representation collapse (\textbf{\hyperlink{representation-health-tracks-architectural-function}{§B.1}}).

Random-weight controls distinguish the contribution of pretraining from information available through the structural input and architecture alone (\textbf{\hyperlink{random-initialisation-and-matched-training-data-controls}{§A.9}}). Pretraining adds most for energy above hull (ΔR² = 0.38--0.41), band gap (0.27--0.30) and crystal system (ΔF1 = 0.19--0.21), but only 0.03--0.06 for density, bulk modulus and magnetic-moment density (\textbf{\hyperlink{fig:3}{Fig.~3a--b}}; \textbf{\hyperlink{tab:S17}{Table~S17}}). For band gap and metal/insulator character, the pretraining advantage grows with depth, while random-network performance remains nearly flat across message-passing layers. Thus, some properties are strongly decodable even without pretraining, whereas others benefit substantially from learned representations.

\begin{figure}[t]
  \hypertarget{fig:3}{}%
  \centering
  \includegraphics{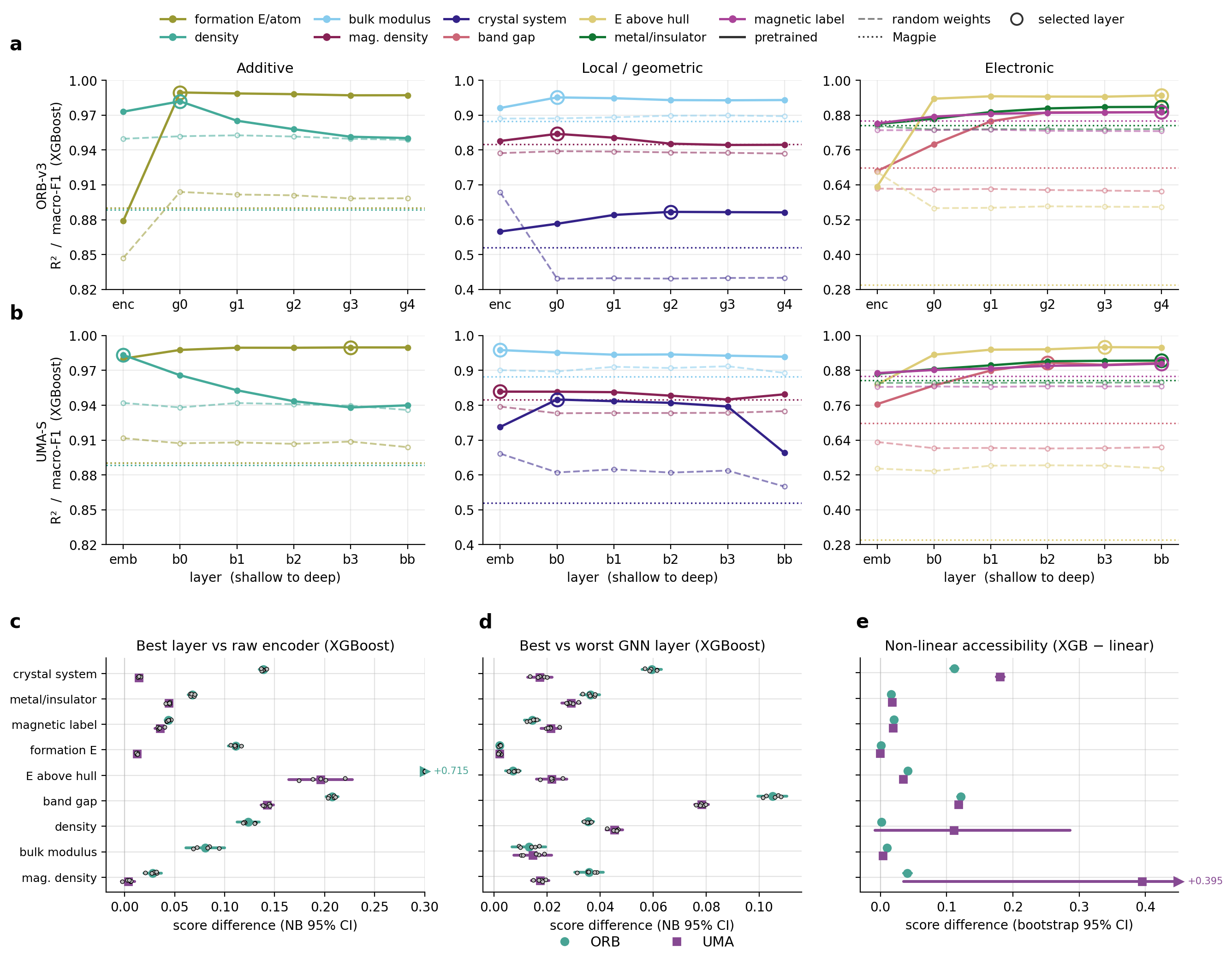}
  \caption*{\textbf{\emph{Figure~3. Property-specific variation across network layers and the contribution of pretraining.}} \emph{\textbf{(a)} ORB-v3 and \textbf{(b)} UMA-S: XGBoost score at every layer of each target's selected read-out and channel view, for the pretrained model (solid), the same model re-initialised with random weights (dashed) and the Magpie composition baseline (dotted); open circles mark the selected layers. Scores are R² for continuous targets and macro-F1 for categorical targets. ORB's backbone read-out equals its final layer (g4) and is omitted. UMA's backbone (bb) is the layer-normalised output of b3 and provides only a node read-out, so where earlier layers use concatenated node and edge features the bb score uses nodes alone; the crystal-system drop at bb therefore also reflects a change of read-out. \textbf{(c)} Best layer versus the raw encoder (ORB) or embedding block (UMA) and \textbf{(d)} best versus worst message-passing layer, both on the node read-out (\hyperlink{statistics}{§2.5}). Grey dots show five retrainings; markers and error bars give the mean and Nadeau--Bengio-corrected 95~\% confidence interval. A missing UMA row in (c) means the embedding block is already the best layer. \textbf{(e)} XGBoost-minus-linear score difference at each target's selected configuration, with its 95~\% paired-bootstrap confidence interval; the large UMA gaps for density and magnetic-moment density come from an unstable linear fit at the embedding layer. Statistics for (c)--(d) are in \textbf{\hyperlink{tab:S7}{Table~S7}}.}}
\end{figure}

\hypertarget{uma-channel-views-and-the-rotation-invariant-crystal-system-gap}{%
\subsection{UMA channel views and the rotation-invariant crystal-system gap}\label{uma-channel-views-and-the-rotation-invariant-crystal-system-gap}}

UMA-S separates scalar (L=0), directional (L=1) and rank-2 angular (L=2) channels. These do not correspond strictly to chemistry and geometry: scalar features can encode local structure, while angular features also depend on the chemical identities of neighbouring atoms. Their predictive contributions differ by property (\textbf{\hyperlink{fig:S1}{Fig.~S1}}). Scalar features nearly match the full tensor for metal/insulator classification, but crystal-system prediction benefits substantially from angular channels. For crystal-system classification, rotation-invariant angular-norm features (F1 0.817) outperform scalar features (0.581) and ORB (0.622), while the full tensor reaches an even higher F1 of 0.897 (\textbf{\hyperlink{fig:4}{Fig.~4a}}). However, the full tensor contains components that rotate with the input frame, raising the question of whether this advantage is robust to rotation.

To test rotational robustness, we randomly rotate 1,000 test structures (atoms and unit cell together), extract their embeddings again and evaluate them with the same trained probes (\textbf{\hyperlink{robustness-controls}{§A.4}}). The full-tensor score falls to F1 = 0.408, below Magpie's full-test-set score (0.520), indicating that the probe exploits the orientation in which Materials Project stores its structures. In contrast, the angular-norm score changes only from 0.834 to 0.813, and ORB is similarly robust (\textbf{\hyperlink{fig:4}{Fig.~4b}}, \textbf{\hyperlink{tab:S13}{Table~S13}}). This contrast does not depend on pretraining: with randomly initialised UMA, the full-tensor score falls from 0.85 to 0.32 after rotation, while the angular-norm score remains nearly unchanged at 0.66 and 0.65, respectively. Feature-level tests on 300 test materials under five independent rotations confirm that the L=0 and angular-norm features change by less than 5 × 10⁻⁴ in relative terms, compared with a median 3--19~\% per layer for the full node tensor (\textbf{\hyperlink{tab:S14}{Table~S14b--c}}, \textbf{\hyperlink{robustness-controls}{§A.4}}). We therefore use angular-norm features for UMA's reported crystal-system scores. These features particularly improve low-symmetry classification: ORB misclassifies 41~\% of triclinic structures as monoclinic, whereas UMA correctly classifies 85~\% of triclinic structures (\textbf{\hyperlink{fig:S3}{Fig.~S3a}}).

\begin{figure}[t]
  \hypertarget{fig:4}{}%
  \centering
  \includegraphics{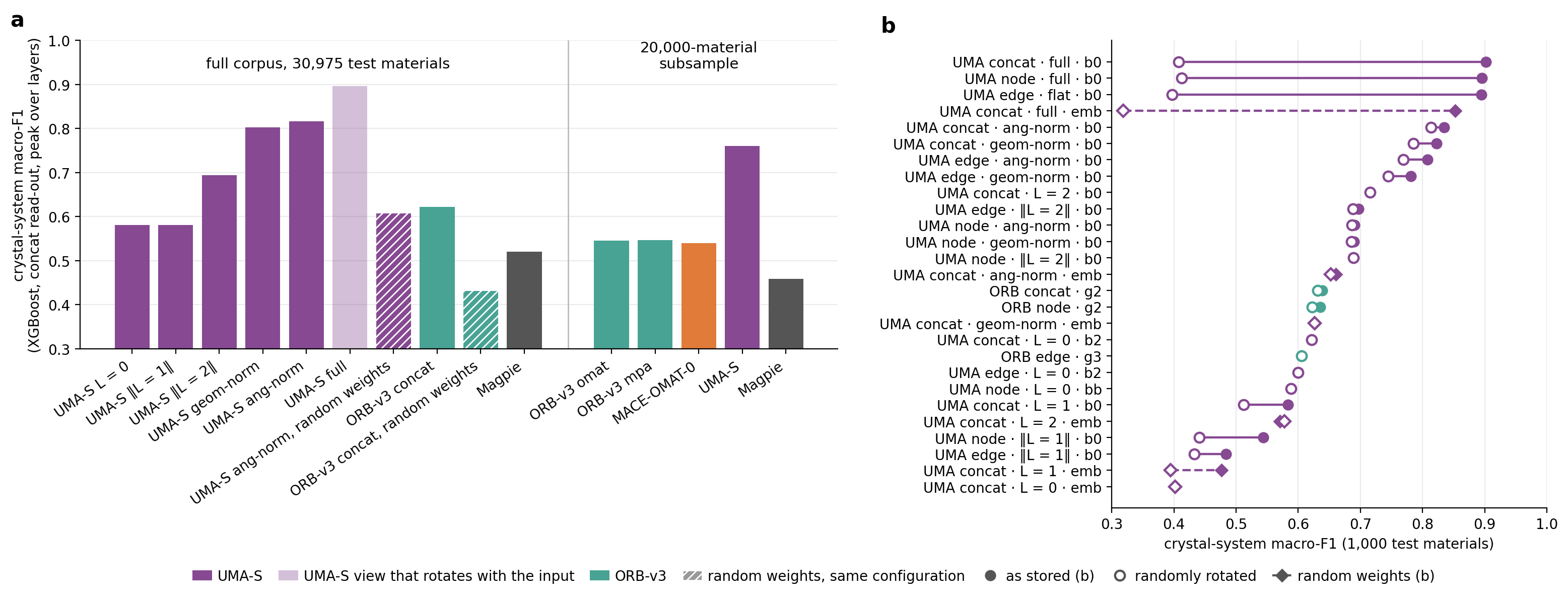}
  \caption*{\textbf{\emph{Figure~4. Crystal-system prediction across UMA-S channels, model controls and rotations.}} \emph{\textbf{(a)} Peak XGBoost macro-F1 across layers on the full test set (left) and the common 20,000-material subset of \textbf{\hyperlink{fig:2}{Fig.~2d}} (right). UMA-S appears once per concatenated channel view (L = 0, ‖L = 1‖, ‖L = 2‖, geom-norm, ang-norm and the full tensor, the last drawn faded because it rotates with the input) and ORB-v3 at its concatenated peak. All embedding bars use concatenated node--edge read-outs except MACE-OMAT-0 (nodes only). Hatched bars are the same networks with randomly re-initialised weights, scored at the read-out, view and layer of the pretrained bar beside them (UMA-S: concat · ang-norm · b0; ORB-v3: concat · g2). \textbf{(b)} Macro-F1 on 1,000 test structures before and after random rotation, evaluated with the same trained probes. Random-weight UMA-S is tested across its six concatenated channel views, each at that network's best layer. Prediction results are in \textbf{\hyperlink{tab:S13}{Table~S13}} and feature-level rotation tests in \textbf{\hyperlink{tab:S14}{Table~S14}}.}}
\end{figure}

UMA's crystal-system advantage cannot be attributed to equivariance alone (\textbf{\hyperlink{random-initialisation-and-matched-training-data-controls}{§A.9}}). Random initialisation also reveals predictive information before pretraining: at the configurations selected for the pretrained models, random-weight UMA scores 0.61 with angular-norm features, compared with 0.43 for random-weight ORB; at their own best configurations, the random networks reach 0.66 and 0.68, respectively (\textbf{\hyperlink{tab:S17}{Table~S17}}). On a common 20,000-material subset, an ORB-v3 variant pretrained on MPtrj and Alexandria differs from the original OMat24-trained ORB-v3 by at most 0.012 across all nine targets. Meanwhile, equivariant MACE-OMAT-0, pretrained on the same OMat24 data as ORB, reaches crystal-system F1 = 0.540, nearly identical to ORB's 0.545 and well below UMA's 0.760 (\textbf{\hyperlink{fig:2}{Fig.~2d}}, \textbf{\hyperlink{fig:4}{Fig.~4a}}; \textbf{\hyperlink{tab:S18}{Table~S18}}). These controls suggest that UMA's advantage depends on characteristics of its particular representation rather than equivariance alone, although they do not isolate the effects of architecture, capacity and training data.

\hypertarget{property-information-is-locally-predictable-without-property-aligned-clusters}{%
\subsection{Property information is locally predictable without property-aligned clusters}\label{property-information-is-locally-predictable-without-property-aligned-clusters}}

Neighbourhood-organisation metrics (k-NN purity and 1−LVR; \textbf{\hyperlink{fig:S4}{Fig.~S4}}) broadly follow the layer-wise probe trends, although their scores are not directly comparable with R² or macro-F1 (\textbf{\hyperlink{unsupervised-label-organisation-metrics-main}{§2.3}}). To test whether these neighbourhoods support prediction, we evaluate a supervised k-NN predictor at each target's selected configuration. It scores within 0.06 of XGBoost in 17 of 18 model--target comparisons; UMA crystal system is the exception, where XGBoost leads by 0.23 macro-F1 (\textbf{\hyperlink{tab:S4}{Table~S4}}). Thus, much of the property information is accessible through local embedding neighbourhoods without fitting a flexible decoder.

Local predictability does not, however, imply distinct property-aligned clusters. HDBSCAN finds little agreement between clusters and property labels (ARI below 0.05 for most targets and layers, at most 0.13; \textbf{\hyperlink{fig:S7}{Fig.~S7b}}, \textbf{\hyperlink{fig:S8}{Fig.~S8b}}). In UMA's b0 representation, band-gap variation is 0.39~eV within clusters versus 1.51~eV overall, despite negligible agreement with property classes (the HDBSCAN clusters are shown in \textbf{\hyperlink{fig:2}{Fig.~2e}}). Element identity, by contrast, forms distinct atom-level clusters (ARI 0.99 for ORB and 0.78 for UMA; \textbf{\hyperlink{fig:S5}{Fig.~S5c}}), visible in the element-coloured UMAP of \textbf{\hyperlink{fig:2}{Fig.~2e}}. Embeddings can therefore place materials with similar properties near one another without separating them into property-specific groups (\textbf{\hyperlink{decodability-clustering-and-nearest-neighbour-retrieval}{§B.2}}).

The two models also need not share the same neighbourhoods to encode similar predictive information. After removing element-identity effects, mutual k-NN agreement between ORB's final layer and UMA's backbone is only 0.248. Yet linear stitching, which maps one model's embeddings into the other's space, preserves probe performance within 0.05 for all nine targets in the UMA-to-ORB direction and within 0.06 for eight of nine in the reverse direction (\textbf{\hyperlink{orb-and-uma-transfer-function-across-layers-but-converge-geometrically-only-in-part}{§B.3}}, \textbf{\hyperlink{tab:S8}{Table~S8a}}, \textbf{\hyperlink{fig:S11}{Fig.~S11}}). Thus, predictive information transfers between the models despite incomplete agreement in their representation geometry.

\hypertarget{limitations}{%
\section{Limitations}\label{limitations}}

ORB-v3 and UMA-S differ in architecture, training data and capacity. Matched-model controls constrain possible explanations for their performance differences but do not isolate the contribution of equivariance. Probe accuracy depends on both the representation and the decoder {[}\citelink{cite:Hewitt2019}{Hewitt \& Liang 2019}{]}; our results establish which properties are decodable, not whether the MLIPs use that information to predict energies or forces. Depth changes both receptive context and non-linear processing, so the observed trends do not establish a causal ordering by physical scale. The main results are evaluated against Materials Project DFT labels rather than experiment, and cross-database transfer is affected by differences in material distributions, reference labels and computational protocols (\textbf{\hyperlink{external-transfer-results}{§B.4}}). The screening experiment evaluates how well the probes identify promising materials from a held-out set with known DFT properties, rather than testing newly discovered candidates. The label-efficiency experiment compares frozen MLIP embeddings with a single small GNN trained from scratch, so its findings do not establish an advantage over all task-specific models.

\hypertarget{conclusion}{%
\section{Conclusion}\label{conclusion}}

Frozen ORB-v3 and UMA-S embeddings predict nine crystal properties, including five outside their training objectives, and outperform composition-only descriptors across all targets. The most informative layers vary by property, with depth trends consistent with a physical-scale ordering. Random-weight controls show that pretraining contributes most to band gap, energy above hull and crystal system. UMA-S's crystal-system advantage is carried by rotation-invariant angular channels, although matched-model controls do not isolate equivariance as its cause. Nearest-neighbour prediction approaches trained-probe performance, despite the absence of property-aligned clusters. For downstream applications, these embeddings provide reusable descriptors without training a new representation from scratch. Their usefulness depends on the required accuracy: band-gap errors of 0.46--0.50~eV can support coarse screening, while energy-above-hull errors of 90--103~meV/atom make further DFT validation important for near-hull stability decisions.

\clearpage
\hypertarget{references}{%
\section{References}\label{references}}

{\small %
\begin{itemize}[label={},leftmargin=0pt,itemindent=0pt,itemsep=3pt]
\tightlist
\item
  \hypertarget{cite:Alain2016}{}Alain, G. \& Bengio, Y. (2016). \emph{Understanding intermediate layers using linear classifier probes.} arXiv:\href{https://arxiv.org/abs/1610.01644}{1610.01644}.
\item
  \hypertarget{cite:Bansal2021}{}Bansal, Y., Nakkiran, P., Barak, B. (2021). \emph{Revisiting model stitching to compare neural representations.} Advances in Neural Information Processing Systems 34; arXiv:\href{https://arxiv.org/abs/2106.07682}{2106.07682}.
\item
  \hypertarget{cite:Batatia2024}{}Batatia, I., Benner, P., Chiang, Y., et al.~(2024). \emph{A foundation model for atomistic materials chemistry (MACE-MP-0).} arXiv:\href{https://arxiv.org/abs/2401.00096}{2401.00096}; Journal of Chemical Physics 163, 184110 (2025). DOI:\href{https://doi.org/10.1063/5.0297006}{10.1063/5.0297006}. MACE-OMAT-0 checkpoints: github.com/ACEsuit/mace-foundations.
\item
  \hypertarget{cite:Benjamini1995}{}Benjamini, Y. \& Hochberg, Y. (1995). \emph{Controlling the false discovery rate: a practical and powerful approach to multiple testing.} Journal of the Royal Statistical Society: Series B 57(1), 289--300. DOI:\href{https://doi.org/10.1111/j.2517-6161.1995.tb02031.x}{10.1111/j.2517-6161.1995.tb02031.x}.
\item
  \hypertarget{cite:Borlido2019}{}Borlido, P., Aull, T., Huran, A. W., Tran, F., Marques, M. A. L., Botti, S. (2019). \emph{Large-scale benchmark of exchange--correlation functionals for the determination of electronic band gaps of solids.} Journal of Chemical Theory and Computation 15(9), 5069--5079. DOI:\href{https://doi.org/10.1021/acs.jctc.9b00322}{10.1021/acs.jctc.9b00322}.
\item
  \hypertarget{cite:Cawley2010}{}Cawley, G. C. \& Talbot, N. L. C. (2010). \emph{On over-fitting in model selection and subsequent selection bias in performance evaluation.} Journal of Machine Learning Research 11, 2079--2107.
\item
  \hypertarget{cite:Chen2022}{}Chen, C. \& Ong, S. P. (2022). \emph{A universal graph deep learning interatomic potential for the periodic table (M3GNet).} Nature Computational Science 2, 718--728. DOI:\href{https://doi.org/10.1038/s43588-022-00349-3}{10.1038/s43588-022-00349-3}.
\item
  \hypertarget{cite:Chen2016}{}Chen, T. \& Guestrin, C. (2016). \emph{XGBoost: a scalable tree boosting system.} KDD 2016, 785--794. DOI:\href{https://doi.org/10.1145/2939672.2939785}{10.1145/2939672.2939785}.
\item
  \hypertarget{cite:Chorna2026}{}Chorna, S., Tisi, D., Malosso, C., How, W. B., Ceriotti, M., Chong, S. (2026). \emph{Comparing the latent features of universal machine-learning interatomic potentials.} Advanced Intelligent Systems 8, e202501497. DOI:\href{https://doi.org/10.1002/aisy.202501497}{10.1002/aisy.202501497}; arXiv:\href{https://arxiv.org/abs/2512.05717}{2512.05717}.
\item
  \hypertarget{cite:Choudhary2020}{}Choudhary, K., Garrity, K. F., Reid, A. C. E., et al.~(2020). \emph{The joint automated repository for various integrated simulations (JARVIS) for data-driven materials design.} npj Computational Materials 6, 173. DOI:\href{https://doi.org/10.1038/s41524-020-00440-1}{10.1038/s41524-020-00440-1}.
\item
  \hypertarget{cite:Cover1967}{}Cover, T. \& Hart, P. (1967). \emph{Nearest neighbor pattern classification.} IEEE Transactions on Information Theory 13, 21--27.
\item
  \hypertarget{cite:Cui2024}{}Cui, H., Wang, C., Maan, H., Pang, K., Luo, F., et al.~(2024). \emph{scGPT: toward building a foundation model for single-cell multi-omics using generative AI.} Nature Methods 21, 1470--1480. DOI:\href{https://doi.org/10.1038/s41592-024-02201-0}{10.1038/s41592-024-02201-0}.
\item
  \hypertarget{cite:DallaTorre2025}{}Dalla-Torre, H., Gonzalez, L., Mendoza-Revilla, J., Lopez Carranza, N., Grzywaczewski, A. H., et al.~(2025). \emph{The Nucleotide Transformer: building and evaluating robust foundation models for human genomics.} Nature Methods 22, 287--297. DOI:\href{https://doi.org/10.1038/s41592-024-02523-z}{10.1038/s41592-024-02523-z}; bioRxiv 2023.01.11.523679.
\item
  \hypertarget{cite:Deng2023}{}Deng, B., Zhong, P., Jun, K., et al.~(2023). \emph{CHGNet as a pretrained universal neural network potential for charge-informed atomistic modelling.} Nature Machine Intelligence 5, 1031--1041. DOI:\href{https://doi.org/10.1038/s42256-023-00716-3}{10.1038/s42256-023-00716-3}; arXiv:\href{https://arxiv.org/abs/2302.14231}{2302.14231}.
\item
  \hypertarget{cite:Efron1979}{}Efron, B. (1979). \emph{Bootstrap methods: another look at the jackknife.} The Annals of Statistics 7(1), 1--26. DOI:\href{https://doi.org/10.1214/aos/1176344552}{10.1214/aos/1176344552}.
\item
  \hypertarget{cite:Gupta2024}{}Gupta, V., Choudhary, K., DeCost, B., Tavazza, F., Campbell, C., Liao, W., et al.~(2024). \emph{Structure-aware GNN-based deep transfer learning for materials property prediction.} npj Computational Materials 10. DOI:\href{https://doi.org/10.1038/s41524-023-01185-3}{10.1038/s41524-023-01185-3}
\item
  \hypertarget{cite:Harris2020}{}Harris, C. R., Millman, K. J., van der Walt, S. J., et al.~(2020). \emph{Array programming with NumPy.} Nature 585, 357--362. DOI:\href{https://doi.org/10.1038/s41586-020-2649-2}{10.1038/s41586-020-2649-2}.
\item
  \hypertarget{cite:Hewitt2019}{}Hewitt, J. \& Liang, P. (2019). \emph{Designing and interpreting probes with control tasks.} EMNLP-IJCNLP 2019, 2733--2743. DOI:\href{https://doi.org/10.18653/v1/D19-1275}{10.18653/v1/D19-1275}.
\item
  \hypertarget{cite:Horton2025}{}Horton, M. K., Huck, P., Yang, R. X., Munro, J. M., Dwaraknath, S., et al.~(2025). \emph{Accelerated data-driven materials science with the Materials Project.} Nature Materials 24, 1522--1532. DOI:\href{https://doi.org/10.1038/s41563-025-02272-0}{10.1038/s41563-025-02272-0}.
\item
  \hypertarget{cite:Huang2025}{}Huang, X., Deng, B., Zhong, P., Kaplan, A. D., Persson, K. A., Ceder, G. (2025). \emph{Cross-functional transferability in foundation machine learning interatomic potentials.} npj Computational Materials 11, 313. DOI:\href{https://doi.org/10.1038/s41524-025-01796-y}{10.1038/s41524-025-01796-y}; arXiv:\href{https://arxiv.org/abs/2504.05565}{2504.05565}.
\item
  \hypertarget{cite:Huh2024}{}Huh, M., Cheung, B., Wang, T., Isola, P. (2024). \emph{The Platonic Representation Hypothesis.} ICML 2024; arXiv:\href{https://arxiv.org/abs/2405.07987}{2405.07987}.
\item
  \hypertarget{cite:Jain2013}{}Jain, A., Ong, S. P., Hautier, G., Chen, W., Richards, W. D., Dacek, S., et al.~(2013). \emph{Commentary: the Materials Project: a materials genome approach to accelerating materials innovation.} APL Materials 1, 011002. DOI:\href{https://doi.org/10.1063/1.4812323}{10.1063/1.4812323}.
\item
  \hypertarget{cite:Kim2025}{}Kim, S. Y., Park, Y. J., Li, J. (2025). \emph{Leveraging neural network interatomic potentials for a foundation model of chemistry.} arXiv:\href{https://arxiv.org/abs/2506.18497}{2506.18497}.
\item
  \hypertarget{cite:Kirklin2015}{}Kirklin, S., Saal, J. E., Meredig, B., Thompson, A., Doak, J. W., Aykol, M., Rühl, S., Wolverton, C. (2015). \emph{The Open Quantum Materials Database (OQMD): assessing the accuracy of DFT formation energies.} npj Computational Materials 1, 15010. DOI:\href{https://doi.org/10.1038/npjcompumats.2015.10}{10.1038/npjcompumats.2015.10}.
\item
  \hypertarget{cite:Klimes2011}{}Klimeš, J., Bowler, D. R., Michaelides, A. (2011). \emph{Van der Waals density functionals applied to solids.} Physical Review B 83, 195131. DOI:\href{https://doi.org/10.1103/PhysRevB.83.195131}{10.1103/PhysRevB.83.195131}.
\item
  \hypertarget{cite:Kornblith2019}{}Kornblith, S., Norouzi, M., Lee, H., Hinton, G. (2019). \emph{Similarity of neural network representations revisited (CKA).} ICML 2019; arXiv:\href{https://arxiv.org/abs/1905.00414}{1905.00414}.
\item
  \hypertarget{cite:Larsen2017}{}Larsen, A. H., Mortensen, J. J., Blomqvist, J., et al.~(2017). \emph{The atomic simulation environment --- a Python library for working with atoms.} Journal of Physics: Condensed Matter 29(27), 273002. DOI:\href{https://doi.org/10.1088/1361-648X/aa680e}{10.1088/1361-648X/aa680e}.
\item
  \hypertarget{cite:Lin2023}{}Lin, Z., Akin, H., Rao, R., Hie, B., Zhu, Z., et al.~(2023). \emph{Evolutionary-scale prediction of atomic-level protein structure with a language model (ESM-2 / ESMFold).} Science 379(6637), 1123--1130. DOI:\href{https://doi.org/10.1126/science.ade2574}{10.1126/science.ade2574}.
\item
  \hypertarget{cite:Li2026}{}Li, Z. \& Walsh, A. (2026). \emph{Platonic representation of foundation machine learning interatomic potentials.} Nature Machine Intelligence 8, 830--840. DOI:\href{https://doi.org/10.1038/s42256-026-01235-7}{10.1038/s42256-026-01235-7}; arXiv:\href{https://arxiv.org/abs/2512.05349}{2512.05349}.
\item
  \hypertarget{cite:MansouriTehrani2018}{}Mansouri Tehrani, A., Oliynyk, A. O., Parry, M., Rizvi, Z., Couper, S., Lin, F., Miyagi, L., Sparks, T. D., Brgoch, J. (2018). \emph{Machine learning directed search for ultraincompressible, superhard materials.} Journal of the American Chemical Society 140(31), 9844--9853. DOI:\href{https://doi.org/10.1021/jacs.8b02717}{10.1021/jacs.8b02717}.
\item
  \hypertarget{cite:McInnes2017}{}McInnes, L., Healy, J., Astels, S. (2017). \emph{hdbscan: hierarchical density based clustering.} Journal of Open Source Software 2(11), 205. DOI:\href{https://doi.org/10.21105/joss.00205}{10.21105/joss.00205}.
\item
  \hypertarget{cite:McInnes2018}{}McInnes, L., Healy, J., Melville, J. (2018). \emph{UMAP: uniform manifold approximation and projection for dimension reduction.} arXiv:\href{https://arxiv.org/abs/1802.03426}{1802.03426}.
\item
  \hypertarget{cite:Meredig2018}{}Meredig, B., Antono, E., Church, C., Hutchinson, M., Ling, J., Paradiso, S., Blaiszik, B., Foster, I., Gibbons, B., Hattrick-Simpers, J., Mehta, A., Ward, L. (2018). \emph{Can machine learning identify the next high-temperature superconductor? Examining extrapolation performance for materials discovery.} Molecular Systems Design \& Engineering 3, 819--825. DOI:\href{https://doi.org/10.1039/C8ME00012C}{10.1039/C8ME00012C}.
\item
  \hypertarget{cite:Nadeau2003}{}Nadeau, C. \& Bengio, Y. (2003). \emph{Inference for the generalization error.} Machine Learning 52, 239--281. DOI:\href{https://doi.org/10.1023/A:1024068626366}{10.1023/A:1024068626366}.
\item
  \hypertarget{cite:Neumann2024}{}Neumann, M., Gin, J., Rhodes, B., et al.~(2024). \emph{Orb: a fast, scalable neural network potential.} arXiv:\href{https://arxiv.org/abs/2410.22570}{2410.22570}.
\item
  \hypertarget{cite:Ong2013}{}Ong, S. P., Richards, W. D., Jain, A., Hautier, G., Kocher, M., Cholia, S., Gunter, D., Chevrier, V. L., Persson, K. A., Ceder, G. (2013). \emph{Python Materials Genomics (pymatgen): a robust, open-source Python library for materials analysis.} Computational Materials Science 68, 314--319. DOI:\href{https://doi.org/10.1016/j.commatsci.2012.10.028}{10.1016/j.commatsci.2012.10.028}.
\item
  \hypertarget{cite:Pedregosa2011}{}Pedregosa, F., et al.~(2011). \emph{Scikit-learn: machine learning in Python.} Journal of Machine Learning Research 12, 2825--2830.
\item
  \hypertarget{cite:Pinto2025}{}Pinto, L. (2025). \emph{Superior molecular representations from intermediate encoder layers.} arXiv:\href{https://arxiv.org/abs/2506.06443}{2506.06443}.
\item
  \hypertarget{cite:Radova2025}{}Radova, M., Stark, W. G., Allen, C. S., Maurer, R. J., Bartók, A. P. (2025). \emph{Fine-tuning foundation models of materials interatomic potentials with frozen transfer learning.} npj Computational Materials 11. DOI:\href{https://doi.org/10.1038/s41524-025-01727-x}{10.1038/s41524-025-01727-x}; arXiv:\href{https://arxiv.org/abs/2502.15582}{2502.15582}.
\item
  \hypertarget{cite:Raghu2017}{}Raghu, M., Gilmer, J., Yosinski, J., Sohl-Dickstein, J. (2017). \emph{SVCCA: singular vector canonical correlation analysis for deep learning dynamics and interpretability.} NeurIPS 2017; arXiv:\href{https://arxiv.org/abs/1706.05806}{1706.05806}.
\item
  \hypertarget{cite:Rao2021}{}Rao, R., Meier, J., Sercu, T., Ovchinnikov, S., Rives, A. (2021). \emph{Transformer protein language models are unsupervised structure learners.} ICLR 2021; bioRxiv 2020.12.15.422761. DOI:\href{https://doi.org/10.1101/2020.12.15.422761}{10.1101/2020.12.15.422761}.
\item
  \hypertarget{cite:Rhodes2025}{}Rhodes, B., Vandenhaute, S., Šimkus, V., Gin, J., Godwin, J., Duignan, T., Neumann, M. (2025). \emph{Orb-v3: atomistic simulation at scale.} arXiv:\href{https://arxiv.org/abs/2504.06231}{2504.06231}.
\item
  \hypertarget{cite:Rives2021}{}Rives, A., Meier, J., Sercu, T., Goyal, S., Lin, Z., et al.~(2021). \emph{Biological structure and function emerge from scaling unsupervised learning to 250 million protein sequences (ESM-1b).} PNAS 118(15), e2016239118. DOI:\href{https://doi.org/10.1073/pnas.2016239118}{10.1073/pnas.2016239118}.
\item
  \hypertarget{cite:Rumelhart1986}{}Rumelhart, D. E., Hinton, G. E., Williams, R. J. (1986). \emph{Learning representations by back-propagating errors.} Nature 323, 533--536.
\item
  \hypertarget{cite:Sun2016}{}Sun, W., Dacek, S. T., Ong, S. P., Hautier, G., Jain, A., Richards, W. D., Gamst, A. C., Persson, K. A., Ceder, G. (2016). \emph{The thermodynamic scale of inorganic crystalline metastability.} Science Advances 2, e1600225. DOI:\href{https://doi.org/10.1126/sciadv.1600225}{10.1126/sciadv.1600225}
\item
  \hypertarget{cite:Theodoris2023}{}Theodoris, C. V., Xiao, L., Chopra, A., Chaffin, M. D., Al Sayed, Z. R., et al.~(2023). \emph{Transfer learning enables predictions in network biology.} Nature 618, 616--624. DOI:\href{https://doi.org/10.1038/s41586-023-06139-9}{10.1038/s41586-023-06139-9}.
\item
  \hypertarget{cite:Virtanen2020}{}Virtanen, P., Gommers, R., Oliphant, T. E., et al.~(2020). \emph{SciPy 1.0: fundamental algorithms for scientific computing in Python.} Nature Methods 17(3), 261--272. DOI:\href{https://doi.org/10.1038/s41592-019-0686-2}{10.1038/s41592-019-0686-2}.
\item
  \hypertarget{cite:Ward2016}{}Ward, L., Agrawal, A., Choudhary, A., Wolverton, C. (2016). \emph{A general-purpose machine learning framework for predicting properties of inorganic materials (Magpie).} npj Computational Materials 2, 16028. DOI:\href{https://doi.org/10.1038/npjcompumats.2016.28}{10.1038/npjcompumats.2016.28}.
\item
  \hypertarget{cite:Ward2018}{}Ward, L., Dunn, A., Faghaninia, A., et al.~(2018). \emph{Matminer: an open source toolkit for materials data mining.} Computational Materials Science 152, 60--69. DOI:\href{https://doi.org/10.1016/j.commatsci.2018.05.018}{10.1016/j.commatsci.2018.05.018}.
\item
  \hypertarget{cite:Wood2025}{}Wood, B. M., Shuaibi, M., et al.~(2025). \emph{UMA: a family of universal models for atoms.} arXiv:\href{https://arxiv.org/abs/2506.23971}{2506.23971}.
\item
  \hypertarget{cite:Yamada2019}{}Yamada, H., Liu, C., Wu, S., Koyama, Y., Ju, S., et al.~(2019). \emph{Predicting materials properties with little data using shotgun transfer learning.} ACS Central Science 5(10). DOI:\href{https://doi.org/10.1021/acscentsci.9b00804}{10.1021/acscentsci.9b00804}
\item
  \hypertarget{cite:Zhuo2018}{}Zhuo, Y., Mansouri Tehrani, A., Brgoch, J. (2018). \emph{Predicting the band gaps of inorganic solids by machine learning.} Journal of Physical Chemistry Letters 9(7), 1668--1673. DOI:\href{https://doi.org/10.1021/acs.jpclett.8b00124}{10.1021/acs.jpclett.8b00124}.
\end{itemize}
}

\hypertarget{supplementary-methods}{%
\appendix
\newpage
\section{Supplementary Methods}\label{supplementary-methods}}

Full technical detail behind the methods of \hyperlink{methods}{§2}.

\hypertarget{read-outs}{%
\subsection{Read-outs}\label{read-outs}}

At each layer, we extract per-atom node and per-bond edge features. ORB-v3 produces 256-dimensional rotation-invariant vectors for both and has 25.5 million parameters. UMA-S instead produces SO(3)-equivariant tensors of shape {[}9, 128{]}, whose nine spherical-harmonic components comprise an L=0 scalar, an L=1 vector and an L=2 rank-2 channel. Its checkpoint contains 291 million parameters, including 289 million in 64 mixture-of-linear-experts weight sets that are merged into one set per structure before the forward pass, leaving approximately 6.6 million active parameters per structure. Both backbones remain frozen throughout the study.

An important difference lies in their initial representations. ORB's encoder processes element embeddings through a node multilayer perceptron (MLP) and bond distances and directions through a separate edge MLP. Its initial node features therefore encode element identity alone; geometric information reaches them only after the first message-passing layer. UMA's embedding block, by contrast, adds a neighbour-dependent term to each atom's element embedding, summing a radial function of bond lengths and neighbouring element identities. Its node features thus already contain local structural information before the first interaction block. This difference helps explain why several local properties peak at UMA's embedding layer but improve after ORB's first message-passing layer (\textbf{\hyperlink{fig:3}{Fig.~3a--b}}). Accordingly, ``before message passing'' for UMA means before its interaction blocks, not before any neighbour information is incorporated (\textbf{\hyperlink{tab:S7}{Table~S7}}).

\textbf{Unpooled channel views.} To examine which parts of UMA's representations carry predictive information, we construct six views from the L=0, L=1 and L=2 channels described above (\textbf{\hyperlink{tab:S1}{Table~S1}}). The \emph{flat} view retains all nine spherical-harmonic slots and is called \emph{full} after pooling. The \emph{L=0} view retains only the scalar channel, while ‖L1‖ and ‖L2‖ take the Euclidean norm of the corresponding angular components. The \emph{ang-norm} view combines L=0, ‖L1‖ and ‖L2‖, whereas \emph{geom-norm} includes only the two angular norms. These operations are applied separately to each atom or bond, preserving one feature vector per site or edge. Node and edge features use the same six views but differ in how their angular components are pooled across a structure, as explained below. The unpooled vectors are used for atom- and bond-level neighbourhood analyses (\textbf{\hyperlink{unsupervised-label-organisation-metrics}{§A.5}}); supervised probes use the material-level pooled representations defined next.

\par\addvspace{0.8\baselineskip}
\begin{table}[H]
\hypertarget{tab:S1}{}%
  \caption*{\textbf{Table~S1 --- UMA-S equivariant read-out family.} \emph{Per-atom node views and their pooled
per-material counterparts. ‖·‖ is the Euclidean norm across spherical harmonic slots (channel-wise magnitude); cat is concatenation; mean is
the atom average. † For nodes the pooled L=1/L=2 views use mean-then-norm (‖L.mean‖, called polar); for edges they use norm-then-mean (‖L‖.mean, called norm).}}
  \centering

  \begin{tabular}{@{}
  >{\raggedright\arraybackslash}p{(\columnwidth - 6\tabcolsep) * \real{0.1034}}
  >{\raggedright\arraybackslash}p{(\columnwidth - 6\tabcolsep) * \real{0.2759}}
  >{\raggedright\arraybackslash}p{(\columnwidth - 6\tabcolsep) * \real{0.2931}}
  >{\raggedright\arraybackslash}p{(\columnwidth - 6\tabcolsep) * \real{0.3276}}@{}}
\toprule\noalign{}
\begin{minipage}[c]{\linewidth}\raggedright
Read-out
\end{minipage} & \begin{minipage}[c]{\linewidth}\raggedright
Per-atom node {[}N, ·{]}
\end{minipage} & \begin{minipage}[c]{\linewidth}\raggedright
What it captures
\end{minipage} & \begin{minipage}[c]{\linewidth}\raggedright
Pooled per material {[}·{]}
\end{minipage} \\
\midrule\noalign{}
\textbf{flat / full} & flatten({[}9,128{]}) → 1152 & all channels; changes under rotation & mean.flatten() → 1152 \\
\textbf{L=0} & {[}:,0,:{]} → 128 & composition / chemistry & L0.mean → 128 \\
\textbf{L=1} † & ‖L1‖ → 128 & how directional the bonding is & ‖L1.mean‖ → 128 \\
\textbf{L=2} † & ‖L2‖ → 128 & how elongated or flattened the coordination shell is & ‖L2.mean‖ → 128 \\
\textbf{ang-norm} & cat(L0, ‖L1‖, ‖L2‖) → 384 & atom identity and how asymmetric its environment is, without the direction & cat(L0, ‖L1‖, ‖L2‖).mean → 384 \\
\textbf{geom-norm} & cat(‖L1‖, ‖L2‖) → 256 & geometry with composition removed & cat(‖L1‖, ‖L2‖).mean → 256 \\
  \bottomrule\end{tabular}
\end{table}
\par\addvspace{0.8\baselineskip}

\textbf{Material-level pooling.} A supervised probe reads one fixed-length vector per material, so the per-atom and per-bond views are averaged over the structure. For ORB-v3 that is a plain mean. UMA-S's equivariant channels can be averaged in two orders, which measure different things.

Norm-then-mean takes each atom's magnitude and then averages, measuring how strong the local geometric asymmetry is at a typical site. The ang-norm and geom-norm views use this order. For edges it is the only option: every bond appears in the neighbour graph in both directions with opposite vectors, so averaging bond vectors first would give identically zero for any structure.

Mean-then-norm averages the vectors and then takes the magnitude, measuring whether the atoms' environments all point the same way. Opposing environments cancel, which is why the output is zero for centrosymmetric structures, where every atom has an inversion partner. The standalone node L=1 and L=2 views use this order, and are called polar (‖mean\_atoms(L)‖).

\hypertarget{probe-configuration}{%
\subsection{Probe configuration}\label{probe-configuration}}

Four supervised probes are used. XGBoost and a linear probe (RidgeCV for regression, logistic regression for classification) are swept over every layer, read-out and view: 18 configurations for ORB-v3 (node, edge and concat read-outs at six layers) and 96 for UMA-S (six channel views at six node layers, five edge layers and five concat layers); the parameter-free k-NN predictor and a multilayer perceptron (MLP) are fitted only at each target's XGBoost-selected configuration. Every setting, including values left at their library default, is listed in \textbf{\hyperlink{tab:S2}{Table~S2}}.

\par\addvspace{0.8\baselineskip}
\hypertarget{tab:S2}{}%
{\small \textbf{Table~S2 --- Hyperparameters and fitted sizes of the four prediction methods (\hyperlink{supervised-probes}{§2.2}, \hyperlink{probe-configuration}{§A.2}).} \emph{The table lists every setting that affects the fit; parameters not shown were left at their library defaults. XGBoost, linear probes and k-nearest-neighbour (k-NN) prediction use fixed settings across all targets, layers, read-outs and channel views. For the multilayer perceptron (MLP), network size and weight decay are selected separately for each target. Fitted sizes are reported as ranges across the 18 model--target configurations in \textbf{\hyperlink{tab:1}{Table~1}}. They represent leaf counts for XGBoost and numbers of weights for linear probes and MLPs; k-NN has no fitted parameters. The Magpie composition-only baseline uses the same XGBoost settings and data split, but replaces embeddings with composition descriptors. Software versions: Python 3.12, XGBoost 3.3.0, scikit-learn 1.8.0, NumPy 2.4.4, matminer 0.10.1 and pymatgen 2026.3.23.}\par}\par\nobreak
{\renewcommand{\arraystretch}{0.93}%
\begin{longtable}{@{}
  >{\raggedright\arraybackslash}p{(\columnwidth - 4\tabcolsep) * \real{0.2667}}
  >{\raggedright\arraybackslash}p{(\columnwidth - 4\tabcolsep) * \real{0.2250}}
  >{\raggedright\arraybackslash}p{(\columnwidth - 4\tabcolsep) * \real{0.5083}}@{}}
\toprule\noalign{}
\begin{minipage}[c]{\linewidth}\raggedright
Component
\end{minipage} & \begin{minipage}[c]{\linewidth}\raggedright
Setting
\end{minipage} & \begin{minipage}[c]{\linewidth}\raggedright
Value
\end{minipage} \\
\midrule\noalign{}
Split & scheme & fixed 80/20 train/test, per target \\
& seed & 42 \\
& stratification & by class (classification targets only) \\
\textbf{Input} & dimension \(d\) & the pooled vector at the selected read-out and layer, 256--768 at the \textbf{\hyperlink{tab:1}{Table~1}} configurations (node 256 or 384; concat, node \(\oplus\) edge at one layer, 512 or 768; L0 concat 256) \\
\textbf{XGBoost} (regressor / classifier) & n\_estimators & 300 \\
& max\_depth & 6 \\
& learning\_rate & 0.05 \\
& subsample & 0.8 \\
& colsample\_bytree & 0.8 \\
& tree\_method & hist \\
& random\_state & 42 \\
& objective & reg:squarederror (reg) / softprob (clf) \\
& reg\_lambda / reg\_alpha & 1.0 / 0.0 (default) \\
& min\_child\_weight / gamma & 1 / 0 (default) \\
& all other parameters & XGBoost 3.3.0 default \\
& fitted size & 15,000--19,000 leaves at the \textbf{\hyperlink{tab:1}{Table~1}} configurations; 109,000--127,000 for the seven-class crystal system (one tree set per class) \\
\textbf{Linear} (regression) & preprocessing & StandardScaler (zero mean, unit variance) \\
& estimator & RidgeCV \\
& alphas & 6 log-spaced values, 10⁻² to 10³ \\
& CV & leave-one-out (RidgeCV default) \\
& fitted size & \(d + 1\) weights: 257--769 \\
\textbf{Linear} (classification) & preprocessing & StandardScaler \\
& estimator & LogisticRegression \\
& penalty / C & ℓ₂ / 1.0 \\
& max\_iter & 2000 \\
& fitted size & \(d + 1\) weights per class: 385--513 for the binary targets, 3,591--5,383 for crystal system \\
\textbf{k-NN predictor} (supervised) & K & 10 \\
& index & 80~\% training split only \\
& features & pooled material-level embedding at each model's best configuration \\
& standardisation & none \\
& distance & Euclidean and cosine, reported separately \\
& fitted size & none \\
\textbf{MLP} (regressor / classifier) & configuration & each target's XGBoost-selected layer, read-out and view \\
& preprocessing & StandardScaler (fitted on the training split) \\
& hidden layers & (256) or (512, 256), selected \\
& alpha (ℓ₂ weight decay) & 10⁻⁴ or 10⁻², selected \\
& selection & best of the 4 settings on a held-out 10~\% of the training split; refit on the full split \\
& activation / solver & ReLU / Adam (default) \\
& batch size / learning rate & 512 / 10⁻³ \\
& max\_iter & 200 \\
& early stopping & on, validation\_fraction 0.1, n\_iter\_no\_change 10 \\
& random\_state & 42 \\
& fitted size & 66,000--527,000 weights at the selected configurations \\
\textbf{Composition baseline} & features & Magpie: 132 elemental-property statistics per formula (matminer ElementProperty, magpie preset) \\
\bottomrule
\end{longtable}
}
\par\addvspace{0.8\baselineskip}

\textbf{k-NN predictor} (supervised). A probe that fits no parameters: the same k-NN graph as the purity statistic of \textbf{\hyperlink{unsupervised-label-organisation-metrics}{§A.5}}, but used to predict rather than to describe. A test material's value is predicted from its nearest training materials, by majority vote for classification and by the neighbour mean for regression:

\[\hat{y}_i = \operatorname*{arg\,max}_c \sum_{j \in N_K(i)} \mathbf{1}[y_j = c] \quad \text{(classification)}, \qquad \hat{y}_i = \frac{1}{K}\sum_{j \in N_K(i)} y_j \quad \text{(regression)}\]

where \(\hat{y}_i\) is the predicted value for test material \(i\), \(N_K(i)\) the set of its \(K\) nearest training materials, \(y_j\) the known value of training material \(j\), \(c\) a candidate class, and \(\mathbf{1}[y_j = c]\) the indicator function, equal to 1 when neighbour \(j\) belongs to class \(c\) and 0 otherwise. Predictions are scored by macro-F1 for classification and \(R^2\) for regression, the same units as the trained probe on the same split.

The index holds only training materials, and scores are computed on test materials, so a queried material is never among its own neighbours; the remaining settings are in \textbf{\hyperlink{tab:S2}{Table~S2}}.

It consumes training labels, unlike the purity statistic of \textbf{\hyperlink{unsupervised-label-organisation-metrics}{§A.5}}, but fits nothing, unlike the XGBoost probe above --- its only inductive bias is that nearby materials share property values.

\textbf{Multilayer perceptron} (supervised). We train a fully connected network with ReLU activations using Adam on standardised features (scikit-learn MLPRegressor / MLPClassifier). We test four configurations: one hidden layer (256 units) or two (512 and 256 units), each with weight decay of 10⁻⁴ or 10⁻². The best configuration is selected using 10~\% of the training data held out for validation (stratified for classification), then refitted on the full training split and evaluated on the test set; each fit uses early stopping on a further internal tenth of its training data. Results are compared with the other three probes in \textbf{\hyperlink{probe-family}{§B.5}} (\textbf{\hyperlink{tab:S4}{Table~S4}}, \textbf{\hyperlink{fig:S2}{Fig.~S2}}).

\textbf{Probe comparison (\hyperlink{frozen-mlip-embeddings-predict-properties-outside-the-training-objective}{§3.1}, \hyperlink{probe-family}{§B.5}).} At each target's test-selected configuration, the one \textbf{\hyperlink{tab:1}{Table~1}} reports, the four probes above are fitted on the seed-42 training split and scored on its test split. Each probe is scored by its primary metric and, for continuous targets, by RMSE in the property's units; the wall time of one fit on 8 cores is recorded. The refitted XGBoost and linear scores reproduce the sweep above to within 0.001. The layer selection is checked against the linear probe without new fits: for every model and target the configuration at which the linear probe scores highest over the same sweep is compared with XGBoost's, counting layers apart in the model's layer sequence (\textbf{\hyperlink{property-decodability-depends-on-network-depth}{§3.3}}, \textbf{\hyperlink{fig:S6}{Fig.~S6}}).

\hypertarget{targets-and-labels}{%
\subsection{Targets and labels}\label{targets-and-labels}}

The nine targets are taken or derived from the Materials Project summary record of each structure (database release 2026.04.13, \textbf{\hyperlink{data-availability-and-licences}{§A.12}}). \textbf{Formation energy per atom}, \textbf{energy above hull}, \textbf{band gap} and \textbf{density} are the summary fields as served. \textbf{Bulk modulus} is the Voigt--Reuss--Hill average from the Materials Project elastic workflow, available for 12,853 of the 154,875 materials; the other targets are available for all of them. \textbf{Magnetic-moment density} is the total magnetisation of the cell divided by its volume, in μ\_B~Å⁻³. \textbf{Crystal system} is the Materials Project symmetry assignment, seven classes from triclinic to cubic. \textbf{Metal/insulator} is metal when the DFT band gap is exactly zero and insulator otherwise, the Materials Project convention. \textbf{Magnetic label} is magnetic when the magnetic-moment density is at least 10⁻⁴~μ\_B~Å⁻³ in magnitude and non-magnetic otherwise. Before probing, values outside physical bounds would have been set to missing (bulk modulus 0--500~GPa, band gap 0--25~eV, formation energy −15 to 10~eV/atom, energy above hull ≥~0, density 0--30~g~cm⁻³); no value in the corpus fell outside them. Splits are drawn per target over the materials with a value, so a material missing a target is absent from that target's split only (\textbf{\hyperlink{tab:S2}{Table~S2}}).

All of these are density-functional-theory (DFT) values and inherit the functional's limitations. Band gaps come from generalised-gradient-approximation functionals (GGA/PBE(+U)), which underestimate experimental gaps by roughly 30--50~\% {[}\citelink{cite:Borlido2019}{Borlido 2019}{]}; the metal/insulator label is therefore operationally ``DFT gap = 0'', which misassigns Mott and charge-transfer insulators. The magnetic label is a binary magnetic / non-magnetic distinction, not the ferromagnetic/antiferromagnetic/ferrimagnetic classification. Reported accuracies are therefore against these DFT labels, not against experiment, and are specific to this functional: the same structures computed with another functional (r²SCAN) give energies that are offset in scale and only weakly correlated with their GGA values {[}\citelink{cite:Huang2025}{Huang 2025}{]}.

\hypertarget{robustness-controls}{%
\subsection{Robustness controls}\label{robustness-controls}}

Every control below runs on the same cached embeddings and the same fixed seed-42 split as the probes of \textbf{\hyperlink{probe-configuration}{§A.2}}, so only the control itself differs.

\textbf{Validation-split selection (\hyperlink{frozen-mlip-embeddings-predict-properties-outside-the-training-objective}{§3.1}).} Picking the best configuration by its test score and then reporting that same score is optimistic: the maximum over many configurations partly fits noise in the test set {[}\citelink{cite:Cawley2010}{Cawley \& Talbot 2010}{]}. A 20~\% validation split is carved from the training data (stratified for classification); each model's peak configuration is selected by validation score (for this ranking, each configuration is trained on at most 50,000 randomly drawn training materials), refit on the full training split, and scored once on the test set. Inflation is the test score of the test-selected configuration minus that of the validation-selected one: how much selecting on the test set flatters the reported number.

\textbf{Grouped splits (\hyperlink{frozen-mlip-embeddings-predict-properties-outside-the-training-objective}{§3.1}).} A random Materials Project split can place polymorphs and near-duplicate stoichiometries on both sides, inflating scores through train/test leakage; grouped cross-validation was introduced to expose this {[}\citelink{cite:Meredig2018}{Meredig 2018}{]}. We therefore split by group rather than at random: materials are first collected into groups, and every material in a group is then assigned to the same side of an 80/20 split (seed 42). We run this twice, with two different definitions of a group, both derived from each material's chemical formula using pymatgen {[}\citelink{cite:Ong2013}{Ong 2013}{]}. Composition-grouped puts every material sharing a reduced formula on the same side, which separates polymorphs. Chemical-system-grouped groups by the set of elements present, so materials built from the same elements never end up on opposite sides.

\textbf{Log-scale bulk modulus (\hyperlink{from-ruxb2-to-expected-downstream-accuracy}{§3.2}).} Bulk modulus in GPa has a long stiff tail (σ = 77~GPa on the test set), so a few very stiff materials carry much of its variance and a probe that places them roughly right earns a high R² whatever it does elsewhere. Following the Matbench elastic task, which predicts log\textsubscript{10}(K\textsubscript{VRH}), the XGBoost probe is refitted at each model's \textbf{\hyperlink{tab:1}{Table~1}} bulk-modulus configuration on the same seed-42 split with log\textsubscript{10}(K\textsubscript{VRH}) as the target, excluding the materials with K \(\le\) 0 (2,569 of the 2,571 test materials remain), and scored by R² on the log scale with the same 2,000-resample bootstrap as the main probes. Raw and log-scale scores are compared in \textbf{\hyperlink{tab:S11}{Table~S11b}}.

\textbf{Band gap on non-metals (\hyperlink{from-ruxb2-to-expected-downstream-accuracy}{§3.2}).} About 47~\% of the test materials are metals with a gap of exactly zero, so a pooled band-gap R² partly rewards separating metals from non-metals. The selected band-gap probes are therefore also scored on the non-metal test materials only (true gap \textgreater{}~0), with the same probe settings and the same seed-42 split and with σ recomputed on that subset (\textbf{\hyperlink{tab:S11}{Table~S11a}}).

\textbf{Permuted labels (\hyperlink{property-decodability-depends-on-network-depth}{§3.3}).} A probe with enough capacity can score well on labels that carry no information about the input, so a high score alone does not show that the representation holds the property {[}\citelink{cite:Hewitt2019}{Hewitt \& Liang 2019}{]}. At each target's peak configuration the XGBoost probe is refitted after the labels have been permuted at random across materials, separately within the training and the test split, and scored on the permuted test labels. Selectivity is the real-label score minus this permuted-label score; the two scores and their difference are in \textbf{\hyperlink{tab:S12}{Table~S12}}.

\textbf{Rotation control (\hyperlink{uma-channel-views-and-the-rotation-invariant-crystal-system-gap}{§3.4}).} Each of the 1,000 test structures in the subsample receives its own random rotation, drawn uniformly from all 3D rotations (any axis, any angle; the Haar measure on SO(3), sampled with scipy.spatial.transform.Rotation.random {[}\citelink{cite:Virtanen2020}{Virtanen 2020}{]} and a fresh seed per material). The same rotation matrix is applied to the atomic positions and to the lattice vectors, so the crystal is rotated rigidly as a whole and nothing but its orientation in space changes. The embedding is then re-extracted through a fresh forward pass and scored with the already-trained probe. The control is repeated for the random-weight UMA-S network of \hyperlink{random-initialisation-and-matched-training-data-controls}{§A.9} on the same 1,000 materials, with its weights re-initialised from the extraction's fixed seed and probes trained on its own features (\textbf{\hyperlink{tab:S13}{Table~S13b}}).

\textbf{Feature-level rotation test (\hyperlink{uma-channel-views-and-the-rotation-invariant-crystal-system-gap}{§3.4}).} For 300 crystal-system test materials and five random rotations each, every pooled read-out --- UMA node and edge views at every layer, ORB node and edge at every layer --- is recomputed from a fresh forward pass and compared with an identity-rotation pass of the same extraction: the relative change ‖f(Rx) − f(x)‖ / ‖f(x)‖, the identity pass against the stored feature cache, the number of graph edges per pass, and the crystal-system F1 of the paper's probes on each rotation.

\textbf{Same-material exclusion (\hyperlink{decodability-clustering-and-nearest-neighbour-retrieval}{§B.2}).} Per-atom purity scores atoms rather than materials, so each atom is given the label of the material it belongs to. An atom's nearest neighbours can then include other atoms of that same material, and those neighbours carry a matching label whatever the embedding has learned, which inflates the score. We recompute the chance-corrected k-nearest-neighbour (k-NN) purity with those same-material neighbours removed from the candidate pool. The index holds all ≈499,000 (ORB) or ≈300,000 (UMA) atoms and K = 10, as in the uncorrected statistic. Purity is defined only for categorical targets, so this control and the whitening control below cover crystal system, metal/insulator and magnetic label; the continuous 1−LVR values are not recomputed.

\textbf{Whitened purity metric (\hyperlink{decodability-clustering-and-nearest-neighbour-retrieval}{§B.2}).} k-NN purity picks each point's neighbours by Euclidean distance, which weights every embedding dimension equally, so a few high-variance directions can decide which points count as neighbours. If the weak purity were an artefact of that, rescaling those directions should recover it. We therefore recompute material-level k-NN purity after principal-component-analysis (PCA) whitening, which rescales the covariance so no direction dominates.

\hypertarget{unsupervised-label-organisation-metrics}{%
\subsection{Unsupervised label-organisation metrics}\label{unsupervised-label-organisation-metrics}}

\textbf{Chance-corrected k-NN purity} (categorical):

\[\text{purity} = \frac{\bar{p} - p_{\text{chance}}}{1 - p_{\text{chance}}}, \quad \bar{p} = \frac{1}{n}\sum_i \frac{1}{K}\sum_{j \in N_K(i)} \mathbf{1}[y_j = y_i], \quad p_{\text{chance}} = \sum_c f_c^2\]

where \(n\) is the number of indexed points, \(N_K(i)\) the set of the \(K\) nearest neighbours of point \(i\), \(y_i\) and \(y_j\) the classes of points \(i\) and \(j\) for the target being scored, \(f_c\) the frequency of class \(c\), and \(\mathbf{1}[y_j = y_i]\) the indicator function, equal to 1 when neighbour \(j\) has the same class as point \(i\) and 0 otherwise. The double sum therefore makes \(\bar{p}\) the mean fraction of a point's neighbours that share its label, and \(p_{\text{chance}}\) the fraction expected from the class frequencies alone. Subtracting that expectation puts chance agreement at zero and leaves the top of the scale for a perfectly pure neighbourhood, so targets with differently balanced classes are comparable.

\textbf{Local Variance Ratio (LVR)} (continuous), reported throughout as 1−LVR:

\[\text{LVR} = \frac{\frac{1}{n}\sum_i \mathrm{SD}_{j \in N_K(i)}(y_j)}{\mathrm{SD}(y)}\]

where \(y_j\) is the value of point \(j\) for the target being scored, \(N_K(i)\) the set of the \(K\) nearest neighbours of point \(i\), \(\mathrm{SD}_{j \in N_K(i)}(y_j)\) the standard deviation of those values across that set, and \(\mathrm{SD}(y)\) is the standard deviation across the whole set of \(n\) points. The numerator averages the neighbourhood standard deviation over every point, so LVR asks how much a target varies locally compared with how much it varies overall. It is small when neighbours hold similar values, and approaches 1 when a neighbourhood is no more homogeneous than the full set. Reporting 1−LVR makes larger mean better organised, as for purity. Working in standard-deviation rather than variance units also keeps LVR in the target's own units, so \(\text{LVR}\cdot\mathrm{SD}(y)\) reads directly as an expected error, on the same footing as the \(\sigma\cdot\sqrt{1-R^2}\) budget of \textbf{\hyperlink{from-ruxb2-to-expected-downstream-accuracy}{§3.2}}.

Purity and LVR use the same neighbourhood \(N_K(i)\), and the neighbourhood size depends on what is indexed. For pooled (material-level) read-outs \(K = \mathrm{round}(\sqrt{n})\) clamped to \([10,\, n/10]\), with \(n\) the number of materials (\(K = 394\) for the corpus). For per-atom read-outs \(K\) is one more than the 75th percentile of atoms per material in the indexed subsample, and for per-bond read-outs one more than the 75th percentile of bonds per bond type, each clamped to \([10,\, n/10]\) with \(n\) the subsample size. \(K\) is calibrated once on the final layer and reused at every layer; on this corpus the rule gives \(K = 394\) for pooled read-outs and \(K = 10\) for the per-atom subsample of ≈499,000 atoms (ORB) and ≈300,000 atoms (UMA), where the 75th percentile of atoms per material is below the floor. The controls of \textbf{\hyperlink{robustness-controls}{§A.4}} (\textbf{\hyperlink{tab:S15}{Tables~S15}} and \textbf{\hyperlink{tab:S16}{S16}}) and the k-NN predictor of \textbf{\hyperlink{tab:S4}{Table~S4}} use a fixed \(K = 10\).

The input for the two further diagnostics (UMAP and HDBSCAN) is the principal components retaining 90~\% of each representation's variance, so the dimensionality is set per layer and read-out by the representation itself (2 to 156 components across ORB's read-outs, 7 to 137 across UMA's node read-outs (UMA's edge read-outs span 5 to the 256-component cap)). The within-cluster band-gap spread quoted in \hyperlink{property-information-is-locally-predictable-without-property-aligned-clusters}{§3.5} is the median, over HDBSCAN clusters with more than 20 materials, of the standard deviation of band gap inside a cluster, set against the standard deviation over all plotted materials.

\textbf{UMAP} (Uniform Manifold Approximation and Projection) {[}\citelink{cite:McInnes2018}{McInnes 2018}{]} projects those components down to two dimensions (15 neighbours throughout, except 50 for ORB's bond-level read-out; minimum distance 0.1, Euclidean, random initialisation, seed 42). We then colour each point by its value for one property, so a property that varies smoothly across the projection is visibly different from one scattered at random. These plots are read by eye and support no numerical claim.

\textbf{HDBSCAN} (Hierarchical Density-Based Spatial Clustering of Applications with Noise) {[}\citelink{cite:McInnes2017}{McInnes 2017}{]} instead finds density-separated clusters without seeing any labels; the clusters are then scored against each property by \textbf{NMI} (Normalised Mutual Information) and \textbf{ARI} (Adjusted Rand Index) {[}\citelink{cite:Pedregosa2011}{Pedregosa 2011}{]}. The minimum cluster size is the 95th percentile of atoms per material, floored at 20 and capped at one tenth of the points clustered, and the minimum sample count is half that, floored at 10. Per-atom and per-bond clustering uses an element- or bond-type-stratified, square-root-weighted sample targeting 100,000 points per layer; pooled material vectors are clustered in full.

\hypertarget{embedding-health-qc}{%
\subsection{Embedding-health QC}\label{embedding-health-qc}}

Because a collapsed or degenerate layer would depress every probe and organisation metric regardless of content, we characterise each layer's geometry as a quality-control signal. Every measure below is computed separately for each layer and read-out. \textbf{\hyperlink{fig:S9}{Fig.~S9}} and \textbf{\hyperlink{fig:S10}{Fig.~S10}} plot ORB's node and edge read-outs and UMA-S's node read-outs respectively; the rest are computed as QC but not shown.

\textbf{Mean pairwise cosine similarity} detects collapse. It is computed over the \(n\) representation vectors of a layer, where \(h_i\) and \(h_j\) are the vectors of points \(i\) and \(j\) and \(\|\cdot\|\) is the Euclidean norm:

\[\bar{c} = \frac{1}{n(n-1)} \sum_{i \neq j} \frac{h_i \cdot h_j}{\|h_i\| \|h_j\|}\]

\(\bar{c}\) near zero means the vectors point in unrelated directions; values approaching 1 mean the layer has collapsed, with every point carrying almost the same representation.

\textbf{Covariance condition number} detects anisotropy: \(\kappa = \lambda_{\max}/\lambda_{\min}'\), with \(\lambda_{\max}\) the largest eigenvalue of the representation's covariance and \(\lambda_{\min}'\) the smallest one above \(10^{-6}\lambda_{\max}\). Eigenvalues below that floor are rounding error rather than variance in the data, so dividing by one would make \(\kappa\) report the precision of the eigendecomposition instead of the shape of the representation. A large \(\kappa\) means the variance is concentrated in a few directions, which can depress neighbourhood metrics built from Euclidean distances (k-NN purity and 1−LVR, \textbf{\hyperlink{unsupervised-label-organisation-metrics}{§A.5}}, and the k-NN Jaccard below), even if no information was lost. The supervised probes can rescale individual axes and therefore are less likely to be affected.

\textbf{PCA-90} measures effective dimensionality: the number of principal components retaining 90~\% of the variance.

\textbf{Layer-to-layer stability.} Linear \textbf{CKA} (Centred Kernel Alignment) {[}\citelink{cite:Kornblith2019}{Kornblith 2019}{]} and \textbf{SVCCA} (Singular Vector Canonical Correlation Analysis) {[}\citelink{cite:Raghu2017}{Raghu 2017}{]} between consecutive layers show whether the representation evolves smoothly from one layer to the next or is re-coded abruptly (global similarity). CKA is computed on a 10,000-point subsample for the atom- and bond-level read-outs and on every material for the pooled ones. SVCCA first reduces each layer to the singular directions carrying 99~\% of its variance, on a 5,000-point subsample throughout (seed 42). \textbf{k-NN Jaccard} compares neighbourhoods, measuring the overlap between each point's neighbourhood at layer \(\ell\) and its neighbourhood at the next layer \(\ell+1\):

\[J = \frac{1}{n}\sum_i \frac{|N_K^{(\ell)}(i) \cap N_K^{(\ell+1)}(i)|}{|N_K^{(\ell)}(i) \cup N_K^{(\ell+1)}(i)|}\]

where \(N_K^{(\ell)}(i)\) is the set of \(K\) nearest neighbours of point \(i\) computed at layer \(\ell\) and \(|\cdot|\) is the size of a set. The fraction is one point's overlap between the two layers, and \(J\) averages it over all \(n\) points. \(J = 1\) means every point keeps exactly its neighbours from one layer to the next. This is the local geometry that purity and LVR depend on.

\hypertarget{cross-model-convergence}{%
\subsection{Cross-model convergence}\label{cross-model-convergence}}

If two architecturally opposite models converge on similar representations, what they encode reflects the physics and training data rather than either architecture. Both models emit per-atom vectors for the same atoms in the same order, so every ORB layer can be scored against every UMA layer. All measures use the same \(N = 30{,}005\) atoms in the same order, alignment verified by element sequence and fractional coordinates. One UMA read-out is used throughout: the rotation-invariant ang-norm view of its node features (384-d), chosen because, like ORB's 256-d node vector, it combines composition and local geometry without orientation-dependent components.

\textbf{Measures.} CKA and SVCCA compare the two spaces globally. \textbf{Mutual k-NN}, the measure behind the Platonic Representation Hypothesis {[}\citelink{cite:Huh2024}{Huh 2024}{]}, is the local test: the fraction of an atom's \(k = 10\) nearest neighbours shared by both spaces, on a 4,000-atom subsample, where the chance level is \(k/n = 0.0025\). Two further measures ask whether one representation can stand in for the other {[}\citelink{cite:Bansal2021}{Bansal 2021}{]}. \textbf{Linear predictivity} asks how well one model's atom embeddings can be mapped onto the other's. A ridge map is fitted from one model to the other on a 70/30 atom split, with standardised inputs and regularisation strength \(\alpha = 100\), and scored by variance-weighted \(R^2\) on the held-out 30~\%. Two references calibrate that score: a map between consecutive layers within a single model, which is an upper bound on what counts as similar, and a map fitted after shuffling the atoms, which is a lower bound. \textbf{Functional stitching} fits the same kind of ridge map at material level (\(\alpha = 100\)), trains a linear probe on the destination model's own pooled embedding --- ridge with \(\alpha = 10\) for regression, logistic regression (\(C = 1\), 300 iterations) for classification --- then applies that probe to the stitched source embedding; both directions are run, and for each target we report Δ = stitched − native, the change in the probe's score when the destination model's own pooled embedding is replaced by the mapped source embedding (80/20 split, 40,000 materials).

\textbf{Controls.} Both models are given atomic numbers, so agreement that merely reflects the periodic table is not learned convergence. Every measure except stitching is therefore computed on element-residualised embeddings, subtracting from each atom's vector the mean vector of its element. Stitching measures whether a probe still works after substitution, and reports the difference between the stitched and the native score. Element identity is present on both sides of that subtraction, so it cannot inflate the difference, and residualising would lower the stitched and the native score alike. A permutation baseline fixes the floor. Pairing each ORB atom with a randomly chosen UMA atom destroys the real correspondence and drops CKA to \(0.002\ [0.002, 0.002]\) and mutual k-NN to \(0.0024\ [0.0021, 0.0029]\).

\hypertarget{statistics-1}{%
\subsection{Statistics}\label{statistics-1}}

\textbf{Bootstrap} {[}\citelink{cite:Efron1979}{Efron 1979}{]}. Comparisons that need no retraining are resampled instead: UMA against ORB, XGBoost against linear, and node against edge against concat. We draw \(B = 2000\) resamples of the fixed seed-42 test set, each one paired, meaning both configurations are scored on the same materials in every resample. The 2.5 and 97.5 percentiles of the resampled scores, and of the per-material differences, give the 95~\% confidence interval (CI). Because each comparison is run once per property, nine tests are made at a time and some would clear a fixed threshold by chance alone; we therefore apply the \textbf{Benjamini--Hochberg false-discovery-rate correction (BH-FDR)} {[}\citelink{cite:Benjamini1995}{Benjamini \& Hochberg 1995}{]} across the nine properties, which controls the expected share of false positives among the differences called significant. Each difference is also graded by its absolute size --- trivial \textless{}~0.005 \textless{} small \textless{}~0.02 \textless{} moderate \textless{}~0.05 \textless{} large --- so that the comparisons rest on effect size as well as significance: with ≈31,000 test materials, even the smallest differences can come out statistically significant.

\textbf{Nadeau--Bengio corrected repeated-split test} {[}\citelink{cite:Nadeau2003}{Nadeau \& Bengio 2003}{]}: \(K = 5\) retrains (seeds 101--105) for the retraining-dependent layer claims --- best vs worst graph-neural-network (GNN) layer, and best vs raw encoder/embed, both on the node read-out; the selected configurations of \textbf{\hyperlink{tab:1}{Table~1}} are mostly concat read-outs, so these means are not the \textbf{\hyperlink{fig:3}{Fig.~3}} curves. Scores in the control tables are seed-42 point estimates unless an interval is printed. Both claims use the \textbf{node} read-out for every target, so a difference between layers cannot be confounded with a change of read-out.

\textbf{Atom resampling} (convergence peak, \textbf{\hyperlink{cross-model-convergence}{§A.7}}). The highest CKA and the highest mutual k-NN over the layer grid are each estimated from one finite sample of atoms, so both their height and the layer pair at which they occur could be an artefact of which atoms happened to be drawn. We therefore repeat the calculation on 25 independent draws of 15,000 atoms from a 45,002-atom pool, each with its own permutation baseline, on the same ang-norm read-out and the same element-residualised footing as the grids of \textbf{\hyperlink{cross-model-convergence}{§A.7}}. For CKA the whole layer grid is recomputed in every draw, and we report the mean peak height with a 95~\% confidence interval across the draws, the layer pair that came out on top in the largest number of draws, and the fraction of draws in which the real peak did not beat its permuted counterpart, which serves as an empirical p-value. Mutual k-NN is resampled at the layer pair already identified as its peak, so it yields a confidence interval on the height alone.

\hypertarget{random-initialisation-and-matched-training-data-controls}{%
\subsection{Random-initialisation and matched-training-data controls}\label{random-initialisation-and-matched-training-data-controls}}

\textbf{Random initialisation (\hyperlink{property-decodability-depends-on-network-depth}{§3.3}).} Each architecture is re-extracted with every parameter re-initialised (a fixed seed shared by all workers, so every material is embedded by the same random network), then passed through the unchanged probing pipeline. The comparison is at the same layer and read-out on both sides: the pretrained model's best configuration against the random network at that identical configuration, with the random network's own best configuration reported as context. Magpie is the untrained composition reference, and the trivial predictor (the training-set mean for regression, the training-set majority class for classification, scored on the test split) is the floor both are read against.

\textbf{Matched data (\hyperlink{uma-channel-views-and-the-rotation-invariant-crystal-system-gap}{§3.4}).} Two pairs on one stratified 20,000-material subsample of the corpus (stratified by crystal system, seed 42), probed with XGBoost at every available read-out on the paper's seed-42 split restricted to the subsample: MACE-OMAT-0 {[}\citelink{cite:Batatia2024}{Batatia 2024}{]}, an equivariant model trained on OMat24, against ORB-v3-omat (trained on OMat24) --- matched data, different architecture; and ORB-v3-mpa (MPtrj + Alexandria) against ORB-v3-omat --- matched architecture, different data. UMA-S and Magpie are probed on the same materials. MACE-OMAT-0's two interaction blocks are read out as the mean over atoms of all channels, of the scalar channels, and of the scalars with the norms of the vector channels (rotation-invariant). The best configuration per arm and target is reported with a paired bootstrap on the shared test rows for the headline pairs.

\hypertarget{external-transfer}{%
\subsection{Cross-database transfer}\label{external-transfer}}

\textbf{Datasets, labels and overlap.} We extract frozen embeddings for JARVIS-DFT 3D {[}\citelink{cite:Choudhary2020}{Choudhary 2020}{]} and OQMD {[}\citelink{cite:Kirklin2015}{Kirklin 2015}{]} using the same pipeline as for Materials Project (MP). JARVIS uses the OptB88vdW functional {[}\citelink{cite:Klimes2011}{Klimeš 2011}{]} and contains 93,902 structures after deduplication. OQMD uses PBE with different computational settings. We sample 39,892 OQMD entries at regular intervals, restricting structures to at most 40 atoms.

We classify external crystals as \emph{shared} with MP if they cite an MP ID in our dataset or if StructureMatcher identifies an MP counterpart with the same reduced formula and site count. This identifies 63,555 shared crystals in JARVIS and 2,088 in OQMD. The remaining 30,346 JARVIS and 37,804 OQMD entries are classified as \emph{absent from MP}. An audit of 1,000 entries per database finds that 95.9~\% of cited MP IDs match geometrically. Relaxing the site-count restriction identifies additional MP matches among 1.9~\% of absent JARVIS entries and 1.1~\% of absent OQMD entries. The absent groups therefore retain a small amount of structural overlap.

We use each database's own reference labels. JARVIS provides formation energy, band gap, energy above hull, magnetisation and elastic tensors, from which we derive magnetic-moment density, magnetic label and bulk modulus (\textbf{\hyperlink{targets-and-labels}{§A.3}}). OQMD provides formation energy, band gap and hull distance, with negative hull values set to zero; it has no bulk-modulus or magnetic labels. Density and crystal system are calculated from the structures, and metal/insulator labels follow MP's convention.

\textbf{Reference-label agreement.} On crystals shared with MP, we compare the two databases' reference labels directly, scoring MP labels against external labels using R² or macro-F1. No predictive model is involved (\textbf{\hyperlink{fig:S12}{Fig.~S12a}}).

\textbf{Prediction transfer.} ORB-v3 and UMA-S remain frozen throughout. For each model and target, we train an XGBoost predictor using the configuration selected on MP and MP's seed-42 training split. We evaluate this MP-trained predictor without retraining on three sets: MP's original test crystals, external crystals absent from MP, and shared crystals whose MP counterparts belong to the MP test split.

For crystals absent from MP, we also retrain the same XGBoost configuration directly on the external database. We use five 80/20 splits and evaluate both the externally trained and MP-trained predictors on the same external test crystals, against the same external labels. Their score difference therefore measures the cost of transferring the MP-trained predictor rather than retraining it on external data (\textbf{\hyperlink{tab:S19}{Table~S19}}).

\textbf{Separating structure and label differences.} We next examine crystals shared with MP whose MP counterparts lie in its test split, so none were used to train the MP predictor. Each crystal has an MP structure and label and an external structure and label. We evaluate the same MP-trained predictor in four settings:

\begin{itemize}[itemsep=1pt,topsep=2pt]
\tightlist
\item S\textsubscript{0}: MP structure, MP label.
\item S\textsubscript{1}: external structure, MP label.
\item S\textsubscript{2}: external structure, external label.
\item S\textsubscript{3}: MP structure, external label.
\end{itemize}

S\textsubscript{0} and S\textsubscript{2} give the scores under the two databases' respective structures and labels. S\textsubscript{1} changes only the structure, while S\textsubscript{3} changes only the label. Because the effect of either change can depend on which is made first, we average over both orders:

\begin{center}
geometry contribution = \(\tfrac{1}{2}\)[(S\textsubscript{0} − S\textsubscript{1}) + (S\textsubscript{3} − S\textsubscript{2})]\\
label contribution = \(\tfrac{1}{2}\)[(S\textsubscript{0} − S\textsubscript{3}) + (S\textsubscript{1} − S\textsubscript{2})]
\end{center}

The two contributions sum to the total score loss, S\textsubscript{0} − S\textsubscript{2}. Positive values indicate lower performance under the external database's structures or labels. We use a fixed class list for macro-F1.

For density and crystal system, reference labels are calculated from the structures themselves, so their label differences also reflect changes in relaxed geometry. Crystals absent from MP additionally differ in material composition and property distributions. The decomposition on shared crystals does not isolate these further effects on the absent group.

\hypertarget{screening-enrichment-and-label-efficiency}{%
\subsection{Screening enrichment and label efficiency}\label{screening-enrichment-and-label-efficiency}}

\textbf{Screening evaluation (\hyperlink{screening-enrichment-results}{§B.6}).} We evaluate how well the selected XGBoost predictors identify promising materials in the fixed seed-42 test set. ORB-v3 and UMA-S use frozen embeddings, while the Magpie composition-only baseline uses the same training and test split.

For formation energy, density, bulk modulus and magnetic-moment density, we use the DFT reference values to identify the best-performing materials in the held-out test set. We then ask how many of these materials would be found by selecting candidates using only the model's predictions. We compare the lowest predicted formation energies and densities and the highest predicted bulk moduli and magnetic-moment densities, selecting the best 1, 2, 5, 10, 20 or 50~\% of materials. Recall is the percentage of reference-best materials recovered in each predicted shortlist. For example, when selecting 1~\% of the test set, random selection would recover only 1~\% of the reference-best group on average. Because the predicted and reference shortlists contain the same number of materials, precision equals recall.

Energy above hull is evaluated differently because approximately 22~\% of test materials lie exactly on the hull. We rank materials by predicted hull energy and report precision: the fraction of selected candidates whose reference hull energy is at most 0, 50 or 100~meV/atom. The prevalence of each threshold in the full test set gives the expected precision under random selection. For band gap, we select materials with predicted gaps between 0.5 and 3.0~eV. Precision measures the fraction of selected materials whose reference gaps fall within this range, while recall measures the fraction of all materials in the true range that were selected.

For categorical targets, we evaluate one-vs-rest area under the receiver operating characteristic curve (AUROC). For each class, AUROC measures how well its predicted probability ranks materials belonging to that class above non-members, across all classification thresholds. An AUROC of 0.5 indicates random ranking and 1.0 perfect separation.

To compare screening performance with random selection and randomly initialised models (\textbf{\hyperlink{fig:S13}{Fig.~S13j}}), we also calculate lift. For the four continuous targets ranked by their best values, lift is recall divided by 0.05 at a 5~\% shortlist. For energy above hull, it is on-hull precision at a 5~\% shortlist divided by the fraction of materials on the hull. For band gap, it is the precision of the predicted 0.5--3.0~eV window divided by the fraction of materials in the true window. For categorical targets, we rank materials by each class's predicted probability, select the top 5~\% for that class, and average precision divided by class prevalence across classes. We include randomly initialised ORB-v3 and UMA-S, each evaluated at its own best configuration (\hyperlink{random-initialisation-and-matched-training-data-controls}{§A.9}).

We estimate 95~\% confidence intervals from 2,000 paired bootstrap resamples where reported. We also repeat the screening analysis using validation-selected configurations when they differ from the original selections.

\textbf{Learning curves (\hyperlink{label-efficiency-results}{§B.7}).} We evaluate how prediction accuracy changes with the number of labelled training materials. For each target, we reserve a fixed test set of 20,000 materials using a seeded random permutation, or half the available labelled materials for bulk modulus. From the remaining training pool, we sample sets of 50, 100, 200, 500, 1,000, 2,000, 5,000, 7,000, 10,000 and 20,000 materials, where sufficient data are available. We repeat sampling 30 times for sizes up to 2,000, 10 times at 5,000, seven times at 7,000 and 10,000, and six times at 20,000. Every sample contains the same materials across the three feature sets: ORB-v3 backbone node embeddings, UMA-S angular-norm backbone node embeddings and Magpie descriptors. This allows paired comparisons using the same XGBoost probe.

For each training size, we report the median test score, its estimated 10th--90th percentile spread across draws, and the difference from the corresponding full-corpus score reported in the main paper. The spread is estimated as 2.5631 times the standard deviation across draws, corresponding to the 10th--90th percentile interval width under a normal approximation. We consider a learning curve stable once this spread remains at most 0.05 for all larger training sizes, and saturated once its score remains within 0.05 of the full-corpus result.

\textbf{From-scratch network.} To compare frozen embeddings with a model trained directly on the target properties, we train a small message-passing neural network from raw crystal structures. It has three message-passing layers with 64-dimensional hidden features, element embeddings, radius-based neighbour graphs constructed with pymatgen, and mean pooling. The network contains approximately 77,500 parameters (77,900 for the seven-class crystal-system classifier). Training uses Adam and early stopping. For each run, we sample a specified number of labelled materials from the paper's seed-42 training split, reserving one fifth of the sample, capped at 500 materials, for validation. The remaining materials are used to fit the network.

We evaluate training sizes of 200, 500, 1,000, 2,000 and 5,000 materials with three independent samples per size, and 10,000 and 20,000 materials with one sample each. We also train three times at each target's saturation size, defined as the smallest training size at which both frozen-embedding probes have saturated (\textbf{\hyperlink{tab:S22}{Table~S22}}). Every network is evaluated on the same fixed seed-42 test split (30,976 materials, or 2,571 for bulk modulus). Sizes exceeding the available training pool are omitted, so bulk modulus is evaluated up to 10,000 materials.

\hypertarget{data-availability-and-licences}{%
\subsection{Data availability and licences}\label{data-availability-and-licences}}

\textbf{Released data.} The processed dataset is released on Hugging Face as two repositories, split by licence: \href{https://huggingface.co/datasets/NatalijaStepurko/orb-mp-embeddings}{huggingface.co/datasets/NatalijaStepurko/orb-mp-embeddings} holds everything derived from Materials Project and ORB-v3 under CC-BY-4.0 and Apache-2.0, and \href{https://huggingface.co/datasets/NatalijaStepurko/uma-mp-embeddings-gated}{huggingface.co/datasets/NatalijaStepurko/uma-mp-embeddings-gated} the UMA-S embeddings under Meta's FAIR Chemistry License. Together they contain:
\begin{itemize}[itemsep=1pt,topsep=2pt]
\tightlist
\item the Materials Project corpus as used: metadata and the nine targets with their seed-42 splits for 154,875 materials, and their structures from the database release stated below;
\item both models' mean-pooled node and edge embeddings at every layer and read-out, UMA-S in all six channel views;
\item the inputs behind the analyses that cannot be rerun from pooled embeddings alone: the atom-aligned subsample of the cross-model comparison, the unrotated and rotated activations of the rotation control, and the Magpie feature matrix;
\item the external databases as used in the transfer test: every JARVIS-DFT 3D entry with its labels as derived, its relation to Materials Project and the matched MP id (the structures themselves come from the versioned jarvis-tools release named below), and the OQMD stride sample with its structures, since that API offers no snapshot;
\item the two result tables that the released code checks a rerun against: every probe score, and the bootstrap metric table of point estimates with their confidence intervals.
\end{itemize}
The other result tables are not deposited; the released code regenerates them from the deposited embeddings, except for the experiments that re-embed structures (the random-weight and matched-model controls, the rotation tests and the cross-database transfer), which it reruns from the deposited structures and the model weights. The full per-atom embeddings and raw per-edge tensors are excluded for size (≈~1 TB) and available on request.

\textbf{Code.} The full code base to reproduce the data analysis pipeline is released under the MIT licence at \href{https://github.com/Natalija-Stepurko/mlip-embedding-probes}{github.com/Natalija-Stepurko/mlip-embedding-probes}.

\textbf{Models.} ORB-v3 {[}\citelink{cite:Neumann2024}{Neumann 2024}; \citelink{cite:Rhodes2025}{Rhodes 2025}{]} is used as released under the Apache-2.0 licence (\href{https://github.com/orbital-materials/orb-models}{github.com/orbital-materials/orb-models}); UMA-S (FairChem) {[}\citelink{cite:Wood2025}{Wood 2025}{]} under Meta's FAIR Chemistry License (\href{https://huggingface.co/facebook/UMA}{huggingface.co/facebook/UMA}). The exact checkpoints are ORB-v3 orb\_v3\_conservative\_inf\_omat and UMA-S uma-s-1p2 loaded with task\_name=``omat''. Their training corpora differ, which is the confound noted in \textbf{\hyperlink{limitations}{§4}}: ORB's omat variant is trained on OMat24, whereas UMA-S is trained jointly across OMat24, OC20, ODAC23, OMC25 and OMol25, spanning catalysts, molecular crystals and molecules as well as bulk materials. The controls of \textbf{\hyperlink{random-initialisation-and-matched-training-data-controls}{§A.9}} use ORB-v3 orb\_v3\_conservative\_inf\_mpa (Apache-2.0; \href{https://github.com/orbital-materials/orb-models}{github.com/orbital-materials/orb-models}) and MACE-OMAT-0 medium-omat-0 {[}\citelink{cite:Batatia2024}{Batatia 2024}{]} (\href{https://github.com/ACEsuit/mace-mp/releases/tag/mace_omat_0}{github.com/ACEsuit/mace-mp}, release mace\_omat\_0) through mace-torch 0.3.16, under the Academic Software License.

\textbf{Analysed data.} All structures and material properties are from the Materials Project {[}\citelink{cite:Jain2013}{Jain 2013}; \citelink{cite:Horton2025}{Horton 2025}{]} under CC-BY 4.0 (database release 2026.04.13, queried through the summary endpoint on 26 May 2026 with the non-deprecated filter). Four mvc-prefixed entries returned by the query are absent from the public dataset release and are excluded throughout, giving the 154,875-material corpus. The transfer test (\textbf{\hyperlink{external-transfer}{§A.10}}) uses JARVIS-DFT 3D {[}\citelink{cite:Choudhary2020}{Choudhary 2020}{]} (\href{https://jarvis.nist.gov}{jarvis.nist.gov}; the dft\_3d release of 2025-09-24 through jarvis-tools 2026.6.12, accessed 2026-09-28), a NIST work not subject to copyright in the United States, and OQMD {[}\citelink{cite:Kirklin2015}{Kirklin 2015}{]} (\href{https://oqmd.org}{oqmd.org}, accessed through its REST API on 2026-09-29) under CC-BY 4.0.

\textbf{Software.} XGBoost 3.3.0 {[}\citelink{cite:Chen2016}{Chen \& Guestrin 2016}{]} (Apache-2.0), scikit-learn 1.8.0 {[}\citelink{cite:Pedregosa2011}{Pedregosa 2011}{]} (BSD-3-Clause), NumPy 2.4.4 {[}\citelink{cite:Harris2020}{Harris 2020}{]} (BSD-3-Clause), SciPy 1.17.1 {[}\citelink{cite:Virtanen2020}{Virtanen 2020}{]} (BSD-3-Clause), pymatgen 2026.3.23 {[}\citelink{cite:Ong2013}{Ong 2013}{]} (MIT), matminer 0.10.1 {[}\citelink{cite:Ward2018}{Ward 2018}{]} (BSD-3-Clause), ASE 3.28.0 {[}\citelink{cite:Larsen2017}{Larsen 2017}{]} (LGPL-2.1), PyTorch 2.8.0 (BSD-3-Clause), orb-models 0.6.2 (Apache-2.0), fairchem-core 2.19.0 (MIT), mace-torch 0.3.16 (MIT), umap-learn 0.5.12 {[}\citelink{cite:McInnes2018}{McInnes 2018}{]} (BSD-3-Clause), hdbscan 0.8.44 {[}\citelink{cite:McInnes2017}{McInnes 2017}{]} (BSD-3-Clause) and jarvis-tools 2026.6.12 (NIST licence); Python 3.12. The exact versions of every package are pinned in the lock file of the released code.

\clearpage
\hypertarget{compute-resources}{%
\subsection{Compute resources}\label{compute-resources}}

Wall-clock cost for every stage of the pipeline is given in \textbf{\hyperlink{tab:S3}{Table~S3}}.

\par\addvspace{0.8\baselineskip}
\hypertarget{tab:S3}{}%
{\small \textbf{Table~S3 --- Compute resources.} \emph{All computation ran on a single CPU-only machine (8-core Intel Xeon, 165 GB RAM). Approximate wall-clock of a full run, grouped by what the time is spent on, with the measured basis of each figure; subtotals per group and the total. Embedding extraction (a) and the random-weight repeat of it dominate; because the per-layer activation caches are reused by every downstream analysis, the results of (b)--(e) reproduce from the caches without re-running extraction.}\par}\par\nobreak
\begin{longtable}{@{}
  >{\raggedright\arraybackslash}p{(\columnwidth - 4\tabcolsep) * \real{0.3833}}
  >{\raggedright\arraybackslash}p{(\columnwidth - 4\tabcolsep) * \real{0.1333}}
  >{\raggedright\arraybackslash}p{(\columnwidth - 4\tabcolsep) * \real{0.4833}}@{}}
\toprule\noalign{}
\begin{minipage}[c]{\linewidth}\raggedright
Stage
\end{minipage} & \begin{minipage}[c]{\linewidth}\raggedright
Wall-clock
\end{minipage} & \begin{minipage}[c]{\linewidth}\raggedright
Basis
\end{minipage} \\
\midrule\noalign{}
\multicolumn{3}{@{}l}{\textbf{(a) Embedding extraction}} \\
ORB-v3, 154,875 structures & ≈~7~h & measured (0.16 s/structure) \\
UMA-S, 154,875 structures & ≈~13~h & measured (0.30 s/structure) \\
Random-weight networks, both models & ≈~16.5~h & measured (4.7 + 11.8~h) \\
Matched models on the 20,000-material subsample (ORB-v3 mpa, MACE-OMAT-0) & ≈~1.3~h & measured (0.8 + 0.5~h) \\
External databases, both models & ≈~13~h & JARVIS-DFT 3D estimated from the per-structure rates ((0.16 + 0.30) s × 93,902); OQMD measured (1.4~h) \\
\textbf{Subtotal} & \textbf{≈~51~h} & \\
\addlinespace[3pt]
\multicolumn{3}{@{}l}{\textbf{(b) Analyses on the cached embeddings}} \\
Unsupervised diagnostics and embedding-health QC, both models & ≈~4--6~h & measured (cold UMAP ≈~16 min per layer × read-out, ≈30 passes over parallel workers) \\
Property-prediction probes, every layer × read-out × channel configuration & ≈~3~h & measured (2 h 45 min) \\
Bootstrap significance and Nadeau--Bengio repeated splits & ≈~2~h & measured (61 + 65 min) \\
Cross-model convergence & ≈~0.2~h & measured (4.7 + 7.7 min) \\
\textbf{Subtotal} & \textbf{≈~10~h} & \\
\addlinespace[3pt]
\multicolumn{3}{@{}l}{\textbf{(c) Controls}} \\
Validation-split re-selection & ≈~5~h & measured (4 h 49 min) \\
Random-weight networks: pooled caches and probes, both models & ≈~31~h & measured (caches 14.5~h; probes 2.5 + 14.3~h) \\
Matched-model probes on the subsample & ≈~2~h & measured (1.9~h) \\
Probe family (linear, XGBoost, k-NN, MLP at the selected configurations) & ≈~2~h & measured (1.9~h) \\
Feature-level rotation test & ≈~0.5~h & measured (33 min) \\
Rotation control, random-weight UMA-S & ≈~1~h & measured (55 min) \\
\textbf{Subtotal} & \textbf{≈~42~h} & \\
\addlinespace[3pt]
\multicolumn{3}{@{}l}{\textbf{(d) Cross-database transfer}} \\
Probes on the crystals absent from MP and the five within-database splits, both databases & ≈~3.5~h & measured (1.9 + 1.6~h) \\
Label and geometry parts on shared crystals; matching audit & ≈~0.7~h & measured (0.6 + 0.1~h) \\
\textbf{Subtotal} & \textbf{≈~4~h} & \\
\addlinespace[3pt]
\multicolumn{3}{@{}l}{\textbf{(e) Screening and label efficiency}} \\
Screening enrichment (both models and Magpie, with the random-weight networks) & ≈~0.5~h & measured (20 + 20 min) \\
Label-efficiency curves (ORB-v3, UMA-S, Magpie) & ≈~4.5~h & measured (1.3 + 2.3 + 0.7~h) \\
From-scratch network, learning curve and saturation point & ≈~5.5~h & measured (5.1 + 0.3~h) \\
\textbf{Subtotal} & \textbf{≈~10.5~h} & \\
\addlinespace[2pt]
\textbf{Total} & \textbf{≈~118~h} & ≈~4.9 days on the one machine \\
\bottomrule
\end{longtable}
\par\addvspace{0.8\baselineskip}

\clearpage
\hypertarget{supplementary-results}{%
\section{Supplementary Results}\label{supplementary-results}}

\hypertarget{representation-health-tracks-architectural-function}{%
\subsection{Representation health tracks architectural function}\label{representation-health-tracks-architectural-function}}

All layers in both models are geometrically healthy --- no collapse, no degenerate directions --- so the depth trends in probe scores (\textbf{\hyperlink{property-decodability-depends-on-network-depth}{§3.3}}) reflect genuine information content rather than representation pathology (metric definitions in \textbf{\hyperlink{embedding-health-qc}{§A.6}}; per-layer trajectories in \textbf{\hyperlink{fig:S9}{Fig.~S9}} and \textbf{\hyperlink{fig:S10}{Fig.~S10}}).

The mean pairwise cosine similarity of ORB's node embeddings rises from 0.165 at gnn\_2 to 0.392 at gnn\_3, then falls to 0.061 at gnn\_4. On its own that spike would suggest collapse at gnn\_3. However, the layer keeps its full effective dimensionality (PCA-90 = 120), which is measured from the mean-subtracted covariance and so would fall if diversity had been lost, and k-NN Jaccard (0.992), adjacent-layer CKA (0.999) and probe scores are all unchanged, therefore indicating that its vectors share a larger common offset rather than having lost diversity. Cosine similarity is measured without centring, so a shared offset raises it even when the spread around that offset is untouched. Across gnn\_2, gnn\_3 and gnn\_4 the mean vector's length is 3.2, 6.0 and 1.9, while the mean spread about it stays at 7.2, 7.5 and 7.5.

Beyond QC, the same metrics localise where each architecture moves from element identity to structural context. ORB's effective dimensionality rises sharply with depth. PCA-90 at the encoder is 9, where element identity is already stored almost losslessly (k-NN purity 1.000). By gnn\_1 it reaches 107, the step where one- and two-hop context arrives and electronic-property scores gain most. That one-hop transition rewrites which atoms are neighbours (Jaccard 0.002) while preserving the principal subspace (CKA 0.82), dissolving the element-cluster geometry into environment-specific representations (\textbf{\hyperlink{fig:S5}{Fig.~S5a}}). UMA reorganises differently. At blk\_0 it compresses its scalar (L=0) channel while populating the angular channels from near-zero. On the flat read-out, which keeps all nine slots (\textbf{\hyperlink{read-outs}{§A.1}}), SVCCA between embed and blk\_0 is 0.64, the lowest of its five adjacent-layer values, so this is where the principal axes change most: from element-type to geometry-aware encoding. A second angular reorganisation follows at blk\_2--blk\_3, where the backbone distils the geometry.

In ORB the deepest layers are the healthiest by every criterion: PCA-90 at its maximum of 120, and the lowest condition number \(\kappa\) (\(5.8 \times 10^3\)) and cosine similarity (0.061) in the network. UMA's deepest layers are healthy too, its flat read-out reaching PCA-90 = 137 with the lowest \(\kappa\), \(2.0 \times 10^4\). Its cosine similarity nonetheless rises with depth, on the ang-norm read-out from 0.293 at embed to 0.689 at the backbone. That rise is the same offset effect, not collapse: the cloud of atom vectors drifts away from the origin until its centre is 1.47 times further out than a typical point is from that centre, up from 0.66 at embed, while PCA-90 rises from 36 to 59. In both models the deepest layers are exactly where local and geometric properties decline and electronic ones peak, so that decline is a genuine redistribution of information, not a pathology. Full per-layer PCA-90, CKA/SVCCA and Jaccard trajectories are in \textbf{\hyperlink{fig:S9}{Fig.~S9}} and \textbf{\hyperlink{fig:S10}{Fig.~S10}}.

The per-channel PCA-90 profiles set UMA's angular channels apart from its scalar one, the same division of labour reported in \textbf{\hyperlink{uma-channel-views-and-the-rotation-invariant-crystal-system-gap}{§3.4}}. The angular channels expand substantially through the first half of the network --- L=1 grows from PCA-90 = 11 at embed to 70 at blk\_2; L=2 from 19 to 66; geometric from 25 to 126 --- before all three contract at blk\_3 (L=1: 33, L=2: 35, geometric: 58). This inverted-U profile indicates that mid-network is when angular information is richest and most diverse, and the backbone distils it. The channels that expand most --- geometric and L=2 --- are those that carry the most crystal-system information, consistent with the F1 ranking. By contrast, the L=0 scalar channel remains at PCA-90 9--13 throughout the main network after the initial compression at blk\_0, reflecting that element-type and composition information are captured in a small number of invariant directions that do not require expansion. That reorganisation shows in the adjacent-layer measures: between blk\_2 and blk\_3 the L=1 channel gives CKA 0.780 and SVCCA 0.496, its lowest CKA of any step. It coincides with the PCA contraction, consistent with high-dimensional geometric context being distilled into a more task-relevant basis.

\hypertarget{decodability-clustering-and-nearest-neighbour-retrieval}{%
\subsection{Decodability, clustering and nearest-neighbour retrieval}\label{decodability-clustering-and-nearest-neighbour-retrieval}}

We compared each target's XGBoost probe score with its unsupervised organisation---how strongly neighbouring embeddings share the same target value---using k-NN purity for categorical properties and \(1-\mathrm{LVR}\) for continuous ones, with higher values indicating stronger organisation in both cases. The two measures order the targets similarly, but they are different quantities on different scales and are not compared numerically (\textbf{\hyperlink{unsupervised-label-organisation-metrics-main}{§2.3}}; \textbf{\hyperlink{tab:S4}{Table~S4}}, \textbf{\hyperlink{fig:S4}{Fig.~S4}}). Crystal system shows the sharpest contrast: per-atom node k-NN purity peaks at only ≈0.45 in both models on the chance-corrected scale, less than halfway from chance (0) to every neighbour sharing the label (1), whereas UMA's XGBoost probe reaches macro-F1 0.82 on the rotation-invariant ang-norm channel (\textbf{\hyperlink{uma-channel-views-and-the-rotation-invariant-crystal-system-gap}{§3.4}}). This weak organisation is not explained by same-material neighbours artificially boosting purity: because such neighbours necessarily share the target label, they bias purity upward. Excluding them however lowers crystal-system purity further, to 0.39 (ORB) and 0.42 (UMA) (\textbf{\hyperlink{robustness-controls}{§A.4}}, \textbf{\hyperlink{tab:S15}{Table~S15}}). Nor is the result caused by high-variance embedding dimensions dominating the distance calculation: after PCA whitening (Mahalanobis distance), crystal-system purity rises only to 0.46 / 0.50 (ORB / UMA), still well below its decodability, while metal/insulator purity decreases in both models (\textbf{\hyperlink{robustness-controls}{§A.4}}, \textbf{\hyperlink{tab:S16}{Table~S16}}).

HDBSCAN shows the same lack of global property organisation. Unlike k-NN, which probes local neighbourhoods, HDBSCAN searches for density-separated regions. Across representations it finds hundreds of clusters with 60--80~\% noise, yet cluster--label agreement stays low (ARI below 0.05 for most targets and layers, at most 0.13; \textbf{\hyperlink{fig:S7}{Figs~S7b}} and \textbf{\hyperlink{fig:S8}{S8b}}). Element identity is the exception (ARI 0.99 for ORB and 0.78 for UMA before any message passing; \textbf{\hyperlink{fig:S5}{Fig.~S5c}}; the element-coloured atom-level UMAP of \textbf{\hyperlink{fig:2}{Fig.~2e}} shows the same structure), showing that the recovered clusters reflect chemical identity rather than emergent material properties (\textbf{\hyperlink{fig:S5}{Fig.~S5a--b}}; per-layer ARI and NMI in \textbf{\hyperlink{fig:S7}{Figs~S7}} and \textbf{\hyperlink{fig:S8}{S8}}).

Despite the lack of property-aligned clusters, a supervised k-NN predictor recovers most of XGBoost's performance: macro-F1 is within 0.035 on five of six classification cases, and R² within 0.06 on every regression target (\textbf{\hyperlink{tab:S4}{Table~S4}}). XGBoost fits 300 trees and can reweight, combine and discard dimensions; k-NN predictor fits no parameters and assumes only that nearby materials have similar target values.

Their close agreement shows that much of the predictive signal is already expressed through local neighbourhood structure. XGBoost nevertheless achieves lower error in every case except ORB crystal system, cutting RMSE by 10--32~\%, with the smallest gain for magnetic moment density (\textbf{\hyperlink{tab:S4}{Table~S4}}). For UMA crystal system, XGBoost reaches 0.817 versus 0.583 for k-NN predictor, consistent with the additional signal carried by UMA's equivariant angular channels (\textbf{\hyperlink{uma-channel-views-and-the-rotation-invariant-crystal-system-gap}{§3.4}}). This gap is not due to the choice of distance metric: cosine and Euclidean retrieval agree to within 0.007 in every case but UMA bulk modulus (0.031). Nor does it simply reflect XGBoost's flexibility, since the same probe trained on permuted labels scores at chance (\textbf{\hyperlink{property-decodability-depends-on-network-depth}{§3.3}}).

The contrast between k-NN predictor and HDBSCAN follows from what each requires of the geometry. Clustering needs separable regions with clear boundaries, whereas nearest-neighbour prediction requires only local smoothness: nearby points must have similar target values. A representation can therefore vary smoothly with a property without forming distinct property clusters. The UMAP projections show this directly, with smooth band-gap gradients across manifolds whose HDBSCAN clusters barely agree with the property labels (\textbf{\hyperlink{fig:S7}{Fig.~S7a}}, \textbf{\hyperlink{fig:S8}{Fig.~S8a}}).

\hypertarget{orb-and-uma-transfer-function-across-layers-but-converge-geometrically-only-in-part}{%
\subsection{ORB and UMA transfer function across layers, but converge geometrically only in part}\label{orb-and-uma-transfer-function-across-layers-but-converge-geometrically-only-in-part}}

We compared the two models layer by layer on 30,005 atom-aligned embeddings, using CKA, SVCCA, mutual k-NN and linear predictivity, and tested functional interchangeability by stitching in both directions (\textbf{\hyperlink{cross-model-convergence}{§A.7}}, \textbf{\hyperlink{tab:S8}{Table~S8}}, \textbf{\hyperlink{fig:S11}{Fig.~S11}}). All measures use UMA's rotation-invariant ang-norm read-out. All are reported after removing element identity from both models, since both are given atomic numbers and agreement on those alone would not be convergence. That correction matters most near the input. Between ORB's encoder and UMA's embed, CKA is 0.862 on the uncorrected embeddings and 0.004 after residualisation. Subtracting each atom's element-mean vector leaves 0.0~\% of ORB's encoder variance and 7.7~\% of UMA's embed variance, against 63.1~\% (ORB) and 18.8~\% (UMA) at depth. The uncorrected values are quoted here for that comparison only; every table and figure reports the residualised ones. The apparent agreement between ORB's encoder and UMA's embed is therefore the shared atomic number, not a shared representation.

At depth the two models agree on more than their shared input: between ORB's gnn\_4 and UMA's backbone, CKA is 0.617 uncorrected and 0.358 residualised, and mutual k-NN 0.333 and 0.248 respectively. Agreement is highest in the middle of both networks, on the residualised measures throughout: CKA peaks at 0.738 for ORB gnn\_0 paired with UMA block\_2, and mutual k-NN at 0.381 for ORB gnn\_1 paired with UMA block\_1, against a chance level of 0.0025. Both peaks are stable under resampling (\textbf{\hyperlink{statistics-1}{§A.8}}): across 25 draws of 15,000 atoms they are \(0.737\ [0.733, 0.742]\) and \(0.377\ [0.370, 0.386]\), with the CKA peak at the same layer pair every time and the permuted baseline below the real value in all 25 draws (empirical p \textless{}~0.04 for both measures), reproducing the full-set values from an independent code path.

The same mid-network peak shows in linear predictivity, which fits a ridge map from one model's atom embeddings to the other's and scores how much of the target it recovers, with element identity residualised out as for the other measures (\textbf{\hyperlink{cross-model-convergence}{§A.7}}). It is a looser test than the geometric measures: any invertible linear transformation of either space leaves it unchanged. Fitted for every layer pair from ORB's gnn\_0 onward, predictivity runs from 0.41 to 0.61 for ORB predicting UMA and from 0.30 to 0.76 for UMA predicting ORB. The two directions peak in different places: UMA predicts ORB best for ORB gnn\_0 paired with UMA blk\_0, recovering 0.764 there against 0.515 the other way, while ORB predicts UMA best for ORB gnn\_1 paired with UMA embed, recovering 0.609 against 0.378. That second peak falls on UMA's layer before any message passing, but it is not the shared input: element identity has been residualised out of both models, so what the map recovers is the variation UMA's embedding block carries beyond element type. Predictivity then falls with depth, as CKA and mutual k-NN do: between ORB's gnn\_4 and UMA's backbone it is 0.503 predicting UMA from ORB and 0.425 the other way, below every consecutive-layer step inside UMA (0.637--0.986) and, in the UMA-to-ORB direction, below every step inside ORB (0.466--0.999, except the encoder-to-gnn\_0 step; \textbf{\hyperlink{tab:S8}{Table~S8}}). CKA, mutual k-NN and linear predictivity therefore agree that cross-model similarity is layer-dependent and greatest in the first half of both networks.

Functional stitching transfers well both ways, scored as Δ = stitched − native, the change in the probe's score when the destination model's own embedding is replaced by the mapped source, in R² for regression and macro-F1 for classification (\textbf{\hyperlink{cross-model-convergence}{§A.7}}). A probe trained on ORB still works on UMA mapped into ORB's space for all nine targets (\textbar Δ\textbar{} ≤~0.05); the reverse holds for eight of nine within 0.06, with crystal system the only larger loss at 0.091. Part of that asymmetry is geometric. UMA's pooled read-out has 384 dimensions and ORB's has 256, so mapping UMA into ORB has more inputs than outputs and can reach the whole of ORB's space, whereas mapping ORB's 256 into UMA's 384 cannot. Unlike the geometric agreement, stitching does not depend on depth: fitted for every layer pair, from ORB's first message-passing layer onward the median Δ over targets is −0.04 (ORB→UMA) and −0.01 (UMA→ORB), with per-layer-pair medians between −0.06 and 0.00. The exception is ORB's encoder, which is too low-dimensional to express UMA's embedding (\textbf{\hyperlink{tab:S8}{Table~S8b}}). A probe trained on one model can therefore be reused on the other by fitting a ridge map, with no retraining and no labels beyond those already used.

The Platonic Representation Hypothesis holds that independently trained models converge on similar representations of the same underlying data {[}\citelink{cite:Huh2024}{Huh 2024}{]}, and Li \& Walsh report such convergence across seven MLIPs, whose latent spaces can be aligned into a common space {[}\citelink{cite:Li2026}{Li \& Walsh 2026}{]}. Our deep-layer results agree, but only on the two measures that fit a linear map from one model's embeddings to the other's, linear predictivity and functional stitching. The measure that asks whether the two models place the same atoms near each other does not: mutual k-NN peaks at 0.381, so most of an atom's neighbours differ between the models. Convergence here is therefore about what can be linearly decoded, not about the geometry (\textbf{\hyperlink{property-information-is-locally-predictable-without-property-aligned-clusters}{§3.5}}).

\hypertarget{external-transfer-results}{%
\subsection{Cross-database transfer}\label{external-transfer-results}}

A probe trained on Materials Project (MP) learns to predict MP's reference values. When applied to JARVIS-DFT 3D or OQMD, it encounters both different crystals and different DFT labels. We examine these two sources of variation separately, on crystals shared with MP and on crystals absent from it (\textbf{\hyperlink{external-transfer}{§A.10}}, \textbf{\hyperlink{tab:S19}{Table~S19}}, \textbf{\hyperlink{fig:S12}{Fig.~S12}}).

\textbf{Agreement between databases.} We first compare the reference labels of crystals shared with MP: 63,555 in JARVIS and 2,088 in OQMD (\textbf{\hyperlink{fig:S12}{Fig.~S12a}}). Density and formation energy agree closely, with R² values of 0.95--0.99 when MP labels are scored against external labels. Bulk modulus reaches R² = 0.92, while categorical targets reach macro-F1 scores of 0.84--0.95. Band gaps show strong rank agreement (Spearman ρ = 0.89--0.91), but differ by 0.5--0.6~eV in root-mean-square terms (R² = 0.86--0.87). Agreement is much weaker for magnetic-moment density (R² = 0.26). Many crystals labelled as magnetic in MP have zero magnetisation in JARVIS. Energy above hull also differs substantially between databases. OQMD agrees well with MP (R² = 0.92), whereas JARVIS does not (R² = −1.6). On the same crystals, median hull energy is 1.78~eV/atom in JARVIS versus 0.015~eV/atom in MP.

\textbf{Transfer to crystals absent from MP.} We next evaluate MP-trained predictors on external crystals with no identified counterpart in MP. For each target, we compare their scores with the same XGBoost configuration retrained directly on the external database, using identical test crystals and external reference labels (\textbf{\hyperlink{fig:S12}{Fig.~S12b}}, \textbf{\hyperlink{tab:S19}{Table~S19}}).

The transfer gap is small for density, formation energy and metal/insulator classification (0.02--0.07 in R² or macro-F1), and for bulk modulus (0.03). Band-gap scores decline by 0.08--0.10 on OQMD and 0.16--0.17 on JARVIS. Larger gaps occur for magnetic label (0.15), crystal system (0.26--0.41) and magnetic-moment density (0.34--0.36). Energy above hull transfers particularly poorly, reaching R² = −1.1 on JARVIS and 0.01 on OQMD. However, even predictors trained directly on OQMD obtain low hull-energy scores: R² = 0.30 ± 0.12 for ORB and 0.16 ± 0.15 for UMA.

\textbf{Separating label and geometry differences on shared crystals.} We next examine crystals present in both databases whose MP counterparts were excluded from probe training. Using MP structures and labels, the predictors score within 0.02 of their full MP test scores in 17 of 30 model--database--target comparisons. The largest differences are −0.10 for OQMD energy above hull and +0.05 for OQMD metal/insulator classification. These shared-crystal scores provide a reference for examining changes in structures and labels.

For each shared crystal, we evaluate the MP-trained predictor using four combinations of MP and external structures and labels (\textbf{\hyperlink{tab:S19}{Table~S19}}). Averaging over both orders of replacement separates the total score loss into label and geometry contributions (\textbf{\hyperlink{external-transfer}{§A.10}}, \textbf{\hyperlink{fig:S12}{Fig.~S12c,d}}). The contributions below are averaged across ORB and UMA.

The largest effect occurs for JARVIS energy above hull. Of the total R² loss of 2.53, reference-label differences contribute 2.32 and geometry differences contribute 0.20. The effect of changing geometry depends on which labels are used: replacing the MP structure with the JARVIS structure reduces R² by 0.38--0.40 against MP labels, but changes it by only about 0.01 against JARVIS labels. This illustrates why the two contributions are averaged over both orders.

Reference-label differences also contribute approximately 0.30 to the JARVIS magnetic-moment-density loss and 0.05--0.06 to band-gap losses in both databases. For the remaining targets, label contributions are at most 0.03. Geometry contributions are generally within ±0.02, except for JARVIS energy above hull (0.20), JARVIS magnetic-moment density (−0.04) and OQMD crystal system (0.05).

\textbf{Interpreting the transfer losses.} The shared-crystal analysis shows that differences in reference labels can substantially change prediction scores, particularly for JARVIS energy above hull. This need not indicate a failure of the learned representations: the same crystal may have very different reference values across databases. For example, median hull energy on shared JARVIS crystals is 1.78~eV/atom, compared with 0.015~eV/atom in MP.

Performance on crystals absent from MP can decline further. For OQMD energy above hull, R² is approximately 0.86 on shared crystals using OQMD structures and labels, but only 0.01 on absent crystals. The latter also have a different hull-energy distribution, with a median of 0.27~eV/atom compared with 0.05~eV/atom in MP. Further declines occur for crystal system on JARVIS (0.22), magnetic-moment density (0.17), and band gap and magnetic label (0.13--0.16).

Overall, transfer performance depends strongly on the target and database. The shared-crystal controls identify how differences in reference labels and relaxed structures affect scores for matched materials. They do not, however, explain the full transfer gap on crystals absent from MP, where compositions and property distributions also change. We therefore cannot attribute that additional loss to unfamiliar structures alone.

\hypertarget{probe-family}{%
\subsection{Probe family}\label{probe-family}}

At each target's selected configuration the MLP stays within 0.035 of XGBoost in 17 of 18 model--target cases; the exception is ORB crystal system, where it leads by 0.075 (\textbf{\hyperlink{fig:S2}{Fig.~S2}}, \textbf{\hyperlink{tab:S4}{Table~S4}}). Small score differences near R² = 1 hide large error differences: the MLP cuts XGBoost's RMSE by 28--71~\% for formation energy, energy above hull and density, and by 3--17~\% for band gap, bulk modulus and magnetic-moment density. The linear probe trails XGBoost by up to 0.18 on crystal system and band gap, and by 0.40 and 0.11 on UMA magnetic-moment density and density, where its fit at the embedding layer is unstable (\textbf{\hyperlink{fig:3}{Fig.~3e}}); elsewhere it stays within 0.05. The probes also differ in cost: a median fit takes 6~s for the linear probe, 21~s for XGBoost and 75~s for the MLP, plus 33--514~s for its four-point hyperparameter grid, while the k-NN predictor fits nothing and needs 14~s to predict the test split.

\hypertarget{screening-enrichment-results}{%
\subsection{Retrospective materials screening}\label{screening-enrichment-results}}

High prediction accuracy does not necessarily mean that a model can identify useful materials. We therefore evaluate how well the frozen-embedding predictors select candidates with desirable properties from the held-out test set (\hyperlink{screening-enrichment-and-label-efficiency}{§A.11}, \textbf{\hyperlink{fig:S13}{Fig.~S13}}, \textbf{\hyperlink{tab:S20}{Table~S20}}).

\textbf{Recovering the best materials.} When selecting the 1~\% of materials with the lowest predicted formation energies, ORB-v3 recovers 88~\% of the best 1~\% of materials, as determined by their DFT reference values; UMA-S recovers 87~\%, and Magpie recovers 76~\%. Random selection would recover only 1~\%. The largest advantage over the composition-only baseline occurs for density: ORB-v3 recovers 84~\% and UMA-S 82~\% of the reference-best 1~\%, compared with 5~\% for Magpie. At the same selection size, ORB-v3 and UMA-S recover 69~\% and 81~\% of the materials with the highest bulk moduli, respectively, and 73~\% and 72~\% of those with the highest magnetic-moment densities. These results show that frozen embeddings can substantially improve the recovery of desirable materials, particularly where composition alone is insufficient.

\textbf{Identifying materials near the stability hull.} We evaluate energy above hull using fixed physical thresholds rather than percentile-based recovery. Among the 1~\% of candidates with the lowest predicted hull energies, 69~\% of ORB-v3's selections and 74~\% of UMA-S's selections are truly on the hull, compared with 61~\% for Magpie. When the criterion is relaxed to within 50~meV/atom of the hull, precision increases to 96~\%, 97~\% and 88~\%, respectively. For comparison, only 22~\% of all test materials lie on the hull, and 51~\% are within 50~meV/atom. The predictors therefore concentrate likely stable materials in a small shortlist. However, their absolute prediction errors remain too large for reliable decisions about individual materials close to the hull (\hyperlink{from-ruxb2-to-expected-downstream-accuracy}{§3.2}).

\textbf{Selecting materials by band gap.} For the 0.5--3.0~eV window, ORB-v3 achieves 73~\% precision (the fraction of selected materials actually in the range) and 87~\% recall (the fraction of all materials in the range successfully identified). UMA-S achieves 76~\% precision and 85~\% recall, while Magpie has substantially lower precision (52~\%) despite similar recall (87~\%). The frozen-embedding predictors therefore identify a comparable fraction of qualifying materials while selecting fewer candidates outside the desired range.

\textbf{Ranking categorical properties.} We assess how well each predicted class probability ranks members of that class above non-members using AUROC. A score of 0.5 indicates random ranking and 1.0 perfect separation. ORB-v3 and UMA-S both achieve approximately 0.97 for metal/insulator classification. For magnetic labels, ORB-v3 reaches 0.95 and UMA-S 0.96. The largest difference occurs for crystal system: UMA-S achieves AUROC values of 0.96--1.00 across individual classes, compared with 0.83--0.97 for ORB-v3.

\textbf{Comparison with random-weight controls.} \textbf{\hyperlink{fig:S13}{Figure~S13j}} summarises screening performance relative to random selection and compares the pretrained predictors with randomly initialised ORB-v3 and UMA-S. This summary uses lift, defined as screening success divided by the corresponding random-selection baseline. A lift of 1 indicates random selection, while larger values indicate greater enrichment. Because the prevalence of qualifying materials differs across screening tasks, the maximum possible lift also differs; values should therefore be compared primarily within each target rather than across targets.

Selecting predictor configurations using validation data instead of the test set changes screening precision and recall by at most 0.048. Overall, frozen embeddings improve retrospective candidate selection over composition-only descriptors, especially for properties where structural information is important. The gains are smaller for the magnetic properties. These results concern materials with known DFT labels and do not establish prospective discovery performance.

\hypertarget{label-efficiency-results}{%
\subsection{Label efficiency and a network trained from scratch}\label{label-efficiency-results}}

\textbf{How many labels are needed?} We evaluate how prediction accuracy changes as the number of labelled training materials increases from 50 to 20,000. ORB, UMA and the composition-only Magpie baseline are trained on the same sampled materials and evaluated on a fixed test set (\textbf{\hyperlink{screening-enrichment-and-label-efficiency}{§A.11}}, \textbf{\hyperlink{fig:S14}{Fig.~S14}}, \textbf{\hyperlink{tab:S21}{Table~S21}}). Previous work has also shown the value of transfer learning when labels are scarce {[}\citelink{cite:Yamada2019}{Yamada 2019}; \citelink{cite:Gupta2024}{Gupta 2024}; \citelink{cite:Radova2025}{Radova 2025}{]}.

We distinguish two stages of learning. A probe becomes \emph{stable} when variation across repeated training samples falls below 0.05. It \emph{saturates} when its score comes within 0.05 of the full-corpus result. Stability is reached relatively quickly: with 50--500 labels for most targets, although ORB needs 1,000 for density and magnetic-moment density, and energy above hull requires 2,000. Saturation takes considerably more data. Formation energy, bulk modulus and metal/insulator classification saturate with 1,000--2,000 labels. Energy above hull and magnetic label need 2,000--10,000. Band gap and density require 7,000, while crystal system and magnetic-moment density require 20,000.

Frozen embeddings do not always outperform composition descriptors when very few labels are available. For density, Magpie performs better up to 1,000 labels, with a clear embedding advantage appearing at 2,000. ORB also trails Magpie for bulk modulus up to 200 labels. Band gap and energy above hull show the opposite pattern: embeddings outperform Magpie in more than 90~\% of paired training samples from 100 labels onward, and from just 50 labels for UMA.

\textbf{Comparison with a network trained from scratch.} We also train a small message-passing network directly on crystal structures (\textbf{\hyperlink{screening-enrichment-and-label-efficiency}{§A.11}}, \textbf{\hyperlink{tab:S22}{Table~S22}}). At the training size where both frozen-embedding probes have saturated, the network performs worse on six of nine targets. Its R² is lower by 0.09--0.10 for formation energy, 0.31--0.33 for band gap and 0.35--0.38 for energy above hull. Its macro-F1 is lower by 0.11--0.12 for metal/insulator, 0.05--0.07 for magnetic label and 0.13--0.20 for crystal system. In contrast, the network performs similarly on bulk modulus and magnetic-moment density. It performs better on density, reaching R² = 0.99 compared with approximately 0.90 for the embedding probes. Density is a special case because it can be calculated directly from the input cell and atomic masses.

\textbf{How the advantage changes with training size.} The network trained from scratch struggles most when labels are scarce. With 200 training materials, it achieves R² = 0.22 for density (mean of three draws ranging from −0.05 to 0.74), compared with 0.60 for the embedding probes, and R² = 0.54 for formation energy, compared with 0.82 for ORB and 0.87 for UMA.

As more labels become available, the network closes some of these gaps. At 20,000 materials, its formation-energy score is within 0.02 of the embedding probes. However, it still trails them by 0.18--0.21 R² for energy above hull, 0.22--0.24 R² for band gap and 0.13--0.20 macro-F1 for crystal system. It matches the embedding probes from 1,000 materials for bulk modulus and from 5,000 for magnetic-moment density. For density, it takes the lead from 500 materials onward, reaching R² = 0.996 at 20,000 labels, compared with 0.93 for ORB and 0.92 for UMA (\textbf{\hyperlink{fig:S14}{Fig.~S14}}, \textbf{\hyperlink{tab:S22}{Tables~S22}} and \textbf{\hyperlink{tab:S21}{S21}}).

Overall, frozen embeddings offer the largest advantage when labelled data are limited. Their benefit remains substantial for band gap, energy above hull and crystal system even as the training set grows. The small network is more competitive for properties closely related to the input structure, particularly density.

\hypertarget{supplementary-tables}{%
\newpage
\section{Supplementary Tables}\label{supplementary-tables}}
\par\addvspace{0.8\baselineskip}
\hypertarget{tab:S4}{}%

{\small \textbf{Table~S4 --- Comparison of prediction methods and embedding-neighbourhood organisation (\hyperlink{frozen-mlip-embeddings-predict-properties-outside-the-training-objective}{§3.1}, \hyperlink{property-information-is-locally-predictable-without-property-aligned-clusters}{§3.5}, \hyperlink{decodability-clustering-and-nearest-neighbour-retrieval}{§B.2}, \hyperlink{probe-family}{§B.5}).} \emph{All prediction methods use the same seed-42 training and test split. For each model and target, we use the peak XGBoost embedding configuration from \textbf{\hyperlink{tab:1}{Table~1}}. UMA-S crystal system uses its best rotation-invariant configuration (concat · ang-norm · b0), since higher-scoring full-tensor configurations are sensitive to rotation (\hyperlink{uma-channel-views-and-the-rotation-invariant-crystal-system-gap}{§3.4}, \hyperlink{tab:S13}{Table~S13}). \textbf{Magpie} is the composition-only XGBoost baseline and is identical for ORB and UMA. \textbf{Linear} regression/classification, \textbf{XGBoost} and the multilayer perceptron (\textbf{MLP}) are trained on the selected embeddings (\hyperlink{probe-configuration}{§A.2}). The \textbf{k-nearest-neighbour} predictor uses the ten closest training materials, with either Euclidean or cosine distance. It predicts continuous properties by averaging their neighbours' values and categorical properties by majority vote. Prediction scores are R² for continuous targets and macro-F1 for categorical targets. Continuous targets also report RMSE in the units shown (\hyperlink{from-ruxb2-to-expected-downstream-accuracy}{§3.2}). The final column measures \textbf{neighbourhood organisation}, not prediction accuracy. It indicates whether nearby embeddings have similar property values or labels, using 1−LVR for continuous targets and chance-corrected k-NN purity for categorical targets. Neighbours are identified without using property labels. Unlike the prediction columns, organisation is reported at the node layer and, for UMA, channel view where this metric is highest. These configurations may differ from the selected prediction configurations. Organisation scores are therefore not directly comparable with R² or macro-F1.}\par}

\noindent\begin{minipage}{\linewidth}
\textbf{(a) ORB-v3 --- continuous properties}\par\nobreak
{\scriptsize\setlength{\tabcolsep}{3pt}\begin{tabular}{@{}l rr rr rr rr rr rr r@{}}
\toprule
\multicolumn{1}{@{}l}{Target (unit)} & \multicolumn{2}{c}{Magpie} & \multicolumn{2}{c}{Linear} & \multicolumn{2}{c}{\shortstack[c]{XGBoost\\probe}} & \multicolumn{2}{c}{\shortstack[c]{k-NN retrieval\\(Euclidean)}} & \multicolumn{2}{c}{\shortstack[c]{k-NN retrieval\\(cosine)}} & \multicolumn{2}{c}{MLP} & \multicolumn{1}{c@{}}{\shortstack[c]{Organi-\\sation}} \\
\cmidrule(lr){2-3} \cmidrule(lr){4-5} \cmidrule(lr){6-7} \cmidrule(lr){8-9} \cmidrule(lr){10-11} \cmidrule(lr){12-13}
 & R² & RMSE & R² & RMSE & R² & RMSE & R² & RMSE & R² & RMSE & R² & RMSE & 1−LVR \\
\midrule
formation E/atom (eV/atom) & 0.890 & 0.397 & 0.989 & 0.128 & 0.990 & 0.123 & 0.982 & 0.163 & 0.981 & 0.164 & 0.995 & 0.085 & 0.850 \\
energy above hull (eV/atom) & 0.296 & 0.378 & 0.906 & 0.138 & 0.948 & 0.103 & 0.914 & 0.132 & 0.919 & 0.128 & 0.978 & 0.068 & 0.837 \\
band gap (eV) & 0.698 & 0.835 & 0.769 & 0.731 & 0.890 & 0.504 & 0.839 & 0.610 & 0.840 & 0.608 & 0.924 & 0.419 & 0.717 \\
density (g~cm⁻³) & 0.888 & 0.919 & 0.980 & 0.385 & 0.982 & 0.370 & 0.960 & 0.550 & 0.961 & 0.541 & 0.998 & 0.116 & 0.807 \\
bulk modulus (GPa) & 0.882 & 26.40 & 0.941 & 18.71 & 0.951 & 17.04 & 0.918 & 21.99 & 0.920 & 21.69 & 0.958 & 15.73 & 0.718 \\
mag. density (μ\_B~Å⁻³) & 0.816 & 0.0126 & 0.805 & 0.0129 & 0.846 & 0.0115 & 0.788 & 0.0135 & 0.789 & 0.0135 & 0.854 & 0.0112 & 0.763 \\
\bottomrule\end{tabular}}
\end{minipage}

\noindent\begin{minipage}{\linewidth}
\textbf{(a) ORB-v3 --- categorical properties}\par\nobreak
\raggedright{\footnotesize\setlength{\tabcolsep}{5pt}\begin{tabular}{@{}l cccccc c@{}}
\toprule
\multicolumn{1}{@{}l}{Target} & \shortstack[c]{Magpie\\F1} & \shortstack[c]{Linear\\F1} & \shortstack[c]{XGBoost\\probe F1} & \shortstack[c]{k-NN retrieval\\F1 (Euclidean)} & \shortstack[c]{k-NN retrieval\\F1 (cosine)} & \shortstack[c]{MLP\\F1} & \multicolumn{1}{c@{}}{\shortstack[c]{Organisation\\(k-NN purity)}} \\
\midrule
crystal system & 0.520 & 0.510 & 0.622 & 0.621 & 0.623 & 0.698 & 0.454 \\
metal/insulator & 0.845 & 0.892 & 0.909 & 0.888 & 0.890 & 0.911 & 0.748 \\
magnetic label & 0.860 & 0.870 & 0.890 & 0.869 & 0.869 & 0.910 & 0.633 \\
\bottomrule\end{tabular}}
\end{minipage}

\noindent\begin{minipage}{\linewidth}
\textbf{(b) UMA-S --- continuous properties}\par\nobreak
{\scriptsize\setlength{\tabcolsep}{3pt}\begin{tabular}{@{}l rr rr rr rr rr rr r@{}}
\toprule
\multicolumn{1}{@{}l}{Target (unit)} & \multicolumn{2}{c}{Magpie} & \multicolumn{2}{c}{Linear} & \multicolumn{2}{c}{\shortstack[c]{XGBoost\\probe}} & \multicolumn{2}{c}{\shortstack[c]{k-NN retrieval\\(Euclidean)}} & \multicolumn{2}{c}{\shortstack[c]{k-NN retrieval\\(cosine)}} & \multicolumn{2}{c}{MLP} & \multicolumn{1}{c@{}}{\shortstack[c]{Organi-\\sation}} \\
\cmidrule(lr){2-3} \cmidrule(lr){4-5} \cmidrule(lr){6-7} \cmidrule(lr){8-9} \cmidrule(lr){10-11} \cmidrule(lr){12-13}
 & R² & RMSE & R² & RMSE & R² & RMSE & R² & RMSE & R² & RMSE & R² & RMSE & 1−LVR \\
\midrule
formation E/atom (eV/atom) & 0.890 & 0.397 & 0.990 & 0.118 & 0.990 & 0.121 & 0.979 & 0.176 & 0.980 & 0.171 & 0.995 & 0.082 & 0.861 \\
energy above hull (eV/atom) & 0.296 & 0.378 & 0.925 & 0.123 & 0.960 & 0.090 & 0.944 & 0.107 & 0.945 & 0.106 & 0.979 & 0.065 & 0.838 \\
band gap (eV) & 0.698 & 0.835 & 0.789 & 0.699 & 0.907 & 0.465 & 0.857 & 0.576 & 0.857 & 0.574 & 0.930 & 0.403 & 0.733 \\
density (g~cm⁻³) & 0.888 & 0.919 & 0.872 & 0.984 & 0.983 & 0.356 & 0.964 & 0.525 & 0.962 & 0.539 & 0.999 & 0.103 & 0.802 \\
bulk modulus (GPa) & 0.882 & 26.40 & 0.954 & 16.41 & 0.958 & 15.67 & 0.916 & 22.23 & 0.885 & 26.03 & 0.963 & 14.76 & 0.711 \\
mag. density (μ\_B~Å⁻³) & 0.816 & 0.0126 & 0.444 & 0.0219 & 0.839 & 0.0118 & 0.802 & 0.0131 & 0.795 & 0.0133 & 0.847 & 0.0115 & 0.774 \\
\bottomrule\end{tabular}}
\end{minipage}

\noindent\begin{minipage}{\linewidth}
\textbf{(b) UMA-S --- categorical properties}\par\nobreak
\raggedright{\footnotesize\setlength{\tabcolsep}{5pt}\begin{tabular}{@{}l cccccc c@{}}
\toprule
\multicolumn{1}{@{}l}{Target} & \shortstack[c]{Magpie\\F1} & \shortstack[c]{Linear\\F1} & \shortstack[c]{XGBoost\\probe F1} & \shortstack[c]{k-NN retrieval\\F1 (Euclidean)} & \shortstack[c]{k-NN retrieval\\F1 (cosine)} & \shortstack[c]{MLP\\F1} & \multicolumn{1}{c@{}}{\shortstack[c]{Organisation\\(k-NN purity)}} \\
\midrule
crystal system & 0.520 & 0.636 & 0.817 & 0.583 & 0.582 & 0.815 & 0.454 \\
metal/insulator & 0.845 & 0.896 & 0.914 & 0.892 & 0.894 & 0.917 & 0.741 \\
magnetic label & 0.860 & 0.884 & 0.903 & 0.869 & 0.870 & 0.914 & 0.643 \\
\bottomrule\end{tabular}}
\end{minipage}

\clearpage
\hypertarget{tab:S5}{}%
{\small \textbf{Table~S5 --- Best XGBoost performance by read-out type and model.} \emph{For each target and read-out (node, edge or concatenated), we report the configuration with the highest R² or macro-F1 across all layers. For UMA-S, the search also includes all equivariant channel views. The \textbf{best configuration} column identifies the selected layer and, for UMA-S, the channel view. Crystal-system results for UMA-S are restricted to rotation-invariant features because higher-scoring full-tensor configurations are sensitive to rotation (\hyperlink{uma-channel-views-and-the-rotation-invariant-crystal-system-gap}{§3.4}, \hyperlink{tab:S13}{Table~S13}). All results use the fixed seed-42 test split. For continuous targets, we report R², mean absolute error (\textbf{MAE}) and root-mean-square error (\textbf{RMSE}). Errors are expressed in the property's native units, and \textbf{σ} is its standard deviation on the test set. For categorical targets, we report \textbf{accuracy} and \textbf{macro-F1}. Accuracy is the proportion of correctly classified materials, whereas macro-F1 gives equal weight to each class. The \textbf{baseline} is a classifier that always predicts the most frequent class. The \textbf{XGB − linear} column shows the difference between XGBoost and the linear probe at the same selected configuration. For UMA-S, \textbf{Difference UMA − ORB} compares the two models on the same target and read-out. The \textbf{Advantage} column classifies the absolute difference as trivial (below 0.005), small (0.005--0.02), moderate (0.02--0.05) or large (at least 0.05). All table entries are point estimates. The 95~\% confidence intervals for peak scores and UMA − ORB differences are shown in \hyperlink{fig:2}{Fig.~2a} and \hyperlink{fig:2}{Fig.~2c}.}\par}

\textbf{node read-out}

\noindent\begin{minipage}{\linewidth}
\emph{node · regression --- ORB-v3}\par\nobreak
\begin{tabular}{@{}
  >{\raggedright\arraybackslash}p{(\columnwidth - 12\tabcolsep) * \real{0.2689}}
  >{\raggedright\arraybackslash}p{(\columnwidth - 12\tabcolsep) * \real{0.0924}}
  >{\raggedright\arraybackslash}p{(\columnwidth - 12\tabcolsep) * \real{0.1092}}
  >{\raggedright\arraybackslash}p{(\columnwidth - 12\tabcolsep) * \real{0.0924}}
  >{\raggedright\arraybackslash}p{(\columnwidth - 12\tabcolsep) * \real{0.2269}}
  >{\raggedright\arraybackslash}p{(\columnwidth - 12\tabcolsep) * \real{0.0840}}
  >{\raggedright\arraybackslash}p{(\columnwidth - 12\tabcolsep) * \real{0.1261}}@{}}
\toprule\noalign{}
\begin{minipage}[c]{\linewidth}\raggedright
target
\end{minipage} & \begin{minipage}[c]{\linewidth}\raggedright
best config
\end{minipage} & \begin{minipage}[c]{\linewidth}\raggedright
MAE
\end{minipage} & \begin{minipage}[c]{\linewidth}\raggedright
RMSE
\end{minipage} & \begin{minipage}[c]{\linewidth}\raggedright
σ
\end{minipage} & \begin{minipage}[c]{\linewidth}\raggedright
R²
\end{minipage} & \begin{minipage}[c]{\linewidth}\raggedright
XGB − linear (R²)
\end{minipage} \\
\midrule\noalign{}
formation E/atom & g0 & 0.0795 & 0.123 & 1.2~eV/atom & 0.99 & 0.000998 \\
energy above hull & g4 & 0.0555 & 0.106 & 0.45~eV/atom & 0.945 & 0.0803 \\
band gap & g2 & 0.293 & 0.507 & 1.52~eV & 0.889 & 0.199 \\
density & g0 & 0.257 & 0.375 & 2.75~g~cm⁻³ & 0.981 & 0.00687 \\
bulk modulus & g0 & 9.07 & 17.1 & 77~GPa & 0.95 & 0.0114 \\
mag. density & g0 & 0.00587 & 0.0115 & 0.0293~μ\_B~Å⁻³ & 0.846 & 0.0472 \\
\bottomrule\end{tabular}
\end{minipage}

\noindent\begin{minipage}{\linewidth}
\emph{node · classification --- ORB-v3}\par\nobreak
\begin{tabular}{@{}
  >{\raggedright\arraybackslash}p{(\columnwidth - 12\tabcolsep) * \real{0.2605}}
  >{\raggedright\arraybackslash}p{(\columnwidth - 12\tabcolsep) * \real{0.1008}}
  >{\raggedright\arraybackslash}p{(\columnwidth - 12\tabcolsep) * \real{0.1345}}
  >{\raggedright\arraybackslash}p{(\columnwidth - 12\tabcolsep) * \real{0.1345}}
  >{\raggedright\arraybackslash}p{(\columnwidth - 12\tabcolsep) * \real{0.1345}}
  >{\raggedright\arraybackslash}p{(\columnwidth - 12\tabcolsep) * \real{0.1345}}
  >{\raggedright\arraybackslash}p{(\columnwidth - 12\tabcolsep) * \real{0.1008}}@{}}
\toprule\noalign{}
\begin{minipage}[c]{\linewidth}\raggedright
target
\end{minipage} & \begin{minipage}[c]{\linewidth}\raggedright
best config
\end{minipage} & \begin{minipage}[c]{\linewidth}\raggedright
accuracy
\end{minipage} & \begin{minipage}[c]{\linewidth}\raggedright
macro-F1
\end{minipage} & \begin{minipage}[c]{\linewidth}\raggedright
baseline accuracy
\end{minipage} & \begin{minipage}[c]{\linewidth}\raggedright
baseline macro-F1
\end{minipage} & \begin{minipage}[c]{\linewidth}\raggedright
XGB − linear (F1)
\end{minipage} \\
\midrule\noalign{}
crystal system & g2 & 0.601 & 0.608 & 0.234 & 0.0542 & 0.136 \\
metal/insulator & g3 & 0.908 & 0.908 & 0.531 & 0.347 & 0.0215 \\
magnetic label & g4 & 0.889 & 0.888 & 0.527 & 0.345 & 0.0308 \\
\bottomrule\end{tabular}
\end{minipage}

\noindent\begin{minipage}{\linewidth}
\emph{node · regression --- UMA-S (Difference/Advantage vs the ORB rows)}\par\nobreak
\begin{tabular}{@{}
  >{\raggedright\arraybackslash}p{(\columnwidth - 16\tabcolsep) * \real{0.1833}}
  >{\raggedright\arraybackslash}p{(\columnwidth - 16\tabcolsep) * \real{0.1500}}
  >{\raggedright\arraybackslash}p{(\columnwidth - 16\tabcolsep) * \real{0.0750}}
  >{\raggedright\arraybackslash}p{(\columnwidth - 16\tabcolsep) * \real{0.0667}}
  >{\raggedright\arraybackslash}p{(\columnwidth - 16\tabcolsep) * \real{0.1500}}
  >{\raggedright\arraybackslash}p{(\columnwidth - 16\tabcolsep) * \real{0.0500}}
  >{\raggedright\arraybackslash}p{(\columnwidth - 16\tabcolsep) * \real{0.0750}}
  >{\raggedright\arraybackslash}p{(\columnwidth - 16\tabcolsep) * \real{0.1083}}
  >{\raggedright\arraybackslash}p{(\columnwidth - 16\tabcolsep) * \real{0.1417}}@{}}
\toprule\noalign{}
\begin{minipage}[c]{\linewidth}\raggedright
target
\end{minipage} & \begin{minipage}[c]{\linewidth}\raggedright
best config
\end{minipage} & \begin{minipage}[c]{\linewidth}\raggedright
MAE
\end{minipage} & \begin{minipage}[c]{\linewidth}\raggedright
RMSE
\end{minipage} & \begin{minipage}[c]{\linewidth}\raggedright
σ
\end{minipage} & \begin{minipage}[c]{\linewidth}\raggedright
R²
\end{minipage} & \begin{minipage}[c]{\linewidth}\raggedright
XGB − linear (R²)
\end{minipage} & \begin{minipage}[c]{\linewidth}\raggedright
Difference UMA − ORB
\end{minipage} & \begin{minipage}[c]{\linewidth}\raggedright
Advantage
\end{minipage} \\
\midrule\noalign{}
formation E/atom & ang-norm · bb & 0.0794 & 0.121 & 1.2~eV/atom & 0.99 & 0.00554 & +0.000 & trivial~/~tie \\
energy above hull & l0 · bb & 0.0486 & 0.0909 & 0.45~eV/atom & 0.959 & 0.0803 & +0.014 & small~UMA \\
band gap & ang-norm · bb & 0.271 & 0.467 & 1.52~eV & 0.906 & 0.144 & +0.017 & small~UMA \\
density & ang-norm · emb & 0.254 & 0.372 & 2.75~g~cm⁻³ & 0.982 & 0.0675 & +0.000 & trivial~/~tie \\
bulk modulus & ang-norm · emb & 8.85 & 16.3 & 77~GPa & 0.955 & 0.00312 & +0.005 & trivial~/~tie \\
mag. density & ang-norm · b1 & 0.00618 & 0.012 & 0.0293~μ\_B~Å⁻³ & 0.832 & 0.0707 & -0.014 & small~ORB \\
\bottomrule\end{tabular}
\end{minipage}

\noindent\begin{minipage}{\linewidth}
\emph{node · classification --- UMA-S (Difference/Advantage vs the ORB rows)}\par\nobreak
\begin{tabular}{@{}
  >{\raggedright\arraybackslash}p{(\columnwidth - 16\tabcolsep) * \real{0.1780}}
  >{\raggedright\arraybackslash}p{(\columnwidth - 16\tabcolsep) * \real{0.1525}}
  >{\raggedright\arraybackslash}p{(\columnwidth - 16\tabcolsep) * \real{0.0932}}
  >{\raggedright\arraybackslash}p{(\columnwidth - 16\tabcolsep) * \real{0.0932}}
  >{\raggedright\arraybackslash}p{(\columnwidth - 16\tabcolsep) * \real{0.0932}}
  >{\raggedright\arraybackslash}p{(\columnwidth - 16\tabcolsep) * \real{0.0932}}
  >{\raggedright\arraybackslash}p{(\columnwidth - 16\tabcolsep) * \real{0.0678}}
  >{\raggedright\arraybackslash}p{(\columnwidth - 16\tabcolsep) * \real{0.1186}}
  >{\raggedright\arraybackslash}p{(\columnwidth - 16\tabcolsep) * \real{0.1102}}@{}}
\toprule\noalign{}
\begin{minipage}[c]{\linewidth}\raggedright
target
\end{minipage} & \begin{minipage}[c]{\linewidth}\raggedright
best config
\end{minipage} & \begin{minipage}[c]{\linewidth}\raggedright
accuracy
\end{minipage} & \begin{minipage}[c]{\linewidth}\raggedright
macro-F1
\end{minipage} & \begin{minipage}[c]{\linewidth}\raggedright
baseline accuracy
\end{minipage} & \begin{minipage}[c]{\linewidth}\raggedright
baseline macro-F1
\end{minipage} & \begin{minipage}[c]{\linewidth}\raggedright
XGB − linear (F1)
\end{minipage} & \begin{minipage}[c]{\linewidth}\raggedright
Difference UMA − ORB
\end{minipage} & \begin{minipage}[c]{\linewidth}\raggedright
Advantage
\end{minipage} \\
\midrule\noalign{}
crystal system & ang-norm · b0 & 0.656 & 0.679 & 0.234 & 0.0542 & 0.136 & +0.071 & large~UMA \\
metal/insulator & ang-norm · bb & 0.914 & 0.914 & 0.531 & 0.347 & 0.0179 & +0.006 & small~UMA \\
magnetic label & ang-norm · bb & 0.904 & 0.903 & 0.527 & 0.345 & 0.0193 & +0.015 & small~UMA \\
\bottomrule\end{tabular}
\end{minipage}

\textbf{edge read-out}

\noindent\begin{minipage}{\linewidth}
\emph{edge · regression --- ORB-v3}\par\nobreak
\begin{tabular}{@{}
  >{\raggedright\arraybackslash}p{(\columnwidth - 12\tabcolsep) * \real{0.2727}}
  >{\raggedright\arraybackslash}p{(\columnwidth - 12\tabcolsep) * \real{0.0992}}
  >{\raggedright\arraybackslash}p{(\columnwidth - 12\tabcolsep) * \real{0.1157}}
  >{\raggedright\arraybackslash}p{(\columnwidth - 12\tabcolsep) * \real{0.0992}}
  >{\raggedright\arraybackslash}p{(\columnwidth - 12\tabcolsep) * \real{0.2314}}
  >{\raggedright\arraybackslash}p{(\columnwidth - 12\tabcolsep) * \real{0.0826}}
  >{\raggedright\arraybackslash}p{(\columnwidth - 12\tabcolsep) * \real{0.0992}}@{}}
\toprule\noalign{}
\begin{minipage}[c]{\linewidth}\raggedright
target
\end{minipage} & \begin{minipage}[c]{\linewidth}\raggedright
best config
\end{minipage} & \begin{minipage}[c]{\linewidth}\raggedright
MAE
\end{minipage} & \begin{minipage}[c]{\linewidth}\raggedright
RMSE
\end{minipage} & \begin{minipage}[c]{\linewidth}\raggedright
σ
\end{minipage} & \begin{minipage}[c]{\linewidth}\raggedright
R²
\end{minipage} & \begin{minipage}[c]{\linewidth}\raggedright
XGB − linear (R²)
\end{minipage} \\
\midrule\noalign{}
formation E/atom & g1 & 0.115 & 0.181 & 1.2~eV/atom & 0.977 & 0.0273 \\
energy above hull & g4 & 0.0708 & 0.143 & 0.45~eV/atom & 0.9 & 0.166 \\
band gap & g4 & 0.344 & 0.561 & 1.52~eV & 0.864 & 0.174 \\
density & g1 & 0.444 & 0.647 & 2.75~g~cm⁻³ & 0.945 & 0.05 \\
bulk modulus & g1 & 10.6 & 18.6 & 77~GPa & 0.941 & 0.0378 \\
mag. density & g1 & 0.00729 & 0.0134 & 0.0293~μ\_B~Å⁻³ & 0.793 & 0.123 \\
\bottomrule\end{tabular}
\end{minipage}

\noindent\begin{minipage}{\linewidth}
\emph{edge · classification --- ORB-v3}\par\nobreak
\begin{tabular}{@{}
  >{\raggedright\arraybackslash}p{(\columnwidth - 12\tabcolsep) * \real{0.2605}}
  >{\raggedright\arraybackslash}p{(\columnwidth - 12\tabcolsep) * \real{0.1008}}
  >{\raggedright\arraybackslash}p{(\columnwidth - 12\tabcolsep) * \real{0.1345}}
  >{\raggedright\arraybackslash}p{(\columnwidth - 12\tabcolsep) * \real{0.1345}}
  >{\raggedright\arraybackslash}p{(\columnwidth - 12\tabcolsep) * \real{0.1345}}
  >{\raggedright\arraybackslash}p{(\columnwidth - 12\tabcolsep) * \real{0.1345}}
  >{\raggedright\arraybackslash}p{(\columnwidth - 12\tabcolsep) * \real{0.1008}}@{}}
\toprule\noalign{}
\begin{minipage}[c]{\linewidth}\raggedright
target
\end{minipage} & \begin{minipage}[c]{\linewidth}\raggedright
best config
\end{minipage} & \begin{minipage}[c]{\linewidth}\raggedright
accuracy
\end{minipage} & \begin{minipage}[c]{\linewidth}\raggedright
macro-F1
\end{minipage} & \begin{minipage}[c]{\linewidth}\raggedright
baseline accuracy
\end{minipage} & \begin{minipage}[c]{\linewidth}\raggedright
baseline macro-F1
\end{minipage} & \begin{minipage}[c]{\linewidth}\raggedright
XGB − linear (F1)
\end{minipage} \\
\midrule\noalign{}
crystal system & g3 & 0.596 & 0.602 & 0.234 & 0.0542 & 0.138 \\
metal/insulator & g4 & 0.897 & 0.896 & 0.531 & 0.347 & 0.022 \\
magnetic label & g4 & 0.87 & 0.869 & 0.527 & 0.345 & 0.0394 \\
\bottomrule\end{tabular}
\end{minipage}

\noindent\begin{minipage}{\linewidth}
\emph{edge · regression --- UMA-S (Difference/Advantage vs the ORB rows)}\par\nobreak
\begin{tabular}{@{}
  >{\raggedright\arraybackslash}p{(\columnwidth - 16\tabcolsep) * \real{0.1849}}
  >{\raggedright\arraybackslash}p{(\columnwidth - 16\tabcolsep) * \real{0.1429}}
  >{\raggedright\arraybackslash}p{(\columnwidth - 16\tabcolsep) * \real{0.0756}}
  >{\raggedright\arraybackslash}p{(\columnwidth - 16\tabcolsep) * \real{0.0672}}
  >{\raggedright\arraybackslash}p{(\columnwidth - 16\tabcolsep) * \real{0.1513}}
  >{\raggedright\arraybackslash}p{(\columnwidth - 16\tabcolsep) * \real{0.0588}}
  >{\raggedright\arraybackslash}p{(\columnwidth - 16\tabcolsep) * \real{0.0672}}
  >{\raggedright\arraybackslash}p{(\columnwidth - 16\tabcolsep) * \real{0.1092}}
  >{\raggedright\arraybackslash}p{(\columnwidth - 16\tabcolsep) * \real{0.1429}}@{}}
\toprule\noalign{}
\begin{minipage}[c]{\linewidth}\raggedright
target
\end{minipage} & \begin{minipage}[c]{\linewidth}\raggedright
best config
\end{minipage} & \begin{minipage}[c]{\linewidth}\raggedright
MAE
\end{minipage} & \begin{minipage}[c]{\linewidth}\raggedright
RMSE
\end{minipage} & \begin{minipage}[c]{\linewidth}\raggedright
σ
\end{minipage} & \begin{minipage}[c]{\linewidth}\raggedright
R²
\end{minipage} & \begin{minipage}[c]{\linewidth}\raggedright
XGB − linear (R²)
\end{minipage} & \begin{minipage}[c]{\linewidth}\raggedright
Difference UMA − ORB
\end{minipage} & \begin{minipage}[c]{\linewidth}\raggedright
Advantage
\end{minipage} \\
\midrule\noalign{}
formation E/atom & ang-norm · b0 & 0.099 & 0.149 & 1.2~eV/atom & 0.985 & 0.0122 & +0.007 & small~UMA \\
energy above hull & l0 · b1 & 0.0619 & 0.12 & 0.45~eV/atom & 0.929 & 0.192 & +0.029 & moderate~UMA \\
band gap & l0 · b2 & 0.291 & 0.508 & 1.52~eV & 0.889 & 0.19 & +0.025 & moderate~UMA \\
density & l0 · b0 & 0.372 & 0.539 & 2.75~g~cm⁻³ & 0.962 & 0.0463 & +0.017 & small~UMA \\
bulk modulus & ang-norm · b0 & 10.8 & 18.7 & 77~GPa & 0.941 & 0.0278 & -0.001 & trivial~/~tie \\
mag. density & l0 · b0 & 0.00641 & 0.0121 & 0.0293~μ\_B~Å⁻³ & 0.829 & 0.084 & +0.036 & moderate~UMA \\
\bottomrule\end{tabular}
\end{minipage}

\noindent\begin{minipage}{\linewidth}
\emph{edge · classification --- UMA-S (Difference/Advantage vs the ORB rows)}\par\nobreak
\begin{tabular}{@{}
  >{\raggedright\arraybackslash}p{(\columnwidth - 16\tabcolsep) * \real{0.1667}}
  >{\raggedright\arraybackslash}p{(\columnwidth - 16\tabcolsep) * \real{0.1500}}
  >{\raggedright\arraybackslash}p{(\columnwidth - 16\tabcolsep) * \real{0.0917}}
  >{\raggedright\arraybackslash}p{(\columnwidth - 16\tabcolsep) * \real{0.0917}}
  >{\raggedright\arraybackslash}p{(\columnwidth - 16\tabcolsep) * \real{0.0917}}
  >{\raggedright\arraybackslash}p{(\columnwidth - 16\tabcolsep) * \real{0.0917}}
  >{\raggedright\arraybackslash}p{(\columnwidth - 16\tabcolsep) * \real{0.0667}}
  >{\raggedright\arraybackslash}p{(\columnwidth - 16\tabcolsep) * \real{0.1167}}
  >{\raggedright\arraybackslash}p{(\columnwidth - 16\tabcolsep) * \real{0.1333}}@{}}
\toprule\noalign{}
\begin{minipage}[c]{\linewidth}\raggedright
target
\end{minipage} & \begin{minipage}[c]{\linewidth}\raggedright
best config
\end{minipage} & \begin{minipage}[c]{\linewidth}\raggedright
accuracy
\end{minipage} & \begin{minipage}[c]{\linewidth}\raggedright
macro-F1
\end{minipage} & \begin{minipage}[c]{\linewidth}\raggedright
baseline accuracy
\end{minipage} & \begin{minipage}[c]{\linewidth}\raggedright
baseline macro-F1
\end{minipage} & \begin{minipage}[c]{\linewidth}\raggedright
XGB − linear (F1)
\end{minipage} & \begin{minipage}[c]{\linewidth}\raggedright
Difference UMA − ORB
\end{minipage} & \begin{minipage}[c]{\linewidth}\raggedright
Advantage
\end{minipage} \\
\midrule\noalign{}
crystal system & ang-norm · b0 & 0.800 & 0.793 & 0.234 & 0.0542 & 0.226 & +0.191 & large~UMA \\
metal/insulator & ang-norm · b3 & 0.906 & 0.905 & 0.531 & 0.347 & 0.0267 & +0.009 & small~UMA \\
magnetic label & ang-norm · b2 & 0.89 & 0.89 & 0.527 & 0.345 & 0.0335 & +0.021 & moderate~UMA \\
\bottomrule\end{tabular}
\end{minipage}

\textbf{concat read-out}

\noindent\begin{minipage}{\linewidth}
\emph{concat · regression --- ORB-v3}\par\nobreak
\begin{tabular}{@{}
  >{\raggedright\arraybackslash}p{(\columnwidth - 12\tabcolsep) * \real{0.2689}}
  >{\raggedright\arraybackslash}p{(\columnwidth - 12\tabcolsep) * \real{0.0924}}
  >{\raggedright\arraybackslash}p{(\columnwidth - 12\tabcolsep) * \real{0.1092}}
  >{\raggedright\arraybackslash}p{(\columnwidth - 12\tabcolsep) * \real{0.0924}}
  >{\raggedright\arraybackslash}p{(\columnwidth - 12\tabcolsep) * \real{0.2269}}
  >{\raggedright\arraybackslash}p{(\columnwidth - 12\tabcolsep) * \real{0.0840}}
  >{\raggedright\arraybackslash}p{(\columnwidth - 12\tabcolsep) * \real{0.1261}}@{}}
\toprule\noalign{}
\begin{minipage}[c]{\linewidth}\raggedright
target
\end{minipage} & \begin{minipage}[c]{\linewidth}\raggedright
best config
\end{minipage} & \begin{minipage}[c]{\linewidth}\raggedright
MAE
\end{minipage} & \begin{minipage}[c]{\linewidth}\raggedright
RMSE
\end{minipage} & \begin{minipage}[c]{\linewidth}\raggedright
σ
\end{minipage} & \begin{minipage}[c]{\linewidth}\raggedright
R²
\end{minipage} & \begin{minipage}[c]{\linewidth}\raggedright
XGB − linear (R²)
\end{minipage} \\
\midrule\noalign{}
formation E/atom & g0 & 0.0794 & 0.123 & 1.2~eV/atom & 0.989 & 2.99e-05 \\
energy above hull & g4 & 0.0539 & 0.103 & 0.45~eV/atom & 0.948 & 0.0412 \\
band gap & g4 & 0.303 & 0.504 & 1.52~eV & 0.89 & 0.121 \\
density & g0 & 0.255 & 0.37 & 2.75~g~cm⁻³ & 0.982 & 0.00144 \\
bulk modulus & g0 & 9.11 & 17 & 77~GPa & 0.951 & 0.0101 \\
mag. density & g0 & 0.00586 & 0.0115 & 0.0293~μ\_B~Å⁻³ & 0.846 & 0.0408 \\
\bottomrule\end{tabular}
\end{minipage}

\noindent\begin{minipage}{\linewidth}
\emph{concat · classification --- ORB-v3}\par\nobreak
\begin{tabular}{@{}
  >{\raggedright\arraybackslash}p{(\columnwidth - 12\tabcolsep) * \real{0.2605}}
  >{\raggedright\arraybackslash}p{(\columnwidth - 12\tabcolsep) * \real{0.1008}}
  >{\raggedright\arraybackslash}p{(\columnwidth - 12\tabcolsep) * \real{0.1345}}
  >{\raggedright\arraybackslash}p{(\columnwidth - 12\tabcolsep) * \real{0.1345}}
  >{\raggedright\arraybackslash}p{(\columnwidth - 12\tabcolsep) * \real{0.1345}}
  >{\raggedright\arraybackslash}p{(\columnwidth - 12\tabcolsep) * \real{0.1345}}
  >{\raggedright\arraybackslash}p{(\columnwidth - 12\tabcolsep) * \real{0.1008}}@{}}
\toprule\noalign{}
\begin{minipage}[c]{\linewidth}\raggedright
target
\end{minipage} & \begin{minipage}[c]{\linewidth}\raggedright
best config
\end{minipage} & \begin{minipage}[c]{\linewidth}\raggedright
accuracy
\end{minipage} & \begin{minipage}[c]{\linewidth}\raggedright
macro-F1
\end{minipage} & \begin{minipage}[c]{\linewidth}\raggedright
baseline accuracy
\end{minipage} & \begin{minipage}[c]{\linewidth}\raggedright
baseline macro-F1
\end{minipage} & \begin{minipage}[c]{\linewidth}\raggedright
XGB − linear (F1)
\end{minipage} \\
\midrule\noalign{}
crystal system & g2 & 0.612 & 0.622 & 0.234 & 0.0542 & 0.112 \\
metal/insulator & g4 & 0.909 & 0.909 & 0.531 & 0.347 & 0.0166 \\
magnetic label & g4 & 0.891 & 0.89 & 0.527 & 0.345 & 0.0203 \\
\bottomrule\end{tabular}
\end{minipage}

\noindent\begin{minipage}{\linewidth}
\emph{concat · regression --- UMA-S (Difference/Advantage vs the ORB rows)}\par\nobreak
\begin{tabular}{@{}
  >{\raggedright\arraybackslash}p{(\columnwidth - 16\tabcolsep) * \real{0.1750}}
  >{\raggedright\arraybackslash}p{(\columnwidth - 16\tabcolsep) * \real{0.1500}}
  >{\raggedright\arraybackslash}p{(\columnwidth - 16\tabcolsep) * \real{0.0750}}
  >{\raggedright\arraybackslash}p{(\columnwidth - 16\tabcolsep) * \real{0.0667}}
  >{\raggedright\arraybackslash}p{(\columnwidth - 16\tabcolsep) * \real{0.1500}}
  >{\raggedright\arraybackslash}p{(\columnwidth - 16\tabcolsep) * \real{0.0500}}
  >{\raggedright\arraybackslash}p{(\columnwidth - 16\tabcolsep) * \real{0.0917}}
  >{\raggedright\arraybackslash}p{(\columnwidth - 16\tabcolsep) * \real{0.1083}}
  >{\raggedright\arraybackslash}p{(\columnwidth - 16\tabcolsep) * \real{0.1333}}@{}}
\toprule\noalign{}
\begin{minipage}[c]{\linewidth}\raggedright
target
\end{minipage} & \begin{minipage}[c]{\linewidth}\raggedright
best config
\end{minipage} & \begin{minipage}[c]{\linewidth}\raggedright
MAE
\end{minipage} & \begin{minipage}[c]{\linewidth}\raggedright
RMSE
\end{minipage} & \begin{minipage}[c]{\linewidth}\raggedright
σ
\end{minipage} & \begin{minipage}[c]{\linewidth}\raggedright
R²
\end{minipage} & \begin{minipage}[c]{\linewidth}\raggedright
XGB − linear (R²)
\end{minipage} & \begin{minipage}[c]{\linewidth}\raggedright
Difference UMA − ORB
\end{minipage} & \begin{minipage}[c]{\linewidth}\raggedright
Advantage
\end{minipage} \\
\midrule\noalign{}
formation E/atom & ang-norm · b3 & 0.0799 & 0.121 & 1.2~eV/atom & 0.99 & -0.00057 & +0.000 & trivial~/~tie \\
energy above hull & l0 · b3 & 0.0491 & 0.09 & 0.45~eV/atom & 0.96 & 0.0347 & +0.012 & small~UMA \\
band gap & ang-norm · b2 & 0.264 & 0.465 & 1.52~eV & 0.907 & 0.118 & +0.016 & small~UMA \\
density & ang-norm · emb & 0.242 & 0.356 & 2.75~g~cm⁻³ & 0.983 & 0.111 & +0.001 & trivial~/~tie \\
bulk modulus & ang-norm · emb & 8.25 & 15.7 & 77~GPa & 0.958 & 0.00403 & +0.008 & small~UMA \\
mag. density & ang-norm · emb & 0.00629 & 0.0118 & 0.0293~μ\_B~Å⁻³ & 0.839 & 0.395 & -0.007 & small~ORB \\
\bottomrule\end{tabular}
\end{minipage}

\noindent\begin{minipage}{\linewidth}
\emph{concat · classification --- UMA-S (Difference/Advantage vs the ORB rows)}\par\nobreak
\begin{tabular}{@{}
  >{\raggedright\arraybackslash}p{(\columnwidth - 16\tabcolsep) * \real{0.1653}}
  >{\raggedright\arraybackslash}p{(\columnwidth - 16\tabcolsep) * \real{0.1488}}
  >{\raggedright\arraybackslash}p{(\columnwidth - 16\tabcolsep) * \real{0.0909}}
  >{\raggedright\arraybackslash}p{(\columnwidth - 16\tabcolsep) * \real{0.0909}}
  >{\raggedright\arraybackslash}p{(\columnwidth - 16\tabcolsep) * \real{0.0909}}
  >{\raggedright\arraybackslash}p{(\columnwidth - 16\tabcolsep) * \real{0.0909}}
  >{\raggedright\arraybackslash}p{(\columnwidth - 16\tabcolsep) * \real{0.0661}}
  >{\raggedright\arraybackslash}p{(\columnwidth - 16\tabcolsep) * \real{0.1074}}
  >{\raggedright\arraybackslash}p{(\columnwidth - 16\tabcolsep) * \real{0.1488}}@{}}
\toprule\noalign{}
\begin{minipage}[c]{\linewidth}\raggedright
target
\end{minipage} & \begin{minipage}[c]{\linewidth}\raggedright
best config
\end{minipage} & \begin{minipage}[c]{\linewidth}\raggedright
accuracy
\end{minipage} & \begin{minipage}[c]{\linewidth}\raggedright
macro-F1
\end{minipage} & \begin{minipage}[c]{\linewidth}\raggedright
baseline accuracy
\end{minipage} & \begin{minipage}[c]{\linewidth}\raggedright
baseline macro-F1
\end{minipage} & \begin{minipage}[c]{\linewidth}\raggedright
XGB − linear (F1)
\end{minipage} & \begin{minipage}[c]{\linewidth}\raggedright
Difference UMA − ORB
\end{minipage} & \begin{minipage}[c]{\linewidth}\raggedright
Advantage
\end{minipage} \\
\midrule\noalign{}
crystal system & ang-norm · b0 & 0.818 & 0.817 & 0.234 & 0.0542 & 0.181 & +0.194 & large~UMA \\
metal/insulator & ang-norm · b3 & 0.914 & 0.913 & 0.531 & 0.347 & 0.012 & +0.005 & trivial~/~tie \\
magnetic label & ang-norm · b3 & 0.9 & 0.9 & 0.527 & 0.345 & 0.0101 & +0.009 & small~UMA \\
\bottomrule\end{tabular}
\end{minipage}

\clearpage
\hypertarget{tab:S6}{}%
{\small \textbf{Table~S6 --- Improvement over the composition-only baseline: score gains and error reduction (\hyperlink{frozen-mlip-embeddings-predict-properties-outside-the-training-objective}{§3.1}).} \emph{We compare the best ORB-v3 and UMA-S embedding probes with the Magpie composition-only baseline. All use XGBoost with identical settings and data splits. \textbf{Comp R²/F1} is the baseline score, while \textbf{ORB Δ} and \textbf{UMA Δ} show each model's improvement over it, with 95~\% paired-bootstrap confidence intervals. For continuous targets, \textbf{comp RMSE}, \textbf{ORB RMSE} and \textbf{UMA RMSE} give prediction errors in the units shown (\hyperlink{from-ruxb2-to-expected-downstream-accuracy}{§3.2}, \hyperlink{tab:S5}{Table~S5}). For categorical targets, the corresponding residual is 1 − macro-F1. \textbf{ORB cut} and \textbf{UMA cut} give the percentage of baseline error removed by each model. All gains are positive across 2,000 paired bootstrap resamples (reported two-sided p and Benjamini--Hochberg-adjusted q \textless{}~0.0005). UMA-S crystal-system results use its best rotation-invariant configuration (concat · ang-norm · b0; \hyperlink{uma-channel-views-and-the-rotation-invariant-crystal-system-gap}{§3.4}, \hyperlink{tab:S13}{Table~S13}).}\par}

\noindent\begin{minipage}{\linewidth}
\textbf{Regression}\par\nobreak
\begin{tabular}{@{}
  >{\raggedright\arraybackslash}p{(\columnwidth - 20\tabcolsep) * \real{0.2000}}
  >{\raggedright\arraybackslash}p{(\columnwidth - 20\tabcolsep) * \real{0.0727}}
  >{\raggedright\arraybackslash}p{(\columnwidth - 20\tabcolsep) * \real{0.0727}}
  >{\raggedright\arraybackslash}p{(\columnwidth - 20\tabcolsep) * \real{0.1091}}
  >{\raggedright\arraybackslash}p{(\columnwidth - 20\tabcolsep) * \real{0.0727}}
  >{\raggedright\arraybackslash}p{(\columnwidth - 20\tabcolsep) * \real{0.1091}}
  >{\raggedright\arraybackslash}p{(\columnwidth - 20\tabcolsep) * \real{0.0909}}
  >{\raggedright\arraybackslash}p{(\columnwidth - 20\tabcolsep) * \real{0.0727}}
  >{\raggedright\arraybackslash}p{(\columnwidth - 20\tabcolsep) * \real{0.0727}}
  >{\raggedright\arraybackslash}p{(\columnwidth - 20\tabcolsep) * \real{0.0636}}
  >{\raggedright\arraybackslash}p{(\columnwidth - 20\tabcolsep) * \real{0.0636}}@{}}
\toprule\noalign{}
\begin{minipage}[c]{\linewidth}\raggedright
Target
\end{minipage} & \begin{minipage}[c]{\linewidth}\raggedright
comp R²
\end{minipage} & \begin{minipage}[c]{\linewidth}\raggedright
ORB Δ
\end{minipage} & \begin{minipage}[c]{\linewidth}\raggedright
ORB 95~\% CI
\end{minipage} & \begin{minipage}[c]{\linewidth}\raggedright
UMA Δ
\end{minipage} & \begin{minipage}[c]{\linewidth}\raggedright
UMA 95~\% CI
\end{minipage} & \begin{minipage}[c]{\linewidth}\raggedright
comp RMSE
\end{minipage} & \begin{minipage}[c]{\linewidth}\raggedright
ORB RMSE
\end{minipage} & \begin{minipage}[c]{\linewidth}\raggedright
UMA RMSE
\end{minipage} & \begin{minipage}[c]{\linewidth}\raggedright
ORB cut, \%
\end{minipage} & \begin{minipage}[c]{\linewidth}\raggedright
UMA cut, \%
\end{minipage} \\
\midrule\noalign{}
formation E/atom & 0.890 & +0.099 & {[}+0.091, +0.108{]} & +0.100 & {[}+0.091, +0.108{]} & 0.397 & 0.123 & 0.121 & 69 & 69 \\
energy above hull & 0.296 & +0.652 & {[}+0.619, +0.680{]} & +0.664 & {[}+0.632, +0.697{]} & 0.378 & 0.103 & 0.0900 & 73 & 76 \\
band gap & 0.698 & +0.192 & {[}+0.184, +0.200{]} & +0.209 & {[}+0.200, +0.217{]} & 0.835 & 0.504 & 0.465 & 40 & 44 \\
density & 0.888 & +0.093 & {[}+0.085, +0.102{]} & +0.095 & {[}+0.087, +0.103{]} & 0.919 & 0.370 & 0.356 & 60 & 61 \\
bulk modulus & 0.882 & +0.069 & {[}+0.053, +0.087{]} & +0.076 & {[}+0.060, +0.094{]} & 26.4 & 17.0 & 15.7 & 35 & 41 \\
mag. density & 0.816 & +0.030 & {[}+0.024, +0.036{]} & +0.023 & {[}+0.017, +0.030{]} & 0.0126 & 0.0115 & 0.0118 & 9 & 6 \\
\bottomrule\end{tabular}
\end{minipage}

\noindent\begin{minipage}{\linewidth}
\textbf{Classification} (residual = 1 − macro-F1)\par\nobreak
\begin{tabular}{@{}
  >{\raggedright\arraybackslash}p{(\columnwidth - 20\tabcolsep) * \real{0.2000}}
  >{\raggedright\arraybackslash}p{(\columnwidth - 20\tabcolsep) * \real{0.0727}}
  >{\raggedright\arraybackslash}p{(\columnwidth - 20\tabcolsep) * \real{0.0727}}
  >{\raggedright\arraybackslash}p{(\columnwidth - 20\tabcolsep) * \real{0.1091}}
  >{\raggedright\arraybackslash}p{(\columnwidth - 20\tabcolsep) * \real{0.0727}}
  >{\raggedright\arraybackslash}p{(\columnwidth - 20\tabcolsep) * \real{0.1091}}
  >{\raggedright\arraybackslash}p{(\columnwidth - 20\tabcolsep) * \real{0.0909}}
  >{\raggedright\arraybackslash}p{(\columnwidth - 20\tabcolsep) * \real{0.0727}}
  >{\raggedright\arraybackslash}p{(\columnwidth - 20\tabcolsep) * \real{0.0727}}
  >{\raggedright\arraybackslash}p{(\columnwidth - 20\tabcolsep) * \real{0.0636}}
  >{\raggedright\arraybackslash}p{(\columnwidth - 20\tabcolsep) * \real{0.0636}}@{}}
\toprule\noalign{}
\begin{minipage}[c]{\linewidth}\raggedright
Target
\end{minipage} & \begin{minipage}[c]{\linewidth}\raggedright
comp F1
\end{minipage} & \begin{minipage}[c]{\linewidth}\raggedright
ORB Δ
\end{minipage} & \begin{minipage}[c]{\linewidth}\raggedright
ORB 95~\% CI
\end{minipage} & \begin{minipage}[c]{\linewidth}\raggedright
UMA Δ
\end{minipage} & \begin{minipage}[c]{\linewidth}\raggedright
UMA 95~\% CI
\end{minipage} & \begin{minipage}[c]{\linewidth}\raggedright
comp residual
\end{minipage} & \begin{minipage}[c]{\linewidth}\raggedright
ORB residual
\end{minipage} & \begin{minipage}[c]{\linewidth}\raggedright
UMA residual
\end{minipage} & \begin{minipage}[c]{\linewidth}\raggedright
ORB cut, \%
\end{minipage} & \begin{minipage}[c]{\linewidth}\raggedright
UMA cut, \%
\end{minipage} \\
\midrule\noalign{}
crystal system & 0.520 & +0.102 & {[}+0.097, +0.108{]} & +0.296 & {[}+0.290, +0.303{]} & 0.480 & 0.377 & 0.183 & 21 & 62 \\
metal/insulator & 0.845 & +0.063 & {[}+0.059, +0.067{]} & +0.069 & {[}+0.065, +0.072{]} & 0.155 & 0.091 & 0.086 & 41 & 44 \\
magnetic label & 0.860 & +0.030 & {[}+0.027, +0.034{]} & +0.043 & {[}+0.040, +0.047{]} & 0.140 & 0.110 & 0.097 & 22 & 31 \\
\bottomrule\end{tabular}
\end{minipage}

\clearpage
\hypertarget{tab:S7}{}%
{\small \textbf{Table~S7 --- Statistical comparisons of prediction scores across network layers (\hyperlink{property-decodability-depends-on-network-depth}{§3.3}).} \emph{Nadeau--Bengio corrected tests compare \textbf{(a)} the best message-passing layer with the initial encoder (ORB-v3) or embedding block (UMA-S), and \textbf{(b)} the best and worst message-passing layers. All comparisons use node features and five retrainings. Scores are R² for continuous targets and macro-F1 for categorical targets. \textbf{Comparison} identifies the two layers being evaluated. \textbf{Mean difference} is the first layer's score minus the second's, averaged over five retrainings. The \textbf{95\% CI} is the Nadeau--Bengio corrected confidence interval; \textbf{p} and \textbf{q} are the unadjusted and Benjamini--Hochberg-adjusted p-values. \textbf{Effect} classifies the absolute mean difference as trivial (below 0.005), small (0.005--0.02), moderate (0.02--0.05) or large (at least 0.05). For UMA-S, \textbf{view} identifies the equivariant channel representation, with crystal-system comparisons restricted to rotation-invariant views (\hyperlink{uma-channel-views-and-the-rotation-invariant-crystal-system-gap}{§3.4}, \hyperlink{tab:S13}{Table~S13}). A dash in (a) indicates that the initial embedding block is already the best layer. Individual retraining differences are shown in \hyperlink{fig:3}{Fig.~3c--d}. The XGBoost--linear score differences and their paired-bootstrap confidence intervals are shown separately in \hyperlink{fig:3}{Fig.~3e}.}\par}

\noindent\begin{minipage}{\linewidth}
\textbf{(a) Best layer vs raw encoder/embed --- ORB-v3}\par\nobreak
\begin{tabular}{@{}
  >{\raggedright\arraybackslash}p{(\columnwidth - 12\tabcolsep) * \real{0.2167}}
  >{\raggedright\arraybackslash}p{(\columnwidth - 12\tabcolsep) * \real{0.1583}}
  >{\raggedright\arraybackslash}p{(\columnwidth - 12\tabcolsep) * \real{0.1333}}
  >{\raggedright\arraybackslash}p{(\columnwidth - 12\tabcolsep) * \real{0.2083}}
  >{\raggedright\arraybackslash}p{(\columnwidth - 12\tabcolsep) * \real{0.0917}}
  >{\raggedright\arraybackslash}p{(\columnwidth - 12\tabcolsep) * \real{0.0917}}
  >{\raggedright\arraybackslash}p{(\columnwidth - 12\tabcolsep) * \real{0.1000}}@{}}
\toprule\noalign{}
\begin{minipage}[c]{\linewidth}\raggedright
target
\end{minipage} & \begin{minipage}[c]{\linewidth}\raggedright
comparison
\end{minipage} & \begin{minipage}[c]{\linewidth}\raggedright
mean difference
\end{minipage} & \begin{minipage}[c]{\linewidth}\raggedright
95\% CI
\end{minipage} & \begin{minipage}[c]{\linewidth}\raggedright
p
\end{minipage} & \begin{minipage}[c]{\linewidth}\raggedright
q
\end{minipage} & \begin{minipage}[c]{\linewidth}\raggedright
effect
\end{minipage} \\
\midrule\noalign{}
formation E/atom & gnn\_0 vs enc & +0.111 & {[}+0.104, +0.118{]} & 1.7e-06 & 4.5e-06 & large \\
energy above hull & gnn\_4 vs enc & +0.715 & {[}+0.694, +0.735{]} & 6.7e-08 & 4.2e-07 & large \\
band gap & gnn\_2 vs enc & +0.208 & {[}+0.201, +0.214{]} & 7.8e-08 & 4.2e-07 & large \\
density & gnn\_0 vs enc & +0.124 & {[}+0.113, +0.135{]} & 6.5e-06 & 1.0e-05 & large \\
bulk modulus & gnn\_0 vs enc & +0.081 & {[}+0.061, +0.100{]} & 3.1e-04 & 3.8e-04 & large \\
mag. density & gnn\_0 vs enc & +0.028 & {[}+0.019, +0.037{]} & 8.6e-04 & 9.1e-04 & moderate \\
crystal system & gnn\_2 vs enc & +0.139 & {[}+0.135, +0.143{]} & 7.0e-08 & 4.2e-07 & large \\
metal/insulator & gnn\_3 vs enc & +0.068 & {[}+0.063, +0.072{]} & 1.4e-06 & 4.5e-06 & large \\
magnetic label & gnn\_4 vs enc & +0.044 & {[}+0.040, +0.047{]} & 3.7e-06 & 6.5e-06 & moderate \\
\bottomrule\end{tabular}
\end{minipage}

\noindent\begin{minipage}{\linewidth}
\textbf{(a) Best layer vs raw encoder/embed --- UMA-S}\par\nobreak
\begin{tabular}{@{}
  >{\raggedright\arraybackslash}p{(\columnwidth - 14\tabcolsep) * \real{0.2017}}
  >{\raggedright\arraybackslash}p{(\columnwidth - 14\tabcolsep) * \real{0.0924}}
  >{\raggedright\arraybackslash}p{(\columnwidth - 14\tabcolsep) * \real{0.1345}}
  >{\raggedright\arraybackslash}p{(\columnwidth - 14\tabcolsep) * \real{0.1176}}
  >{\raggedright\arraybackslash}p{(\columnwidth - 14\tabcolsep) * \real{0.1933}}
  >{\raggedright\arraybackslash}p{(\columnwidth - 14\tabcolsep) * \real{0.0840}}
  >{\raggedright\arraybackslash}p{(\columnwidth - 14\tabcolsep) * \real{0.0840}}
  >{\raggedright\arraybackslash}p{(\columnwidth - 14\tabcolsep) * \real{0.0924}}@{}}
\toprule\noalign{}
\begin{minipage}[c]{\linewidth}\raggedright
target
\end{minipage} & \begin{minipage}[c]{\linewidth}\raggedright
view
\end{minipage} & \begin{minipage}[c]{\linewidth}\raggedright
comparison
\end{minipage} & \begin{minipage}[c]{\linewidth}\raggedright
mean difference
\end{minipage} & \begin{minipage}[c]{\linewidth}\raggedright
95\% CI
\end{minipage} & \begin{minipage}[c]{\linewidth}\raggedright
p
\end{minipage} & \begin{minipage}[c]{\linewidth}\raggedright
q
\end{minipage} & \begin{minipage}[c]{\linewidth}\raggedright
effect
\end{minipage} \\
\midrule\noalign{}
formation E/atom & ang-norm & bb vs emb & +0.012 & {[}+0.011, +0.013{]} & 2.7e-06 & 5.4e-06 & small \\
energy above hull & l0 & bb vs emb & +0.196 & {[}+0.164, +0.228{]} & 6.9e-05 & 9.2e-05 & large \\
band gap & ang-norm & bb vs emb & +0.143 & {[}+0.136, +0.149{]} & 3.6e-07 & 1.4e-06 & large \\
density & ang-norm & emb is best & --- & --- & --- & --- & --- \\
bulk modulus & ang-norm & emb is best & --- & --- & --- & --- & --- \\
mag. density & ang-norm & b1 vs emb & +0.003 & {[}-0.003, +0.010{]} & 2.3e-01 & 2.3e-01 & trivial \\
crystal system & ang-norm & b0 vs emb & +0.014 & {[}+0.012, +0.017{]} & 7.0e-05 & 8.2e-05 & small \\
metal/insulator & ang-norm & bb vs emb & +0.044 & {[}+0.041, +0.047{]} & 2.4e-06 & 5.4e-06 & moderate \\
magnetic label & ang-norm & bb vs emb & +0.035 & {[}+0.030, +0.040{]} & 4.2e-05 & 6.0e-05 & moderate \\
\bottomrule\end{tabular}
\end{minipage}

\noindent\begin{minipage}{\linewidth}
\textbf{(b) Best vs worst GNN layer --- ORB-v3}\par\nobreak
\begin{tabular}{@{}
  >{\raggedright\arraybackslash}p{(\columnwidth - 12\tabcolsep) * \real{0.2167}}
  >{\raggedright\arraybackslash}p{(\columnwidth - 12\tabcolsep) * \real{0.1750}}
  >{\raggedright\arraybackslash}p{(\columnwidth - 12\tabcolsep) * \real{0.1250}}
  >{\raggedright\arraybackslash}p{(\columnwidth - 12\tabcolsep) * \real{0.2000}}
  >{\raggedright\arraybackslash}p{(\columnwidth - 12\tabcolsep) * \real{0.0917}}
  >{\raggedright\arraybackslash}p{(\columnwidth - 12\tabcolsep) * \real{0.0917}}
  >{\raggedright\arraybackslash}p{(\columnwidth - 12\tabcolsep) * \real{0.1000}}@{}}
\toprule\noalign{}
\begin{minipage}[c]{\linewidth}\raggedright
target
\end{minipage} & \begin{minipage}[c]{\linewidth}\raggedright
comparison
\end{minipage} & \begin{minipage}[c]{\linewidth}\raggedright
mean difference
\end{minipage} & \begin{minipage}[c]{\linewidth}\raggedright
95\% CI
\end{minipage} & \begin{minipage}[c]{\linewidth}\raggedright
p
\end{minipage} & \begin{minipage}[c]{\linewidth}\raggedright
q
\end{minipage} & \begin{minipage}[c]{\linewidth}\raggedright
effect
\end{minipage} \\
\midrule\noalign{}
formation E/atom & gnn\_0 vs gnn\_3 & +0.002 & {[}+0.002, +0.003{]} & 3.2e-04 & 4.4e-04 & trivial \\
energy above hull & gnn\_4 vs gnn\_0 & +0.007 & {[}+0.004, +0.010{]} & 2.3e-03 & 2.6e-03 & small \\
band gap & gnn\_2 vs gnn\_0 & +0.105 & {[}+0.100, +0.111{]} & 7.5e-07 & 6.0e-06 & large \\
density & gnn\_0 vs gnn\_4 & +0.036 & {[}+0.033, +0.038{]} & 1.1e-06 & 6.0e-06 & moderate \\
bulk modulus & gnn\_0 vs gnn\_3 & +0.013 & {[}+0.007, +0.020{]} & 4.5e-03 & 4.8e-03 & small \\
mag. density & gnn\_0 vs gnn\_3 & +0.036 & {[}+0.030, +0.041{]} & 5.0e-05 & 1.1e-04 & moderate \\
crystal system & gnn\_2 vs gnn\_0 & +0.060 & {[}+0.056, +0.063{]} & 1.3e-06 & 6.0e-06 & large \\
metal/insulator & gnn\_3 vs gnn\_0 & +0.036 & {[}+0.033, +0.040{]} & 9.2e-06 & 2.8e-05 & moderate \\
magnetic label & gnn\_4 vs gnn\_0 & +0.014 & {[}+0.012, +0.017{]} & 1.5e-04 & 2.3e-04 & small \\
\bottomrule\end{tabular}
\end{minipage}

\noindent\begin{minipage}{\linewidth}
\textbf{(b) Best vs worst GNN layer --- UMA-S}\par\nobreak
\begin{tabular}{@{}
  >{\raggedright\arraybackslash}p{(\columnwidth - 14\tabcolsep) * \real{0.2083}}
  >{\raggedright\arraybackslash}p{(\columnwidth - 14\tabcolsep) * \real{0.1000}}
  >{\raggedright\arraybackslash}p{(\columnwidth - 14\tabcolsep) * \real{0.1167}}
  >{\raggedright\arraybackslash}p{(\columnwidth - 14\tabcolsep) * \real{0.1167}}
  >{\raggedright\arraybackslash}p{(\columnwidth - 14\tabcolsep) * \real{0.1917}}
  >{\raggedright\arraybackslash}p{(\columnwidth - 14\tabcolsep) * \real{0.0833}}
  >{\raggedright\arraybackslash}p{(\columnwidth - 14\tabcolsep) * \real{0.0833}}
  >{\raggedright\arraybackslash}p{(\columnwidth - 14\tabcolsep) * \real{0.1000}}@{}}
\toprule\noalign{}
\begin{minipage}[c]{\linewidth}\raggedright
target
\end{minipage} & \begin{minipage}[c]{\linewidth}\raggedright
view
\end{minipage} & \begin{minipage}[c]{\linewidth}\raggedright
comparison
\end{minipage} & \begin{minipage}[c]{\linewidth}\raggedright
mean difference
\end{minipage} & \begin{minipage}[c]{\linewidth}\raggedright
95\% CI
\end{minipage} & \begin{minipage}[c]{\linewidth}\raggedright
p
\end{minipage} & \begin{minipage}[c]{\linewidth}\raggedright
q
\end{minipage} & \begin{minipage}[c]{\linewidth}\raggedright
effect
\end{minipage} \\
\midrule\noalign{}
formation E/atom & ang-norm & bb vs b0 & +0.002 & {[}+0.001, +0.003{]} & 1.2e-03 & 1.4e-03 & trivial \\
energy above hull & l0 & bb vs b0 & +0.022 & {[}+0.016, +0.027{]} & 4.6e-04 & 5.9e-04 & moderate \\
band gap & ang-norm & bb vs b0 & +0.078 & {[}+0.076, +0.081{]} & 1.3e-07 & 2.4e-06 & large \\
density & ang-norm & emb vs b3 & +0.045 & {[}+0.042, +0.048{]} & 2.1e-06 & 7.7e-06 & moderate \\
bulk modulus & ang-norm & emb vs b3 & +0.015 & {[}+0.007, +0.022{]} & 4.8e-03 & 4.8e-03 & small \\
mag. density & ang-norm & b1 vs b2 & +0.017 & {[}+0.014, +0.021{]} & 1.3e-04 & 2.1e-04 & small \\
crystal system & ang-norm & b0 vs b3 & +0.017 & {[}+0.013, +0.022{]} & 5.0e-04 & 8.0e-04 & small \\
metal/insulator & ang-norm & bb vs b0 & +0.029 & {[}+0.026, +0.032{]} & 2.0e-05 & 5.2e-05 & moderate \\
magnetic label & ang-norm & bb vs b0 & +0.021 & {[}+0.018, +0.025{]} & 8.4e-05 & 1.7e-04 & moderate \\
\bottomrule\end{tabular}
\end{minipage}

\clearpage
\hypertarget{tab:S8}{}%
{\small \textbf{Table~S8 --- Cross-model representation transfer between ORB-v3 and UMA-S (\hyperlink{orb-and-uma-transfer-function-across-layers-but-converge-geometrically-only-in-part}{§B.3}).} \emph{\textbf{(a--b) Functional stitching.} We test whether a property predictor trained on one model can use embeddings from the other. First, a ridge regression maps the source model's pooled node embeddings into the destination model's embedding space. A linear probe is trained on the destination model's original embeddings, then evaluated using the mapped source embeddings without retraining. Both directions use 40,000 materials with an 80/20 train/test split. \textbf{UMA native} and \textbf{ORB native} are scores using each model's original embeddings; \textbf{stitched ORB→UMA} and \textbf{stitched UMA→ORB} are scores after replacing them with mapped embeddings. \textbf{Δ} is the stitched score minus the native score, so values near zero indicate little loss in predictive performance. Scores are R² for regression and macro-F1 for classification. \textbf{(a)} Deep-layer comparisons. \textbf{(b)} Near-input comparisons, where the low effective dimensionality of ORB's encoder (9 versus 36 for UMA's embedding layer; \hyperlink{representation-health-tracks-architectural-function}{§B.1}) limits its ability to reproduce UMA's representation. \textbf{(c) Linear predictivity.} We separately test how well a ridge regression reconstructs one model's atom embeddings from the other's, without using property labels. The map is trained on 70~\% of 30,005 aligned atoms and evaluated on the remaining 30~\%, after removing element-identity effects. \textbf{Comparison} indicates the mapping direction; \textbf{Source dim} and \textbf{Predicted dim} give the respective embedding dimensions. \textbf{Variance-weighted R²} measures how accurately the destination embeddings are reconstructed. Results cover both directions across all 36 layer pairs (\hyperlink{fig:S11}{Fig.~S11e--f}), with consecutive layers within each model and shuffled atom correspondences as reference comparisons. Summary ranges exclude ORB's encoder, whose variance is almost entirely removed by the element-identity correction.}\par}

\noindent\begin{minipage}{\linewidth}
\textbf{(a) Deep layers: ORB g4 ↔ UMA backbone, both directions}\par\nobreak
\begin{tabular}{@{}
  >{\raggedright\arraybackslash}p{(\columnwidth - 14\tabcolsep) * \real{0.2583}}
  >{\raggedright\arraybackslash}p{(\columnwidth - 14\tabcolsep) * \real{0.1250}}
  >{\raggedright\arraybackslash}p{(\columnwidth - 14\tabcolsep) * \real{0.0917}}
  >{\raggedright\arraybackslash}p{(\columnwidth - 14\tabcolsep) * \real{0.1250}}
  >{\raggedright\arraybackslash}p{(\columnwidth - 14\tabcolsep) * \real{0.0917}}
  >{\raggedright\arraybackslash}p{(\columnwidth - 14\tabcolsep) * \real{0.0917}}
  >{\raggedright\arraybackslash}p{(\columnwidth - 14\tabcolsep) * \real{0.1250}}
  >{\raggedright\arraybackslash}p{(\columnwidth - 14\tabcolsep) * \real{0.0917}}@{}}
\toprule\noalign{}
\begin{minipage}[c]{\linewidth}\raggedright
Target
\end{minipage} & \begin{minipage}[c]{\linewidth}\raggedright
Metric
\end{minipage} & \begin{minipage}[c]{\linewidth}\raggedright
UMA native
\end{minipage} & \begin{minipage}[c]{\linewidth}\raggedright
stitched ORB→UMA
\end{minipage} & \begin{minipage}[c]{\linewidth}\raggedright
Δ
\end{minipage} & \begin{minipage}[c]{\linewidth}\raggedright
ORB native
\end{minipage} & \begin{minipage}[c]{\linewidth}\raggedright
stitched UMA→ORB
\end{minipage} & \begin{minipage}[c]{\linewidth}\raggedright
Δ
\end{minipage} \\
\midrule\noalign{}
formation E/atom & R² & 0.985 & 0.974 & -0.011 & 0.978 & 0.978 & 0.000 \\
energy above hull & R² & 0.912 & 0.856 & -0.056 & 0.873 & 0.877 & +0.004 \\
band gap & R² & 0.763 & 0.704 & -0.059 & 0.730 & 0.721 & -0.009 \\
density & R² & 0.885 & 0.867 & -0.018 & 0.902 & 0.855 & -0.047 \\
bulk modulus & R² & 0.915 & 0.881 & -0.034 & 0.912 & 0.889 & -0.023 \\
mag. density & R² & 0.721 & 0.685 & -0.036 & 0.717 & 0.676 & -0.041 \\
crystal system & macro-F1 & 0.523 & 0.432 & -0.091 & 0.471 & 0.437 & -0.035 \\
metal/insulator & macro-F1 & 0.889 & 0.871 & -0.019 & 0.882 & 0.875 & -0.006 \\
magnetic label & macro-F1 & 0.883 & 0.844 & -0.039 & 0.857 & 0.864 & +0.007 \\
\bottomrule\end{tabular}
\end{minipage}

\noindent\begin{minipage}{\linewidth}
\textbf{(b) Near-input layers: ORB encoder ↔ UMA embed, both directions}\par\nobreak
\begin{tabular}{@{}
  >{\raggedright\arraybackslash}p{(\columnwidth - 14\tabcolsep) * \real{0.2583}}
  >{\raggedright\arraybackslash}p{(\columnwidth - 14\tabcolsep) * \real{0.1250}}
  >{\raggedright\arraybackslash}p{(\columnwidth - 14\tabcolsep) * \real{0.0917}}
  >{\raggedright\arraybackslash}p{(\columnwidth - 14\tabcolsep) * \real{0.1250}}
  >{\raggedright\arraybackslash}p{(\columnwidth - 14\tabcolsep) * \real{0.0917}}
  >{\raggedright\arraybackslash}p{(\columnwidth - 14\tabcolsep) * \real{0.0917}}
  >{\raggedright\arraybackslash}p{(\columnwidth - 14\tabcolsep) * \real{0.1250}}
  >{\raggedright\arraybackslash}p{(\columnwidth - 14\tabcolsep) * \real{0.0917}}@{}}
\toprule\noalign{}
\begin{minipage}[c]{\linewidth}\raggedright
Target
\end{minipage} & \begin{minipage}[c]{\linewidth}\raggedright
Metric
\end{minipage} & \begin{minipage}[c]{\linewidth}\raggedright
UMA native
\end{minipage} & \begin{minipage}[c]{\linewidth}\raggedright
stitched ORB→UMA
\end{minipage} & \begin{minipage}[c]{\linewidth}\raggedright
Δ
\end{minipage} & \begin{minipage}[c]{\linewidth}\raggedright
ORB native
\end{minipage} & \begin{minipage}[c]{\linewidth}\raggedright
stitched UMA→ORB
\end{minipage} & \begin{minipage}[c]{\linewidth}\raggedright
Δ
\end{minipage} \\
\midrule\noalign{}
formation E/atom & R² & 0.958 & 0.673 & -0.285 & 0.673 & 0.690 & +0.017 \\
energy above hull & R² & 0.794 & 0.073 & -0.721 & 0.074 & 0.086 & +0.012 \\
band gap & R² & 0.587 & 0.353 & -0.235 & 0.358 & 0.363 & +0.005 \\
density & R² & 0.981 & 0.764 & -0.217 & 0.765 & 0.776 & +0.011 \\
bulk modulus & R² & 0.954 & 0.622 & -0.332 & 0.776 & 0.780 & +0.003 \\
mag. density & R² & 0.763 & 0.647 & -0.116 & 0.651 & 0.658 & +0.008 \\
crystal system & macro-F1 & 0.495 & 0.253 & -0.243 & 0.301 & 0.294 & -0.007 \\
metal/insulator & macro-F1 & 0.835 & 0.772 & -0.063 & 0.780 & 0.785 & +0.006 \\
magnetic label & macro-F1 & 0.842 & 0.779 & -0.063 & 0.786 & 0.784 & -0.002 \\
\bottomrule\end{tabular}
\end{minipage}

\noindent\begin{minipage}{\linewidth}
\textbf{(c) Linear predictivity across layer pairs, both directions (UMA ang-norm channel)}\par\nobreak
\begin{tabular}{@{}
  >{\raggedright\arraybackslash}p{(\columnwidth - 6\tabcolsep) * \real{0.5917}}
  >{\raggedright\arraybackslash}p{(\columnwidth - 6\tabcolsep) * \real{0.0750}}
  >{\raggedright\arraybackslash}p{(\columnwidth - 6\tabcolsep) * \real{0.1167}}
  >{\raggedright\arraybackslash}p{(\columnwidth - 6\tabcolsep) * \real{0.2167}}@{}}
\toprule\noalign{}
\begin{minipage}[c]{\linewidth}\raggedright
Comparison
\end{minipage} & \begin{minipage}[c]{\linewidth}\raggedright
Source dim
\end{minipage} & \begin{minipage}[c]{\linewidth}\raggedright
Predicted dim
\end{minipage} & \begin{minipage}[c]{\linewidth}\raggedright
Variance-weighted R²
\end{minipage} \\
\midrule\noalign{}
ORB g4 predicting UMA backbone (deepest pair) & 256 & 384 & 0.503 \\
UMA backbone predicting ORB g4 (deepest pair) & 384 & 256 & 0.425 \\
ORB predicting UMA, best layer pair (g1 × emb) & 256 & 384 & 0.609 \\
UMA predicting ORB, best layer pair (b0 × g0) & 384 & 256 & 0.764 \\
ORB predicting UMA, all layer pairs (range) & 256 & 384 & 0.41--0.61 \\
UMA predicting ORB, all layer pairs (range) & 384 & 256 & 0.30--0.76 \\
within-ORB consecutive layers (reference range) & 256 & 256 & 0.47--1.00 \\
within-UMA consecutive layers (reference range) & 384 & 384 & 0.64--0.99 \\
shuffled-atom map (floor) & 256 & 384 & −0.013 \\
\bottomrule\end{tabular}
\end{minipage}

\par\addvspace{0.8\baselineskip}
\noindent\begin{minipage}{\linewidth}
\hypertarget{tab:S9}{}%
{\small \textbf{Table~S9 --- Validation-split selection control (\hyperlink{frozen-mlip-embeddings-predict-properties-outside-the-training-objective}{§3.1}).} \emph{Each model's peak configuration is normally chosen on the test set; here it is instead chosen on a validation split (20~\% of the training data), retrained on the full training set, and scored once on the untouched test set. \textbf{test-sel} is the score of the test-selected configuration (as in \textbf{\hyperlink{tab:1}{Table~1}}); \textbf{val-sel} is the test score of the validation-selected configuration; the \textbf{inflation} columns are their difference, test-sel − val-sel. The last two columns give the UMA − ORB difference under each selection rule. Regression scored by R², classification by macro-F1. For crystal system, UMA-S's configuration is selected among the rotation-invariant ones, since the non-invariant configurations collapse under random rotation (\hyperlink{uma-channel-views-and-the-rotation-invariant-crystal-system-gap}{§3.4}, \hyperlink{tab:S13}{Table~S13}).}\par}\par\nobreak
\begin{tabular}{@{}
  >{\raggedright\arraybackslash}p{(\columnwidth - 16\tabcolsep) * \real{0.2521}}
  >{\raggedright\arraybackslash}p{(\columnwidth - 16\tabcolsep) * \real{0.0924}}
  >{\raggedright\arraybackslash}p{(\columnwidth - 16\tabcolsep) * \real{0.0756}}
  >{\raggedright\arraybackslash}p{(\columnwidth - 16\tabcolsep) * \real{0.1008}}
  >{\raggedright\arraybackslash}p{(\columnwidth - 16\tabcolsep) * \real{0.0924}}
  >{\raggedright\arraybackslash}p{(\columnwidth - 16\tabcolsep) * \real{0.0756}}
  >{\raggedright\arraybackslash}p{(\columnwidth - 16\tabcolsep) * \real{0.1008}}
  >{\raggedright\arraybackslash}p{(\columnwidth - 16\tabcolsep) * \real{0.1092}}
  >{\raggedright\arraybackslash}p{(\columnwidth - 16\tabcolsep) * \real{0.1008}}@{}}
\toprule\noalign{}
\begin{minipage}[c]{\linewidth}\raggedright
Target
\end{minipage} & \begin{minipage}[c]{\linewidth}\raggedright
ORB test-sel
\end{minipage} & \begin{minipage}[c]{\linewidth}\raggedright
ORB val-sel
\end{minipage} & \begin{minipage}[c]{\linewidth}\raggedright
ORB inflation
\end{minipage} & \begin{minipage}[c]{\linewidth}\raggedright
UMA test-sel
\end{minipage} & \begin{minipage}[c]{\linewidth}\raggedright
UMA val-sel
\end{minipage} & \begin{minipage}[c]{\linewidth}\raggedright
UMA inflation
\end{minipage} & \begin{minipage}[c]{\linewidth}\raggedright
UMA − ORB (test-sel)
\end{minipage} & \begin{minipage}[c]{\linewidth}\raggedright
UMA − ORB (val-sel)
\end{minipage} \\
\midrule\noalign{}
crystal system (F1) & 0.622 & 0.622 & +0.000 & 0.817 & 0.817 & +0.000 & +0.194 & +0.194 \\
metal/insulator (F1) & 0.909 & 0.908 & +0.001 & 0.914 & 0.914 & +0.000 & +0.005 & +0.006 \\
magnetic label (F1) & 0.890 & 0.890 & +0.000 & 0.903 & 0.903 & +0.000 & +0.013 & +0.013 \\
formation E/atom (R²) & 0.990 & 0.990 & +0.000 & 0.990 & 0.990 & +0.000 & +0.000 & +0.000 \\
energy above hull (R²) & 0.948 & 0.948 & +0.000 & 0.960 & 0.959 & +0.002 & +0.012 & +0.011 \\
band gap (R²) & 0.890 & 0.890 & +0.000 & 0.907 & 0.907 & +0.000 & +0.016 & +0.016 \\
density (R²) & 0.982 & 0.982 & +0.000 & 0.983 & 0.983 & +0.000 & +0.001 & +0.001 \\
bulk modulus (R²) & 0.951 & 0.951 & +0.000 & 0.958 & 0.958 & +0.000 & +0.008 & +0.008 \\
mag. density (R²) & 0.846 & 0.846 & +0.000 & 0.839 & 0.839 & +0.000 & −0.007 & −0.007 \\
\bottomrule\end{tabular}
\end{minipage}

\par\addvspace{0.8\baselineskip}
\hypertarget{tab:S10}{}%
{\small \textbf{Table~S10 --- Effect of composition- and chemical-system-grouped splits on prediction scores (\hyperlink{frozen-mlip-embeddings-predict-properties-outside-the-training-objective}{§3.1}).} \emph{Each model's selected configuration from \textbf{\hyperlink{tab:1}{Table~1}} is refitted and evaluated under three 80/20 splits. \textbf{Random} is the original seed-42 split and reproduces the reported scores within 0.001. \textbf{Comp-grouped} keeps materials with the same reduced formula in one split, while \textbf{chemsys-grouped} keeps materials containing the same set of elements together. Each \textbf{Δ} is the random-split score minus the corresponding grouped-split score; positive values indicate lower performance under grouping. Scores are R² for continuous targets and macro-F1 for categorical targets. UMA-S crystal-system results use its best rotation-invariant configuration (concat · ang-norm · b0; \hyperlink{uma-channel-views-and-the-rotation-invariant-crystal-system-gap}{§3.4}, \hyperlink{tab:S13}{Table~S13}).}\par}

\noindent\begin{minipage}{\linewidth}
\textbf{ORB-v3}\par\nobreak
\begin{tabular}{@{}
  >{\raggedright\arraybackslash}p{(\columnwidth - 10\tabcolsep) * \real{0.2400}}
  >{\raggedright\arraybackslash}p{(\columnwidth - 10\tabcolsep) * \real{0.1000}}
  >{\raggedright\arraybackslash}p{(\columnwidth - 10\tabcolsep) * \real{0.1300}}
  >{\raggedright\arraybackslash}p{(\columnwidth - 10\tabcolsep) * \real{0.2050}}
  >{\raggedright\arraybackslash}p{(\columnwidth - 10\tabcolsep) * \real{0.1200}}
  >{\raggedright\arraybackslash}p{(\columnwidth - 10\tabcolsep) * \real{0.2050}}@{}}
\toprule\noalign{}
\begin{minipage}[c]{\linewidth}\raggedright
Target
\end{minipage} & \begin{minipage}[c]{\linewidth}\raggedright
random
\end{minipage} & \begin{minipage}[c]{\linewidth}\raggedright
comp-grouped
\end{minipage} & \begin{minipage}[c]{\linewidth}\raggedright
Δ (random − comp-grouped)
\end{minipage} & \begin{minipage}[c]{\linewidth}\raggedright
chemsys-grouped
\end{minipage} & \begin{minipage}[c]{\linewidth}\raggedright
Δ (random − chemsys-grouped)
\end{minipage} \\
\midrule\noalign{}
crystal system (F1) & 0.623 & 0.608 & 0.015 & 0.585 & 0.038 \\
metal/insulator (F1) & 0.909 & 0.903 & 0.005 & 0.909 & 0.000 \\
magnetic label (F1) & 0.890 & 0.887 & 0.003 & 0.877 & 0.013 \\
formation E/atom (R²) & 0.990 & 0.988 & 0.002 & 0.989 & 0.000 \\
energy above hull (R²) & 0.948 & 0.929 & 0.019 & 0.944 & 0.003 \\
band gap (R²) & 0.890 & 0.884 & 0.006 & 0.869 & 0.021 \\
density (R²) & 0.982 & 0.979 & 0.003 & 0.978 & 0.004 \\
bulk modulus (R²) & 0.951 & 0.949 & 0.002 & 0.959 & −0.008 \\
mag. density (R²) & 0.846 & 0.827 & 0.019 & 0.816 & 0.030 \\
\bottomrule\end{tabular}
\end{minipage}

\noindent\begin{minipage}{\linewidth}
\textbf{UMA-S}\par\nobreak
\begin{tabular}{@{}
  >{\raggedright\arraybackslash}p{(\columnwidth - 10\tabcolsep) * \real{0.2400}}
  >{\raggedright\arraybackslash}p{(\columnwidth - 10\tabcolsep) * \real{0.1000}}
  >{\raggedright\arraybackslash}p{(\columnwidth - 10\tabcolsep) * \real{0.1300}}
  >{\raggedright\arraybackslash}p{(\columnwidth - 10\tabcolsep) * \real{0.2050}}
  >{\raggedright\arraybackslash}p{(\columnwidth - 10\tabcolsep) * \real{0.1200}}
  >{\raggedright\arraybackslash}p{(\columnwidth - 10\tabcolsep) * \real{0.2050}}@{}}
\toprule\noalign{}
\begin{minipage}[c]{\linewidth}\raggedright
Target
\end{minipage} & \begin{minipage}[c]{\linewidth}\raggedright
random
\end{minipage} & \begin{minipage}[c]{\linewidth}\raggedright
comp-grouped
\end{minipage} & \begin{minipage}[c]{\linewidth}\raggedright
Δ (random − comp-grouped)
\end{minipage} & \begin{minipage}[c]{\linewidth}\raggedright
chemsys-grouped
\end{minipage} & \begin{minipage}[c]{\linewidth}\raggedright
Δ (random − chemsys-grouped)
\end{minipage} \\
\midrule\noalign{}
crystal system (F1) & 0.817 & 0.819 & −0.002 & 0.800 & 0.016 \\
metal/insulator (F1) & 0.914 & 0.912 & 0.002 & 0.915 & −0.002 \\
magnetic label (F1) & 0.903 & 0.901 & 0.002 & 0.893 & 0.010 \\
formation E/atom (R²) & 0.990 & 0.988 & 0.002 & 0.988 & 0.002 \\
energy above hull (R²) & 0.960 & 0.949 & 0.011 & 0.959 & 0.001 \\
band gap (R²) & 0.907 & 0.905 & 0.002 & 0.891 & 0.016 \\
density (R²) & 0.983 & 0.981 & 0.002 & 0.980 & 0.004 \\
bulk modulus (R²) & 0.959 & 0.955 & 0.004 & 0.963 & −0.004 \\
mag. density (R²) & 0.839 & 0.820 & 0.020 & 0.805 & 0.034 \\
\bottomrule\end{tabular}
\end{minipage}

\clearpage
\hypertarget{tab:S11}{}%
{\small \textbf{Table~S11 --- Two scale controls on the headline band-gap and bulk-modulus scores (\hyperlink{robustness-controls}{§A.4}, \hyperlink{from-ruxb2-to-expected-downstream-accuracy}{§3.2}).} \emph{Both blocks use the probe settings and the seed-42 split of \textbf{\hyperlink{tab:1}{Table~1}}. \textbf{(a)} re-scores band gap on the non-metal test subset only (true gap \textgreater{}~0), since roughly 47~\% of the materials studied here are metals with a gap of exactly zero and a pooled R² therefore partly rewards separating metals from non-metals; the \emph{full} rows reproduce \textbf{\hyperlink{tab:1}{Table~1}}, \textbf{n} is the number of test materials scored and \textbf{σ} is recomputed on whichever set the row uses. \textbf{(b)} refits the bulk-modulus probe with log\textsubscript{10}(K\textsubscript{VRH}) as the target, the Matbench convention; the GPa score uses all 2,571 test materials, the log score the 2,569 with K \textgreater{}~0, both with 95~\% bootstrap confidence intervals from 2,000 resamples of the test set, and \textbf{Difference} is the log-scale R² minus the GPa one.}\par}

\noindent\begin{minipage}{\linewidth}
\textbf{(a) Band gap restricted to non-metals}\par\nobreak
\begin{tabular}{@{}
  >{\raggedright\arraybackslash}p{(\columnwidth - 10\tabcolsep) * \real{0.1429}}
  >{\raggedright\arraybackslash}p{(\columnwidth - 10\tabcolsep) * \real{0.3613}}
  >{\raggedright\arraybackslash}p{(\columnwidth - 10\tabcolsep) * \real{0.1429}}
  >{\raggedright\arraybackslash}p{(\columnwidth - 10\tabcolsep) * \real{0.1176}}
  >{\raggedright\arraybackslash}p{(\columnwidth - 10\tabcolsep) * \real{0.1176}}
  >{\raggedright\arraybackslash}p{(\columnwidth - 10\tabcolsep) * \real{0.1176}}@{}}
\toprule\noalign{}
\begin{minipage}[c]{\linewidth}\raggedright
Model
\end{minipage} & \begin{minipage}[c]{\linewidth}\raggedright
Test set
\end{minipage} & \begin{minipage}[c]{\linewidth}\raggedright
n
\end{minipage} & \begin{minipage}[c]{\linewidth}\raggedright
R²
\end{minipage} & \begin{minipage}[c]{\linewidth}\raggedright
RMSE (eV)
\end{minipage} & \begin{minipage}[c]{\linewidth}\raggedright
σ (eV)
\end{minipage} \\
\midrule\noalign{}
ORB-v3 & full & 30,976 & 0.890 & 0.504 & 1.521 \\
ORB-v3 & non-metals only & 16,452 & 0.859 & 0.589 & 1.568 \\
\midrule
UMA-S & full & 30,974 & 0.907 & 0.465 & 1.521 \\
UMA-S & non-metals only & 16,451 & 0.881 & 0.542 & 1.568 \\
\bottomrule\end{tabular}
\end{minipage}

\noindent\begin{minipage}{\linewidth}
\textbf{(b) Bulk modulus re-scored on log\textsubscript{10}(K\textsubscript{VRH})}\par\nobreak
\begin{tabular}{@{}
  >{\raggedright\arraybackslash}p{(\columnwidth - 10\tabcolsep) * \real{0.1200}}
  >{\raggedright\arraybackslash}p{(\columnwidth - 10\tabcolsep) * \real{0.2600}}
  >{\raggedright\arraybackslash}p{(\columnwidth - 10\tabcolsep) * \real{0.1400}}
  >{\raggedright\arraybackslash}p{(\columnwidth - 10\tabcolsep) * \real{0.1800}}
  >{\raggedright\arraybackslash}p{(\columnwidth - 10\tabcolsep) * \real{0.1800}}
  >{\raggedright\arraybackslash}p{(\columnwidth - 10\tabcolsep) * \real{0.1200}}@{}}
\toprule\noalign{}
\begin{minipage}[c]{\linewidth}\raggedright
Model
\end{minipage} & \begin{minipage}[c]{\linewidth}\raggedright
Configuration
\end{minipage} & \begin{minipage}[c]{\linewidth}\raggedright
n (GPa / log)
\end{minipage} & \begin{minipage}[c]{\linewidth}\raggedright
R² on K (GPa)
\end{minipage} & \begin{minipage}[c]{\linewidth}\raggedright
R² on log\textsubscript{10} K
\end{minipage} & \begin{minipage}[c]{\linewidth}\raggedright
Difference
\end{minipage} \\
\midrule\noalign{}
ORB-v3 & concat · g0 & 2,571 / 2,569 & 0.951 {[}0.938, 0.962{]} & 0.935 {[}0.922, 0.945{]} & −0.016 \\
UMA-S & concat · ang-norm · emb & 2,571 / 2,569 & 0.958 {[}0.948, 0.968{]} & 0.939 {[}0.924, 0.951{]} & −0.020 \\
\bottomrule\end{tabular}
\end{minipage}

\par\addvspace{0.8\baselineskip}
\noindent\begin{minipage}{\linewidth}
\hypertarget{tab:S12}{}%
{\small \textbf{Table~S12 --- Probe-selectivity control task (\hyperlink{property-decodability-depends-on-network-depth}{§3.3}).} \emph{The probe at each target's peak configuration is trained on the real labels (\textbf{real}) and, separately, on randomly permuted labels (\textbf{control}); selectivity = real − control. A probe that merely exploited its own capacity would score highly on permuted labels. Regression scored by R², classification by macro-F1; permuted-label control R² is ≈~0 (the probe cannot fit random continuous targets) and permuted-label control macro-F1 sits at the class-frequency chance level. Selectivity is therefore close to the full reported score for the continuous targets (it slightly exceeds it for bulk modulus, whose permuted-label R² is negative) and is the part of the score above chance for the categorical ones, confirming that decodability reflects the representation, not the probe. For crystal system, UMA-S's peak is reported as its best rotation-invariant configuration (concat · ang-norm · b0), since the higher-scoring non-invariant configurations collapse under random rotation (\hyperlink{uma-channel-views-and-the-rotation-invariant-crystal-system-gap}{§3.4}, \hyperlink{tab:S13}{Table~S13}).}\par}\par\nobreak
\begin{tabular}{@{}
  >{\raggedright\arraybackslash}p{(\columnwidth - 12\tabcolsep) * \real{0.3277}}
  >{\raggedright\arraybackslash}p{(\columnwidth - 12\tabcolsep) * \real{0.0756}}
  >{\raggedright\arraybackslash}p{(\columnwidth - 12\tabcolsep) * \real{0.1008}}
  >{\raggedright\arraybackslash}p{(\columnwidth - 12\tabcolsep) * \real{0.1597}}
  >{\raggedright\arraybackslash}p{(\columnwidth - 12\tabcolsep) * \real{0.0756}}
  >{\raggedright\arraybackslash}p{(\columnwidth - 12\tabcolsep) * \real{0.1008}}
  >{\raggedright\arraybackslash}p{(\columnwidth - 12\tabcolsep) * \real{0.1597}}@{}}
\toprule\noalign{}
\begin{minipage}[c]{\linewidth}\raggedright
Target
\end{minipage} & \begin{minipage}[c]{\linewidth}\raggedright
ORB real
\end{minipage} & \begin{minipage}[c]{\linewidth}\raggedright
ORB control
\end{minipage} & \begin{minipage}[c]{\linewidth}\raggedright
ORB selectivity
\end{minipage} & \begin{minipage}[c]{\linewidth}\raggedright
UMA real
\end{minipage} & \begin{minipage}[c]{\linewidth}\raggedright
UMA control
\end{minipage} & \begin{minipage}[c]{\linewidth}\raggedright
UMA selectivity
\end{minipage} \\
\midrule\noalign{}
crystal system (F1) & 0.622 & 0.093 & 0.530 & 0.817 & 0.090 & 0.726 \\
metal/insulator (F1) & 0.909 & 0.471 & 0.437 & 0.914 & 0.465 & 0.449 \\
magnetic label (F1) & 0.890 & 0.471 & 0.419 & 0.903 & 0.470 & 0.433 \\
formation E/atom (R²) & 0.990 & −0.010 & 0.999 & 0.990 & −0.007 & 0.997 \\
energy above hull (R²) & 0.948 & −0.018 & 0.966 & 0.960 & −0.014 & 0.974 \\
band gap (R²) & 0.890 & −0.010 & 0.900 & 0.907 & −0.008 & 0.915 \\
density (R²) & 0.982 & −0.010 & 0.991 & 0.983 & −0.008 & 0.991 \\
bulk modulus (R²) & 0.951 & −0.075 & 1.026 & 0.958 & −0.053 & 1.011 \\
mag. density (R²) & 0.846 & −0.009 & 0.855 & 0.839 & −0.010 & 0.849 \\
\bottomrule\end{tabular}
\end{minipage}

\clearpage
\noindent\begin{minipage}{\linewidth}
\hypertarget{tab:S13}{}%
{\small \textbf{Table~S13 --- Effect of rotation on crystal-system classification (\hyperlink{uma-channel-views-and-the-rotation-invariant-crystal-system-gap}{§3.4}).} \emph{Macro-F1 scores for ORB-v3 and UMA-S before and after rotating 1,000 test crystals (atoms and unit cell rotated together). Each row identifies the read-out, channel view (for UMA-S) and best-performing layer. \textbf{Test-set F1} is the score on the full test set (30,976 materials). \textbf{Control (unrot.)} and \textbf{Control (rot.)} are scores on the same 1,000 materials before and after rotation. \textbf{Δ} is their difference (unrotated minus rotated); larger values indicate greater sensitivity to orientation. \textbf{(a)} Pretrained models. \textbf{(b)} Random-weight UMA-S, with probes trained on its own embeddings. Full-tensor features lose accuracy under rotation, whereas rotation-invariant channels remain largely stable. The paper therefore uses the ang-norm view for UMA-S crystal-system classification (\textbf{\hyperlink{tab:1}{Table~1}}; \hyperlink{fig:4}{Fig.~4}).}\par}\par\nobreak
\textbf{(a) Pretrained networks}\par\nobreak
\begin{tabular}{@{}
  >{\raggedright\arraybackslash}p{(\columnwidth - 10\tabcolsep) * \real{0.0909}}
  >{\raggedright\arraybackslash}p{(\columnwidth - 10\tabcolsep) * \real{0.3967}}
  >{\raggedright\arraybackslash}p{(\columnwidth - 10\tabcolsep) * \real{0.1405}}
  >{\raggedright\arraybackslash}p{(\columnwidth - 10\tabcolsep) * \real{0.1405}}
  >{\raggedright\arraybackslash}p{(\columnwidth - 10\tabcolsep) * \real{0.1240}}
  >{\raggedright\arraybackslash}p{(\columnwidth - 10\tabcolsep) * \real{0.1074}}@{}}
\toprule\noalign{}
\begin{minipage}[c]{\linewidth}\raggedright
Model
\end{minipage} & \begin{minipage}[c]{\linewidth}\raggedright
Read-out · channel · layer
\end{minipage} & \begin{minipage}[c]{\linewidth}\raggedright
Test-set F1
\end{minipage} & \begin{minipage}[c]{\linewidth}\raggedright
Control (unrot.)
\end{minipage} & \begin{minipage}[c]{\linewidth}\raggedright
Control (rot.)
\end{minipage} & \begin{minipage}[c]{\linewidth}\raggedright
Δ
\end{minipage} \\
\midrule\noalign{}
UMA & concat · full · b0 & 0.897 & 0.902 & 0.408 & \textbf{+0.494} \\
UMA & edge · flat · b0 & 0.892 & 0.894 & 0.397 & +0.497 \\
UMA & node · full · b0 & 0.888 & 0.895 & 0.412 & +0.483 \\
UMA & concat · ang-norm · b0 & 0.817 & 0.834 & 0.813 & +0.021 \\
UMA & concat · geom-norm · b0 & 0.802 & 0.822 & 0.785 & +0.037 \\
UMA & edge · ang-norm · b0 & 0.793 & 0.808 & 0.769 & +0.039 \\
UMA & edge · geom-norm · b0 & 0.771 & 0.781 & 0.745 & +0.037 \\
UMA & concat · L=2 · b0 & 0.694 & 0.716 & 0.715 & +0.001 \\
UMA & node · ang-norm · b0 & 0.679 & 0.690 & 0.687 & +0.004 \\
UMA & edge · L=2 norm · b0 & 0.672 & 0.697 & 0.688 & +0.009 \\
UMA & node · geom-norm · b0 & 0.672 & 0.689 & 0.685 & +0.004 \\
UMA & node · L=2 polar · b0 & 0.662 & 0.688 & 0.689 & −0.001 \\
ORB & concat · g2 & 0.622 & 0.639 & 0.631 & +0.007 \\
ORB & node · g2 & 0.608 & 0.635 & 0.622 & +0.013 \\
ORB & edge · g3 & 0.602 & 0.606 & 0.605 & +0.000 \\
UMA & concat · L=1 · b0 & 0.581 & 0.584 & 0.512 & +0.072 \\
UMA & concat · L=0 · b2 & 0.581 & 0.621 & 0.621 & +0.000 \\
UMA & node · L=0 · bb & 0.577 & 0.588 & 0.588 & +0.000 \\
UMA & edge · L=0 · b2 & 0.570 & 0.600 & 0.600 & +0.000 \\
UMA & node · L=1 polar · b0 & 0.544 & 0.544 & 0.441 & +0.103 \\
UMA & edge · L=1 norm · b0 & 0.487 & 0.484 & 0.433 & +0.051 \\
\bottomrule\end{tabular}
\end{minipage}

\noindent\begin{minipage}{\linewidth}
\textbf{(b) UMA-S with random weights (\hyperlink{random-initialisation-and-matched-training-data-controls}{§A.9}), each read-out at its own best layer}\par\nobreak
\begin{tabular}{@{}
  >{\raggedright\arraybackslash}p{(\columnwidth - 10\tabcolsep) * \real{0.0909}}
  >{\raggedright\arraybackslash}p{(\columnwidth - 10\tabcolsep) * \real{0.3967}}
  >{\raggedright\arraybackslash}p{(\columnwidth - 10\tabcolsep) * \real{0.1405}}
  >{\raggedright\arraybackslash}p{(\columnwidth - 10\tabcolsep) * \real{0.1405}}
  >{\raggedright\arraybackslash}p{(\columnwidth - 10\tabcolsep) * \real{0.1240}}
  >{\raggedright\arraybackslash}p{(\columnwidth - 10\tabcolsep) * \real{0.1074}}@{}}
\toprule\noalign{}
\begin{minipage}[c]{\linewidth}\raggedright
Model
\end{minipage} & \begin{minipage}[c]{\linewidth}\raggedright
Read-out · channel · layer
\end{minipage} & \begin{minipage}[c]{\linewidth}\raggedright
Test-set F1
\end{minipage} & \begin{minipage}[c]{\linewidth}\raggedright
Control (unrot.)
\end{minipage} & \begin{minipage}[c]{\linewidth}\raggedright
Control (rot.)
\end{minipage} & \begin{minipage}[c]{\linewidth}\raggedright
Δ
\end{minipage} \\
\midrule\noalign{}
UMA & node · full · b1 & 0.845 & 0.854 & 0.325 & +0.529 \\
UMA & concat · full · emb & 0.858 & 0.853 & 0.318 & \textbf{+0.534} \\
UMA & edge · flat · emb & 0.845 & 0.847 & 0.321 & +0.525 \\
UMA & concat · ang-norm · emb & 0.661 & 0.660 & 0.652 & +0.008 \\
UMA & edge · ang-norm · emb & 0.627 & 0.628 & 0.629 & −0.001 \\
UMA & concat · geom-norm · emb & 0.629 & 0.626 & 0.626 & −0.000 \\
UMA & concat · L=2 · emb & 0.580 & 0.570 & 0.578 & −0.007 \\
UMA & edge · geom-norm · emb & 0.579 & 0.570 & 0.535 & +0.035 \\
UMA & node · geom-norm · b3 & 0.543 & 0.557 & 0.551 & +0.005 \\
UMA & node · ang-norm · b3 & 0.571 & 0.552 & 0.552 & +0.000 \\
UMA & node · L=2 polar · emb & 0.554 & 0.541 & 0.541 & +0.000 \\
UMA & edge · L=2 norm · emb & 0.516 & 0.513 & 0.412 & +0.101 \\
UMA & concat · L=1 · emb & 0.447 & 0.476 & 0.394 & +0.082 \\
UMA & node · L=1 polar · b0 & 0.431 & 0.444 & 0.389 & +0.054 \\
UMA & concat · L=0 · emb & 0.426 & 0.402 & 0.402 & +0.000 \\
UMA & edge · L=0 · emb & 0.407 & 0.396 & 0.396 & +0.000 \\
UMA & node · L=0 · b3 & 0.424 & 0.395 & 0.395 & +0.000 \\
UMA & edge · L=1 norm · emb & 0.325 & 0.327 & 0.280 & +0.047 \\
\bottomrule\end{tabular}
\end{minipage}

\clearpage
\hypertarget{tab:S14}{}%
{\small \textbf{Table~S14 --- Effect of rotation on frozen model features (\hyperlink{uma-channel-views-and-the-rotation-invariant-crystal-system-gap}{§3.4}).} \emph{Relative changes (dimensionless) in pooled embeddings after five random rotations of 300 test crystals (atoms and unit cell rotated together; 1,500 comparisons per layer), measured as ‖f(Rx) − f(x)‖ / ‖f(x)‖. \textbf{(a)} ORB-v3 and \textbf{(b)} UMA-S: median and maximum relative changes across layers for each read-out and channel view. \textbf{Layers} gives the number of layers evaluated, while \textbf{Identity vs cache} measures numerical differences between freshly extracted unrotated features and stored embeddings. Large maxima for the L=1 and L=2 norm views come from materials whose pooled vector in these channels is zero or nearly zero (for example highly symmetric crystals, where these components cancel), so the ratio divides by almost nothing; medians are more representative. \textbf{(c)} Median relative change in UMA-S full-tensor node features at each layer. Rotation changes the neighbour-graph edge count for only one of the 300 structures in each model, by at most two edges.}\par}

\noindent\begin{minipage}{\linewidth}
\textbf{(a) ORB-v3 --- relative change of each read-out}\par\nobreak
\begin{tabular}{@{}
  >{\raggedright\arraybackslash}p{(\columnwidth - 8\tabcolsep) * \real{0.1754}}
  >{\raggedright\arraybackslash}p{(\columnwidth - 8\tabcolsep) * \real{0.1228}}
  >{\raggedright\arraybackslash}p{(\columnwidth - 8\tabcolsep) * \real{0.2281}}
  >{\raggedright\arraybackslash}p{(\columnwidth - 8\tabcolsep) * \real{0.2281}}
  >{\raggedright\arraybackslash}p{(\columnwidth - 8\tabcolsep) * \real{0.2456}}@{}}
\toprule\noalign{}
\begin{minipage}[c]{\linewidth}\raggedright
Read-out
\end{minipage} & \begin{minipage}[c]{\linewidth}\raggedright
Layers
\end{minipage} & \begin{minipage}[c]{\linewidth}\raggedright
Median relative change
\end{minipage} & \begin{minipage}[c]{\linewidth}\raggedright
Max relative change
\end{minipage} & \begin{minipage}[c]{\linewidth}\raggedright
Identity vs cache (max)
\end{minipage} \\
\midrule\noalign{}
node & 6 & 1.18 × 10\textsuperscript{−2} & 4.71 × 10\textsuperscript{−1} & 3.60 × 10\textsuperscript{−4} \\
edge & 6 & 1.48 × 10\textsuperscript{−3} & 2.07 × 10\textsuperscript{−1} & 3.60 × 10\textsuperscript{−4} \\
concat & 6 & 5.96 × 10\textsuperscript{−3} & 2.02 × 10\textsuperscript{−1} & 3.08 × 10\textsuperscript{−4} \\
\bottomrule\end{tabular}
\end{minipage}

\noindent\begin{minipage}{\linewidth}
\textbf{(b) UMA-S --- relative change of each read-out and view}\par\nobreak
\begin{tabular}{@{}
  >{\raggedright\arraybackslash}p{(\columnwidth - 10\tabcolsep) * \real{0.1343}}
  >{\raggedright\arraybackslash}p{(\columnwidth - 10\tabcolsep) * \real{0.1642}}
  >{\raggedright\arraybackslash}p{(\columnwidth - 10\tabcolsep) * \real{0.1045}}
  >{\raggedright\arraybackslash}p{(\columnwidth - 10\tabcolsep) * \real{0.1940}}
  >{\raggedright\arraybackslash}p{(\columnwidth - 10\tabcolsep) * \real{0.1940}}
  >{\raggedright\arraybackslash}p{(\columnwidth - 10\tabcolsep) * \real{0.2090}}@{}}
\toprule\noalign{}
\begin{minipage}[c]{\linewidth}\raggedright
Read-out
\end{minipage} & \begin{minipage}[c]{\linewidth}\raggedright
View
\end{minipage} & \begin{minipage}[c]{\linewidth}\raggedright
Layers
\end{minipage} & \begin{minipage}[c]{\linewidth}\raggedright
Median relative change
\end{minipage} & \begin{minipage}[c]{\linewidth}\raggedright
Max relative change
\end{minipage} & \begin{minipage}[c]{\linewidth}\raggedright
Identity vs cache (max)
\end{minipage} \\
\midrule\noalign{}
node & full & 6 & 5.82 × 10\textsuperscript{−2} & 1.18 & 4.33 × 10\textsuperscript{−4} \\
node & L0 & 6 & 4.98 × 10\textsuperscript{−7} & 4.36 × 10\textsuperscript{−4} & 4.36 × 10\textsuperscript{−4} \\
node & ang-norm & 6 & 4.59 × 10\textsuperscript{−5} & 3.89 × 10\textsuperscript{−4} & 3.74 × 10\textsuperscript{−4} \\
node & ‖L1‖ polar & 6 & 7.91 × 10\textsuperscript{−2} & 1.10 × 10\textsuperscript{2} & 9.78 × 10\textsuperscript{1} \\
node & ‖L2‖ polar & 6 & 1.77 × 10\textsuperscript{−4} & 6.64 × 10\textsuperscript{2} & 1.51 × 10\textsuperscript{1} \\
node & geom-norm & 6 & 6.56 × 10\textsuperscript{−5} & 3.14 & 3.12 × 10\textsuperscript{−1} \\
edge & flat & 5 & 1.53 × 10\textsuperscript{−1} & 1.67 & 1.42 × 10\textsuperscript{−4} \\
edge & L0 & 5 & 9.91 × 10\textsuperscript{−6} & 3.30 × 10\textsuperscript{−4} & 1.49 × 10\textsuperscript{−4} \\
edge & ang-norm & 5 & 5.55 × 10\textsuperscript{−5} & 3.31 × 10\textsuperscript{−4} & 1.39 × 10\textsuperscript{−4} \\
edge & ‖L1‖ norm & 5 & 4.06 × 10\textsuperscript{−1} & 1.00 × 10\textsuperscript{12} & 1.00 × 10\textsuperscript{12} \\
edge & ‖L2‖ norm & 5 & 3.01 × 10\textsuperscript{−4} & 1.38 & 1.15 \\
edge & geom-norm & 5 & 3.94 × 10\textsuperscript{−4} & 1.38 & 1.15 \\
\bottomrule\end{tabular}
\end{minipage}

\noindent\begin{minipage}{\linewidth}
\textbf{(c) UMA-S full node tensor --- relative change by layer}\par\nobreak
\begin{tabular}{@{}
  >{\raggedright\arraybackslash}p{(\columnwidth - 12\tabcolsep) * \real{0.2200}}
  >{\raggedright\arraybackslash}p{(\columnwidth - 12\tabcolsep) * \real{0.1300}}
  >{\raggedright\arraybackslash}p{(\columnwidth - 12\tabcolsep) * \real{0.1300}}
  >{\raggedright\arraybackslash}p{(\columnwidth - 12\tabcolsep) * \real{0.1300}}
  >{\raggedright\arraybackslash}p{(\columnwidth - 12\tabcolsep) * \real{0.1300}}
  >{\raggedright\arraybackslash}p{(\columnwidth - 12\tabcolsep) * \real{0.1300}}
  >{\raggedright\arraybackslash}p{(\columnwidth - 12\tabcolsep) * \real{0.1300}}@{}}
\toprule\noalign{}
Layer & emb & b0 & b1 & b2 & b3 & bb \\
\midrule\noalign{}
Median relative change & 0.050 & 0.034 & 0.051 & 0.065 & 0.173 & 0.195 \\
\bottomrule\end{tabular}
\end{minipage}

\par\addvspace{0.8\baselineskip}
\noindent\begin{minipage}{\linewidth}
\hypertarget{tab:S15}{}%
{\small \textbf{Table~S15 --- Effect of excluding same-material neighbours on atom-level k-NN purity (\hyperlink{decodability-clustering-and-nearest-neighbour-retrieval}{§B.2}).} \emph{Each atom inherits its material's label, so nearest neighbours from the same material share that label by construction. We compare chance-corrected purity using all neighbours (\textbf{standard}) and after excluding atoms from the same material (\textbf{same-material excluded}). Both use K = 10 and 30,000 query atoms, drawn from indices of approximately 499,000 ORB-v3 atoms and 300,000 UMA-S atoms. \textbf{Channel · layer} identifies the configuration with the highest standard purity for each model and target. \textbf{Δ} is standard purity minus purity after exclusion, while \textbf{Same-material neighbours} gives the percentage of the original 10 neighbours belonging to the same material. For UMA-S, configurations are selected from the geom-norm, ang-norm, ‖L2‖ and full views, excluding L0; crystal-system selection is restricted to rotation-invariant views (\hyperlink{uma-channel-views-and-the-rotation-invariant-crystal-system-gap}{§3.4}, \hyperlink{tab:S13}{Table~S13}). Standard scores differ slightly from \hyperlink{tab:S4}{Table~S4} because of query subsampling (at most 0.004 for ORB-v3 and 0.005 for UMA-S crystal system). For UMA-S metal/insulator and magnetic label, differences of up to 0.012 also reflect the exclusion of the L0 view selected in \hyperlink{tab:S4}{Table~S4}.}\par}\par\nobreak
\begin{tabular}{@{}
  >{\raggedright\arraybackslash}p{(\columnwidth - 12\tabcolsep) * \real{0.0750}}
  >{\raggedright\arraybackslash}p{(\columnwidth - 12\tabcolsep) * \real{0.1833}}
  >{\raggedright\arraybackslash}p{(\columnwidth - 12\tabcolsep) * \real{0.1917}}
  >{\raggedright\arraybackslash}p{(\columnwidth - 12\tabcolsep) * \real{0.1333}}
  >{\raggedright\arraybackslash}p{(\columnwidth - 12\tabcolsep) * \real{0.1833}}
  >{\raggedright\arraybackslash}p{(\columnwidth - 12\tabcolsep) * \real{0.0667}}
  >{\raggedright\arraybackslash}p{(\columnwidth - 12\tabcolsep) * \real{0.1667}}@{}}
\toprule\noalign{}
\begin{minipage}[c]{\linewidth}\raggedright
Model
\end{minipage} & \begin{minipage}[c]{\linewidth}\raggedright
Channel · layer
\end{minipage} & \begin{minipage}[c]{\linewidth}\raggedright
Target
\end{minipage} & \begin{minipage}[c]{\linewidth}\raggedright
Purity (standard)
\end{minipage} & \begin{minipage}[c]{\linewidth}\raggedright
Purity (same-material excluded)
\end{minipage} & \begin{minipage}[c]{\linewidth}\raggedright
Δ
\end{minipage} & \begin{minipage}[c]{\linewidth}\raggedright
Same-material neighbours
\end{minipage} \\
\midrule\noalign{}
ORB & g4 & crystal system & 0.454 & 0.390 & 0.064 & 9.4~\% \\
ORB & g3 & metal/insulator & 0.752 & 0.720 & 0.032 & 9.2~\% \\
ORB & g3 & magnetic label & 0.630 & 0.592 & 0.038 & 9.3~\% \\
\midrule
UMA & geom-norm · b0 & crystal system & 0.459 & 0.416 & 0.043 & 6.4~\% \\
UMA & ang-norm · b3 & metal/insulator & 0.730 & 0.710 & 0.020 & 7.0~\% \\
UMA & ang-norm · b2 & magnetic label & 0.631 & 0.597 & 0.034 & 8.0~\% \\
\bottomrule\end{tabular}
\end{minipage}

\par\addvspace{0.8\baselineskip}
\noindent\begin{minipage}{\linewidth}
\hypertarget{tab:S16}{}%
{\small \textbf{Table~S16 --- Metric control: Euclidean vs PCA-whitened k-NN purity (\hyperlink{decodability-clustering-and-nearest-neighbour-retrieval}{§B.2}).} \emph{Material-level chance-corrected k-NN purity (K = 10) for the three categorical targets. \textbf{Euclidean} is the purity under the raw Euclidean metric, the one used everywhere else in the paper. \textbf{Whitened} is the same purity under a PCA-whitened (Mahalanobis) metric, which rescales the covariance so that no single high-variance direction dominates distances. \textbf{Gain} is \textbf{Whitened} minus \textbf{Euclidean}. If poor Euclidean retrieval were merely a wrong-metric artefact, whitening would recover it; instead it moves purity by at most 0.079 and lowers it in three of the six cases, so the weak neighbourhood organisation is a property of the representation, not the distance function. For crystal system, UMA-S's peak is reported as its best rotation-invariant configuration (concat · ang-norm · b0), since the higher-scoring non-invariant configurations collapse under random rotation (\hyperlink{uma-channel-views-and-the-rotation-invariant-crystal-system-gap}{§3.4}, \hyperlink{tab:S13}{Table~S13}).}\par}\par\nobreak
\begin{tabular}{@{}
  >{\raggedright\arraybackslash}p{(\columnwidth - 8\tabcolsep) * \real{0.1333}}
  >{\raggedright\arraybackslash}p{(\columnwidth - 8\tabcolsep) * \real{0.3417}}
  >{\raggedright\arraybackslash}p{(\columnwidth - 8\tabcolsep) * \real{0.2083}}
  >{\raggedright\arraybackslash}p{(\columnwidth - 8\tabcolsep) * \real{0.1833}}
  >{\raggedright\arraybackslash}p{(\columnwidth - 8\tabcolsep) * \real{0.1333}}@{}}
\toprule\noalign{}
\begin{minipage}[c]{\linewidth}\raggedright
Model
\end{minipage} & \begin{minipage}[c]{\linewidth}\raggedright
Target
\end{minipage} & \begin{minipage}[c]{\linewidth}\raggedright
Euclidean
\end{minipage} & \begin{minipage}[c]{\linewidth}\raggedright
Whitened
\end{minipage} & \begin{minipage}[c]{\linewidth}\raggedright
Gain
\end{minipage} \\
\midrule\noalign{}
ORB & crystal system & 0.458 & 0.464 & +0.006 \\
ORB & metal/insulator & 0.724 & 0.698 & −0.026 \\
ORB & magnetic label & 0.642 & 0.641 & −0.001 \\
\midrule
UMA & crystal system & 0.424 & 0.503 & +0.079 \\
UMA & metal/insulator & 0.739 & 0.689 & −0.050 \\
UMA & magnetic label & 0.659 & 0.663 & +0.004 \\
\bottomrule\end{tabular}
\end{minipage}

\par\addvspace{0.8\baselineskip}
\hypertarget{tab:S17}{}%
{\small \textbf{Table~S17 --- Effect of random initialisation on prediction performance (\hyperlink{property-decodability-depends-on-network-depth}{§3.3}).} \emph{ORB-v3 and UMA-S are compared with identical architectures but randomly reinitialised weights, using the same embedding extraction, data split and XGBoost settings (\hyperlink{random-initialisation-and-matched-training-data-controls}{§A.9}). \textbf{Best config (pretrained)} identifies each pretrained model's optimal read-out, channel view and layer (\textbf{\hyperlink{tab:1}{Table~1}}). \textbf{Pretrained score} and \textbf{Random, same config} compare the pretrained and random-weight networks at that configuration. \textbf{Best config (random)} and \textbf{Random best config score} report the random-weight network's own optimal configuration and score. The Magpie composition-only baseline and a trivial predictor (training-set mean for regression, majority class for classification) are included for reference. Scores are R² for continuous targets and macro-F1 for categorical targets.}\par}

\noindent\begin{minipage}{\linewidth}
\textbf{(a) ORB-v3}\par\nobreak
\begin{tabular}{@{}
  >{\raggedright\arraybackslash}p{(\columnwidth - 14\tabcolsep) * \real{0.1850}}
  >{\raggedright\arraybackslash}p{(\columnwidth - 14\tabcolsep) * \real{0.2350}}
  >{\raggedright\arraybackslash}p{(\columnwidth - 14\tabcolsep) * \real{0.0700}}
  >{\raggedright\arraybackslash}p{(\columnwidth - 14\tabcolsep) * \real{0.0700}}
  >{\raggedright\arraybackslash}p{(\columnwidth - 14\tabcolsep) * \real{0.2350}}
  >{\raggedright\arraybackslash}p{(\columnwidth - 14\tabcolsep) * \real{0.0700}}
  >{\raggedright\arraybackslash}p{(\columnwidth - 14\tabcolsep) * \real{0.0675}}
  >{\raggedright\arraybackslash}p{(\columnwidth - 14\tabcolsep) * \real{0.0675}}@{}}
\toprule\noalign{}
\begin{minipage}[c]{\linewidth}\raggedright
Target
\end{minipage} & \begin{minipage}[c]{\linewidth}\raggedright
Best config (pretrained)
\end{minipage} & \begin{minipage}[c]{\linewidth}\raggedright
Pretrained score
\end{minipage} & \begin{minipage}[c]{\linewidth}\raggedright
Random, same config
\end{minipage} & \begin{minipage}[c]{\linewidth}\raggedright
Best config (random)
\end{minipage} & \begin{minipage}[c]{\linewidth}\raggedright
Random best config score
\end{minipage} & \begin{minipage}[c]{\linewidth}\raggedright
Magpie
\end{minipage} & \begin{minipage}[c]{\linewidth}\raggedright
Trivial
\end{minipage} \\
\midrule\noalign{}
formation E/atom & node · g0 & 0.990 & 0.904 & concat · enc & 0.920 & 0.890 & −0.000 \\
energy above hull & concat · g4 & 0.948 & 0.564 & concat · enc & 0.684 & 0.296 & −0.000 \\
band gap & concat · g4 & 0.890 & 0.618 & concat · enc & 0.627 & 0.698 & −0.000 \\
density & concat · g0 & 0.982 & 0.952 & concat · g1 & 0.953 & 0.888 & −0.000 \\
bulk modulus & concat · g0 & 0.951 & 0.891 & concat · g3 & 0.899 & 0.882 & −0.002 \\
mag. density & concat · g0 & 0.846 & 0.797 & concat · g0 & 0.797 & 0.816 & −0.000 \\
crystal system & concat · g2 & 0.622 & 0.431 & edge · enc & 0.681 & 0.520 & 0.054 \\
metal/insulator & concat · g4 & 0.909 & 0.832 & concat · enc & 0.842 & 0.845 & 0.347 \\
magnetic label & concat · g4 & 0.890 & 0.825 & concat · g1 & 0.830 & 0.860 & 0.345 \\
\bottomrule\end{tabular}
\end{minipage}

\noindent\begin{minipage}{\linewidth}
\textbf{(b) UMA-S}\par\nobreak
\begin{tabular}{@{}
  >{\raggedright\arraybackslash}p{(\columnwidth - 14\tabcolsep) * \real{0.1850}}
  >{\raggedright\arraybackslash}p{(\columnwidth - 14\tabcolsep) * \real{0.2350}}
  >{\raggedright\arraybackslash}p{(\columnwidth - 14\tabcolsep) * \real{0.0700}}
  >{\raggedright\arraybackslash}p{(\columnwidth - 14\tabcolsep) * \real{0.0700}}
  >{\raggedright\arraybackslash}p{(\columnwidth - 14\tabcolsep) * \real{0.2350}}
  >{\raggedright\arraybackslash}p{(\columnwidth - 14\tabcolsep) * \real{0.0700}}
  >{\raggedright\arraybackslash}p{(\columnwidth - 14\tabcolsep) * \real{0.0675}}
  >{\raggedright\arraybackslash}p{(\columnwidth - 14\tabcolsep) * \real{0.0675}}@{}}
\toprule\noalign{}
\begin{minipage}[c]{\linewidth}\raggedright
Target
\end{minipage} & \begin{minipage}[c]{\linewidth}\raggedright
Best config (pretrained)
\end{minipage} & \begin{minipage}[c]{\linewidth}\raggedright
Pretrained score
\end{minipage} & \begin{minipage}[c]{\linewidth}\raggedright
Random, same config
\end{minipage} & \begin{minipage}[c]{\linewidth}\raggedright
Best config (random)
\end{minipage} & \begin{minipage}[c]{\linewidth}\raggedright
Random best config score
\end{minipage} & \begin{minipage}[c]{\linewidth}\raggedright
Magpie
\end{minipage} & \begin{minipage}[c]{\linewidth}\raggedright
Trivial
\end{minipage} \\
\midrule\noalign{}
formation E/atom & concat · ang-norm · b3 & 0.990 & 0.909 & concat · ang-norm · emb & 0.912 & 0.890 & −0.000 \\
energy above hull & concat · L0 · b3 & 0.960 & 0.552 & node · full · b3 & 0.669 & 0.296 & −0.000 \\
band gap & concat · ang-norm · b2 & 0.907 & 0.611 & concat · ang-norm · emb & 0.634 & 0.698 & −0.000 \\
density & concat · ang-norm · emb & 0.983 & 0.942 & concat · L0 · b1 & 0.945 & 0.888 & −0.000 \\
bulk modulus & concat · ang-norm · emb & 0.958 & 0.901 & concat · ang-norm · b3 & 0.912 & 0.882 & −0.002 \\
mag. density & concat · ang-norm · emb & 0.839 & 0.796 & concat · ang-norm · emb & 0.796 & 0.816 & −0.000 \\
crystal system & concat · ang-norm · b0 & 0.817 & 0.607 & concat · ang-norm · emb & 0.661 & 0.520 & 0.054 \\
metal/insulator & node · ang-norm · bb & 0.914 & 0.839 & concat · ang-norm · emb & 0.839 & 0.845 & 0.347 \\
magnetic label & node · ang-norm · bb & 0.903 & 0.826 & concat · ang-norm · emb & 0.830 & 0.860 & 0.345 \\
\bottomrule\end{tabular}
\end{minipage}

\clearpage
\hypertarget{tab:S18}{}%
{\small \textbf{Table~S18 --- Comparison of models with matched training data or architecture (\hyperlink{uma-channel-views-and-the-rotation-invariant-crystal-system-gap}{§3.4}).} \emph{We compare four models on the same 20,000-material subset, stratified by crystal system and using the paper's seed-42 training and test split restricted to these materials (\hyperlink{random-initialisation-and-matched-training-data-controls}{§A.9}). For each model and target, we evaluate XGBoost across all available layers and read-outs and report the best-performing configuration. MACE-OMAT-0 has two message-passing layers, each ending in a product block that combines the neighbour messages into many-body atomic features; p0 and p1 are the atom features after the first and second of these blocks (p0: 128 scalar and 128 vector channels; p1: 128 scalar channels), the MACE counterparts of the ORB and UMA layer labels. Scores are R² for continuous targets and macro-F1 for categorical targets; the Magpie column of block (a) is the composition-only baseline on the same split, shown for reference. Block (c) reports pairwise score differences between models, with 95~\% confidence intervals from paired bootstrap resampling of the shared test materials.}\par}

\noindent\begin{minipage}{\linewidth}
\textbf{(a) Best score per model}\par\nobreak
\begin{tabular}{@{}
  >{\raggedright\arraybackslash}p{(\columnwidth - 10\tabcolsep) * \real{0.2131}}
  >{\raggedright\arraybackslash}p{(\columnwidth - 10\tabcolsep) * \real{0.1639}}
  >{\raggedright\arraybackslash}p{(\columnwidth - 10\tabcolsep) * \real{0.1639}}
  >{\raggedright\arraybackslash}p{(\columnwidth - 10\tabcolsep) * \real{0.1803}}
  >{\raggedright\arraybackslash}p{(\columnwidth - 10\tabcolsep) * \real{0.1639}}
  >{\raggedright\arraybackslash}p{(\columnwidth - 10\tabcolsep) * \real{0.1148}}@{}}
\toprule\noalign{}
\begin{minipage}[c]{\linewidth}\raggedright
Target
\end{minipage} & \begin{minipage}[c]{\linewidth}\raggedright
MACE-OMAT-0 (equivariant, OMat24)
\end{minipage} & \begin{minipage}[c]{\linewidth}\raggedright
ORB-v3 omat (OMat24)
\end{minipage} & \begin{minipage}[c]{\linewidth}\raggedright
ORB-v3 mpa (MPtrj + Alexandria)
\end{minipage} & \begin{minipage}[c]{\linewidth}\raggedright
UMA-S (five datasets)
\end{minipage} & \begin{minipage}[c]{\linewidth}\raggedright
Magpie
\end{minipage} \\
\midrule\noalign{}
formation E/atom & 0.978 & 0.986 & 0.986 & 0.986 & 0.882 \\
energy above hull & 0.845 & 0.909 & 0.915 & 0.935 & 0.198 \\
band gap & 0.776 & 0.863 & 0.874 & 0.885 & 0.652 \\
density & 0.934 & 0.975 & 0.972 & 0.972 & 0.873 \\
bulk modulus & 0.904 & 0.928 & 0.931 & 0.914 & 0.866 \\
mag. density & 0.704 & 0.798 & 0.787 & 0.779 & 0.760 \\
crystal system & 0.540 & 0.545 & 0.546 & 0.760 & 0.458 \\
metal/insulator & 0.867 & 0.894 & 0.899 & 0.904 & 0.826 \\
magnetic label & 0.839 & 0.876 & 0.868 & 0.886 & 0.834 \\
\bottomrule\end{tabular}
\end{minipage}

\noindent\begin{minipage}{\linewidth}
\textbf{(b) Best configuration per model}\par\nobreak
\begin{tabular}{@{}
  >{\raggedright\arraybackslash}p{(\columnwidth - 8\tabcolsep) * \real{0.2000}}
  >{\raggedright\arraybackslash}p{(\columnwidth - 8\tabcolsep) * \real{0.2000}}
  >{\raggedright\arraybackslash}p{(\columnwidth - 8\tabcolsep) * \real{0.1667}}
  >{\raggedright\arraybackslash}p{(\columnwidth - 8\tabcolsep) * \real{0.1833}}
  >{\raggedright\arraybackslash}p{(\columnwidth - 8\tabcolsep) * \real{0.2500}}@{}}
\toprule\noalign{}
\begin{minipage}[c]{\linewidth}\raggedright
Target
\end{minipage} & \begin{minipage}[c]{\linewidth}\raggedright
MACE-OMAT-0 (equivariant, OMat24)
\end{minipage} & \begin{minipage}[c]{\linewidth}\raggedright
ORB-v3 omat (OMat24)
\end{minipage} & \begin{minipage}[c]{\linewidth}\raggedright
ORB-v3 mpa (MPtrj + Alexandria)
\end{minipage} & \begin{minipage}[c]{\linewidth}\raggedright
UMA-S (five datasets)
\end{minipage} \\
\midrule\noalign{}
formation E/atom & node · scalar · p0 & node · g0 & node · g0 & concat · ang-norm · b1 \\
energy above hull & node · ang-norm · p1 & concat · g1 & concat · g4 & node · L0 · b3 \\
band gap & node · ang-norm · p1 & concat · g3 & concat · g2 & node · ang-norm · bb \\
density & node · scalar · p0 & node · g0 & node · g0 & concat · ang-norm · emb \\
bulk modulus & node · ang-norm · p0 & node · g0 & concat · g0 & concat · ang-norm · emb \\
mag. density & node · scalar · p0 & concat · g0 & concat · g0 & concat · ang-norm · emb \\
crystal system & node · ang-norm · p0 & concat · g3 & concat · g2 & concat · ang-norm · b0 \\
metal/insulator & node · ang-norm · p0 & concat · g2 & node · g3 & node · ang-norm · b3 \\
magnetic label & node · ang-norm · p0 & node · g2 & node · g3 & node · L0 · bb \\
\bottomrule\end{tabular}
\end{minipage}

\noindent\begin{minipage}{\linewidth}
\textbf{(c) Paired differences on the shared test rows (difference and 95~\% CI)}\par\nobreak
\begin{tabular}{@{}
  >{\raggedright\arraybackslash}p{(\columnwidth - 6\tabcolsep) * \real{0.2075}}
  >{\raggedright\arraybackslash}p{(\columnwidth - 6\tabcolsep) * \real{0.2642}}
  >{\raggedright\arraybackslash}p{(\columnwidth - 6\tabcolsep) * \real{0.2642}}
  >{\raggedright\arraybackslash}p{(\columnwidth - 6\tabcolsep) * \real{0.2642}}@{}}
\toprule\noalign{}
\begin{minipage}[c]{\linewidth}\raggedright
Target
\end{minipage} & \begin{minipage}[c]{\linewidth}\raggedright
Matched data: MACE-OMAT-0 − ORB-v3 omat
\end{minipage} & \begin{minipage}[c]{\linewidth}\raggedright
Matched architecture: ORB-v3 mpa − ORB-v3 omat
\end{minipage} & \begin{minipage}[c]{\linewidth}\raggedright
Paper's pair: UMA-S − ORB-v3 omat
\end{minipage} \\
\midrule\noalign{}
formation E/atom & −0.008 {[}−0.011, −0.005{]} & −0.000 {[}−0.001, +0.001{]} & −0.000 {[}−0.002, +0.001{]} \\
energy above hull & −0.064 {[}−0.093, −0.042{]} & 0.006 {[}−0.005, +0.018{]} & 0.026 {[}+0.011, +0.041{]} \\
band gap & −0.087 {[}−0.103, −0.072{]} & 0.012 {[}+0.001, +0.027{]} & 0.022 {[}+0.015, +0.030{]} \\
density & −0.040 {[}−0.046, −0.035{]} & −0.002 {[}−0.003, −0.001{]} & −0.002 {[}−0.004, +0.000{]} \\
bulk modulus & −0.024 {[}−0.041, −0.008{]} & 0.003 {[}−0.004, +0.012{]} & −0.014 {[}−0.030, +0.006{]} \\
mag. density & −0.094 {[}−0.118, −0.072{]} & −0.011 {[}−0.021, −0.002{]} & −0.019 {[}−0.032, −0.005{]} \\
crystal system & −0.005 {[}−0.020, +0.012{]} & 0.001 {[}−0.011, +0.013{]} & 0.215 {[}+0.198, +0.232{]} \\
metal/insulator & −0.028 {[}−0.036, −0.019{]} & 0.004 {[}−0.003, +0.012{]} & 0.009 {[}+0.002, +0.017{]} \\
magnetic label & −0.037 {[}−0.046, −0.027{]} & −0.008 {[}−0.016, +0.001{]} & 0.010 {[}+0.002, +0.019{]} \\
\bottomrule\end{tabular}
\end{minipage}

\clearpage
\hypertarget{tab:S19}{}%
{\small \textbf{Table~S19 --- Cross-database transfer from Materials Project (MP) to JARVIS-DFT 3D and OQMD (\hyperlink{external-transfer-results}{§B.4}).} \emph{Each panel compares one model--database pair. ORB-v3 and UMA-S remain frozen; XGBoost predictors use the configurations selected on MP. Results are grouped by evaluation set. \textbf{MP reference:} \textbf{MP test} is the score of the MP-trained predictor on MP's original test set. \textbf{Crystals absent from MP:} \textbf{n} is the number of external crystals with no identified MP counterpart and an external label for the target. \textbf{External refit} gives the score of the same predictor configuration retrained on the external database using five 80/20 splits. \textbf{MP-trained} gives the score of the original MP-trained predictor on the same external test crystals, against the same external labels. \textbf{Transfer gap} is MP-trained minus external refit; negative values indicate lower performance after transfer. Scores are reported as mean ± standard deviation across splits. \textbf{Crystals shared with MP:} \textbf{n} is the number of external crystals whose MP counterparts belong to MP's held-out test set and that carry the target's label in both databases. The four columns show scores of the MP-trained predictor using different combinations of input structures and reference labels: \textbf{S\textsubscript{0}} (MP structure, MP label), \textbf{S\textsubscript{1}} (external structure, MP label), \textbf{S\textsubscript{3}} (MP structure, external label) and \textbf{S\textsubscript{2}} (external structure, external label). These scores allow the contributions of geometry and label differences to be separated (\hyperlink{external-transfer}{§A.10}); they do not directly decompose the transfer gap on MP-absent crystals. Scores are R² for continuous targets and macro-F1 for categorical targets. OQMD has no bulk-modulus or magnetic labels.}\par}\par\nobreak

\noindent\textbf{(a) ORB-v3 --- JARVIS-DFT 3D}\par\nobreak
{\setlength{\tabcolsep}{3pt}\setlength{\LTpre}{2pt}\setlength{\LTpost}{6pt}
\begin{longtable}{@{}
  >{\raggedright\arraybackslash}p{(\columnwidth - 20\tabcolsep) * \real{0.1608}}
  >{\raggedright\arraybackslash}p{(\columnwidth - 20\tabcolsep) * \real{0.0697}}
  >{\raggedright\arraybackslash}p{(\columnwidth - 20\tabcolsep) * \real{0.0804}}
  >{\raggedright\arraybackslash}p{(\columnwidth - 20\tabcolsep) * \real{0.1018}}
  >{\raggedright\arraybackslash}p{(\columnwidth - 20\tabcolsep) * \real{0.1018}}
  >{\raggedright\arraybackslash}p{(\columnwidth - 20\tabcolsep) * \real{0.1018}}
  >{\raggedright\arraybackslash}p{(\columnwidth - 20\tabcolsep) * \real{0.0750}}
  >{\raggedright\arraybackslash}p{(\columnwidth - 20\tabcolsep) * \real{0.0772}}
  >{\raggedright\arraybackslash}p{(\columnwidth - 20\tabcolsep) * \real{0.0772}}
  >{\raggedright\arraybackslash}p{(\columnwidth - 20\tabcolsep) * \real{0.0772}}
  >{\raggedright\arraybackslash}p{(\columnwidth - 20\tabcolsep) * \real{0.0772}}@{}}
\toprule\noalign{}
 & \multicolumn{1}{@{}l@{}}{\textit{\shortstack[l]{MP\\reference}}} & \multicolumn{4}{@{}l@{}}{\textit{\shortstack[l]{Crystals absent from MP}}} & \multicolumn{5}{@{}l@{}}{\textit{\shortstack[l]{Crystals shared with MP (structure / label)}}} \\
\cmidrule(r){2-2}\cmidrule(r){3-6}\cmidrule(r){7-11}
\begin{minipage}[c]{\linewidth}\raggedright
Target
\end{minipage} & \begin{minipage}[c]{\linewidth}\raggedright
MP test
\end{minipage} & \begin{minipage}[c]{\linewidth}\raggedright
n
\end{minipage} & \begin{minipage}[c]{\linewidth}\raggedright
External refit
\end{minipage} & \begin{minipage}[c]{\linewidth}\raggedright
MP-trained
\end{minipage} & \begin{minipage}[c]{\linewidth}\raggedright
Transfer gap
\end{minipage} & \begin{minipage}[c]{\linewidth}\raggedright
n
\end{minipage} & \begin{minipage}[c]{\linewidth}\raggedright
S\textsubscript{0}\newline MP / MP
\end{minipage} & \begin{minipage}[c]{\linewidth}\raggedright
S\textsubscript{1}\newline Ext. / MP
\end{minipage} & \begin{minipage}[c]{\linewidth}\raggedright
S\textsubscript{3}\newline MP / Ext.
\end{minipage} & \begin{minipage}[c]{\linewidth}\raggedright
S\textsubscript{2}\newline Ext. / Ext.
\end{minipage} \\
\midrule\noalign{}
\endhead
\bottomrule
\endlastfoot
formation E/atom & 0.990 & 30,346 & 0.969\newline ±\,0.002 & 0.896\newline ±\,0.004 & −0.073\newline ±\,0.003 & 12,769 & 0.989 & 0.962 & 0.938 & 0.968 \\
energy above hull & 0.948 & 30,335 & 0.951\newline ±\,0.004 & −1.117\newline ±\,0.015 & −2.068\newline ±\,0.014 & 12,769 & 0.918 & 0.536 & −1.585 & −1.599 \\
band gap & 0.890 & 30,335 & 0.848\newline ±\,0.016 & 0.693\newline ±\,0.008 & −0.155\newline ±\,0.015 & 12,769 & 0.882 & 0.872 & 0.817 & 0.824 \\
density & 0.982 & 30,335 & 0.978\newline ±\,0.001 & 0.955\newline ±\,0.002 & −0.023\newline ±\,0.001 & 12,769 & 0.980 & 0.926 & 0.917 & 0.977 \\
bulk modulus & 0.951 & 14,500 & 0.914\newline ±\,0.008 & 0.884\newline ±\,0.009 & −0.030\newline ±\,0.001 & 1,536 & 0.942 & 0.935 & 0.924 & 0.927 \\
mag. density & 0.846 & 30,176 & 0.742\newline ±\,0.015 & 0.379\newline ±\,0.036 & −0.363\newline ±\,0.036 & 11,738 & 0.804 & 0.755 & 0.417 & 0.541 \\
crystal system & 0.623 & 30,335 & 0.794\newline ±\,0.005 & 0.383\newline ±\,0.003 & −0.412\newline ±\,0.006 & 12,827 & 0.610 & 0.590 & 0.599 & 0.588 \\
metal/insulator & 0.909 & 30,335 & 0.908\newline ±\,0.003 & 0.885\newline ±\,0.005 & −0.024\newline ±\,0.005 & 12,868 & 0.930 & 0.921 & 0.895 & 0.908 \\
magnetic label & 0.890 & 30,176 & 0.861\newline ±\,0.006 & 0.715\newline ±\,0.004 & −0.146\newline ±\,0.004 & 11,606 & 0.884 & 0.877 & 0.857 & 0.863 \\
\end{longtable}}

\noindent\textbf{(b) ORB-v3 --- OQMD}\par\nobreak
{\setlength{\tabcolsep}{3pt}\setlength{\LTpre}{2pt}\setlength{\LTpost}{6pt}
\begin{longtable}{@{}
  >{\raggedright\arraybackslash}p{(\columnwidth - 20\tabcolsep) * \real{0.1608}}
  >{\raggedright\arraybackslash}p{(\columnwidth - 20\tabcolsep) * \real{0.0697}}
  >{\raggedright\arraybackslash}p{(\columnwidth - 20\tabcolsep) * \real{0.0804}}
  >{\raggedright\arraybackslash}p{(\columnwidth - 20\tabcolsep) * \real{0.1018}}
  >{\raggedright\arraybackslash}p{(\columnwidth - 20\tabcolsep) * \real{0.1018}}
  >{\raggedright\arraybackslash}p{(\columnwidth - 20\tabcolsep) * \real{0.1018}}
  >{\raggedright\arraybackslash}p{(\columnwidth - 20\tabcolsep) * \real{0.0750}}
  >{\raggedright\arraybackslash}p{(\columnwidth - 20\tabcolsep) * \real{0.0772}}
  >{\raggedright\arraybackslash}p{(\columnwidth - 20\tabcolsep) * \real{0.0772}}
  >{\raggedright\arraybackslash}p{(\columnwidth - 20\tabcolsep) * \real{0.0772}}
  >{\raggedright\arraybackslash}p{(\columnwidth - 20\tabcolsep) * \real{0.0772}}@{}}
\toprule\noalign{}
 & \multicolumn{1}{@{}l@{}}{\textit{\shortstack[l]{MP\\reference}}} & \multicolumn{4}{@{}l@{}}{\textit{\shortstack[l]{Crystals absent from MP}}} & \multicolumn{5}{@{}l@{}}{\textit{\shortstack[l]{Crystals shared with MP (structure / label)}}} \\
\cmidrule(r){2-2}\cmidrule(r){3-6}\cmidrule(r){7-11}
\begin{minipage}[c]{\linewidth}\raggedright
Target
\end{minipage} & \begin{minipage}[c]{\linewidth}\raggedright
MP test
\end{minipage} & \begin{minipage}[c]{\linewidth}\raggedright
n
\end{minipage} & \begin{minipage}[c]{\linewidth}\raggedright
External refit
\end{minipage} & \begin{minipage}[c]{\linewidth}\raggedright
MP-trained
\end{minipage} & \begin{minipage}[c]{\linewidth}\raggedright
Transfer gap
\end{minipage} & \begin{minipage}[c]{\linewidth}\raggedright
n
\end{minipage} & \begin{minipage}[c]{\linewidth}\raggedright
S\textsubscript{0}\newline MP / MP
\end{minipage} & \begin{minipage}[c]{\linewidth}\raggedright
S\textsubscript{1}\newline Ext. / MP
\end{minipage} & \begin{minipage}[c]{\linewidth}\raggedright
S\textsubscript{3}\newline MP / Ext.
\end{minipage} & \begin{minipage}[c]{\linewidth}\raggedright
S\textsubscript{2}\newline Ext. / Ext.
\end{minipage} \\
\midrule\noalign{}
\endhead
\bottomrule
\endlastfoot
formation E/atom & 0.990 & 37,797 & 0.976\newline ±\,0.004 & 0.931\newline ±\,0.004 & −0.045\newline ±\,0.000 & 439 & 0.988 & 0.987 & 0.976 & 0.975 \\
energy above hull & 0.948 & 37,804 & 0.301\newline ±\,0.122 & 0.011\newline ±\,0.009 & −0.290\newline ±\,0.129 & 439 & 0.860 & 0.839 & 0.861 & 0.864 \\
band gap & 0.890 & 37,693 & 0.818\newline ±\,0.034 & 0.722\newline ±\,0.031 & −0.096\newline ±\,0.011 & 436 & 0.929 & 0.921 & 0.864 & 0.875 \\
density & 0.982 & 37,804 & 0.971\newline ±\,0.001 & 0.940\newline ±\,0.002 & −0.031\newline ±\,0.002 & 439 & 0.981 & 0.978 & 0.975 & 0.982 \\
crystal system & 0.623 & 37,804 & 0.806\newline ±\,0.005 & 0.454\newline ±\,0.004 & −0.351\newline ±\,0.005 & 418 & 0.556 & 0.521 & 0.530 & 0.529 \\
metal/insulator & 0.909 & 37,693 & 0.938\newline ±\,0.003 & 0.903\newline ±\,0.004 & −0.035\newline ±\,0.004 & 461 & 0.963 & 0.946 & 0.923 & 0.920 \\
\end{longtable}}

\clearpage
\noindent\textbf{(c) UMA-S --- JARVIS-DFT 3D}\par\nobreak
{\setlength{\tabcolsep}{3pt}\setlength{\LTpre}{2pt}\setlength{\LTpost}{6pt}
\begin{longtable}{@{}
  >{\raggedright\arraybackslash}p{(\columnwidth - 20\tabcolsep) * \real{0.1608}}
  >{\raggedright\arraybackslash}p{(\columnwidth - 20\tabcolsep) * \real{0.0697}}
  >{\raggedright\arraybackslash}p{(\columnwidth - 20\tabcolsep) * \real{0.0804}}
  >{\raggedright\arraybackslash}p{(\columnwidth - 20\tabcolsep) * \real{0.1018}}
  >{\raggedright\arraybackslash}p{(\columnwidth - 20\tabcolsep) * \real{0.1018}}
  >{\raggedright\arraybackslash}p{(\columnwidth - 20\tabcolsep) * \real{0.1018}}
  >{\raggedright\arraybackslash}p{(\columnwidth - 20\tabcolsep) * \real{0.0750}}
  >{\raggedright\arraybackslash}p{(\columnwidth - 20\tabcolsep) * \real{0.0772}}
  >{\raggedright\arraybackslash}p{(\columnwidth - 20\tabcolsep) * \real{0.0772}}
  >{\raggedright\arraybackslash}p{(\columnwidth - 20\tabcolsep) * \real{0.0772}}
  >{\raggedright\arraybackslash}p{(\columnwidth - 20\tabcolsep) * \real{0.0772}}@{}}
\toprule\noalign{}
 & \multicolumn{1}{@{}l@{}}{\textit{\shortstack[l]{MP\\reference}}} & \multicolumn{4}{@{}l@{}}{\textit{\shortstack[l]{Crystals absent from MP}}} & \multicolumn{5}{@{}l@{}}{\textit{\shortstack[l]{Crystals shared with MP (structure / label)}}} \\
\cmidrule(r){2-2}\cmidrule(r){3-6}\cmidrule(r){7-11}
\begin{minipage}[c]{\linewidth}\raggedright
Target
\end{minipage} & \begin{minipage}[c]{\linewidth}\raggedright
MP test
\end{minipage} & \begin{minipage}[c]{\linewidth}\raggedright
n
\end{minipage} & \begin{minipage}[c]{\linewidth}\raggedright
External refit
\end{minipage} & \begin{minipage}[c]{\linewidth}\raggedright
MP-trained
\end{minipage} & \begin{minipage}[c]{\linewidth}\raggedright
Transfer gap
\end{minipage} & \begin{minipage}[c]{\linewidth}\raggedright
n
\end{minipage} & \begin{minipage}[c]{\linewidth}\raggedright
S\textsubscript{0}\newline MP / MP
\end{minipage} & \begin{minipage}[c]{\linewidth}\raggedright
S\textsubscript{1}\newline Ext. / MP
\end{minipage} & \begin{minipage}[c]{\linewidth}\raggedright
S\textsubscript{3}\newline MP / Ext.
\end{minipage} & \begin{minipage}[c]{\linewidth}\raggedright
S\textsubscript{2}\newline Ext. / Ext.
\end{minipage} \\
\midrule\noalign{}
\endhead
\bottomrule
\endlastfoot
formation E/atom & 0.990 & 30,346 & 0.969\newline ±\,0.003 & 0.919\newline ±\,0.004 & −0.050\newline ±\,0.001 & 12,768 & 0.988 & 0.960 & 0.935 & 0.967 \\
energy above hull & 0.960 & 30,346 & 0.948\newline ±\,0.003 & −1.071\newline ±\,0.026 & −2.018\newline ±\,0.026 & 12,768 & 0.937 & 0.534 & −1.587 & −1.598 \\
band gap & 0.907 & 30,346 & 0.875\newline ±\,0.014 & 0.708\newline ±\,0.029 & −0.167\newline ±\,0.022 & 12,768 & 0.895 & 0.886 & 0.832 & 0.845 \\
density & 0.983 & 30,346 & 0.980\newline ±\,0.001 & 0.953\newline ±\,0.001 & −0.027\newline ±\,0.001 & 12,768 & 0.982 & 0.926 & 0.919 & 0.979 \\
bulk modulus & 0.959 & 14,500 & 0.921\newline ±\,0.008 & 0.896\newline ±\,0.008 & −0.025\newline ±\,0.003 & 1,536 & 0.955 & 0.948 & 0.927 & 0.932 \\
mag. density & 0.839 & 30,187 & 0.701\newline ±\,0.020 & 0.359\newline ±\,0.054 & −0.342\newline ±\,0.045 & 11,737 & 0.807 & 0.751 & 0.398 & 0.546 \\
crystal system & 0.817 & 30,346 & 0.873\newline ±\,0.005 & 0.580\newline ±\,0.005 & −0.293\newline ±\,0.008 & 12,827 & 0.822 & 0.797 & 0.797 & 0.823 \\
metal/insulator & 0.914 & 30,346 & 0.922\newline ±\,0.004 & 0.877\newline ±\,0.008 & −0.045\newline ±\,0.005 & 12,868 & 0.936 & 0.923 & 0.898 & 0.911 \\
magnetic label & 0.903 & 30,187 & 0.891\newline ±\,0.007 & 0.737\newline ±\,0.002 & −0.154\newline ±\,0.005 & 11,605 & 0.899 & 0.886 & 0.864 & 0.872 \\
\end{longtable}}

\noindent\textbf{(d) UMA-S --- OQMD}\par\nobreak
{\setlength{\tabcolsep}{3pt}\setlength{\LTpre}{2pt}\setlength{\LTpost}{6pt}
\begin{longtable}{@{}
  >{\raggedright\arraybackslash}p{(\columnwidth - 20\tabcolsep) * \real{0.1608}}
  >{\raggedright\arraybackslash}p{(\columnwidth - 20\tabcolsep) * \real{0.0697}}
  >{\raggedright\arraybackslash}p{(\columnwidth - 20\tabcolsep) * \real{0.0804}}
  >{\raggedright\arraybackslash}p{(\columnwidth - 20\tabcolsep) * \real{0.1018}}
  >{\raggedright\arraybackslash}p{(\columnwidth - 20\tabcolsep) * \real{0.1018}}
  >{\raggedright\arraybackslash}p{(\columnwidth - 20\tabcolsep) * \real{0.1018}}
  >{\raggedright\arraybackslash}p{(\columnwidth - 20\tabcolsep) * \real{0.0750}}
  >{\raggedright\arraybackslash}p{(\columnwidth - 20\tabcolsep) * \real{0.0772}}
  >{\raggedright\arraybackslash}p{(\columnwidth - 20\tabcolsep) * \real{0.0772}}
  >{\raggedright\arraybackslash}p{(\columnwidth - 20\tabcolsep) * \real{0.0772}}
  >{\raggedright\arraybackslash}p{(\columnwidth - 20\tabcolsep) * \real{0.0772}}@{}}
\toprule\noalign{}
 & \multicolumn{1}{@{}l@{}}{\textit{\shortstack[l]{MP\\reference}}} & \multicolumn{4}{@{}l@{}}{\textit{\shortstack[l]{Crystals absent from MP}}} & \multicolumn{5}{@{}l@{}}{\textit{\shortstack[l]{Crystals shared with MP (structure / label)}}} \\
\cmidrule(r){2-2}\cmidrule(r){3-6}\cmidrule(r){7-11}
\begin{minipage}[c]{\linewidth}\raggedright
Target
\end{minipage} & \begin{minipage}[c]{\linewidth}\raggedright
MP test
\end{minipage} & \begin{minipage}[c]{\linewidth}\raggedright
n
\end{minipage} & \begin{minipage}[c]{\linewidth}\raggedright
External refit
\end{minipage} & \begin{minipage}[c]{\linewidth}\raggedright
MP-trained
\end{minipage} & \begin{minipage}[c]{\linewidth}\raggedright
Transfer gap
\end{minipage} & \begin{minipage}[c]{\linewidth}\raggedright
n
\end{minipage} & \begin{minipage}[c]{\linewidth}\raggedright
S\textsubscript{0}\newline MP / MP
\end{minipage} & \begin{minipage}[c]{\linewidth}\raggedright
S\textsubscript{1}\newline Ext. / MP
\end{minipage} & \begin{minipage}[c]{\linewidth}\raggedright
S\textsubscript{3}\newline MP / Ext.
\end{minipage} & \begin{minipage}[c]{\linewidth}\raggedright
S\textsubscript{2}\newline Ext. / Ext.
\end{minipage} \\
\midrule\noalign{}
\endhead
\bottomrule
\endlastfoot
formation E/atom & 0.990 & 37,797 & 0.976\newline ±\,0.005 & 0.958\newline ±\,0.005 & −0.018\newline ±\,0.001 & 439 & 0.987 & 0.985 & 0.979 & 0.977 \\
energy above hull & 0.960 & 37,805 & 0.158\newline ±\,0.146 & 0.014\newline ±\,0.010 & −0.144\newline ±\,0.155 & 439 & 0.859 & 0.832 & 0.857 & 0.848 \\
band gap & 0.907 & 37,694 & 0.814\newline ±\,0.023 & 0.732\newline ±\,0.033 & −0.082\newline ±\,0.015 & 436 & 0.939 & 0.936 & 0.871 & 0.890 \\
density & 0.983 & 37,805 & 0.973\newline ±\,0.001 & 0.937\newline ±\,0.002 & −0.036\newline ±\,0.002 & 439 & 0.977 & 0.971 & 0.966 & 0.975 \\
crystal system & 0.817 & 37,805 & 0.908\newline ±\,0.007 & 0.647\newline ±\,0.005 & −0.261\newline ±\,0.004 & 418 & 0.804 & 0.668 & 0.734 & 0.692 \\
metal/insulator & 0.914 & 37,694 & 0.943\newline ±\,0.004 & 0.907\newline ±\,0.004 & −0.036\newline ±\,0.001 & 461 & 0.967 & 0.951 & 0.932 & 0.930 \\
\end{longtable}}

\clearpage
\hypertarget{tab:S20}{}%
{\small \textbf{Table~S20 --- Retrospective materials screening with frozen embeddings and composition descriptors (\hyperlink{screening-enrichment-results}{§B.6}).} \emph{Selected XGBoost predictors using ORB-v3, UMA-S or Magpie features are evaluated on the same seed-42 test set. \textbf{(a)} Recovery of the best 1~\%, 5~\% and 10~\% of materials, as determined by their DFT reference values. For each selection size, we compare the materials selected using model predictions with those identified using the reference values. \textbf{Recall (\%)} is the percentage of these reference-best materials successfully selected. Lower values are preferred for formation energy and density; higher values for bulk modulus and magnetic-moment density. Brackets give 95~\% confidence intervals from 2,000 paired bootstrap resamples. \textbf{(b)} Energy above hull: \textbf{Precision (\%)} is the percentage of selected candidates whose reference hull energy is at most 0, 50 or 100~meV/atom. \textbf{Top k} is the percentage selected, while \textbf{Base rate} is the percentage of all test materials meeting each threshold, equivalent to the expected precision under random selection. \textbf{(c)} Band gap: precision is the percentage of materials predicted to lie in the 0.5--3.0~eV window that truly do; recall is the percentage of all materials in the true window that are selected. \textbf{True in window} gives the number of qualifying test materials, and Spearman ρ measures agreement between predicted and reference band-gap rankings. \textbf{(d)} Categorical targets: one-vs-rest AUROC measures how well each class's predicted probability ranks its members above non-members, across classification thresholds (0.5 = random ranking; 1 = perfect separation). \textbf{Base rate} gives the proportion of test materials belonging to each class.}\par}

\noindent\begin{minipage}{\linewidth}
\textbf{(a) Regression targets: recall (\%) of the reference-best 1, 5 and 10 \%}\par\nobreak
\begin{tabular}{@{}
  >{\raggedright\arraybackslash}p{(\columnwidth - 8\tabcolsep) * \real{0.2453}}
  >{\raggedright\arraybackslash}p{(\columnwidth - 8\tabcolsep) * \real{0.1132}}
  >{\raggedright\arraybackslash}p{(\columnwidth - 8\tabcolsep) * \real{0.2138}}
  >{\raggedright\arraybackslash}p{(\columnwidth - 8\tabcolsep) * \real{0.2138}}
  >{\raggedright\arraybackslash}p{(\columnwidth - 8\tabcolsep) * \real{0.2139}}@{}}
\toprule\noalign{}
\begin{minipage}[c]{\linewidth}\raggedright
Target
\end{minipage} & \begin{minipage}[c]{\linewidth}\raggedright
Reference best
\end{minipage} & \begin{minipage}[c]{\linewidth}\raggedright
ORB-v3 recall (\%)
\end{minipage} & \begin{minipage}[c]{\linewidth}\raggedright
UMA-S recall (\%)
\end{minipage} & \begin{minipage}[c]{\linewidth}\raggedright
Magpie recall (\%)
\end{minipage} \\
\midrule\noalign{}
formation E/atom & 1\% & 87.7 {[}85.2, 91.3{]} & 86.8 {[}84.8, 90.7{]} & 75.8 {[}72.3, 79.7{]} \\
 & 5\% & 92.2 {[}91.2, 93.1{]} & 92.9 {[}92.2, 94.0{]} & 80.6 {[}78.8, 81.9{]} \\
 & 10\% & 93.3 {[}92.6, 94.0{]} & 93.1 {[}92.6, 94.0{]} & 85.3 {[}84.4, 86.2{]} \\
density & 1\% & 83.9 {[}80.7, 86.8{]} & 82.3 {[}78.1, 85.2{]} & 4.8 {[}2.9, 8.1{]} \\
 & 5\% & 88.6 {[}87.4, 89.8{]} & 89.0 {[}87.8, 90.1{]} & 60.4 {[}58.1, 62.9{]} \\
 & 10\% & 88.2 {[}87.5, 89.1{]} & 89.4 {[}88.6, 90.3{]} & 74.9 {[}73.7, 76.3{]} \\
bulk modulus & 1\% & 69.2 {[}53.9, 84.6{]} & 80.8 {[}61.5, 88.5{]} & 65.4 {[}46.1, 76.9{]} \\
 & 5\% & 86.8 {[}83.0, 92.2{]} & 87.6 {[}83.0, 91.4{]} & 80.6 {[}75.2, 86.8{]} \\
 & 10\% & 91.4 {[}87.2, 93.4{]} & 91.0 {[}87.9, 94.2{]} & 86.0 {[}82.1, 89.1{]} \\
mag. density & 1\% & 73.2 {[}68.7, 77.1{]} & 71.9 {[}67.4, 75.2{]} & 70.0 {[}66.1, 74.5{]} \\
 & 5\% & 81.9 {[}80.5, 83.8{]} & 80.2 {[}78.7, 82.2{]} & 79.5 {[}77.8, 81.6{]} \\
 & 10\% & 84.6 {[}83.6, 85.6{]} & 83.0 {[}82.0, 84.1{]} & 81.8 {[}80.7, 82.9{]} \\
\bottomrule\end{tabular}
\end{minipage}

\noindent\begin{minipage}{\linewidth}
\textbf{(b) Energy above hull: precision (\%) of the predicted top k}\par\nobreak
\begin{tabular}{@{}
  >{\raggedright\arraybackslash}p{(\columnwidth - 10\tabcolsep) * \real{0.3000}}
  >{\raggedright\arraybackslash}p{(\columnwidth - 10\tabcolsep) * \real{0.1300}}
  >{\raggedright\arraybackslash}p{(\columnwidth - 10\tabcolsep) * \real{0.0900}}
  >{\raggedright\arraybackslash}p{(\columnwidth - 10\tabcolsep) * \real{0.1600}}
  >{\raggedright\arraybackslash}p{(\columnwidth - 10\tabcolsep) * \real{0.1600}}
  >{\raggedright\arraybackslash}p{(\columnwidth - 10\tabcolsep) * \real{0.1600}}@{}}
\toprule\noalign{}
\begin{minipage}[c]{\linewidth}\raggedright
Criterion
\end{minipage} & \begin{minipage}[c]{\linewidth}\raggedright
Base rate (\%)
\end{minipage} & \begin{minipage}[c]{\linewidth}\raggedright
Top k
\end{minipage} & \begin{minipage}[c]{\linewidth}\raggedright
ORB-v3 precision (\%)
\end{minipage} & \begin{minipage}[c]{\linewidth}\raggedright
UMA-S precision (\%)
\end{minipage} & \begin{minipage}[c]{\linewidth}\raggedright
Magpie precision (\%)
\end{minipage} \\
\midrule\noalign{}
true E-hull ≤ 0.00 eV/atom & 22.0 & 1\% & 68.7 & 73.9 & 60.7 \\
 &  & 5\% & 66.6 & 70.1 & 54.9 \\
 &  & 10\% & 63.9 & 65.8 & 48.4 \\
true E-hull ≤ 0.05 eV/atom & 50.6 & 1\% & 96.5 & 97.1 & 88.1 \\
 &  & 5\% & 97.4 & 97.7 & 87.5 \\
 &  & 10\% & 97.2 & 96.7 & 83.5 \\
true E-hull ≤ 0.10 eV/atom & 67.2 & 1\% & 98.7 & 98.4 & 95.8 \\
 &  & 5\% & 99.2 & 99.3 & 94.0 \\
 &  & 10\% & 99.0 & 99.1 & 93.2 \\
\bottomrule\end{tabular}
\end{minipage}

\noindent\begin{minipage}{\linewidth}
\textbf{(c) Band-gap window 0.5–3.0 eV: precision and recall (\%)}\par\nobreak
\begin{tabular}{@{}
  >{\raggedright\arraybackslash}p{(\columnwidth - 8\tabcolsep) * \real{0.2000}}
  >{\raggedright\arraybackslash}p{(\columnwidth - 8\tabcolsep) * \real{0.2000}}
  >{\raggedright\arraybackslash}p{(\columnwidth - 8\tabcolsep) * \real{0.2000}}
  >{\raggedright\arraybackslash}p{(\columnwidth - 8\tabcolsep) * \real{0.2000}}
  >{\raggedright\arraybackslash}p{(\columnwidth - 8\tabcolsep) * \real{0.2000}}@{}}
\toprule\noalign{}
\begin{minipage}[c]{\linewidth}\raggedright
Model
\end{minipage} & \begin{minipage}[c]{\linewidth}\raggedright
True in window
\end{minipage} & \begin{minipage}[c]{\linewidth}\raggedright
Precision (\%)
\end{minipage} & \begin{minipage}[c]{\linewidth}\raggedright
Recall (\%)
\end{minipage} & \begin{minipage}[c]{\linewidth}\raggedright
Spearman ρ
\end{minipage} \\
\midrule\noalign{}
ORB-v3 & 8,939 & 72.6 & 86.8 & 0.871 \\
UMA-S & 8,939 & 75.9 & 85.4 & 0.882 \\
Magpie & 8,939 & 51.6 & 87.0 & 0.782 \\
\bottomrule\end{tabular}
\end{minipage}

\noindent\begin{minipage}{\linewidth}
\textbf{(d) Categorical targets: one-vs-rest AUROC per class}\par\nobreak
\begin{tabular}{@{}
  >{\raggedright\arraybackslash}p{(\columnwidth - 10\tabcolsep) * \real{0.2063}}
  >{\raggedright\arraybackslash}p{(\columnwidth - 10\tabcolsep) * \real{0.1587}}
  >{\raggedright\arraybackslash}p{(\columnwidth - 10\tabcolsep) * \real{0.1587}}
  >{\raggedright\arraybackslash}p{(\columnwidth - 10\tabcolsep) * \real{0.1587}}
  >{\raggedright\arraybackslash}p{(\columnwidth - 10\tabcolsep) * \real{0.1587}}
  >{\raggedright\arraybackslash}p{(\columnwidth - 10\tabcolsep) * \real{0.1587}}@{}}
\toprule\noalign{}
\begin{minipage}[c]{\linewidth}\raggedright
Target
\end{minipage} & \begin{minipage}[c]{\linewidth}\raggedright
Class
\end{minipage} & \begin{minipage}[c]{\linewidth}\raggedright
Base rate
\end{minipage} & \begin{minipage}[c]{\linewidth}\raggedright
ORB-v3 AUROC
\end{minipage} & \begin{minipage}[c]{\linewidth}\raggedright
UMA-S AUROC
\end{minipage} & \begin{minipage}[c]{\linewidth}\raggedright
Magpie AUROC
\end{minipage} \\
\midrule\noalign{}
crystal system & Cubic & 0.135 & 0.968 & 1.000 & 0.926 \\
 & Hexagonal & 0.069 & 0.938 & 0.981 & 0.903 \\
 & Monoclinic & 0.234 & 0.833 & 0.956 & 0.809 \\
 & Orthorhombic & 0.210 & 0.843 & 0.967 & 0.788 \\
 & Tetragonal & 0.113 & 0.911 & 0.973 & 0.869 \\
 & Triclinic & 0.156 & 0.900 & 0.974 & 0.884 \\
 & Trigonal & 0.084 & 0.901 & 0.983 & 0.863 \\
metal/insulator & Insulator & 0.531 & 0.970 & 0.973 & 0.924 \\
 & Metal & 0.469 & 0.970 & 0.973 & 0.924 \\
magnetic label & Magnetic & 0.473 & 0.954 & 0.962 & 0.928 \\
 & Non-magnetic & 0.527 & 0.954 & 0.962 & 0.928 \\
\bottomrule\end{tabular}
\end{minipage}

\clearpage
\hypertarget{tab:S21}{}%
{\small \textbf{Table~S21 --- Label efficiency of frozen-embedding predictors and the composition-only baseline (\hyperlink{label-efficiency-results}{§B.7}).} \emph{XGBoost predictors are trained on frozen ORB-v3 and UMA-S backbone node embeddings or Magpie composition descriptors, using the same randomly sampled training materials at each label count. Evaluation uses a fixed test set of 20,000 materials per target, or half the labelled materials for bulk modulus. Scores are R² for continuous targets and macro-F1 for categorical targets, reported as medians across training draws. \textbf{(a)} Prediction scores with 100, 1,000 and 20,000 labelled training materials. \textbf{(b)} Minimum training size at which each predictor becomes \textbf{stable} (score variation across training draws remains at most 0.05) and at which it becomes \textbf{saturated} (score remains within 0.05 of its full-data performance). \textbf{Not reached} indicates that saturation was not achieved at the largest training size. Bulk modulus has no 20,000-material training sample.}\par}

\noindent\begin{minipage}{\linewidth}
\textbf{(a) Median score at 100, 1,000 and 20,000 training materials}\par\nobreak
\begin{tabular}{@{}
  >{\raggedright\arraybackslash}p{(\columnwidth - 8\tabcolsep) * \real{0.2889}}
  >{\raggedright\arraybackslash}p{(\columnwidth - 8\tabcolsep) * \real{0.1778}}
  >{\raggedright\arraybackslash}p{(\columnwidth - 8\tabcolsep) * \real{0.1778}}
  >{\raggedright\arraybackslash}p{(\columnwidth - 8\tabcolsep) * \real{0.1778}}
  >{\raggedright\arraybackslash}p{(\columnwidth - 8\tabcolsep) * \real{0.1778}}@{}}
\toprule\noalign{}
\begin{minipage}[c]{\linewidth}\raggedright
Target
\end{minipage} & \begin{minipage}[c]{\linewidth}\raggedright
n train
\end{minipage} & \begin{minipage}[c]{\linewidth}\raggedright
ORB-v3
\end{minipage} & \begin{minipage}[c]{\linewidth}\raggedright
UMA-S
\end{minipage} & \begin{minipage}[c]{\linewidth}\raggedright
Magpie
\end{minipage} \\
\midrule\noalign{}
formation E/atom & 100 & 0.739 & 0.803 & 0.702 \\
 & 1000 & 0.926 & 0.946 & 0.813 \\
 & 20000 & 0.980 & 0.985 & 0.866 \\
energy above hull & 100 & 0.385 & 0.500 & −0.323 \\
 & 1000 & 0.763 & 0.834 & −0.021 \\
 & 20000 & 0.918 & 0.945 & 0.232 \\
band gap & 100 & 0.438 & 0.560 & 0.260 \\
 & 1000 & 0.728 & 0.768 & 0.450 \\
 & 20000 & 0.862 & 0.883 & 0.655 \\
density & 100 & 0.460 & 0.483 & 0.647 \\
 & 1000 & 0.775 & 0.782 & 0.794 \\
 & 20000 & 0.927 & 0.918 & 0.860 \\
bulk modulus & 100 & 0.613 & 0.665 & 0.652 \\
 & 1000 & 0.857 & 0.868 & 0.823 \\
 & 20000 & -- & -- & -- \\
mag. density & 100 & 0.206 & 0.369 & 0.297 \\
 & 1000 & 0.578 & 0.620 & 0.638 \\
 & 20000 & 0.763 & 0.782 & 0.783 \\
crystal system & 100 & 0.223 & 0.293 & 0.222 \\
 & 1000 & 0.372 & 0.440 & 0.327 \\
 & 20000 & 0.554 & 0.617 & 0.474 \\
metal/insulator & 100 & 0.787 & 0.819 & 0.743 \\
 & 1000 & 0.863 & 0.870 & 0.784 \\
 & 20000 & 0.897 & 0.905 & 0.835 \\
magnetic label & 100 & 0.698 & 0.743 & 0.684 \\
 & 1000 & 0.812 & 0.846 & 0.775 \\
 & 20000 & 0.877 & 0.894 & 0.844 \\
\bottomrule\end{tabular}
\end{minipage}

\noindent\begin{minipage}{\linewidth}
\textbf{(b) Training size needed}\par\nobreak
\begin{tabular}{@{}
  >{\raggedright\arraybackslash}p{(\columnwidth - 12\tabcolsep) * \real{0.2131}}
  >{\raggedright\arraybackslash}p{(\columnwidth - 12\tabcolsep) * \real{0.1311}}
  >{\raggedright\arraybackslash}p{(\columnwidth - 12\tabcolsep) * \real{0.1311}}
  >{\raggedright\arraybackslash}p{(\columnwidth - 12\tabcolsep) * \real{0.1311}}
  >{\raggedright\arraybackslash}p{(\columnwidth - 12\tabcolsep) * \real{0.1311}}
  >{\raggedright\arraybackslash}p{(\columnwidth - 12\tabcolsep) * \real{0.1311}}
  >{\raggedright\arraybackslash}p{(\columnwidth - 12\tabcolsep) * \real{0.1311}}@{}}
\toprule\noalign{}
\begin{minipage}[c]{\linewidth}\raggedright
Target
\end{minipage} & \begin{minipage}[c]{\linewidth}\raggedright
ORB-v3 stable
\end{minipage} & \begin{minipage}[c]{\linewidth}\raggedright
UMA-S stable
\end{minipage} & \begin{minipage}[c]{\linewidth}\raggedright
Magpie stable
\end{minipage} & \begin{minipage}[c]{\linewidth}\raggedright
ORB-v3 saturated
\end{minipage} & \begin{minipage}[c]{\linewidth}\raggedright
UMA-S saturated
\end{minipage} & \begin{minipage}[c]{\linewidth}\raggedright
Magpie saturated
\end{minipage} \\
\midrule\noalign{}
formation E/atom & 500 & 500 & 500 & 2000 & 1000 & 5000 \\
energy above hull & 2000 & 2000 & 10000 & 10000 & 5000 & not reached \\
band gap & 500 & 500 & 500 & 7000 & 7000 & 20000 \\
density & 1000 & 500 & 500 & 7000 & 7000 & 7000 \\
bulk modulus & 500 & 500 & 500 & 2000 & 2000 & 2000 \\
mag. density & 1000 & 500 & 2000 & 20000 & 20000 & 20000 \\
crystal system & 500 & 50 & 50 & 20000 & 20000 & 20000 \\
metal/insulator & 100 & 100 & 100 & 1000 & 1000 & 2000 \\
magnetic label & 100 & 200 & 200 & 5000 & 2000 & 5000 \\
\bottomrule\end{tabular}
\end{minipage}

\clearpage
\hypertarget{tab:S22}{}%
{\small \textbf{Table~S22 --- Comparison of a network trained from scratch with frozen-embedding predictors (\hyperlink{label-efficiency-results}{§B.7}).} \emph{A three-layer message-passing network (64 hidden features, approximately 77,500 parameters) is trained from random initialisation on crystal structures and compared with XGBoost predictors using frozen ORB-v3 and UMA-S embeddings or Magpie composition descriptors. Scores are R² for continuous targets and macro-F1 for categorical targets. \textbf{(a)} Performance at the smallest training size (\textbf{n train}) at which both the ORB-v3 and UMA-S frozen-embedding predictors have saturated: their median score remains within 0.05 of their full-corpus score at all larger training sizes (\hyperlink{screening-enrichment-and-label-efficiency}{§A.11}). \textbf{GNN mean} and \textbf{GNN sd} report the mean and standard deviation across three independent training runs. The ORB-v3, UMA-S and Magpie columns give the corresponding median learning-curve scores at the same training size. \textbf{(b)} Performance of the network trained from scratch across increasing training sizes. Scores are averaged over three runs up to 5,000 materials and reported from a single run at larger sizes, except at the saturation sizes in (a), where three runs are used. Dashes indicate sizes not evaluated. The from-scratch network is scored on the seed-42 test set (30,976 materials; 2,571 for bulk modulus), while the frozen-embedding predictors and Magpie use the separate fixed test sets of the label-efficiency experiment (\hyperlink{screening-enrichment-and-label-efficiency}{§A.11}).}\par}

\noindent\begin{minipage}{\linewidth}
\textbf{(a) At the training size where both embedding probes have saturated}\par\nobreak
\begin{tabular}{@{}
  >{\raggedright\arraybackslash}p{(\columnwidth - 14\tabcolsep) * \real{0.1900}}
  >{\raggedright\arraybackslash}p{(\columnwidth - 14\tabcolsep) * \real{0.1000}}
  >{\raggedright\arraybackslash}p{(\columnwidth - 14\tabcolsep) * \real{0.0900}}
  >{\raggedright\arraybackslash}p{(\columnwidth - 14\tabcolsep) * \real{0.1100}}
  >{\raggedright\arraybackslash}p{(\columnwidth - 14\tabcolsep) * \real{0.1000}}
  >{\raggedright\arraybackslash}p{(\columnwidth - 14\tabcolsep) * \real{0.1400}}
  >{\raggedright\arraybackslash}p{(\columnwidth - 14\tabcolsep) * \real{0.1400}}
  >{\raggedright\arraybackslash}p{(\columnwidth - 14\tabcolsep) * \real{0.1300}}@{}}
\toprule\noalign{}
\begin{minipage}[c]{\linewidth}\raggedright
Target
\end{minipage} & \begin{minipage}[c]{\linewidth}\raggedright
Metric
\end{minipage} & \begin{minipage}[c]{\linewidth}\raggedright
n train
\end{minipage} & \begin{minipage}[c]{\linewidth}\raggedright
GNN mean
\end{minipage} & \begin{minipage}[c]{\linewidth}\raggedright
GNN sd
\end{minipage} & \begin{minipage}[c]{\linewidth}\raggedright
ORB-v3 probe
\end{minipage} & \begin{minipage}[c]{\linewidth}\raggedright
UMA-S probe
\end{minipage} & \begin{minipage}[c]{\linewidth}\raggedright
Magpie
\end{minipage} \\
\midrule\noalign{}
formation E/atom & R² & 2000 & 0.861 & 0.008 & 0.948 & 0.963 & 0.829 \\
energy above hull & R² & 10000 & 0.555 & 0.077 & 0.902 & 0.934 & 0.180 \\
band gap & R² & 7000 & 0.528 & 0.015 & 0.836 & 0.858 & 0.599 \\
density & R² & 7000 & 0.992 & 0.000 & 0.897 & 0.890 & 0.843 \\
bulk modulus & R² & 2000 & 0.885 & 0.001 & 0.893 & 0.905 & 0.850 \\
mag. density & R² & 20000 & 0.774 & 0.021 & 0.763 & 0.782 & 0.783 \\
crystal system & macro-F1 & 20000 & 0.421 & 0.011 & 0.554 & 0.617 & 0.474 \\
metal/insulator & macro-F1 & 1000 & 0.755 & 0.012 & 0.863 & 0.870 & 0.784 \\
magnetic label & macro-F1 & 5000 & 0.803 & 0.006 & 0.854 & 0.877 & 0.820 \\
\bottomrule\end{tabular}
\end{minipage}

\noindent\begin{minipage}{\linewidth}
\textbf{(b) Score of the from-scratch network by training size}\par\nobreak
\begin{tabular}{@{}
  >{\raggedright\arraybackslash}p{(\columnwidth - 16\tabcolsep) * \real{0.1688}}
  >{\raggedright\arraybackslash}p{(\columnwidth - 16\tabcolsep) * \real{0.1039}}
  >{\raggedright\arraybackslash}p{(\columnwidth - 16\tabcolsep) * \real{0.1039}}
  >{\raggedright\arraybackslash}p{(\columnwidth - 16\tabcolsep) * \real{0.1039}}
  >{\raggedright\arraybackslash}p{(\columnwidth - 16\tabcolsep) * \real{0.1039}}
  >{\raggedright\arraybackslash}p{(\columnwidth - 16\tabcolsep) * \real{0.1039}}
  >{\raggedright\arraybackslash}p{(\columnwidth - 16\tabcolsep) * \real{0.1039}}
  >{\raggedright\arraybackslash}p{(\columnwidth - 16\tabcolsep) * \real{0.1039}}
  >{\raggedright\arraybackslash}p{(\columnwidth - 16\tabcolsep) * \real{0.1039}}@{}}
\toprule\noalign{}
\begin{minipage}[c]{\linewidth}\raggedright
Target
\end{minipage} & \begin{minipage}[c]{\linewidth}\raggedright
n=200
\end{minipage} & \begin{minipage}[c]{\linewidth}\raggedright
n=500
\end{minipage} & \begin{minipage}[c]{\linewidth}\raggedright
n=1,000
\end{minipage} & \begin{minipage}[c]{\linewidth}\raggedright
n=2,000
\end{minipage} & \begin{minipage}[c]{\linewidth}\raggedright
n=5,000
\end{minipage} & \begin{minipage}[c]{\linewidth}\raggedright
n=7,000
\end{minipage} & \begin{minipage}[c]{\linewidth}\raggedright
n=10,000
\end{minipage} & \begin{minipage}[c]{\linewidth}\raggedright
n=20,000
\end{minipage} \\
\midrule\noalign{}
formation E/atom & 0.544 & 0.710 & 0.800 & 0.861 & 0.923 & -- & 0.939 & 0.969 \\
energy above hull & −0.043 & 0.164 & 0.183 & 0.394 & 0.504 & -- & 0.555 & 0.736 \\
band gap & 0.196 & 0.239 & 0.354 & 0.390 & 0.514 & 0.528 & 0.540 & 0.642 \\
density & 0.216 & 0.886 & 0.947 & 0.977 & 0.990 & 0.992 & 0.992 & 0.996 \\
bulk modulus & 0.317 & 0.777 & 0.863 & 0.885 & 0.917 & -- & 0.931 & -- \\
mag. density & 0.057 & 0.385 & 0.556 & 0.633 & 0.725 & -- & 0.761 & 0.774 \\
crystal system & 0.166 & 0.213 & 0.216 & 0.286 & 0.330 & -- & 0.364 & 0.421 \\
metal/insulator & 0.706 & 0.744 & 0.755 & 0.776 & 0.792 & -- & 0.801 & 0.813 \\
magnetic label & 0.646 & 0.702 & 0.731 & 0.763 & 0.803 & -- & 0.823 & 0.834 \\
\bottomrule\end{tabular}
\end{minipage}

\hypertarget{supplementary-figures}{%
\section{Supplementary Figures}\label{supplementary-figures}}

\begin{figure}[H]
  \hypertarget{fig:S1}{}%
  \centering
  \includegraphics{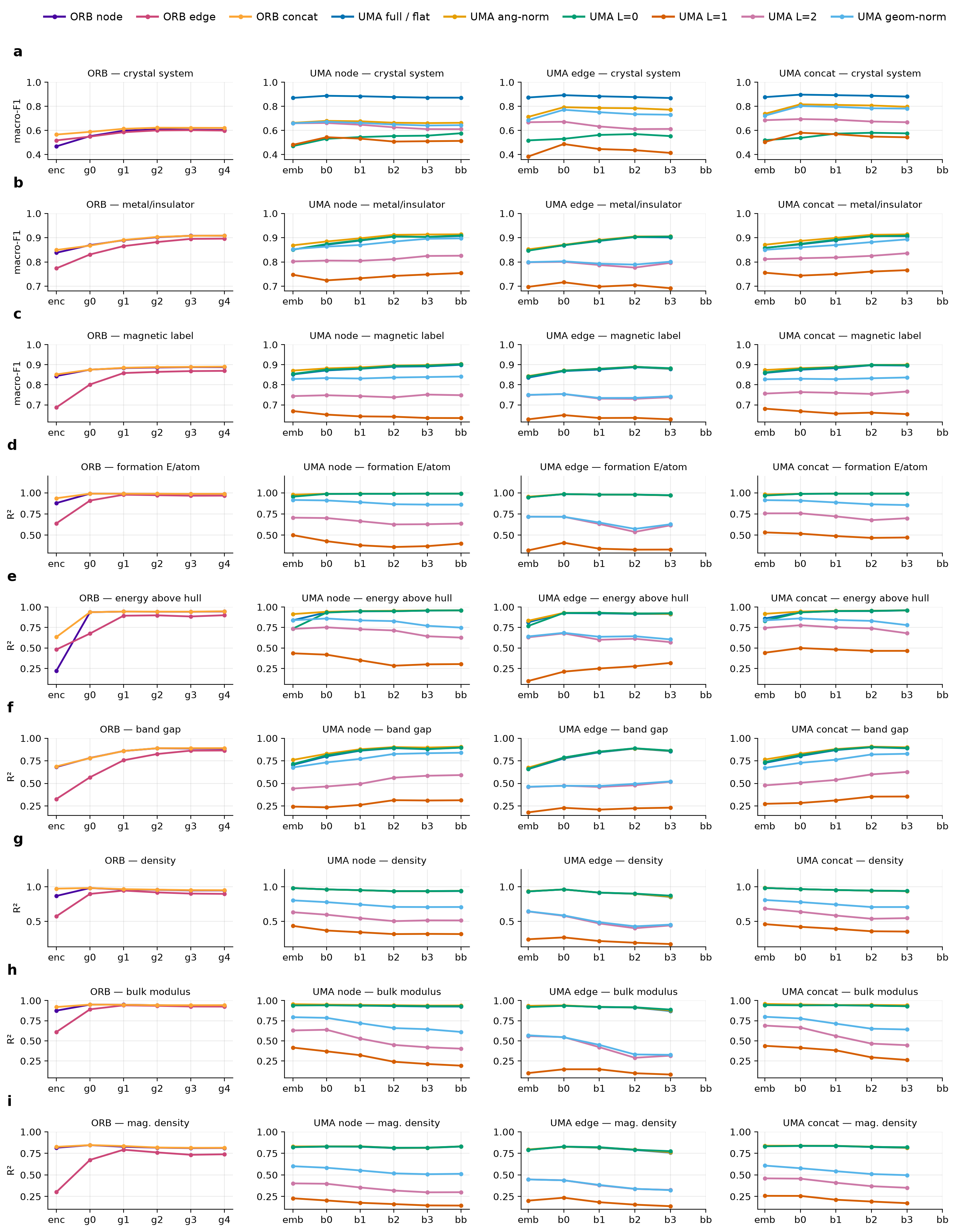}
  \caption*{\textbf{\emph{Figure~S1. XGBoost score across all layers, read-outs, and equivariant channel families for all nine properties.}} \emph{Each row corresponds to one property: \textbf{(a)} crystal system, \textbf{(b)} metal/insulator, \textbf{(c)} magnetic label, \textbf{(d)} formation E/atom, \textbf{(e)} energy above hull, \textbf{(f)} band gap, \textbf{(g)} density, \textbf{(h)} bulk modulus, \textbf{(i)} mag. density. Within each row the leftmost panel shows ORB-v3 layer curves for the three read-outs (node, edge, concat); the remaining three panels show UMA-S layer curves for the node, edge, and concat read-out groups, with each equivariant channel family (full/flat, ang-norm, L=0, L=1, L=2, geom-norm) as a separate line. The x-axis runs from the earliest to the deepest layer (ORB: enc--g4; UMA: emb--bb). Panels a--c use macro-F1; panels d--i use R². Legend at the top identifies read-out colours for ORB and channel-family colours for UMA. Best-configuration metrics per read-out are in \hyperlink{tab:S5}{Table~S5}.}}
\end{figure}

\begin{figure}[H]
  \hypertarget{fig:S2}{}%
  \centering
  \includegraphics{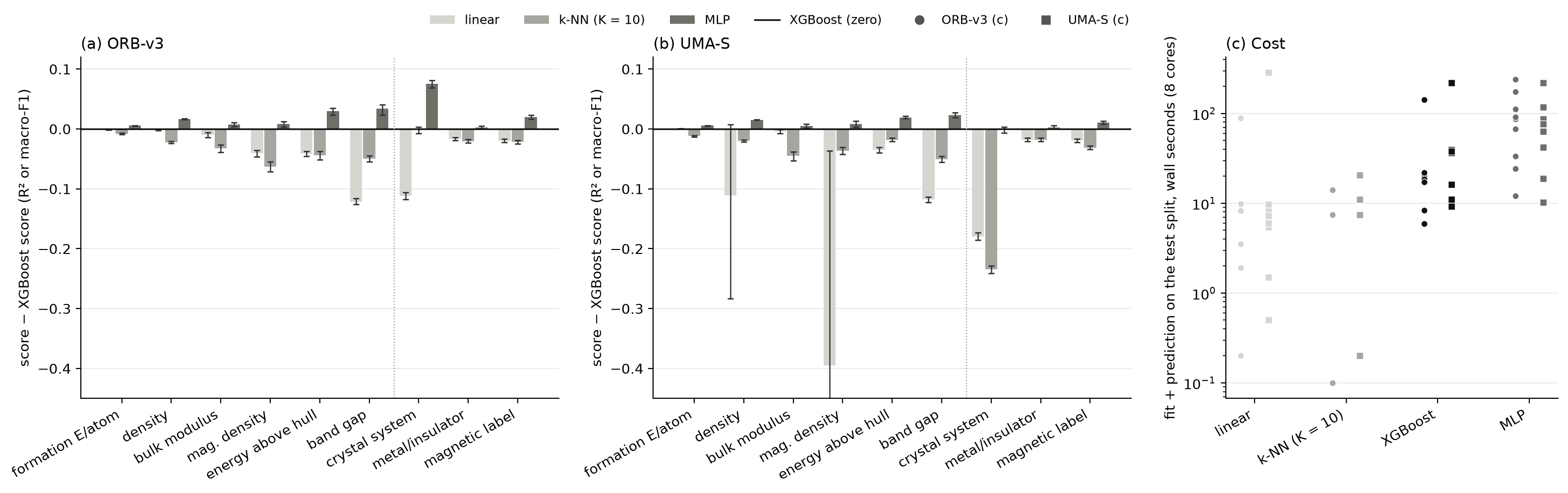}
  \caption*{\textbf{\emph{Figure~S2. Comparison of prediction accuracy and runtime across four supervised methods (\hyperlink{frozen-mlip-embeddings-predict-properties-outside-the-training-objective}{§3.1}).}} \emph{All methods use the same training and test materials and, for each target, the embedding configuration at which XGBoost scores highest (\textbf{\hyperlink{tab:1}{Table~1}}). \textbf{(a--b)} Differences in prediction score relative to XGBoost, for ORB-v3 and UMA-S, respectively. Zero indicates equal performance; negative values indicate lower accuracy than XGBoost. Error bars show 95~\% confidence intervals from paired bootstrap resampling of the test set. The vertical axis is clipped at −0.45; the linear probe's interval for UMA-S magnetic-moment density extends to −0.9. Scores are R² for continuous targets and macro-F1 for categorical targets. \textbf{(c)} Wall-clock time for one model fit and prediction on the test set, using eight CPU cores. Numerical scores are in \hyperlink{tab:S4}{Table~S4} and runtime details in \hyperlink{probe-family}{§B.5}.}}
\end{figure}

\begin{figure}[H]
  \hypertarget{fig:S3}{}%
  \centering
  \includegraphics{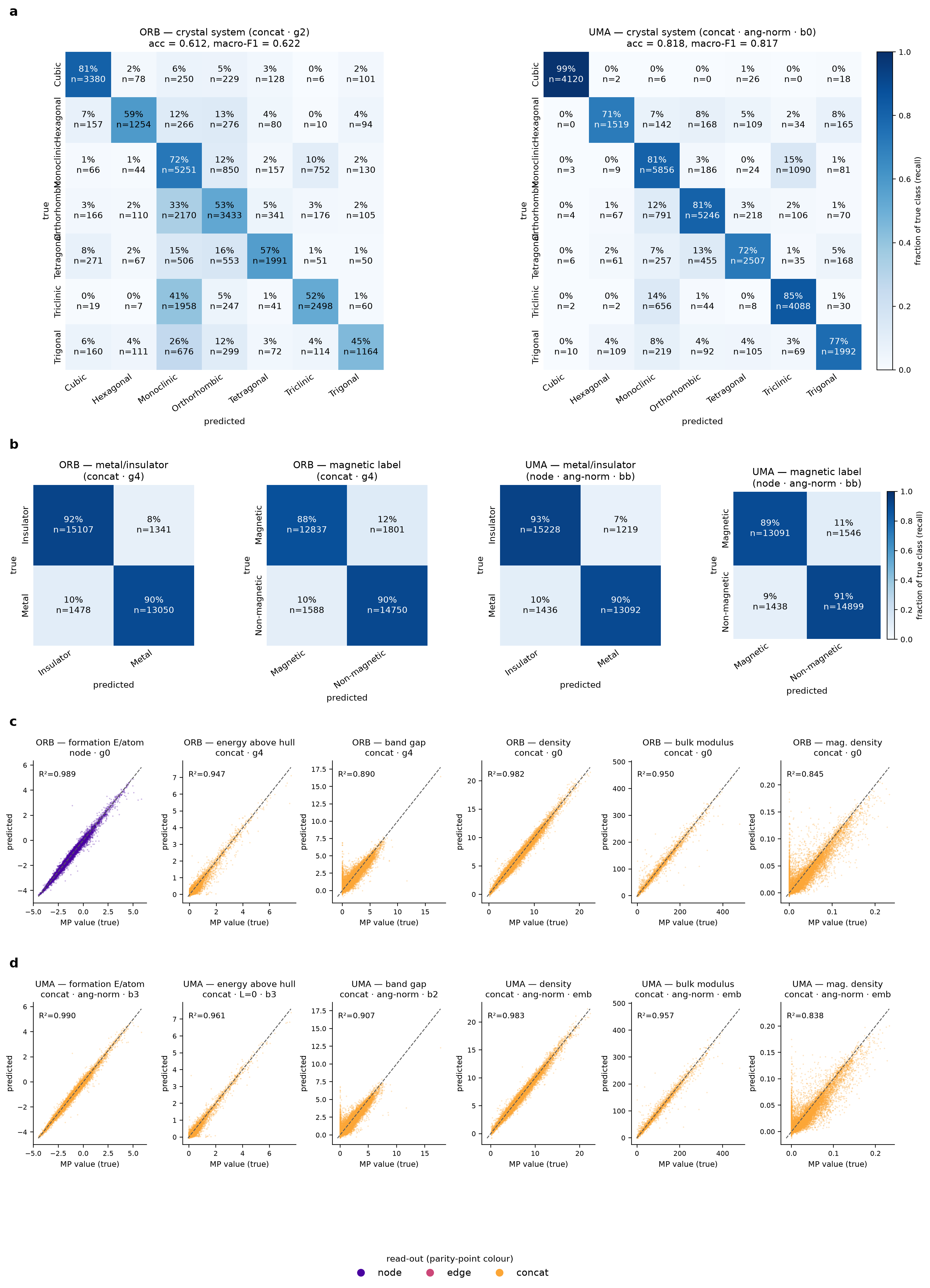}
  \caption*{\textbf{\emph{Figure~S3. Confusion matrices and parity plots at each target's selected configuration.}} \emph{\textbf{(a)} Crystal system, ORB-v3 (concat · g2) and UMA-S (concat · ang-norm · b0). \textbf{(b)} Metal/insulator and magnetic label, ORB-v3 (concat · g4) and UMA-S (node · ang-norm · bb). Cells give row-normalised recall and the test count. \textbf{(c)} ORB-v3 and \textbf{(d)} UMA-S parity plots for the six continuous targets, with the configuration in each panel title; dashed line, y = x. Metrics in \textbf{\hyperlink{tab:S5}{Table~S5}}.}}
\end{figure}

\begin{figure}[H]
  \hypertarget{fig:S4}{}%
  \centering
  \includegraphics[height=0.86\textheight]{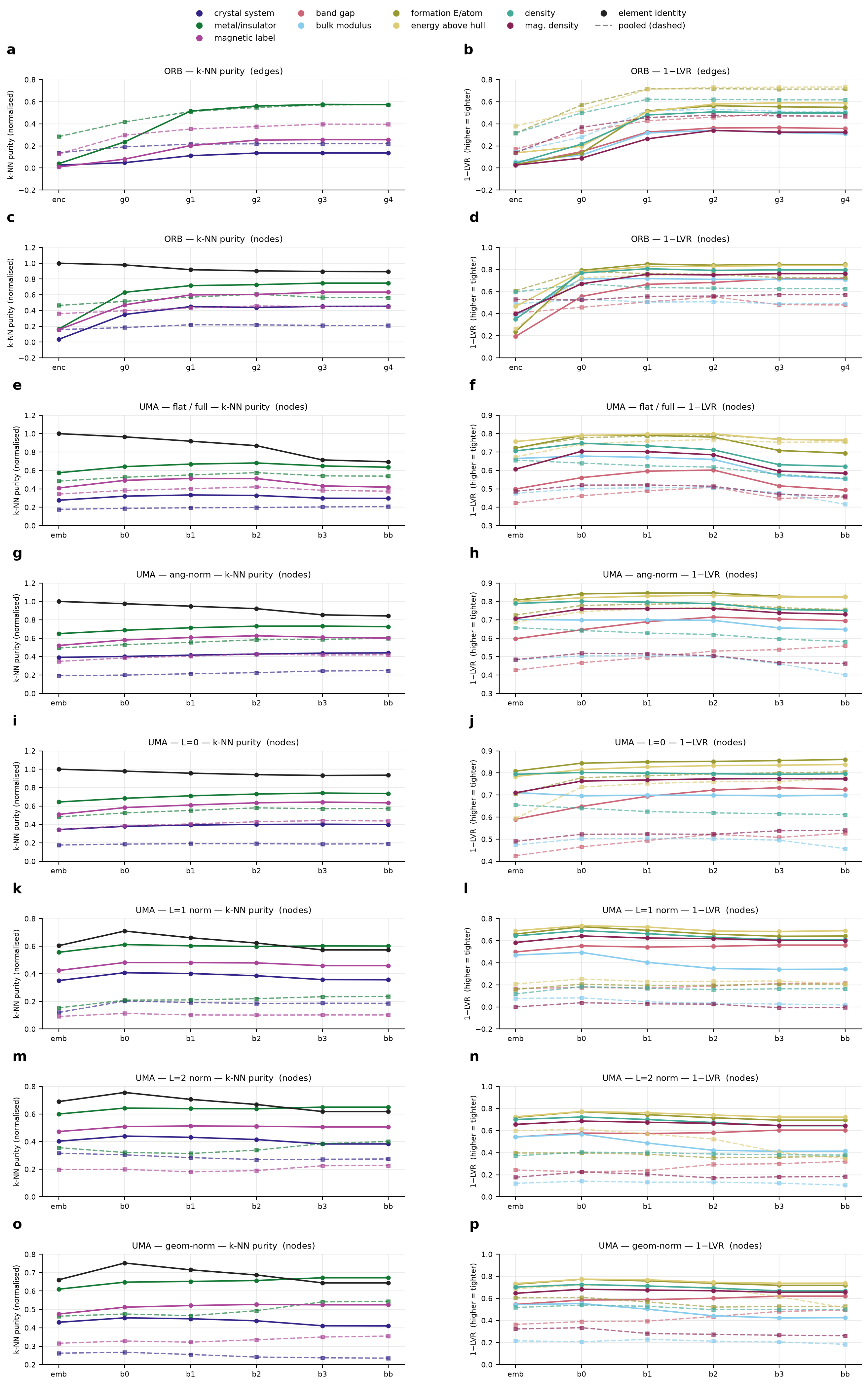}
  \caption*{\textbf{\emph{Figure~S4. k-NN purity and 1−LVR depth curves for ORB-v3 edge and node read-outs and all UMA-S node channel families.}} \emph{k-NN purity for the categorical properties (K = 10; higher is better class separation) and the local variance ratio for the continuous ones, reported as 1−LVR (higher is better within-class compactness), as a function of network layer. Solid lines: individual (atom/bond level); dashed: mean-pooled (material level). Panels run in pairs, k-NN purity (left) then 1−LVR (right): \textbf{(a--b)} ORB-v3 edge; \textbf{(c--d)} ORB-v3 node; \textbf{(e--f)} UMA-S flat/full; \textbf{(g--h)} ang-norm; \textbf{(i--j)} L=0; \textbf{(k--l)} L=1 norm; \textbf{(m--n)} L=2 norm; \textbf{(o--p)} geom-norm. Element identity is included as a reference class wherever it is defined, which is the node read-outs but not ORB's edges. Supports \hyperlink{property-decodability-depends-on-network-depth}{§3.3} and \hyperlink{property-information-is-locally-predictable-without-property-aligned-clusters}{§3.5}.}}
\end{figure}

\begin{figure}[H]
  \hypertarget{fig:S5}{}%
  \centering
  \includegraphics{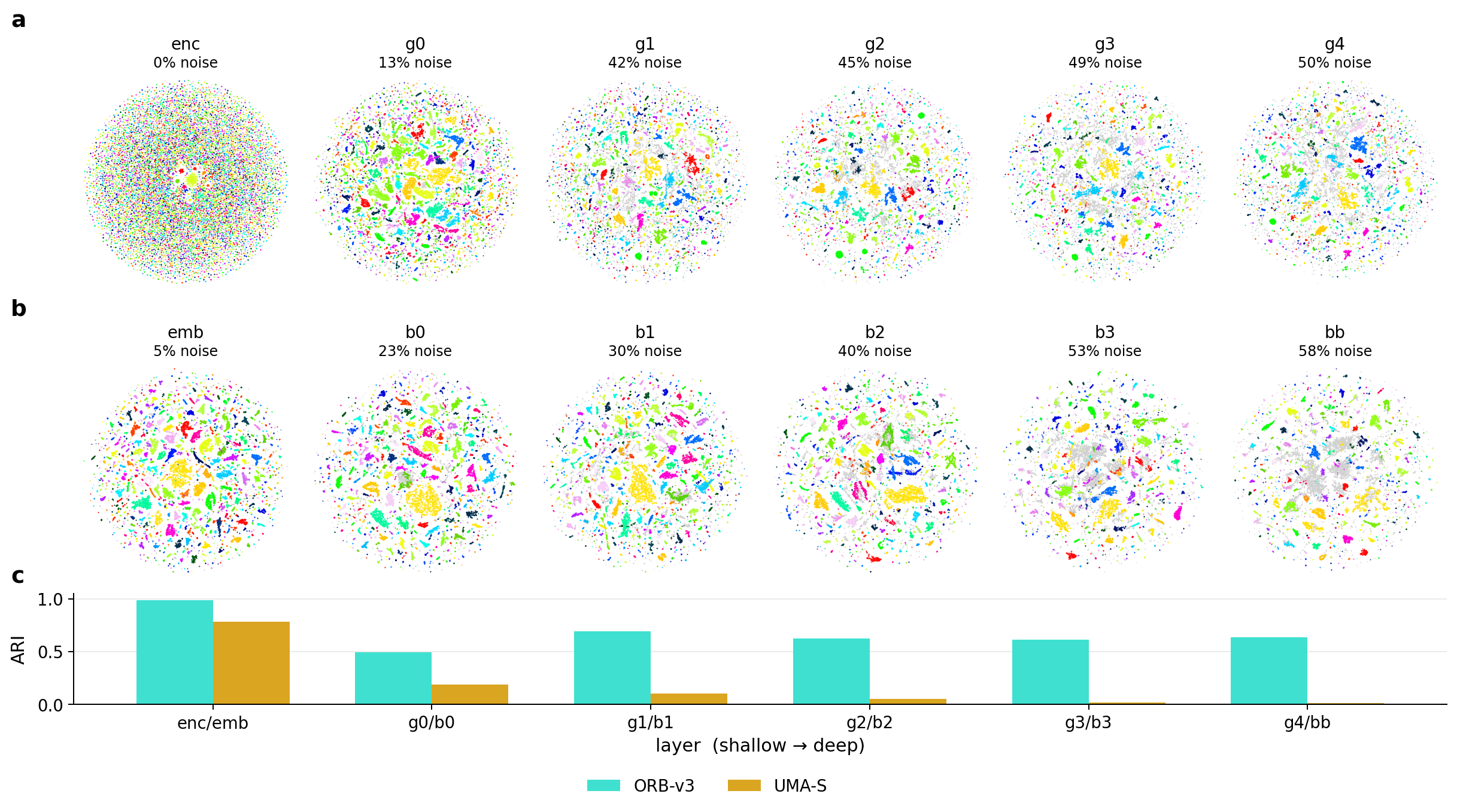}
  \caption*{\textbf{\emph{Figure~S5. Element-identity organisation in per-atom embeddings across all layers.}}
\emph{\textbf{(a)} ORB-v3: HDBSCAN cluster maps for all six layers (enc to g4, left to right), coloured by the majority element of each cluster; grey points are noise. Element clusters are dense and chemically coherent at the encoder, where almost every atom is assigned to one cluster (0.1~\% noise), and dissolve as message passing mixes neighbourhood context into the representations: by g4 half the atoms belong to no cluster at all. \textbf{(b)} UMA-S: same, for all six layers (emb to bb), on the ang-norm read-out. The dissolution pattern mirrors ORB: noise rises from 5.4~\% at emb to 58.3~\% at bb. \textbf{(c)} Adjusted Rand Index (ARI) between HDBSCAN cluster labels and element-identity labels, at each layer for both models. At the input layer HDBSCAN recovers element identity almost exactly for ORB (0.99) and closely for UMA (0.78). The two then diverge: UMA falls to 0.015 by the backbone, while ORB settles near 0.63, its remaining clusters still element-coherent even though half its atoms are unclustered by g4.}}
\end{figure}

\begin{figure}[H]
  \hypertarget{fig:S6}{}%
  \centering
  \includegraphics{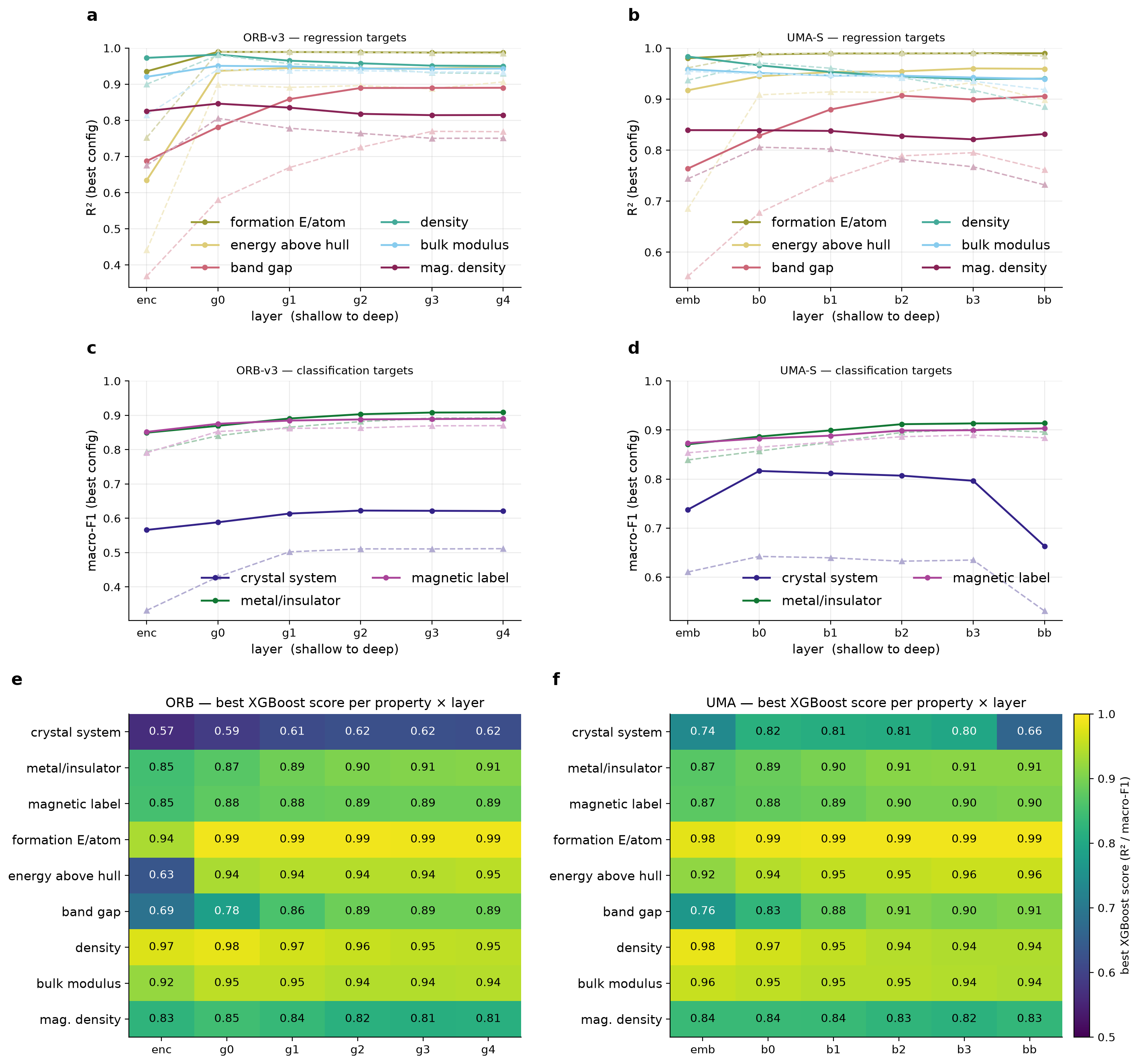}
  \caption*{\textbf{\emph{Figure~S6. Where information lives: the best XGBoost probe score at each layer for all nine material properties and both models, with the linear probe alongside it.}} \emph{\textbf{(a,b)} Per-layer best XGBoost score (dark solid) and linear probe score (light dashed), taking the best read-out at each layer. Layers run from earliest to deepest: ORB-v3 enc to g0--g4 \textbf{(a)}; UMA-S emb to b0--b3 to bb \textbf{(b)}. \textbf{(c,d)} Same as (a,b) for classification targets. UMA-S's edge features stop at b3 --- its backbone aggregates to nodes (\hyperlink{models-and-layer-wise-embedding-extraction}{§2.1}) --- so the best read-out at bb is node-only where every earlier layer can also use node-plus-edge. For eight of the nine targets that costs at most 0.003, node and concatenated read-outs being close to equivalent there; for crystal system, the one target whose signal is partly carried by the bonds, the score falls from 0.797 at b3 to 0.663 at bb. Held to the node read-out throughout, UMA-S's crystal-system trajectory is flat (0.661--0.679), so that fall is a change of available read-out rather than a loss of information with depth. \textbf{(e,f)} The same scores shown as heatmaps, one row per property and one column per layer, for ORB-v3 \textbf{(e)} and UMA-S \textbf{(f)}: R² for the regression targets and macro-F1 for the classification ones. Both panels share one colour scale, and the bar on the right applies to each. For crystal system, UMA-S's maximum is taken over the rotation-invariant channels only, throughout this figure, since the non-invariant configurations collapse under random rotation (\hyperlink{uma-channel-views-and-the-rotation-invariant-crystal-system-gap}{§3.4}, \hyperlink{tab:S13}{Table~S13}). Full layer-depth significance statistics (Nadeau--Bengio, both claim families, all nine material properties) are in \hyperlink{tab:S7}{Table~S7} (supports \hyperlink{property-decodability-depends-on-network-depth}{§3.3}).}}
\end{figure}

\begin{figure}[H]
  \hypertarget{fig:S7}{}%
  \centering
  \includegraphics[height=0.78\textheight]{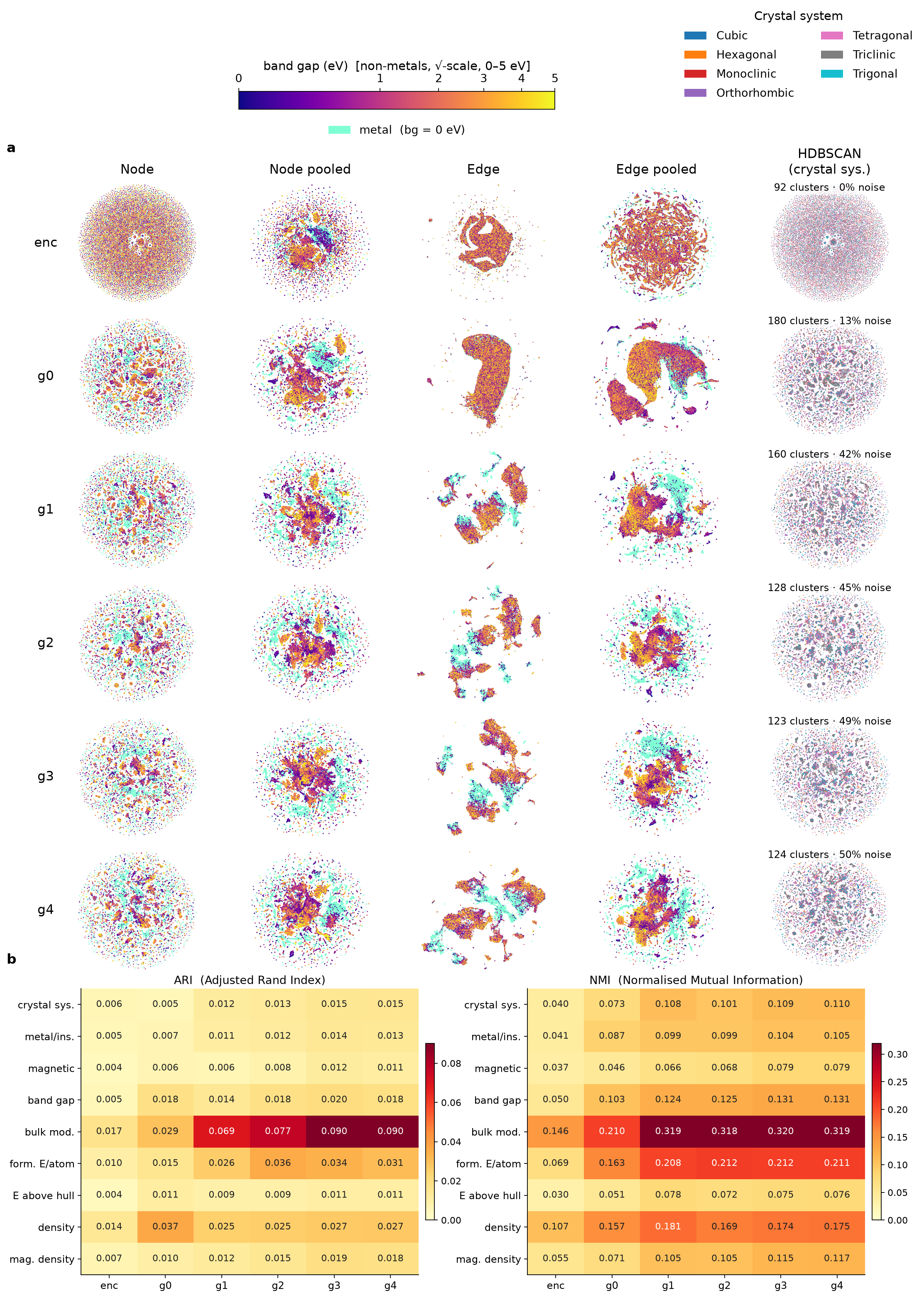}
  \caption*{\textbf{\emph{Figure~S7. ORB-v3 unsupervised geometry across all six layers (enc, g0--g4).}} \emph{\textbf{(a)} UMAP projections and HDBSCAN clusters, one row per ORB layer. The first four columns are the read-outs --- node (atom-level), node-pooled (material-level), edge (bond-level) and edge-pooled (material-level) --- coloured by band-gap magnitude (plasma scale, power-normalised γ = 0.5; metals aquamarine, missing data grey). The last column shows the node representation on the subsample HDBSCAN was run on, coloured by crystal system (grey: no crystal-system label), with the number of HDBSCAN clusters and the noise fraction printed above each panel. \textbf{(b)} Adjusted Rand Index (ARI, left) and Normalised Mutual Information (NMI, right) compare HDBSCAN clusters of node embeddings with material-property labels at each layer. Each atom carries its material's label; continuous targets are binned into quartiles, and HDBSCAN noise points are excluded. \textbf{ARI} measures whether pairs of atoms grouped together by their property labels are also grouped together by HDBSCAN, correcting for chance (0 = chance agreement; 1 = perfect agreement). \textbf{NMI} measures how much knowing an atom's cluster helps identify its property label (0 = no association; 1 = perfect agreement). Unlike ARI, NMI is not corrected for chance. Supports \hyperlink{property-information-is-locally-predictable-without-property-aligned-clusters}{§3.5}.}}
\end{figure}

\begin{figure}[H]
  \hypertarget{fig:S8}{}%
  \centering
  \includegraphics[height=510pt]{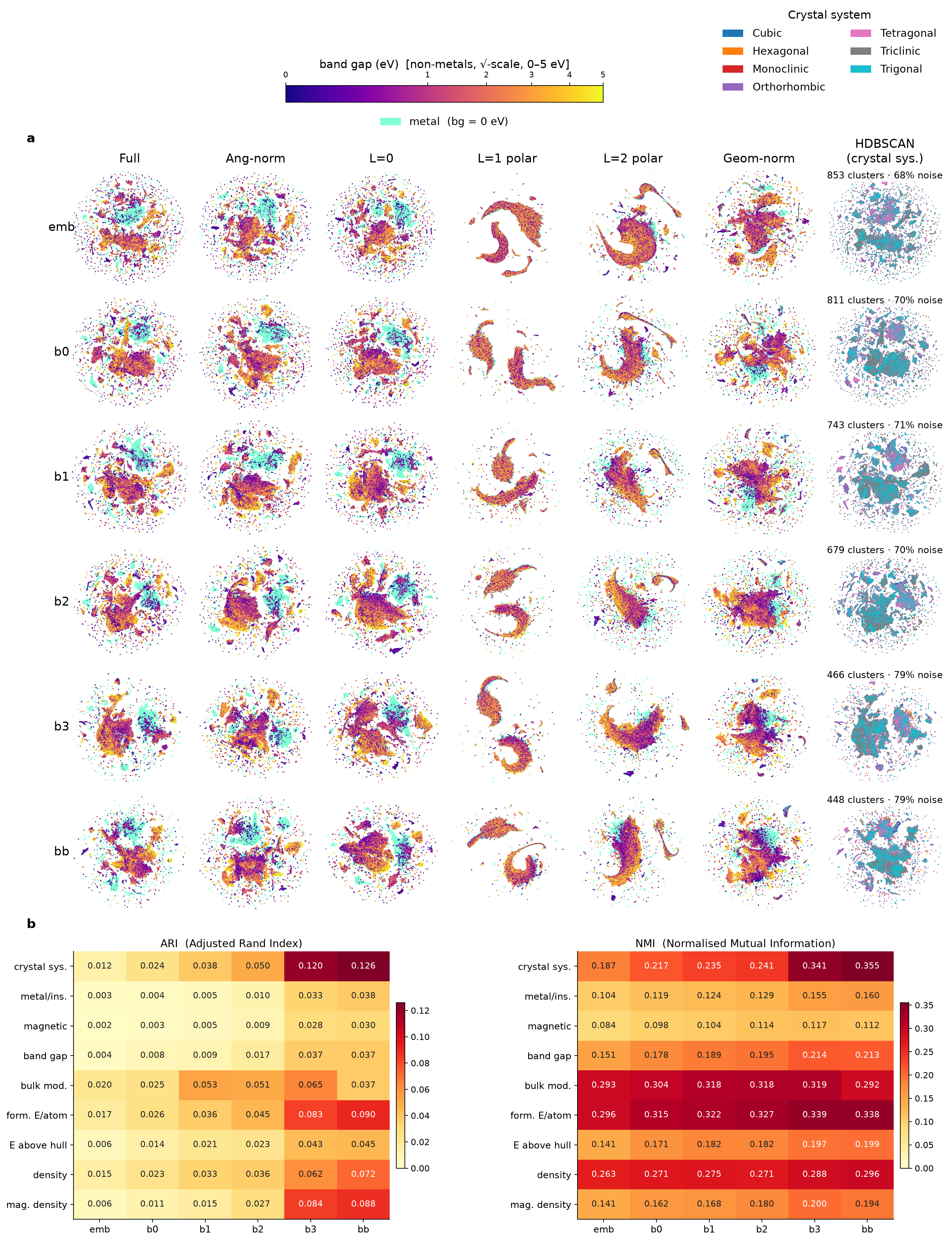}
  \caption*{\textbf{\emph{Figure~S8. UMA-S unsupervised geometry across all six layers (emb, b0--b3, bb).}} \emph{\textbf{(a)} UMAP projections and HDBSCAN clusters, one row per UMA-S layer. The first six columns are the pooled equivariant-channel families --- full, ang-norm, L=0, L=1 polar, L=2 polar and geom-norm --- and the last column shows HDBSCAN clusters of the pooled full channel. UMAPs are coloured by band-gap magnitude (plasma scale, power-normalised γ = 0.5; metals aquamarine, missing data grey); the HDBSCAN column is coloured by crystal system (grey: no crystal-system label), with the number of HDBSCAN clusters and the noise fraction printed above each panel. \textbf{(b)} Adjusted Rand Index (ARI, left) and Normalised Mutual Information (NMI, right) compare HDBSCAN clusters of the pooled full-channel embeddings with material-property labels at each layer. Continuous targets are binned into quartiles, and HDBSCAN noise points are excluded. \textbf{ARI} measures whether pairs of materials grouped together by their property labels are also grouped together by HDBSCAN, correcting for chance (0 = chance agreement; 1 = perfect agreement). \textbf{NMI} measures how much knowing a material's cluster helps identify its property label (0 = no association; 1 = perfect agreement). Unlike ARI, NMI is not corrected for chance. Supports \hyperlink{property-information-is-locally-predictable-without-property-aligned-clusters}{§3.5}.}}
\end{figure}

\begin{figure}[H]
  \hypertarget{fig:S9}{}%
  \centering
  \includegraphics{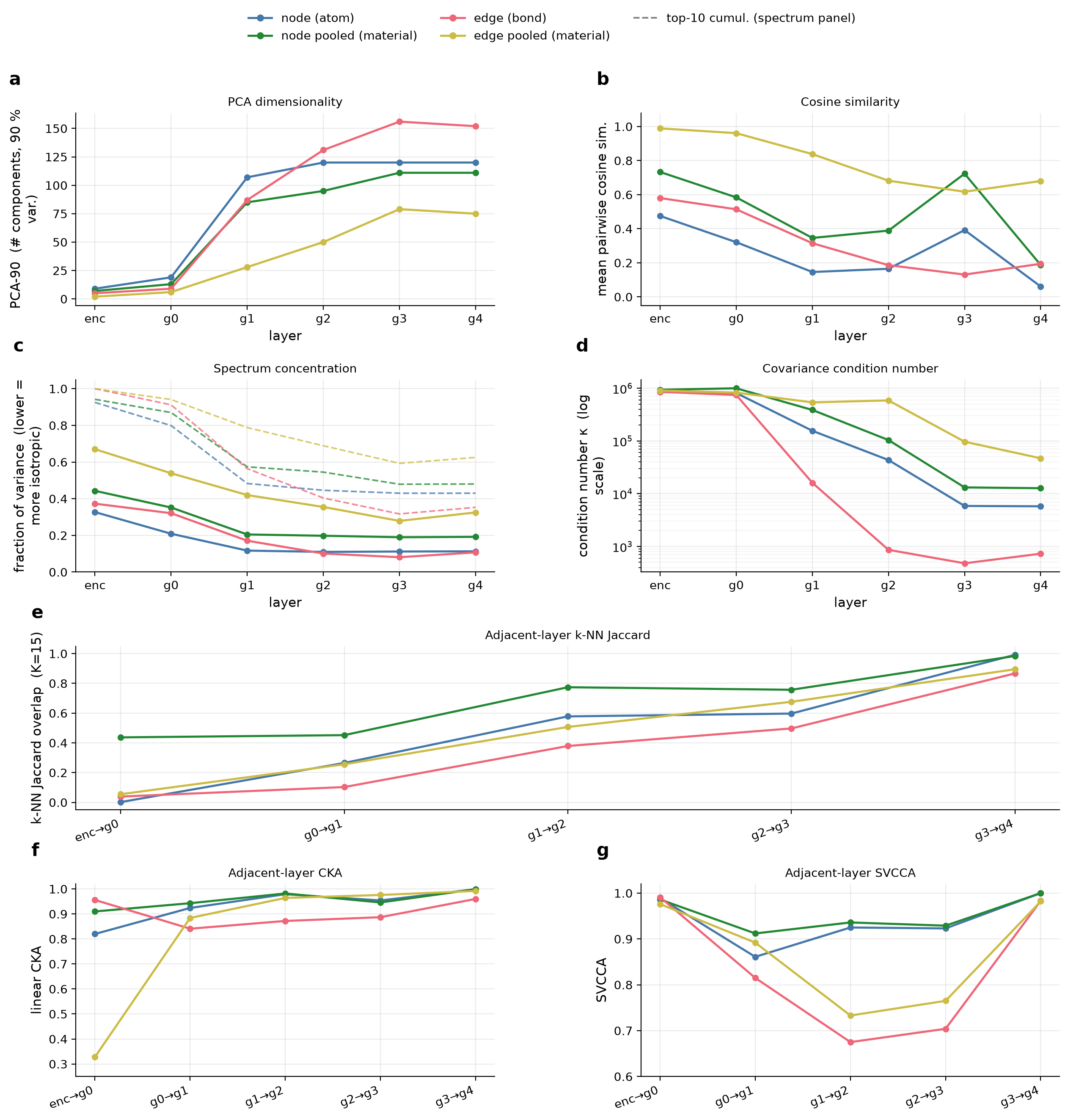}
  \caption*{\textbf{\emph{Figure~S9. ORB-v3 embedding health QC across all layers (enc, g0--g4).}} \emph{Every panel plots the same four read-outs as separate lines, keyed by the legend at the top: node (atom-level), node-pooled (material-level), edge (bond-level) and edge-pooled (material-level). \textbf{(a)} PCA-90 dimensionality (number of components needed to explain 90~\% of variance; higher means richer, less collapsed representation). \textbf{(b)} Mean pairwise cosine similarity (collapse indicator; lower is healthier). \textbf{(c)} Spectral concentration --- how much of the embedding's variance sits in a few directions: the share carried by the single largest principal direction (solid) and by the ten largest together (dashed). A high share means the cloud is stretched along a few axes and the remaining directions carry little; lower is closer to isotropic, and healthier. \textbf{(d)} Condition number κ, the ratio of the largest eigenvalue to the smallest one exceeding a millionth of it (\hyperlink{embedding-health-qc}{§A.6}); lower means better-conditioned. \textbf{(e)} Adjacent-layer k-NN Jaccard overlap (neighbourhood stability; higher means more stable neighbourhoods across layers). \textbf{(f)} Adjacent-layer CKA: whether pairs of inputs that are similar in one layer stay similar in the next, which captures the overall pairwise structure of the representation. \textbf{(g)} Adjacent-layer SVCCA: whether the principal axes of variance align between layers, which focuses on the main directions of variance and ignores minor noise dimensions. In both, higher means greater similarity between consecutive layers. Supports \hyperlink{representation-health-tracks-architectural-function}{§B.1}.}}
\end{figure}

\begin{figure}[H]
  \hypertarget{fig:S10}{}%
  \centering
  \includegraphics[height=0.82\textheight]{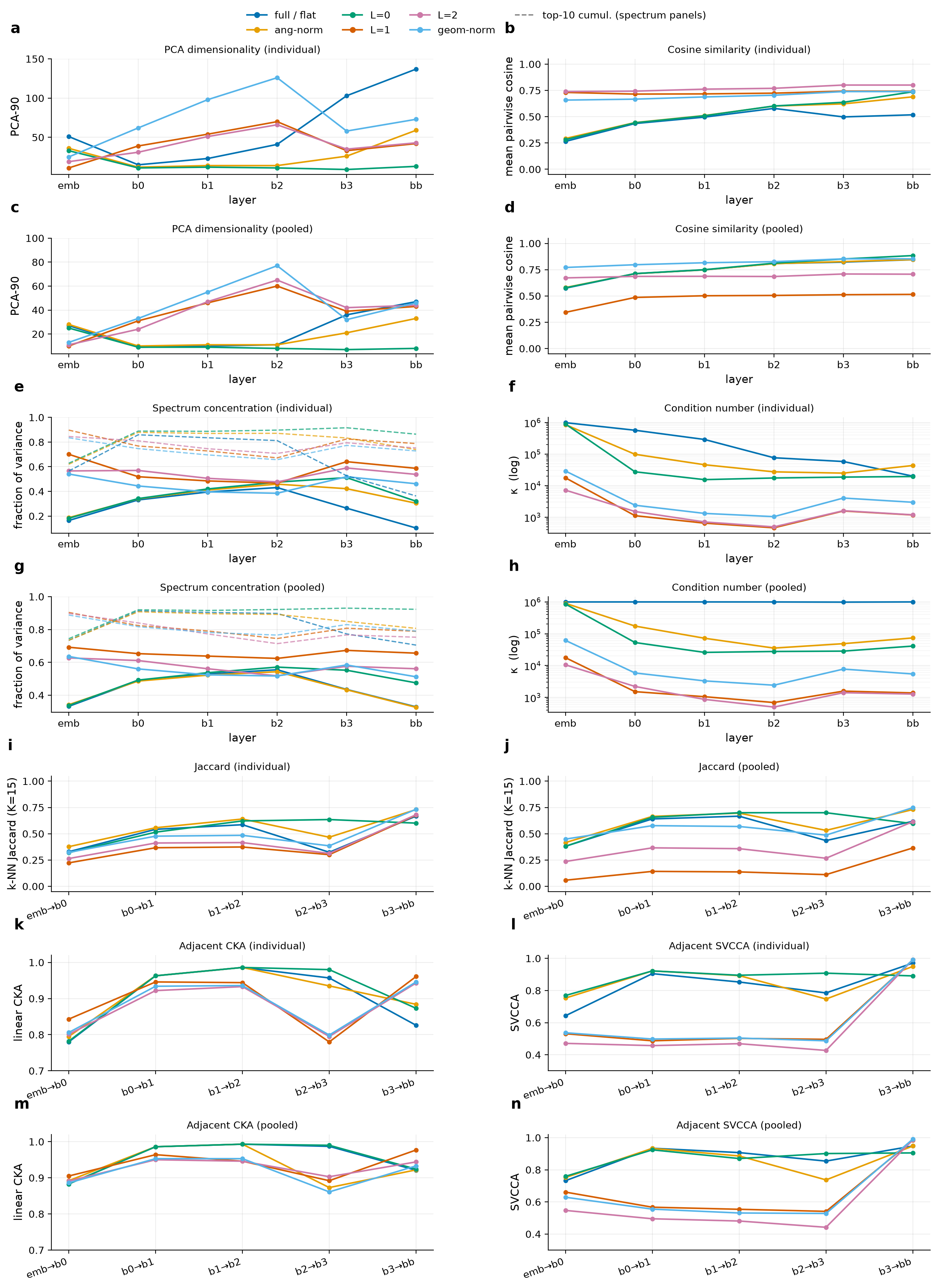}
  \caption*{\textbf{\emph{Figure~S10. UMA-S embedding health QC across all node read-outs (flat, ang-norm, L=0, L=1, L=2, geom-norm) at all layers (emb, b0--b3, bb).}} \emph{Identical QC metrics to \hyperlink{fig:S9}{Fig.~S9}, colour-coded by UMA-S equivariant channel family (legends at the top); \hyperlink{fig:S9}{Fig.~S9} defines each metric. Individual panels are atom-level, pooled panels material-level. \textbf{(a)} PCA-90 dimensionality, individual (higher means richer). \textbf{(b)} Mean pairwise cosine similarity, individual (lower is healthier). \textbf{(c)} PCA-90, pooled. \textbf{(d)} Cosine similarity, pooled. \textbf{(e)} Spectral concentration, for the largest and the ten largest principal directions, individual (lower means more isotropic). \textbf{(f)} Condition number κ, individual, log scale (lower means better-conditioned). \textbf{(g)} Spectral concentration, pooled. \textbf{(h)} Condition number, pooled. \textbf{(i)} Adjacent-layer k-NN Jaccard overlap, individual (higher means more stable neighbourhoods). \textbf{(j)} k-NN Jaccard, pooled. \textbf{(k)} Adjacent-layer CKA, individual (higher means greater similarity). \textbf{(l)} Adjacent-layer SVCCA, individual. \textbf{(m)} CKA, pooled. \textbf{(n)} SVCCA, pooled. Supports \hyperlink{representation-health-tracks-architectural-function}{§B.1}.}}
\end{figure}

\begin{figure}[H]
  \hypertarget{fig:S11}{}%
  \centering
  \includegraphics[width=0.94\linewidth]{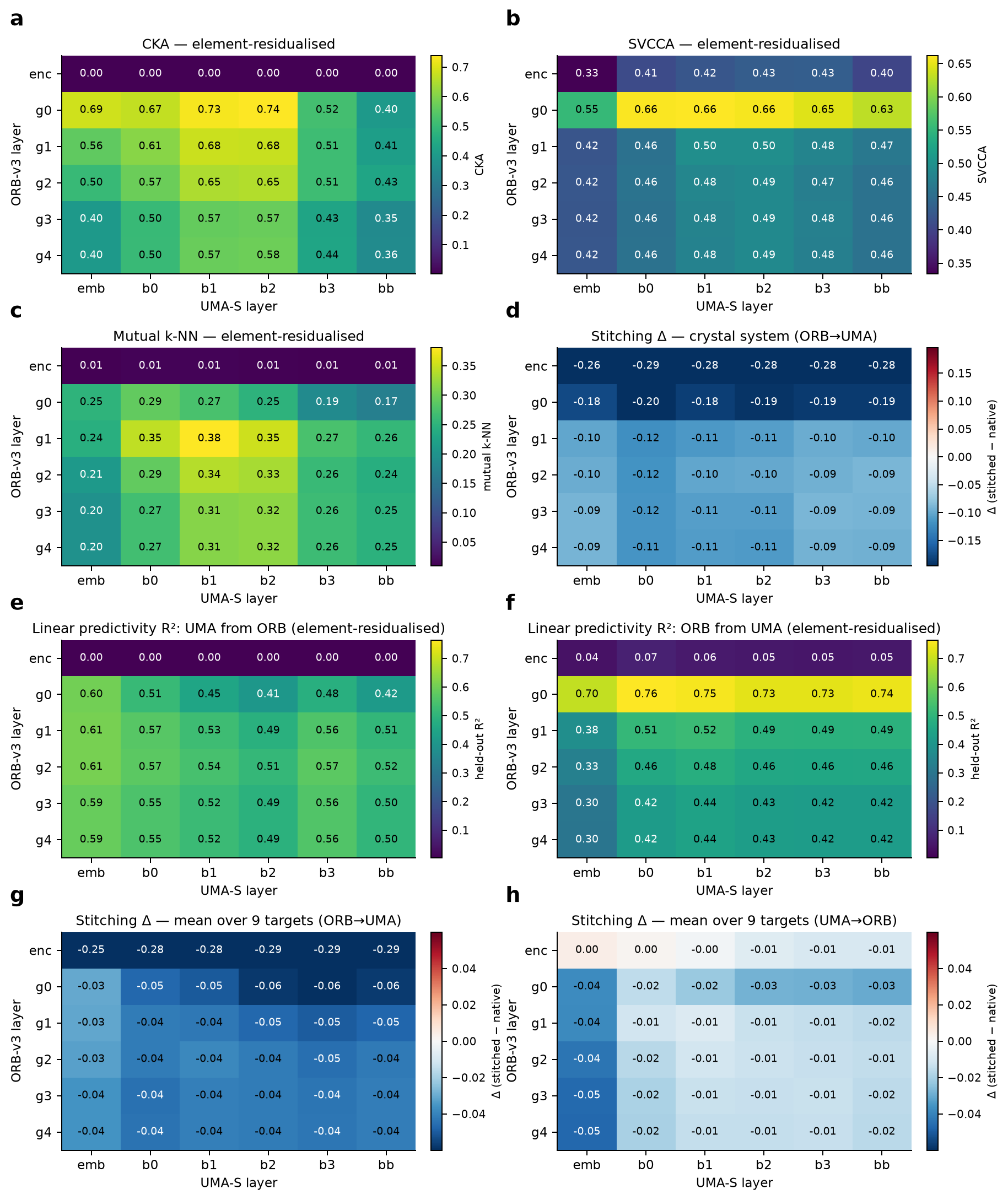}
  \caption*{\textbf{\emph{Figure~S11. Cross-model comparison between ORB-v3 (rows: enc to g4) and UMA-S (columns: emb to bb, in depth order), over every layer pair.}} \emph{All panels compare per-atom node embeddings: ORB's node vector (256-d) and UMA's rotation-invariant ang-norm view (384-d), chosen because, like ORB's node vector, it combines composition and local geometry without orientation-dependent components (\hyperlink{cross-model-convergence}{§A.7}). \textbf{(a--c)} Structural similarity on atom-aligned embeddings: \textbf{(a)} centered kernel alignment (CKA), measuring similarity between the two models' overall representation patterns; \textbf{(b)} singular vector canonical correlation analysis (SVCCA), measuring how closely their representation subspaces align; and \textbf{(c)} mutual k-nearest-neighbour (k-NN) overlap, measuring the fraction of the K = 10 nearest neighbours shared between the models' atom representations. Higher values indicate greater agreement for all three metrics. CKA and SVCCA use all 30,005 atoms; mutual k-NN uses a 4,000-atom subsample. \textbf{(d)} Functional stitching for crystal system, the target with the largest transfer loss: Δ = stitched − native macro-F1 when ORB is mapped into UMA's space and scored by UMA's own probe. \textbf{(e,f)} Linear predictivity, the variance-weighted R² of a ridge map from one model's atom embeddings to the other's (\textbf{\hyperlink{cross-model-convergence}{§A.7}}): \textbf{(e)} UMA predicted from ORB and \textbf{(f)} the reverse, sharing a colour scale. \textbf{(g,h)} Functional stitching averaged over all nine targets: \textbf{(g)} ORB mapped into UMA and \textbf{(h)} UMA mapped into ORB, sharing a colour scale. Similarity panels (a--c, e--f) are computed after subtracting each element's mean embedding from both models, so they show agreement beyond shared element identity (\hyperlink{cross-model-convergence}{§A.7}); the correction leaves 0.0~\% of ORB's encoder variance and 7.7~\% of UMA's embed variance, against 63.1~\% (ORB) and 18.8~\% (UMA) at depth. Stitching panels (d, g, h) are uncorrected, since Δ is a difference between two scores that both carry element identity (\hyperlink{cross-model-convergence}{§A.7}). Δ panels use a diverging scale centred on zero, set from the gnn\_stacks rows. Supports \hyperlink{orb-and-uma-transfer-function-across-layers-but-converge-geometrically-only-in-part}{§B.3}.}}
\end{figure}

\begin{figure}[H]
  \hypertarget{fig:S12}{}%
  \centering
  \includegraphics{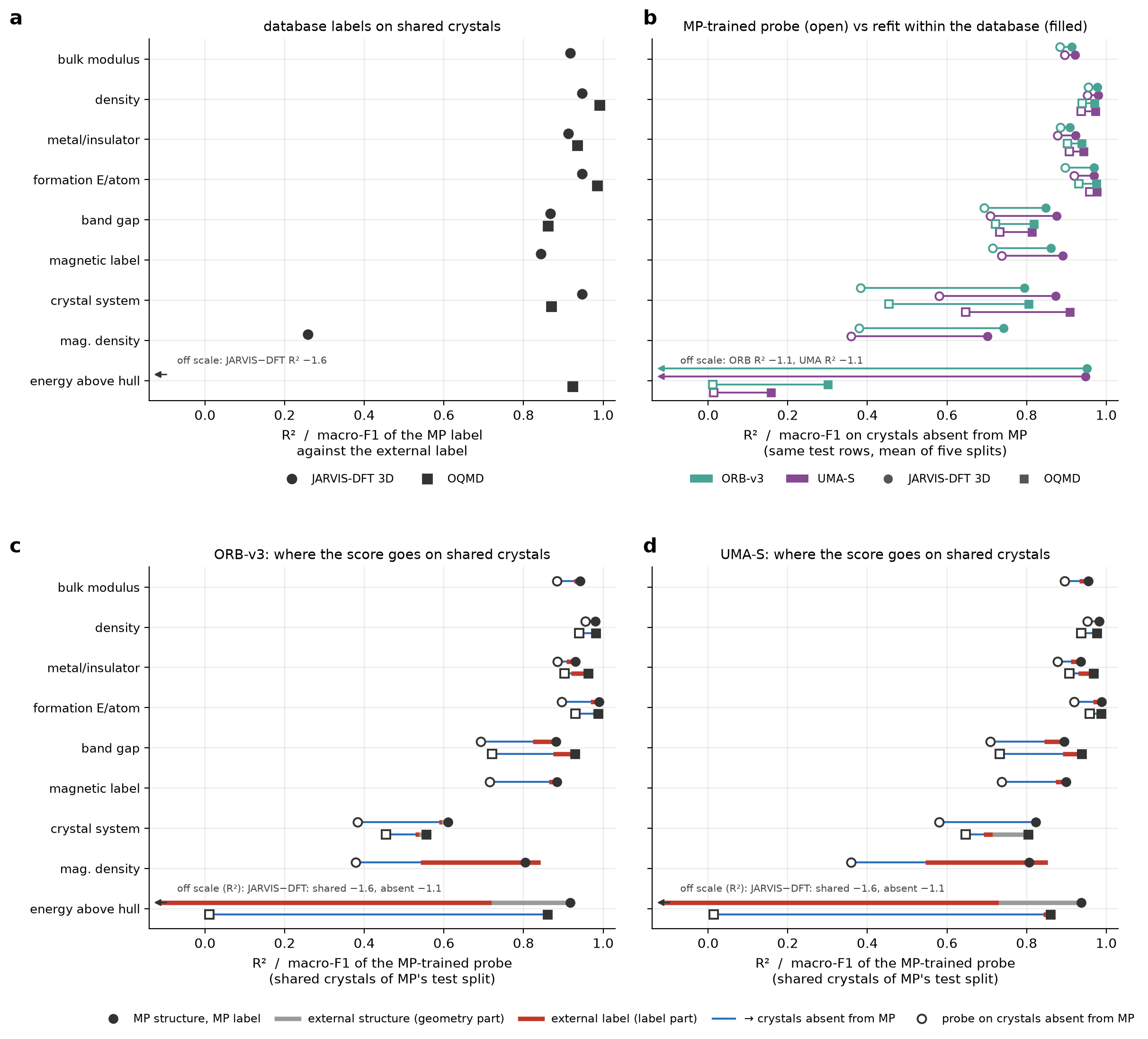}
  \caption*{\textbf{\emph{Figure~S12. Cross-database transfer from Materials Project (MP) to JARVIS-DFT 3D and OQMD (\hyperlink{external-transfer-results}{§B.4}).}} \emph{Each row shows a property, ordered by decreasing transfer performance. Circles represent JARVIS and squares OQMD. Scores are R² for continuous targets and macro-F1 for categorical targets. \textbf{(a) Reference-label agreement:} MP reference labels are scored against external-database labels for crystals present in both databases, without a predictive model. \textbf{(b) Transfer to crystals absent from MP:} Open markers show scores of MP-trained ORB-v3 (green) and UMA-S (purple) predictors. Filled markers show the same predictor configurations retrained on the external database. Both are evaluated on identical test crystals across five 80/20 splits. \textbf{(c--d) Sources of score changes on shared crystals:} Results for ORB-v3 (c) and UMA-S (d), using only crystals whose MP counterparts belong to MP's held-out test set. Filled black markers show the MP-trained predictor evaluated on MP structures against MP labels. Grey and red segments show the contributions of replacing the structures and reference labels with their external-database counterparts, averaged over both orders of replacement (\hyperlink{external-transfer}{§A.10}). Together, these give the score using external structures and labels. Blue segments then connect this score to performance on crystals absent from MP (open markers), where material and property distributions also differ. JARVIS energy above hull extends beyond the plotted range and is annotated. Numerical results for (b--d) are in \textbf{\hyperlink{tab:S19}{Table~S19}}.}}
\end{figure}

\begin{figure}[H]
  \hypertarget{fig:S13}{}%
  \centering
  \includegraphics[height=0.64\textheight]{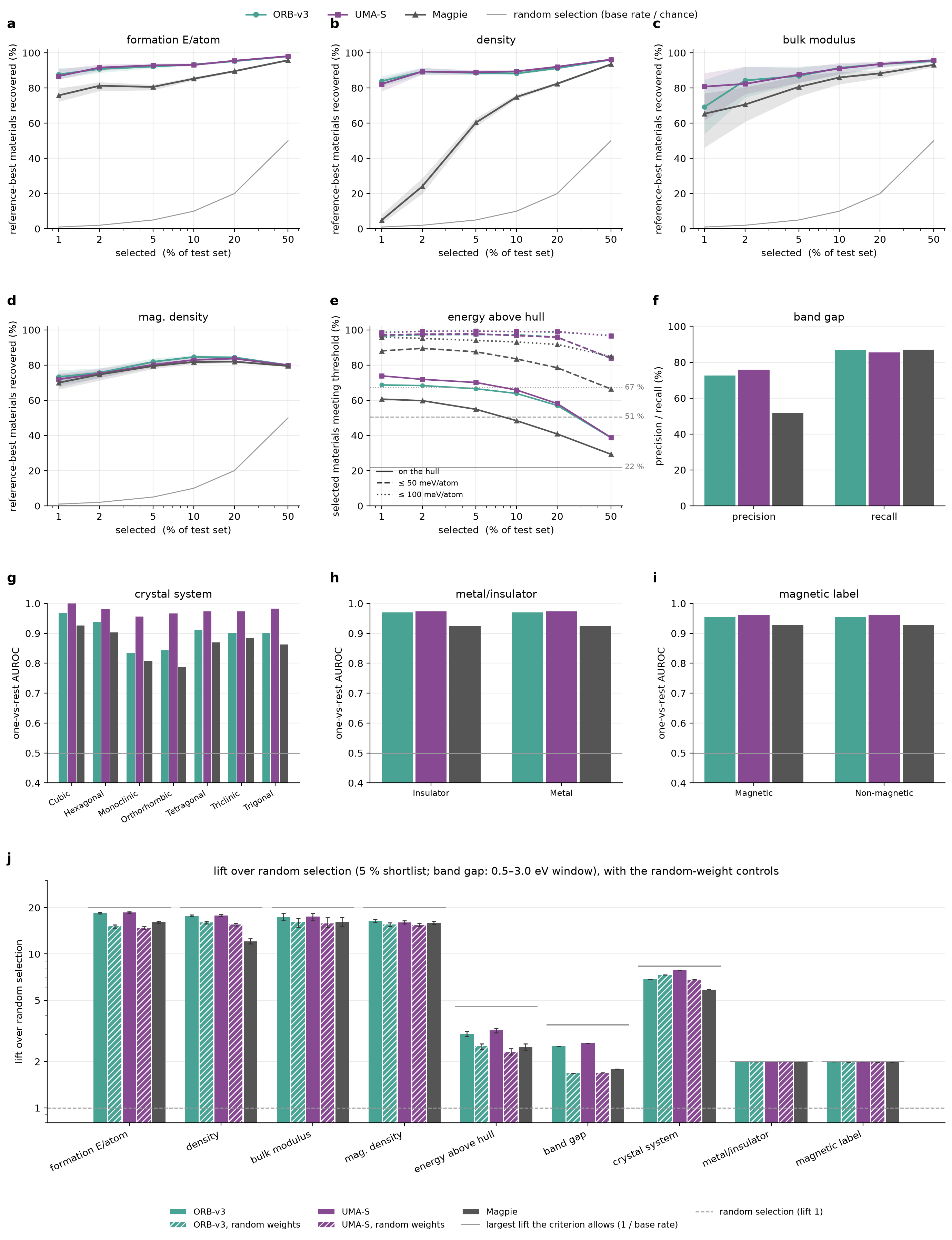}
  \caption*{\textbf{\emph{Figure~S13. Materials screening performance across nine properties and comparison with random-weight controls (\hyperlink{screening-enrichment-results}{§B.6}).}} \emph{Selected XGBoost predictors use frozen ORB-v3 and UMA-S embeddings or Magpie composition descriptors (grey) on the fixed seed-42 test set. \textbf{(a--d)} Percentage of the best materials, as determined by their DFT reference values, recovered as the fraction selected increases: lowest formation energy \textbf{(a)} and density \textbf{(b)}, and highest bulk modulus \textbf{(c)} and magnetic-moment density \textbf{(d)}. The grey line shows expected recovery under random selection. \textbf{(e)} Energy above hull: percentage of selected candidates whose reference hull energy is at most 0, 50 or 100~meV/atom (different line styles). Matching grey lines show the prevalence of each threshold in the full test set, equivalent to random-selection precision. \textbf{(f)} Band gap: precision and recall for selecting materials with predicted gaps between 0.5 and 3.0~eV. Precision is the percentage of selected materials in the true window; recall is the percentage of all qualifying materials recovered. \textbf{(g--i)} Crystal system, metal/insulator and magnetic label: one-vs-rest AUROC for each class. AUROC measures how well predicted class probabilities rank members above non-members, with 0.5 indicating random ranking and 1.0 perfect separation. \textbf{(j)} Lift relative to random selection for the pretrained predictors, Magpie and randomly initialised ORB-v3 and UMA-S (hatched bars, each at its own best configuration; \hyperlink{random-initialisation-and-matched-training-data-controls}{§A.9}). For the continuous targets in (a--d), lift is recall divided by 0.05 when selecting the best 5~\%. For energy above hull, lift is on-hull precision at a 5~\% shortlist divided by the fraction of materials on the hull; for the band-gap window, it is precision divided by the fraction of materials in the true window. For categorical targets, it is mean class-wise lift when selecting the top 5~\% by predicted class probability. The dashed line marks random selection (lift = 1); the short grey line above each target marks the maximum possible lift, which differs across targets. Shaded bands in (a--d) and error bars in (j) show 95~\% paired-bootstrap confidence intervals where available. Numerical results for (a--i) are in \textbf{\hyperlink{tab:S20}{Table~S20}}; the lift values and random-weight comparisons in (j) are shown only here.}}
\end{figure}

\begin{figure}[H]
  \hypertarget{fig:S14}{}%
  \centering
  \includegraphics[width=\linewidth]{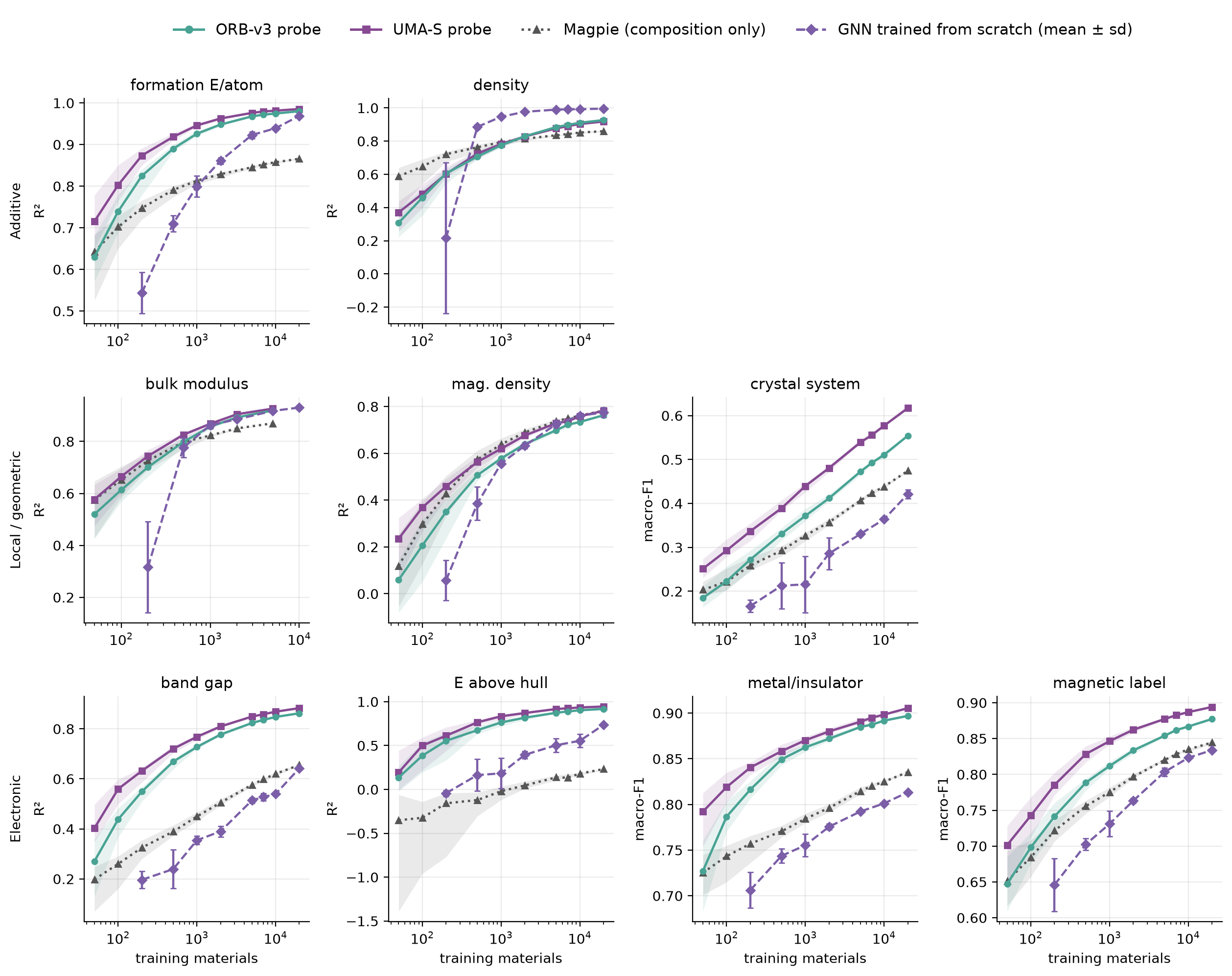}
\end{figure}
\vspace{-0.5\baselineskip}{\small\noindent \textbf{\emph{Figure~S14. Effect of training-set size on materials property prediction (\hyperlink{label-efficiency-results}{§B.7}).}} \emph{Prediction scores as a function of the number of labelled training materials, shown separately for each of the nine properties. Frozen ORB-v3 (circles) and UMA-S (squares) embeddings and Magpie composition descriptors (grey dotted triangles) are evaluated using XGBoost on the same training samples and a fixed test set of 20,000 materials per target (half the available labelled materials for bulk modulus). Lines show median scores across repeated training draws, with shaded bands indicating the 10th--90th percentile range across draws. The message-passing network trained from scratch (dashed diamonds) is evaluated on the paper's seed-42 test split; its scores are means across three runs up to 5,000 training materials and single-run results at larger sizes, except at the saturation sizes where three runs are used. Scores are R² for continuous targets and macro-F1 for categorical targets. Numerical results are reported in \textbf{\hyperlink{tab:S22}{Tables~S22}} and \textbf{\hyperlink{tab:S21}{S21}}.}\par}

\end{document}